\documentclass[twocolumn]{aastex62}
\usepackage{amsfonts}
\usepackage{graphicx}
\usepackage{color}
\usepackage{scrextend}
\usepackage{multirow}

\usepackage{bm}
\usepackage{amsmath}
\def\dif{\mathrm{d}}
\newcommand\E[1]{\left\langle #1\right\rangle}

\usepackage{aas_macros}

\newcommand{\changes}[1]{#1}

\def\CatalogFigureComponentMassesAll{4}   
\def\CatalogFigureQChieffAll{5}   

\usepackage{ci_macros}
\usepackage{ci_macros_manual}

\newcommand\MPRIMARYimrpNoSpinVTQuadCboundninetynine{41.8}

\newcommand\APPDimrpNoSpinVTQuadCboundfifty{0.27}
\newcommand\APPDimrpNoSpinVTQuadCboundninety{0.55}

\newcommand\LNBFMassModelABIMRPhenomP{-1.42}
\newcommand\LNBFMassModelACIMRPhenomP{-2.28}
\newcommand\LNBFMassModelBCIMRPhenomP{-0.86}
\newcommand\SDDRDeltaMvsCIMRPhenomP{0.14}
\newcommand\SDDRLambdavsCIMRPhenomP{-1.92}
\newcommand\SDDRGaussianvsMixtureIMRPhenomP{0.15}

\newcommand{ \bbhPlawCombR }{\ensuremath{ 57^{+40}_{-25} } }
\newcommand{ \bbhUlmCombR }{\ensuremath{ 19^{+13}_{-8.2} } }

\newcommand{\MTOTSCOMPACTOneSevenZeroSevenTwoNineCat}{\ensuremath{85.2_{-11.2}^{+15.4}}} 

\newcommand{\MTOTSCOMPACTOneSevenZeroSixZeroEightCat}{\ensuremath{18.7_{-0.7}^{+3.3}}} 
\newcommand{\REDSHIFTCOMPACTOneSevenZeroSevenTwoNineCat}{\ensuremath{0.48_{-0.20}^{+0.19}}} 

\def\Msol{\ensuremath{\mathit{M_\odot}}}

\def\chieff{\ensuremath{\chi_{\textrm{eff}}}{}}

\def\mmax{\ensuremath{m_{\textrm{max}}}{}}
\def\mmin{\ensuremath{m_{\textrm{min}}}{}}

\def\mminVal{5}

\def\Vc{\ensuremath{V_c}{}}
\def\VT{\ensuremath{\langle VT \rangle}{}}

\newcommand{\msource}[1]{m_{#1}^{\mathrm{source}}}
\newcommand{\mdet}[1]{m_{#1}^{\mathrm{det}}}

\newcommand{\dd}{\mathrm{d}}

\def\invstvol{Gpc$^{-3}$ yr$^{-1}$}

\begin{document}

\received{2018 December 15}
\revised{2019 July 11}
\accepted{2019 July 21}
\published{2019 September 9}

\title{Binary Black Hole Population Properties Inferred from the First and Second Observing Runs\\of Advanced LIGO and Advanced Virgo\\---\\This is the Accepted Manuscript version of an article accepted for publication in\\The Astrophysical Journal Letters and published as ApJL, 882, L24 (2019).\\IOP Publishing Ltd is not responsible for any errors or omissions in this\\version of the manuscript or any version derived from it. The Version of\\Record is available online at https://doi.org/10.3847/2041-8213/ab3800.}

\AuthorCollaborationLimit=3000  

\author{B.~P.~Abbott}
\affiliation{LIGO, California Institute of Technology, Pasadena, CA 91125, USA}
\author{R.~Abbott}
\affiliation{LIGO, California Institute of Technology, Pasadena, CA 91125, USA}
\author{T.~D.~Abbott}
\affiliation{Louisiana State University, Baton Rouge, LA 70803, USA}
\author{S.~Abraham}
\affiliation{Inter-University Centre for Astronomy and Astrophysics, Pune 411007, India}
\author{F.~Acernese}
\affiliation{Universit\`a di Salerno, Fisciano, I-84084 Salerno, Italy}
\affiliation{INFN, Sezione di Napoli, Complesso Universitario di Monte S.Angelo, I-80126 Napoli, Italy}
\author{K.~Ackley}
\affiliation{OzGrav, School of Physics \& Astronomy, Monash University, Clayton 3800, Victoria, Australia}
\author{C.~Adams}
\affiliation{LIGO Livingston Observatory, Livingston, LA 70754, USA}
\author{R.~X.~Adhikari}
\affiliation{LIGO, California Institute of Technology, Pasadena, CA 91125, USA}
\author{V.~B.~Adya}
\affiliation{Max Planck Institute for Gravitational Physics (Albert Einstein Institute), D-30167 Hannover, Germany}
\affiliation{Leibniz Universit\"at Hannover, D-30167 Hannover, Germany}
\author{C.~Affeldt}
\affiliation{Max Planck Institute for Gravitational Physics (Albert Einstein Institute), D-30167 Hannover, Germany}
\affiliation{Leibniz Universit\"at Hannover, D-30167 Hannover, Germany}
% ...

\collaboration{The LIGO Scientific Collaboration and the Virgo Collaboration}


\begin{abstract} We present results on the mass, spin, and redshift
distributions \changes{with phenomenological population models using} the ten binary black hole mergers detected in the first and second observing runs completed by Advanced LIGO and Advanced Virgo. We constrain properties of
the binary black hole (BBH) mass spectrum using models with a range of
parameterizations of the BBH mass and spin distributions. We find that the mass
distribution of the more massive black hole in such binaries is well
approximated by models with no more than 1\% of black holes more massive than 45
\Msol, and a power law index of $\alpha = \ALPHAMimrpNoSpinVTQuadBpm$ (90\%
credibility). We also show that BBHs are unlikely to be composed of black holes
with large spins aligned to the orbital angular momentum. Modelling the
evolution of the BBH merger rate with redshift, we show that it is flat or increasing
with redshift with $\FRACLambdaGTZeroP{}$ probability. Marginalizing over
uncertainties in the BBH population, we find robust estimates of the BBH merger
rate density of $R = \RATEimrpNoSpinVTQuadBpm$ \invstvol (90\% credibility). As
the BBH catalog grows in future observing runs, we expect that uncertainties in
the population model parameters will shrink, potentially providing insights into
the formation of black holes via supernovae, binary interactions of massive
stars, stellar cluster dynamics, and the formation history of black holes across
cosmic time. \end{abstract}

\section{Introduction} \label{sec:intro}
The second LIGO/Virgo observing run (O2) spanned nine months between November 2016 through August 2017, building upon the first, four-month run (O1) in 2015.
The LIGO/Virgo
gravitational-wave (GW) interferometer network is comprised of two instruments
in the United States (LIGO)~\citep{2015CQGra..32g4001L,2016PhRvL.116m1103A} and
a third in Europe (Virgo)~\citep{2015CQGra..32b4001A}, the latter joining the run in the
summer of 2017. In total, ten binary black hole (BBH) mergers have been detected
to date~\citep{o2catalog}. 
The BBHs detected
possess a wide range of physical properties. The lightest so far is
GW170608~\citep{2017ApJ...851L..35A} with an inferred total mass of
\MTOTSCOMPACTOneSevenZeroSixZeroEightCat \Msol.
GW170729~\citep{o2catalog}---exceptional in several ways---is likely to be the
heaviest BBH to date, having total mass \MTOTSCOMPACTOneSevenZeroSevenTwoNineCat
\Msol, as well as the most distant, at redshift \REDSHIFTCOMPACTOneSevenZeroSevenTwoNineCat. Both GW151226 and GW170729 show evidence for at least one black hole with a spin greater than zero~\citep{2016PhRvL.116x1103A,o2catalog}.

By measuring the distributions of mass, spin, and merger redshift in the BBH
population, we may make inferences about the physics of binary mergers and
better understand the origin of these systems.  \changes{We employ Bayesian inference
and modelling~\citep{gelman2004bayesian,2010PhRvD..81h4029M,2014ApJ...795...64F,hilbe_de_souza_ishida_2017,asensio_ramos_arregui_2018} which, when applied to parameterized models of the population},
is able to infer population-level parameters --- sometimes called
hyperparameters to distinguish them from the event-level parameters --- while
properly accounting for the uncertainty in the measurements of each event's
parameters~\citep{2010PhRvD..81h4029M,2010ApJ...725.2166H}.

The structure and parameterization of BBH populations models are guided by the physical processes and evolutionary environments in which BBH are expected to form and merge. Several BBH formation channels have been proposed in the literature, each of them involving a specific environment and a number of physical processes. For example, BBHs might form from isolated massive binaries
in the galactic field through common-envelope
evolution~\citep{1998ApJ...506..780B,1998A&A...332..173P,2002ApJ...572..407B,2003MNRAS.342.1169V,2006MNRAS.368.1742D,2007ApJ...662..504B,2008ApJS..174..223B,2013ApJ...779...72D,2014ApJ...789..120B,2014A&A...564A.134M,2015MNRAS.451.4086S,2017ApJ...846..170T,2016MNRAS.462.3302E,2017NatCo...814906S,2018MNRAS.474.2937C,2017MNRAS.472.2422M,2018MNRAS.474.2959G,2018MNRAS.479.4391M,2018MNRAS.481.1908K,2018MNRAS.480.2011G}
or via chemically homogeneous
evolution~\citep{2016A&A...588A..50M,2016MNRAS.460.3545D,2016MNRAS.458.2634M}.
Alternatively, BBHs might form via dynamical processes in stellar
clusters~\citep{2000ApJ...528L..17P,1993Natur.364..421K,1993Natur.364..423S,2006NatPh...2..116G,2006ApJ...637..937O,2008ApJ...676.1162S,2008MNRAS.386..553I,2010MNRAS.407.1946D,2011MNRAS.416..133D,2013MNRAS.428.3618C,2014MNRAS.441.3703Z,2015PhRvL.115e1101R,
2016PhRvD..93h4029R,2016MNRAS.459.3432M,2017MNRAS.464L..36A,2017MNRAS.467..524B,2017ApJ...836L..26C}
and galactic nuclei~\citep{2012ApJ...757...27A,
2016ApJ...831..187A,2017ApJ...846..146P}, evolution of hierarchical triple
systems~\citep{2014ApJ...781...45A,2016MNRAS.463.2443K,2017ApJ...841...77A,2018ApJ...863...68L}, gas
drag and stellar scattering in accretion disks surrounding super-massive
black holes~\citep{2012MNRAS.425..460M,2017ApJ...835..165B,2017MNRAS.464..946S}.
Finally, BBHs might originate as part of a primordial black hole population in the early
Universe~\citep{1974MNRAS.168..399C,2016PhRvD..94h3504C,2016PhRvL.117f1101S,2017PhRvD..95l3510I,2016MNRAS.461.2722I,2016PhRvL.116t1301B,2017PhRvD..96l3523A,2017PDU....15..142C,2018ApJ...864...61C,2018PhRvD..97l3512A}, \changes{where their mass spectrum is typically proposed as having power law behavior, but spanning a much wider range of masses than stellar mass BH.}
Each channel contributes differently to the distributions of
the mass, spin, distance, and orbital characteristics of BBHs.

There are several processes common to most pathways through stellar evolution which affect the properties of the resultant BBH system. Examples include mass
loss~\citep{2001A&A...369..574V,2005A&A...442..587V,2008A&A...482..945G} and
supernovae~\citep{2011ApJ...730...70O,2012ApJ...749...91F,2012ARNPS..62..407J,2012ApJ...757...69U,2016ApJ...818..124E,2016ApJ...821...38S}.
The mass of the compact object left after the supernova is directly
related to its pre-supernova mass and the supernova mechanism itself.
Metallicity has been shown~\citep{2000ARA&A..38..613K,2001A&A...369..574V,2011A&A...530A.115B} to
have important effects on stellar mass loss through winds --- line-driven winds
are quenched in metal-poor progenitors, enabling large black holes to form through
direct collapse or post-supernova mass fallback
\citep{2003ApJ...591..288H,2009MNRAS.395L..71M,2010ApJ...714.1217B,2015MNRAS.451.4086S}.
This also, in turn, might suppress supernova kicks~\citep{2012ApJ...749...91F}
and hence enhance the number of binaries which are not disrupted.

Theoretical and phenomenological models of BBH formation are explored by population synthesis. This requires modelling not only of stellar evolution but also the influence of their evolutionary environments.
For instance, isolated
evolution in galactic fields requires prescriptions for binary interactions,
such as common envelope
physics, as well as mass transfer episodes (see reviews in~\cite{2007PhR...442...75K,2009NewAR..53...27V,2014LRR....17....3P}), and more recently, the effects of rapid rotation~\cite{2009A&A...497..243D,2016MNRAS.458.2634M,2016A&A...588A..50M}. Meanwhile, BBH formation in dense
stellar
clusters~\citep{2014MNRAS.441.3703Z,2015PhRvL.115e1101R,2016PhRvD..93h4029R,2016MNRAS.459.3432M,2017MNRAS.464L..36A,2017MNRAS.467..524B}
is impacted primarily by dynamical interactions within the
cluster~\citep{2004PhDT.......185F,2013ApJ...763L..15M}, but also by cluster
size and initial mass
functions~\citep{2007A&A...469..925S,2010ARA&A..48..431P,2019ApJ...871...38K}. GW observations provide an alternative to sharpen our understanding of those processes.

Electromagnetic observations and modeling of systems containing black holes have
led to speculation about the existence of potential gaps in the black hole mass
spectrum.  Both gaps may be probed using data from current ground-based
gravitational-wave interferometers, and as such, have been the target of
parametric
studies.
At low masses, observations of X-ray binaries (XRB) combined via Bayesian
population
modeling~\citep{1998ApJ...499..367B,2010ApJ...725.1918O,2011ApJ...741..103F}
suggest a minimum black hole mass well above the largest neutron star masses.
While the existence and nature of this gap is still
uncertain~\citep{2012ApJ...757...36K}, it is proposed to exist between the most
massive neutron
stars~\citep{2016ARA&A..54..401O,2008ApJ...675..670F,2017ApJ...850L..19M} ($2.1
- 2.5 \Msol$) and the lightest black holes $\sim 5 \Msol$.
It is possible to constrain the existence of this lower mass gap with GW observations~\citep{2015ApJ...807L..24L,2015MNRAS.450L..85M,2017PhRvD..95j3010K,2017MNRAS.465.3254M}. In
Section~\ref{sec:mass_distr}, we find our current GW observations do not inform
the upper edge of this gap, inferring a minimum mass on the primary black hole
at $\mmin \lesssim 9$ \Msol. Our volumetric sensitivity to BBH systems with
masses less than 5 \Msol{} is small enough that we expect (and observe) no
events in the lower gap region. Thus, our ability to place constraints in this
region is severely limited.

Recently, there have been claims of an upper cutoff in the BBH mass spectrum based on the first few LIGO detections~\citep{2017ApJ...851L..25F,2018ApJ...856..173T,2018arXiv180506442W,2018arXiv180204909B,2019MNRAS.tmp..231R}.
This might be expected as a consequence of a different
supernova type, called the (pulsational) pair-instability
supernova~\citep{2002ApJ...567..532H,2016A&A...594A..97B,2017ApJ...836..244W,2017MNRAS.470.4739S,2018arXiv181013412M}.
\changes{
Evolved stars with a Helium core mass $\gtrsim{}30 \Msol$ are expected to become unstable, because efficient pair production softens their equation of state. For Helium core mass $\sim{}30$ -- $64 \Msol$, the star undergoes a sequence of pulsations, losing mass until stability is reestablished~\citep{2007Natur.450..390W}. The enhanced mass loss during pulsational pair instability is expected to affect the final collapse of the star, leading to smaller black hole masses. The fate of a star with He core mass $\sim{}64$ -- $135 \Msol$ is more dramatic: the entire star is disrupted by a pair instability supernova, leaving no remnant~\citep{1964ApJS....9..201F,1967PhRvL..18..379B,1967ApJ...148..803R}. From the combination of pair instability and pulsational pair instability, it is expected that
pair-instability supernovae should leave no black hole remnants between $\sim 50$ -- $150 \Msol$ because the progenitor star is partially or entirely disrupted by the explosion.
}
\changes{It is also possible that contributions from the merger of previous merger products --- second generation mergers~\citep{2016ApJ...824L..12O,2017PhRvD..95l4046G,2017ApJ...840L..24F,2018PhRvL.120o1101R} --- could occupy this gap. Primordial BHs could also span numerous decades of the mass spectrum~\citep{2017JHEP...09..138G}, but their number density in either mass gap is dependent on the behavior of fluctuations in the early Universe~\citep{2018JCAP...08..041B}.}
Nonetheless, consistent with prior work, we find that all our mass models have almost no merging black holes above $\sim 45$ \Msol.

Observational constraints on the BBH merger
rate~\citep{2016PhRvX...6d1015A,o2catalog} generally assume a rate density which is
uniform in the comoving volume.
As first shown in~\citet{2018ApJ...863L..41F}, it is also possible to search for redshift evolution in the rate density using current data.
Different redshift-dependent evolutionary
behavior is
possible~\citep{2013ApJ...779...72D,2016MNRAS.458.2634M,2016PhRvD..93h4029R,2017MNRAS.472.2422M,2018ApJ...866L...5R}
with different environments and stellar evolution
scenarios~\citep{2010ApJ...716..615O,2016Natur.534..512B}. For instance, theoretical models of
isolated evolution through common envelope lead to a distribution of times to merger $p(t_{{\rm GW}})
\propto t_{{\rm GW}}^{-1}$~\citep{2012ApJ...759...52D,2016Natur.534..512B}.
This would imply that many isolated binaries
will coalesce near their formation redshift and produce a BBH merger
rate that approximately tracks the star formation rate, peaking near
$z\sim2$.
We find in Section~\ref{sec:rates} that the current sample of BBH mergers does
not provide enough information to confidently constrain any but the most
extreme models.  While we place more posterior mass on merger rates that
increase with increasing redshift than those that decrease, the scenario of a
uniform rate in comoving volume is comfortably within our constraints.

Black hole spin measurements also provide a powerful tool to discriminate
between different channels of BBH
formation~\citep{2010CQGra..27k4007M,2016ApJ...818L..22A,2016ApJ...832L...2R,2017CQGra..34cLT01V,2017PhRvD..95l4046G,2017Natur.548..426F,2018ApJ...854L...9F,2018PhRvD..98h4036G}.
For example, BBHs formed in a dynamic environment will have no preferred
direction for alignment, producing isotropically oriented
spins~\citep{1993Natur.364..423S,2000ApJ...528L..17P,2010CQGra..27k4007M,2015PhRvL.115e1101R,2016ApJ...832L...2R,2017MNRAS.464..946S}. \changes{However, some evidence has been presented for correlation in spin direction due to the natal environment of the progenitor stars within the cluster~\citep{2017NatAs...1E..64C}.}
In contrast, isolated binaries are expected to preferentially produce mergers
with alignment between the spins of the constituent black holes and the orbital
angular momentum of the
system~\citep{1993MNRAS.260..675T,2000ApJ...541..319K,2004PhRvD..69j2002G,2016Natur.534..512B,2016ApJ...832L...2R,2016MNRAS.458.2634M,2016A&A...588A..50M,2017NatCo...814906S,2017PhRvL.119a1101O,2018PhRvD..98h4036G}.
Other effects occurring in stellar systems like hierarchical triples could also
produce a weak preference for certain spin-orbit
misalignments~\citep{2018ApJ...863....7R}. All of our parameterized models point
to preferences against high spin magnitudes when the spin tilts are aligned with the orbital angular momentum. In Section~\ref{sec:spin_distr}, we find
that the dimensionless spin magnitude inference prefers distributions which
decline as the spin magnitude increases from zero, but our ability to
distinguish between assumed distributions of spin orientation is very limited.

GW170817, the first binary neutron star merger observed through GW emission~\citep{2017PhRvL.119p1101A}, was detected by GW observatories and associated with a short GRB~\citep{2017ApJ...848L..13A}) in August of 2017.
A subsequent
post-merger transient (AT 2017gfo) was observed across the electromagnetic spectrum, from
radio~\citep{2017ApJ...848L..21A},
NIR/optical~\citep{2017Sci...358.1556C,2017ApJ...848L..16S,2017ApJ...848L..19C,2017ApJ...848L..17C,2017ApJ...848L..18N,2017Natur.551...67P},
to X-ray~\citep{2017Natur.551...71T,2017ApJ...848L..20M} and
$\gamma$-ray~\citep{2017ApJ...848L..13A,2017ApJ...848L..14G,2017ApJ...848L..15S}.
Unfortunately, with only one confident detection, it is not yet possible to
infer details of binary neutron star populations more than to note that the
gravitational-wave measurement is mostly compatible with the observed Galactic
population~\citep{2012ApJ...757...55O}. However, if GW170817 did form a black
hole, it would also occupy the lower mass gap described previously.

We structure the paper as follows. First, notation and models are
established in Section~\ref{sec:notation}. Section~\ref{sec:mass_distr}
describes our modeling of the black hole mass distribution, followed by rate
distributions and evolution in Section~\ref{sec:rates}. The black hole spin magnitude
and orientation distributions are discussed in Section~\ref{sec:spin_distr}. We conclude
in Section~\ref{sec:conclude}. Studies of various systematics are presented in
Appendix~\ref{sec:systematics}. In Appendix~\ref{sec:alt_spins} we present
additional studies of spin distributions with model selection for a number of
zero-parameter spin models and mixtures of spin orientations. To motivate and
enable more detailed studies, we have established a repository of our samples
and other derived products\footnote{The data release for this work can be found at \url{https://dcc.ligo.org/LIGO-P1800324/public}.}.

\section{Data, Notation, and Models} \label{sec:notation}
In this work, we analyze the population of 10 BBH merger events confidently
identified in the first and second observing run (O1 and O2)~\citep{o2catalog}.
We do not include marginal detections, but these likely have a minimal impact
our conclusions here~\citep{2019MNRAS.tmp..230G}. Ordered roughly from smallest
to most massive by source-frame chirp mass, the mergers considered in this paper
are GW170608,\, GW151226,\, GW151012,\, GW170104,\, GW170814,\, GW170809,\,
GW170818,\, GW150914,\, GW170823,\, and GW170729.

The individual properties of those 10 sources were inferred using a Bayesian framework, with results summarized in \cite{o2catalog}.
For BBH systems, two waveform models have been used, both calibrated to numerical relativity simulations and incorporating spin effects, albeit differently: {\tt IMRPhenomPv2}~\citep{2014PhRvL.113o1101H, 2016PhRvD..93d4006H, 2016PhRvD..93d4007K}, which includes an effective representation~\citep{2015PhRvD..91b4043S} of precession effects, and {\tt SEOBNRv3}~\citep{2014PhRvD..89h4006P, 2014PhRvD..89f1502T, 2017PhRvD..95b4010B}, which incorporates all spin degrees of freedom.
The results presented in this work use {\tt IMRPhenomPv2}; a discussion of potential systematic biases in our inference are discussed in Appendix~\ref{sec:systematics}.
We also refer to Appendix B in~\citep{o2catalog} for more details on comparisons between those two waveform families.

To assess the stability of our results to statistical effects and systematic error we focus on one modestly exceptional event.
Both GW151226 and GW170729 exhibit evidence for measurable black hole spin, but GW170729 in particular is an outlier by several other metrics as well.
In addition to spins, it is also more massive and more distant than any of the other events in the catalog.
All events used in the population analysis have confident probabilities of astrophysical origin, but GW170729 is the least significant, having the smallest odds ratio of astrophysical versus noise origin \citep{o2catalog}.
As we describe in Sections~\ref{sec:mass_distr} and \ref{sec:rates}, this event has an impact on our inferred merger rate versus both mass and redshift.
To demonstrate the robustness of our result, we present these analyses twice: once using every event, and again omitting GW170729.

\subsection{Binary Parameters}

A coalescing compact binary in a quasi-circular orbit can be completely characterized by its eight intrinsic
parameters, namely its component masses $m_i$ and spins $\bm{S}_i$, and its seven extrinsic parameters: right
ascension, declination, luminosity distance, coalescence time, and three Euler angles characterizing its orientation
(e.g., inclination, orbital phase, and polarization).
Binary eccentricity is also a potentially observable quantity in BBH mergers, \changes{with several channels having imprints on eccentricity distributions, e.g.~\citep{1987ApJ...321..199Q,2012PhRvD..85l3005K,2014ApJ...784...71S,2018ApJ...856...92F,2018PhRvD..98l3005R}}.
However, our ability to parameterize~\citep{2014PhRvD..90h4016H,2017PhRvD..95b4038H,2018PhRvD..98j4043K,2018PhRvD..98d4015H} and measure~\citep{2015PhRvD..91f3004C,2016ApJ...818L..22A,2017CQGra..34j4002A,2018PhRvD..98h3028L} eccentricity is an area of active development. For low to moderate eccentricity at formation, binaries are expected to circularize~\citep{1964PhRv..136.1224P,2008PhRvD..77h1502H} before entering the bandwidth of ground-based GW interferometers.  We therefore assume zero eccentricity in our models.

In this work, we define the mass ratio as $q = m_2 / m_1$ where $m_1 \geq m_2$. The frequency of gravitational wave emission is directly related to the component masses.  However, due to the expansion of spacetime as the gravitational wave is propagating, the frequencies measured by the instrument are redshifted relative to those emitted at the source~\citep{1983grr..proc....1T}. We capture these effects by distinguishing between masses as they would be measured in the source frame, denoted as above, and the redshifted masses, $(1+z) m_i$, which are measured in the detector frame. Meanwhile, the amplitude of the wave scales inversely with the luminosity distance~\citep{Misner1973Gravitation}. We use the GW measurement of the luminosity distance to obtain the cosmological redshift and therefore convert between detector-frame and source-frame masses.
We assume a fixed Planck 2015~\citep{2016A&A...594A..13P} cosmology throughout to convert between a source's luminosity distance and its redshift~\citep{1999astro.ph..5116H}.

We characterize black hole spins using the dimensionless spin parameter $\bm{\chi}_{i} = \bm{S}_{i} / m_{i}^{2}$. Of particular interest are the magnitude of the dimensionless spin, $a_i = |{\bm \chi}_i|$, and the tilt angle with respect to the orbital angular momentum, $\hat{\bm{L}}$, given by $\cos t_i = \hat{\bm{L}}\cdot \hat{{\bm \chi}}_i$.
We also define an overall effective spin, $\chieff$\ \citep{2001PhRvD..64l4013D,2008PhRvD..78d4021R,2011PhRvL.106x1101A}, which is a combination of the individual spin components along to orbital angular momentum:
\begin{equation}
  \chieff =
  \frac{(\bm{\chi}_{1} + q \bm{\chi}_{2}) \cdot \hat{\bm{L}}}{1 + q}.
%
  \label{eq:chieff}
\end{equation}
\noindent \chieff{} is approximately proportional to the lowest order contribution to the GW waveform phase that contains spin for systems with similar masses. Additionally, \chieff{} is conserved throughout the binary evolution to high accuracy~\citep{2008PhRvD..78d4021R,2015PhRvD..92f4016G}.

\subsection{Model Features} \label{ssec:model-features}

\changes{
The current sample is not sufficient to allow for a high-fidelity comparison with models (e.g., population synthesis) which include more detailed descriptions of stellar evolution and environmental influences.
As such, we adopt the union of the parameterizations presented in \cite{2017PhRvD..96b3012T,2017ApJ...851L..25F,2018arXiv180506442W,2018ApJ...856..173T,2018ApJ...863L..41F}. This allows for better facilitation of comparison between models, and the ability to vary the subsets of parameters influencing the mass and spin distributions while leaving others fixed.
}

The general model family has \textbf{8} parameters to characterize the mass model; \textbf{3} to characterize each black hole's spin distribution; one parameter describing the local merger rate, $\mathcal{R}_0$; and one parameter characterizing redshift dependence.
We refer to the set of these population parameters as $\theta$.
All of the population parameters introduced in this section are summarised in Table~\ref{tab:params}.

\begin{table*}[htbp]
\centering
\begin{tabular}{|c|l|} \hline
$\alpha$ & Spectral index of $m_1$ for the power-law distributed component of the mass spectrum. \\ \hline
$m_{\max}$ & Maximum mass of the power-law distributed component of the mass spectrum. \\ \hline
$m_{\min}$ & Minimum black hole mass. \\ \hline
$\beta_{q}$ & Spectral index of the mass ratio distribution. \\ \hline
$\lambda_m$ & Fraction of binary black holes in the Gaussian component. \\  \hline
$\mu_{m}$ & Mean mass of black holes in the Gaussian component. \\ \hline
$\sigma_{m}$ & Standard deviation of masses of black holes in the Gaussian component. \\ \hline
$\delta m$ & Mass range over which black hole mass spectrum turns on. \\ \hline \hline
$\zeta$ & Fraction of binaries with isotropic spin orientations. \\ \hline
$\sigma_i$ & Width of the preferentially aligned component of the distribution of black hole spin orientations. \\ \hline
$\mathbb{E}[a]$ & Mean of the Beta distribution of spin magnitudes. \\ \hline
$\mathrm{Var}[a]$ & Variance of the Beta distribution of spin magnitudes. \\ \hline \hline
$\lambda$ & How the merger rate evolves with redshift. \\ \hline
\end{tabular}
\caption{Parameters describing the binary black hole population.
See the text for a more thorough discussion and the functional forms of the models.
}
\label{tab:params}
\end{table*}

\subsection{Parameterized Mass Models} \label{sec:mass_models}

The power-law distribution considered
previously~\citep{2016PhRvX...6d1015A,2017PhRvL.118v1101A} modeled the BBH
primary mass distribution as a one-parameter power-law, with fixed limits on the
minimum and maximum allowed black hole mass. With our sample of ten binaries, we extend
this analysis by considering three increasingly complex models for the
distribution of black hole masses. The first extension, Model A (derived
from~\cite{2017ApJ...851L..25F,2018arXiv180506442W}), allows the maximum black
hole mass $\mmax$ and the power-law index $\alpha$
to vary. In Model B (derived
from~\cite{2017PhRvD..95j3010K,2017ApJ...851L..25F,2018ApJ...856..173T}) the
minimum black hole mass $\mmin$ and the mass ratio power-law index $\beta_q$
are also free parameters. However, the priors on Model B and C enforce a minimum of 5 \Msol{} on \mmin{} --- see Table~\ref{tab:mass_priors}. Explicitly, the mass distribution in Model A and Model
B takes the form
\begin{widetext}
\begin{equation}
\label{eq:power-law:concrete}
p(m_1, m_2 \vert \mmin, \mmax, \alpha, \beta_q) \propto
\begin{cases}
        C(m_1) m_1^{-\alpha} q^{\beta_q} &
                \text{if}\ \mmin \leq m_2 \leq m_1 \leq \mmax \\
        0 &
                \text{otherwise}
\end{cases},
\end{equation}
\end{widetext}
where $C\left( m_1\right)$ is chosen so that the marginal distribution is a
power law in $m_1$: $p\left( m_1 \vert \mmin, \mmax, \alpha, \beta_q\right) =
m_1^{-\alpha}$.

Model A fixes $\mmin = \mminVal \Msol$ and $\beta_q = 0$, whereas Model B fits for all four parameters.
Equation~\ref{eq:power-law:concrete} implies that the conditional mass ratio distribution is a power-law with $p(q \mid m_1)\propto q^{\beta_q}$.
When $\beta_q=0$, $C(m_1) \propto 1/(m_1-\mmin)$, as assumed in~\cite{2016PhRvX...6d1015A,2017PhRvL.118v1101A}.

Model C (from~\cite{2018ApJ...856..173T}) further builds upon the mass distribution in Equation~\ref{eq:power-law:concrete} by allowing for a second, Gaussian component at high mass, as well as introducing
smoothing scales $\delta m$, which taper the hard edges of the low- and high-mass cutoffs of the primary and secondary
mass power-law.
The second Gaussian component is designed to capture a possible build-up of high-mass black holes created from pulsational pair instability supernovae. The tapered low-mass smoothing reflects the fact that parameters such as metallicity probably blur the edge of the lower mass gap, if it exists.
Model C therefore introduces four additional model parameters, the mean, $\mu_{m}$, and standard deviation, $\sigma_{m}$, of the Gaussian component, $\lambda_{m}$, the fraction of primary black holes in this Gaussian component, and $\delta m$ the smoothing scale at the low mass end of the distribution.

\begin{figure*}[htbp]
\includegraphics[width=\textwidth]{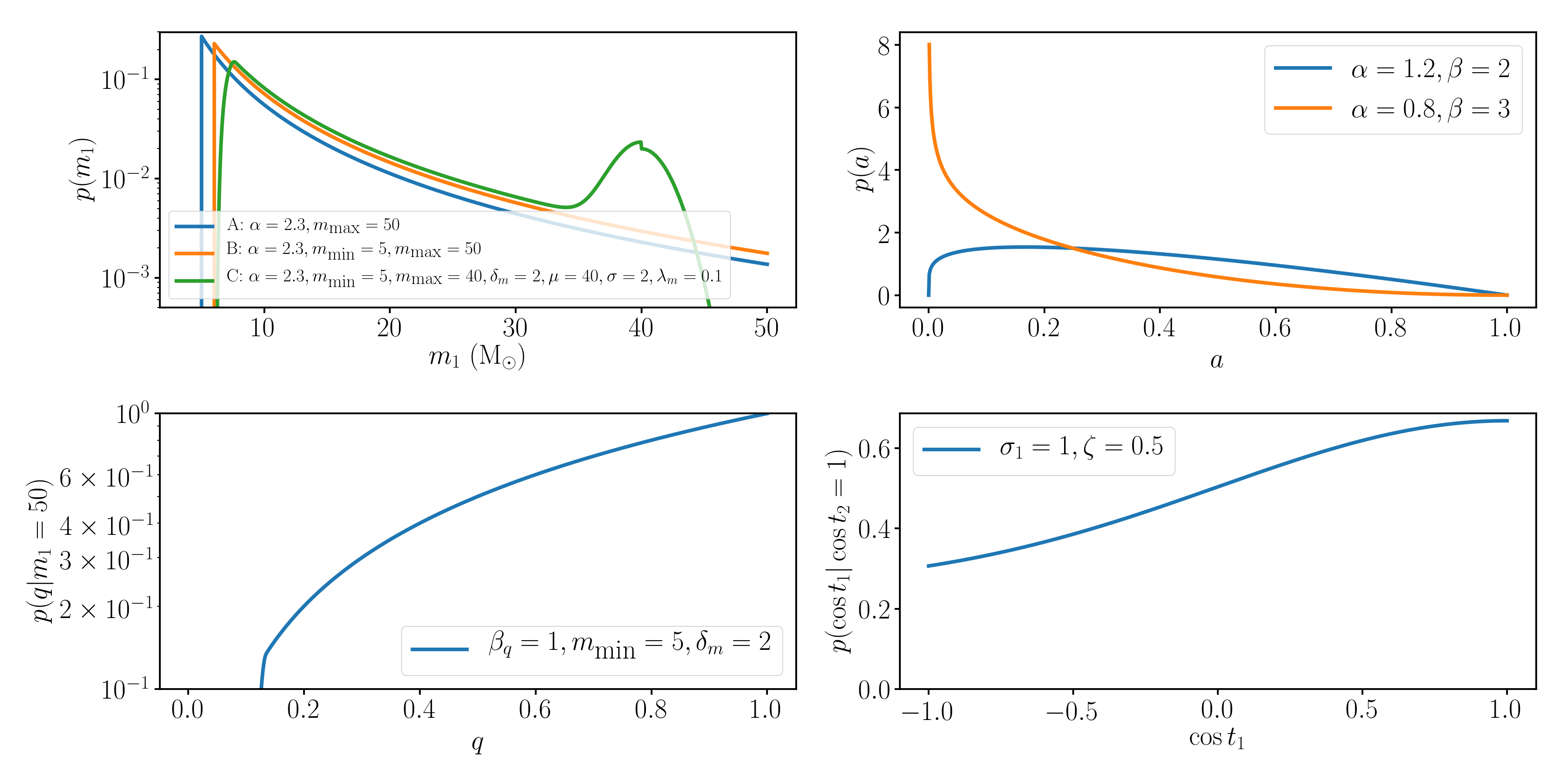}
\caption{The probability distributions for models encoded by Equations \ref{eq:mass-model-full}, \ref{eq:spin_mag}, and \ref{eq:spin-alignment}, are shown in the left panels, upper right panel, and lower right panel, respectively. In each, the legend indicates the parameter values for the models plotted. In the case of the lower left ($q$) distribution, we condition the value of $m_1=40\Msol$ rather than marginalizing for simplicity.}
\label{fig:model_demo}
\end{figure*}

The full form of this distribution is
\begin{widetext}
\begin{equation}
\begin{split}
\label{eq:mass-model-full}
    &p(m_1 \vert \theta) =
    \left[ (1 - \lambda_{m}) A(\theta) m_{1}^{-\alpha} \Theta(m_{\max}-m_1) + \lambda_{m} B(\theta) \exp\left(-\frac{(m_1-\mu_{m})^2}{2\sigma_{m}^2}\right) \right] S(m_1 ,m_{\min},\delta m), \\
    &p(q \vert m_1, \theta) =
    C(m_1, \theta) q^{\beta_{q}} S(m_2,m_{\min},\delta m).
\end{split}
\end{equation}
\end{widetext}
The factors $A$, $B$, and $C$ ensure each of the power-law component, Gaussian component, and mass ratio distributions are correctly normalized.
$S$ is a smoothing function which rises from zero at $m_{\min}$ to one at $m_{\min} + \delta m$ as defined in~\cite{2018ApJ...856..173T}.
$\Theta$ is the Heaviside step function. \changes{Models A, B, and C are displayed with a selection of parameters for demonstration purposes in the left panels of Figure \ref{fig:model_demo}.}

\subsection{Parameterized Spin Models} \label{sec:spin_models}

The black hole spin distribution is decomposed into independent models of spin magnitudes, $a$, and orientations, $t$.
For simplicity and lacking compelling evidence to the contrary, we assume both black hole spin magnitudes in a binary, $a_i$,
are drawn from a beta distribution \citep{2018arXiv180506442W}:
\begin{equation}
    p(a_i | \alpha_a, \beta_a) = \frac{a_i^{\alpha_a-1}(1-a_i)^{\beta_a-1}}{\mathrm{B}(\alpha_a, \beta_a)}.
\label{eq:spin_mag}
\end{equation}
\changes{This distribution is a convenient and flexible parameterization for describing values on the unit interval~\citep{BetaStats}.} \changes{Two examples of this distribution are shown in the upper right hand panel of Figure \ref{fig:model_demo}.} We choose to model the moments of the beta distribution using the  mean ($\mathbb{E}[a]$) and variance ($\mathrm{Var}[a]$), given by

\begin{eqnarray}\label{eq:beta_distr_params}
  \mathbb{E}[a] & = &
  \frac{\alpha_a}{\alpha_a+\beta_a}; \nonumber \\
  \mathrm{Var}[a] & = &
  \frac{\alpha_a\beta_a}{(\alpha_a+\beta_a)^2(\alpha_a+\beta_a+1)}.
\end{eqnarray}

\noindent \changes{We adopt a prior on the spin magnitude model parameters which are uniform over the values of $\mathbb{E}[a]$ and $\mathrm{Var}[a]$ which satisfy $\alpha_a, \beta_a \ge 1$. This choice of which values to sample avoids numerically-challenging singular spin distributions.}

To describe the spin orientation, we assume that the tilt angles between each black hole spin and the orbital angular momentum, $t_{i}$, are drawn from a mixture of two distributions: an isotropic component, and a preferentially aligned component, represented by a truncated Gaussian distribution in $\cos t_i$ peaked at $\cos t_i=1$~\citep{2017PhRvD..96b3012T}

\begin{align}
\label{eq:spin-alignment}
    p(\cos t_1, \cos t_2 | & {} \sigma_1, \sigma_2, \zeta) =  \frac{(1-\zeta)}{4} \nonumber \\
 & {} + \frac{2 \zeta}{\pi} \prod_{i \in \{1, 2\}} \frac{\exp\left(-(1-\cos t_i)^2 / (2\sigma_i^2)\right)}{\sigma_i \text{erf}(\sqrt{2}/\sigma_i)}.
\end{align}

\noindent \changes{We choose to parameterize the cosine of the tilt angles, rather than the angles themselves. This choice prompts the selection of a Gaussian (or uniform) model, rather than a wrapped distribution which would be more appropriate for an angular variable.} \changes{An example of the Mixture distribution is displayed in the lower right panel of Figure \ref{fig:model_demo}.}

The parameter $\zeta$ denotes the fraction of binaries which are preferentially aligned with the orbital angular momentum; $\zeta = 1$ implies all black hole spins are preferentially aligned and $\zeta = 0$ is an isotropic distribution of spin orientations. The typical degree of spin misalignment is represented by the $\sigma_i$. For spin orientations we explore two parameterized families of models:

\begin{itemize}
    \item Gaussian (G): $\zeta$ = 1.
    \item Mixture (M): $0 \leq \zeta \leq 1$.
\end{itemize}

The Gaussian model is motivated by formation in isolated binary evolution, with significant natal misalignment, while the mixture scenarios allow for an arbitrary combination of this scenario and randomly oriented spins, which arise naturally in dynamical formation.

\subsection{Redshift Evolution Models} \label{sec:redshift_models}

The previous two subsections described the probability distributions of intrinsic parameters $p(\xi)$ (i.e. masses and spins) that characterize the population of BBHs.
In addition, we also measure the value of one extrinsic parameter of the population: the overall merger rate density $R$.
The models described in the previous two subsections assume that the distribution of intrinsic parameters is independent of cosmological redshift $z$,  at least over the redshift range accessible to the LIGO and Virgo interferometers during the first two observing runs ($z \lesssim 1$).
However, we consider an additional model in which the overall event rate evolves with redshift.
We follow~\cite{2018ApJ...863L..41F} by parameterizing the evolving merger rate density $R(z)$ in the comoving frame by
\begin{equation}
\label{eq:Rz}
R(z | \lambda) = R_0 \left( 1+z \right)^{\lambda},
\end{equation}
where $R_0$ is the rate density at $z=0$.
In this model, $\lambda = 0$ corresponds to a merger rate density that is uniform in comoving volume and source-frame time,
while $\lambda \sim 3$ corresponds to a merger rate that approximately follows the star-formation rate in the redshift range relevant to the detections in O1 and O2~\citep{2014ARA&A..52..415M}.
Various BBH formation channels predict different merger rate histories, ranging from rate densities that will peak in the future ($\lambda < 0$) to rate densities that peak earlier than the star-formation rate ($\lambda \gtrsim 3$). These depend on the formation rate history and the distribution of delay times between formation and redshift.
In cases where we do not explicitly write the event rate density as $R(z)$, it is assumed that the rate density $R$ is constant in comoving volume and source-frame time.

The general model family, including the distributions of masses, spins and merger redshift, is therefore given by the distribution
\begin{equation}
\label{eq:dN}
\frac{\mathrm{d}N}{\mathrm{d}\xi \mathrm{d}z}(\theta) = R\left(z\right) \left[\frac{\mathrm{d} V_c}{\mathrm{d} z}\left(z\right)\right]\frac{T_\mathrm{obs}}{1+z} p(\xi | \theta),
\end{equation}
where $N$ is the total number of mergers that occur within the detection horizon (i.e. the maximum redshift considered) over the total observing time, $T_\mathrm{obs}$, as measured in the detector-frame, $\theta$ is the collection of all hyper-parameters that characterize the distribution, and
$d\Vc / dz$ is the differential comoving volume per unit redshift.
The merger rate density $R(z)$ is related to $N$ by
\begin{equation}
R(z) = \frac{\mathrm{d}N}{\mathrm{d}V_c \mathrm{d}t}\left( z \right),
\end{equation}
where $t$ is the time in the source-frame,
so that Eq.~\ref{eq:dN} can be written equivalently in terms of the merger rate density:
\begin{equation}
\frac{\mathrm{d}R}{\mathrm{d}\xi} \left( z | \theta \right) = R_0 p(\xi | \theta) (1 + z)^\lambda.
\end{equation}

\subsection{Hierarchical Population Model}

We perform a hierarchical Bayesian analysis, accounting for measurement
uncertainty and selection effects
\citep{2004AIPC..735..195L,2016PhRvX...6d1015A,2018arXiv180506442W,2018ApJ...863L..41F,2018arXiv180902063M,2018arXiv181111723M}.
\changes{We model the occurrence rate of events through a Poisson process with a mean dependent on the parameter distribution of the binaries\footnote{While this assumption is embedded~\citep{2015PhRvD..91b3005F} in the selection of events used in this work, studies of event count per time do not show significant evidence for deviations from Poissonian statistics~\citep{o2catalog}.}.}
The likelihood of the observed GW data given the population hyperparameters $\theta$
that describe the general astrophysical distribution, $dN / d\xi dz$, is given
by the inhomogeneous Poisson likelihood:
\begin{align}
\label{eq:hyperparameter-likelihood}
  \mathcal{L} & {} (\{d_n\} | \theta) \propto \nonumber \\
  & {} e^{-\mu(\theta)}
  \prod_{n=1}^{N_{\mathrm{obs}}}
    \int
      \mathcal{L}(d_n | \xi, z) \, \frac{\mathrm{d}N}{\mathrm{d}\xi \mathrm{d}z} \left(\theta \right) \,
    \dif \xi \dif z,
\end{align}
\changes{where $\mu(\theta)$ is the rate constant describing the mean number of events as a function of the
population hyper-parameters}, $N_\mathrm{obs}$ is the number of detections, and
$\mathcal{L}(d_{n} | \xi, z)$ is the individual-event likelihood for the $n$th
detection having parameters $\xi, z$.

In order to calculate the expected number of detections $\mu(\theta)$, we must understand the selection effects of our detectors.
The sensitivity of GW detectors is a strong function of the binary masses and distance, and also varies with spin.
For any binary, we define the sensitive spacetime volume $VT(\xi)$ of a network with a given sensitivity to be
\begin{equation} \label{eqn:sens_vol}
VT(\xi) = T_\mathrm{obs} \int_0^{\infty} f(z|\xi) \frac{\mathrm{d}\Vc}{\mathrm{d}z} \frac{1}{1+z} \dif z,
\end{equation}
where the sensitivity is assumed to be constant over the observing time, $T_\mathrm{obs}$, as measured in the detector-frame and $f(z|\xi)$ is the detection probability of a BBH with the given parameter set $\xi$ at
redshift $z$~\citep{2010ApJ...716..615O}, \changes{averaged over the extrinsic binary orientation parameters~\citep{1993PhRvD..47.2198F}}.
The factor of $1/(1+z)$ arises from the difference in clocks timed between the source frame and the detector frame.
For a given population with hyper-parameters $\theta$, we can calculate the total observed spacetime volume
\begin{equation}
    \VT_{\theta} = \int_{\xi} p(\xi | \theta) VT(\xi) \dif\xi,
\end{equation}
where $p(\xi | \theta)$ describes the underlying distribution of the intrinsic parameters. We performed large scale simulation runs wherein the spacetime volume in the above equation is estimated by Monte-Carlo integration~\citep{2018CQGra..35n5009T} --- these runs are restricted to have no BH less massive than 5 \Msol. We then use a semi-analytic prescription, calibrated to the simulation results, to derive the $\VT_{\theta}$ for specific hyper-parameters.

Allowing the merger rate to evolve with redshift, the expected number of detections is given by
\begin{widetext}
\begin{equation}
    \mu(\theta) = T_\mathrm{obs} \int_\xi \int_0^{\infty} p(\xi | \theta) f(z \mid \xi) R(z) \frac{\mathrm{d}\Vc}{\mathrm{d}z} \frac{1}{1+z} \dif z \dif\xi.
\end{equation}
\end{widetext}
If the merger rate does not evolve with redshift, i.e., $R(z) = R_0$, this reduces to $\mu(\theta) = R_0 \VT_{\theta}$.

We note that the hyperparameter likelihood given by Eq.~\ref{eq:hyperparameter-likelihood} reduces to the likelihood used in the O1 mass distribution analysis \citep[Eq. D10 of][]{2016PhRvX...6d1015A}, which fit only for the shape, not the rate / normalization of the mass distribution, if one marginalizes over the rate parameter with a flat-in-log prior $p(R_0) \propto 1/R_0$~\citep{2018ApJ...863L..41F,2018arXiv180902063M}.
For consistency with previous analyses, we adopt a flat-in-log prior on the rate parameter throughout this work.

\subsection{Statistical Framework and Prior Choices}

In practice, we sample the likelihood
$\mathcal{L}(d_{n} | \xi, z)$ using the parameter estimation pipeline
LALInference~\citep{2015PhRvD..91d2003V}. \changes{Since LALInference gives us a set of
posterior samples for each event, we first divide out the priors used in the
individual-event analyses before applying
Eq.~\ref{eq:hyperparameter-likelihood}
\citep{2010ApJ...725.2166H,2010PhRvD..81h4029M} (see Appendix \ref{sec:reweighting}).}

Where not fixed, we adopt uniform priors on population parameters describing the models. Unless otherwise noted, for the event rate distribution we use a log-uniform distribution in $R_0$, bounded between $[10^{-1}, 10^{3}]$. While this is a different form than the priors adopted in~\cite{o2catalog}, we note that similar results are obtained on the rates (see Sec.~\ref{sec:rates}), indicating that the choice of prior does not strongly influence the posterior distributions.
We provide specific limits on all priors when the priors for a given model are introduced. \changes{Unless otherwise
stated all posterior credible intervals are 90\% intervals, symmetric in the quantiles around the median.}
\changes{The MCMC-based analyses presented in this work have approximately $10^4$ effective samples, after
thinning by their autocorrelation time.}

The normalization factor of the posterior density in Bayes' theorem is the evidence --- it is the probability of the data given the model. We are interested in the preferences of the data for one model versus another. This preference is encoded in the Bayes factor, or the ratio of evidences. The odds ratio is the Bayes factor multiplied by their ratio of the model prior probabilities. In all cases presented here, the prior model probabilities are assumed to be equal, and odds ratios are equivalent to Bayes factors.

We often present the posterior population distribution (PPD) of various quantities. The PPD is the expected distribution of new mergers conditioned on previously obtained observations. It integrates the distribution of values (e.g., $\xi$, such as the masses and spins) conditioned on the model parameters (e.g, the power law index) over the posteriors obtained for the model parameters:

\begin{equation} \label{eqn:ppd}
p(\xi_{\textrm{new}}|\xi_{\textrm{observed}}) = \int p(\xi_{\textrm{new}}|\theta) p(\theta|\xi_{\textrm{observed}}) d\theta
\end{equation}

\noindent It is a predictor for future merger values $\xi_{\textrm{new}}$ given observed data $\xi_{\textrm{observed}}$ and factors in the uncertainties imposed by the posterior on the model parameters.  Note that the PPD does not incorporate the detector sensitivity, and therefore is not a straightforward predictor of the properties of future \emph{observed} mergers.

\section{The Mass Distribution} \label{sec:mass_distr}
For context, Figure \CatalogFigureComponentMassesAll{} in \cite{o2catalog} illustrates the inferred masses for all of the significant BBH observations
identified in our GW surveys in O1 and O2.
Despite at least moderate sensitivity to total masses between 0.1 -- 500 \Msol, current observations occupy only a portion of the binary mass
parameter space.
Notably, we have not yet observed a pair of very massive (e.g., 100 \Msol) black holes, a binary which is bounded away from equal mass in its posterior, or a binary with a component mass confidently below 5 \Msol.
In our survey, we also find a preponderance of observations at higher masses: six with significant posterior support above $30 \Msol$.
In this section, we attempt to reconstruct the binary black hole merger rate as a function of the component masses using parameterized models.
Table \ref{tab:mass_priors} summarizes the mass models adopted from Section~\ref{sec:mass_models} and the prior distributions for each of the parameters in those models. \changes{We present results for three increasingly general mass and spin models, the most complex of which ranges over the full set of model parameters in Section \ref{sec:notation} with the exception of rate dependence of rate on redshift. The interdependence of the mass and redshift distribution is explored more fully in Section \ref{sec:rates}.}

\begin{table*}[htbp]
\centering
\begin{tabular}{|c|cc|cc|cccc|cccc|} \hline
 & \multicolumn{8}{c|}{Mass Parameters} & \multicolumn{4}{c|}{Spin Parameters} \\
Model & $\alpha$ & $m_{\max}$ & $m_{\min}$ & $\beta_q$ & $\lambda_{m}$ & $\mu_{m}$ & $\sigma_{m}$ & $\delta m$ & $\mathbb{E}[a]$ & $\mathrm{Var}[a]$ & $\zeta$ & $\sigma_i$ \\ \hline
A & [-4, 12] & [30, 100] & \textbf{5} & \textbf{0} & \textbf{0} & N/A & N/A & N/A & [0, 1] & [0, 0.25] & \textbf{1} & [0, 4] \\
B & [-4, 12] & [30, 100] & [5, 10] & [-4, 12] & \textbf{0} & N/A & N/A & N/A & [0, 1] & [0, 0.25] & \textbf{1} & [0, 4] \\
C & [-4, 12] & [30, 100] & [5, 10] & [-4, 12] & [0, 1] & [20, 50] & (0, 10] & [0, 10] & [0, 1] & [0, 0.25] & [0, 1] & [0, 4] \\ \hline
\end{tabular}%
\caption{Summary of models used in Sections~\ref{sec:mass_distr},
\ref{sec:rates}, and \ref{sec:spin_distr}, with the prior ranges for the
population parameters. The fixed parameters are in
bold. Each of these distributions is uniform over the stated range. All models
in this Section assume rates which are uniform in the comoving volume ($\lambda
= 0$).  The lower limit on $\mmin$ is chosen to be consistent with
\cite{o2catalog}.}
\label{tab:mass_priors}
\end{table*}

\subsection{Parameterized Modeling Results}
\label{ssec:postmass}
\begin{figure*}[htbp]
\includegraphics[width=2\columnwidth]{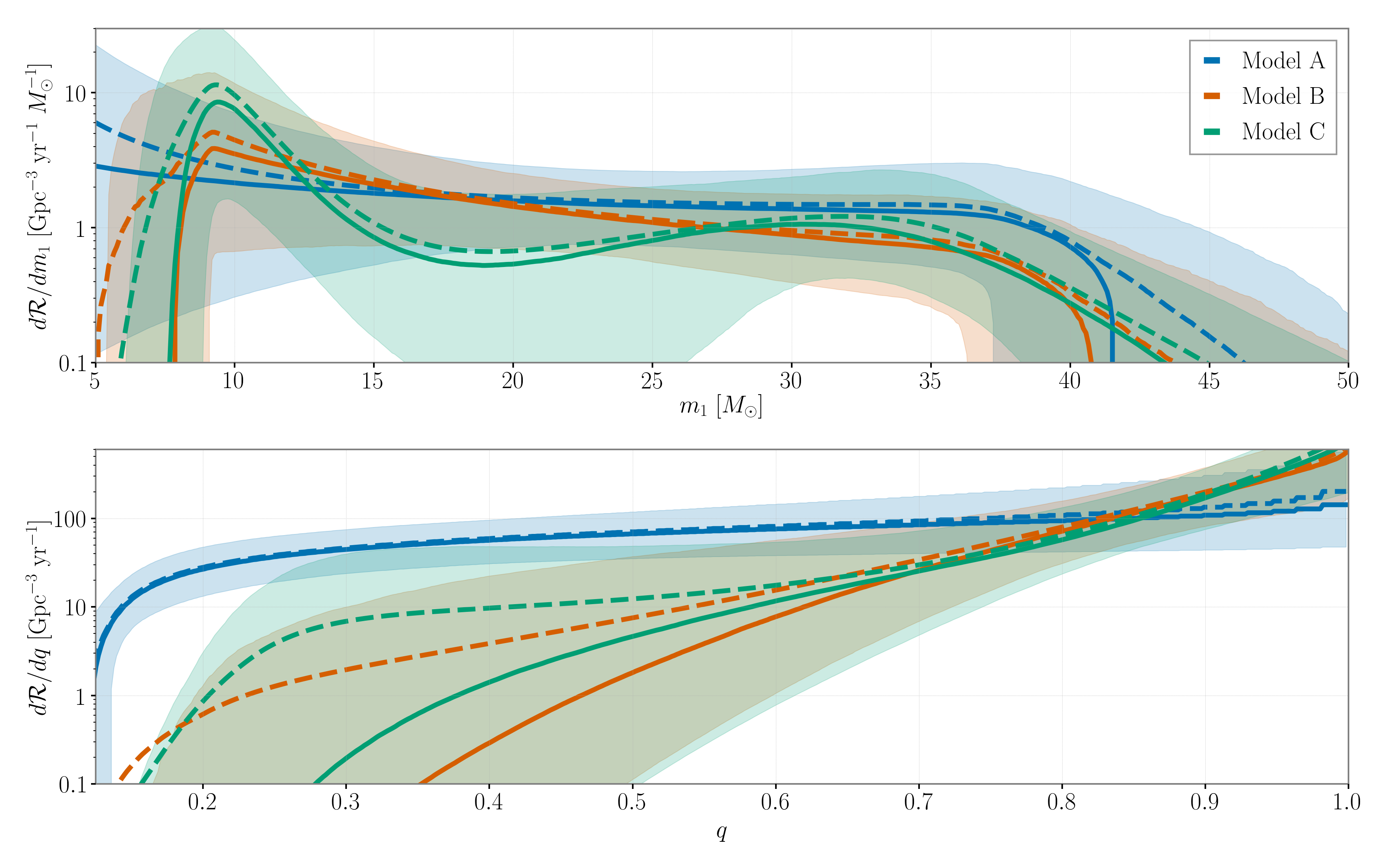}
\caption{Inferred differential merger
rate as a function of primary mass, $m_{1}$, and mass ratio, $q$, for three
different assumptions. For each of the three increasingly complex assumptions A,
B, C described in the text we show the PPD
(dashed) and median (solid), \changes{and the 90\% symmetric credible intervals
(shaded regions), for the differential rate.} The results shown marginalize over
the spin distribution model.  The falloff at small masses in models B and C is
driven by our choice of the prior limits on the $\mmin$ parameter (see Table\
\ref{tab:mass_priors}).  All three models give consistent mass distributions
within their 90\% credible intervals over a broad range of masses, consistent
with their near-unity evidence ratios (Table \ref{tab:mass_bayes_factors}); in
particular, the peaks and trough seen in Model C, while suggestive, are not
identified at high credibility in the mass distribution.}
\label{fig:dndm}
\end{figure*}

\begin{figure*}[htbp]
\includegraphics[width=2\columnwidth]{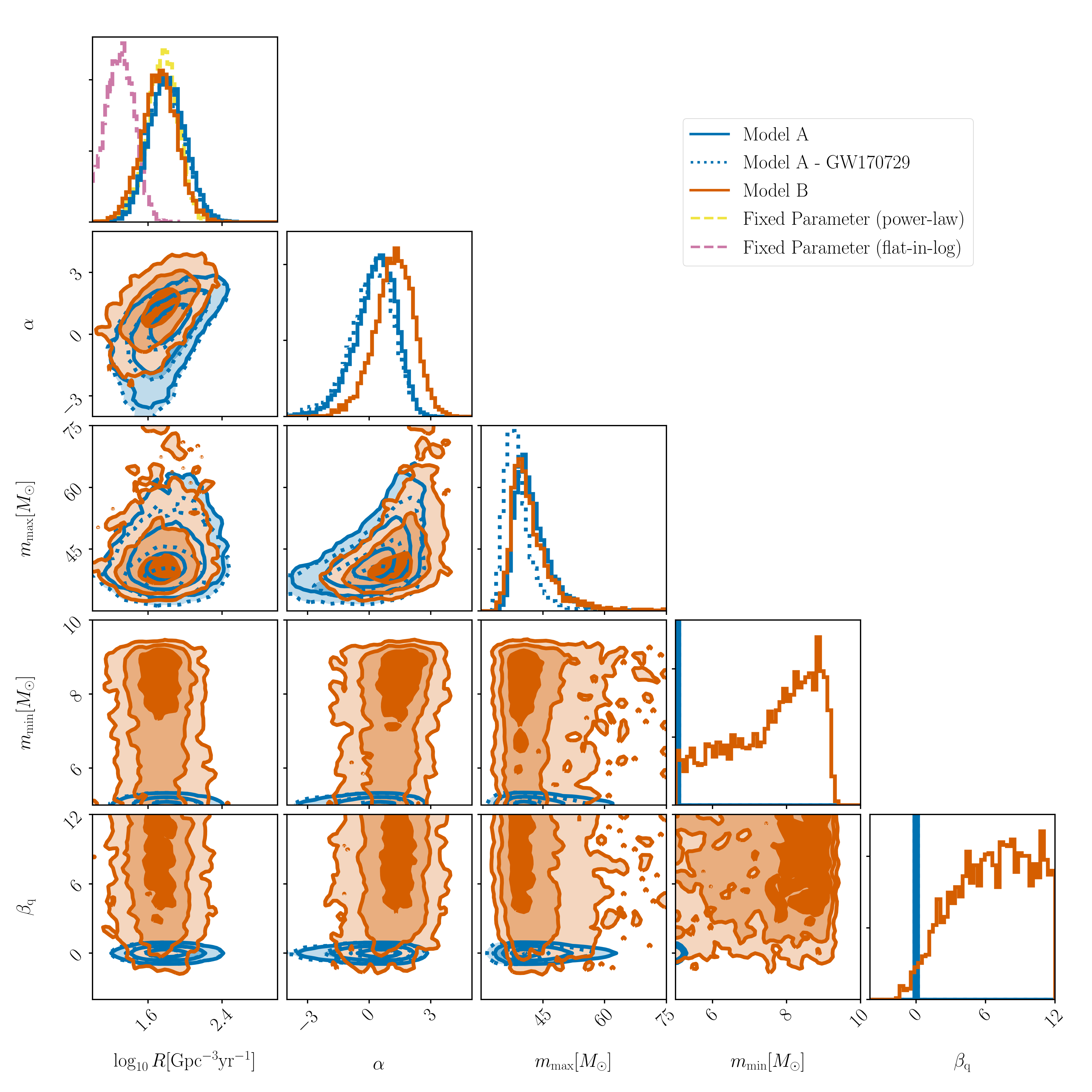}
\caption{
    One- and two-dimensional posterior distributions for the hyperparameters describing Models A and B.
    Large values of $\alpha$ correspond to a mass distribution which rapidly decays with increasing mass.
    Large values of $\beta$ correspond to a mass-ratio distribution which prefers equal mass binaries.
    Also shown is the one-dimensional posterior distribution for the merger rate discussed in~\cite{o2catalog}, and the stability of Model A to the removal of the GW170729 event.
}
\label{fig:dndm:hyperparameters_ab}
\end{figure*}

\begin{figure*}[htbp]
\includegraphics[width=\textwidth]{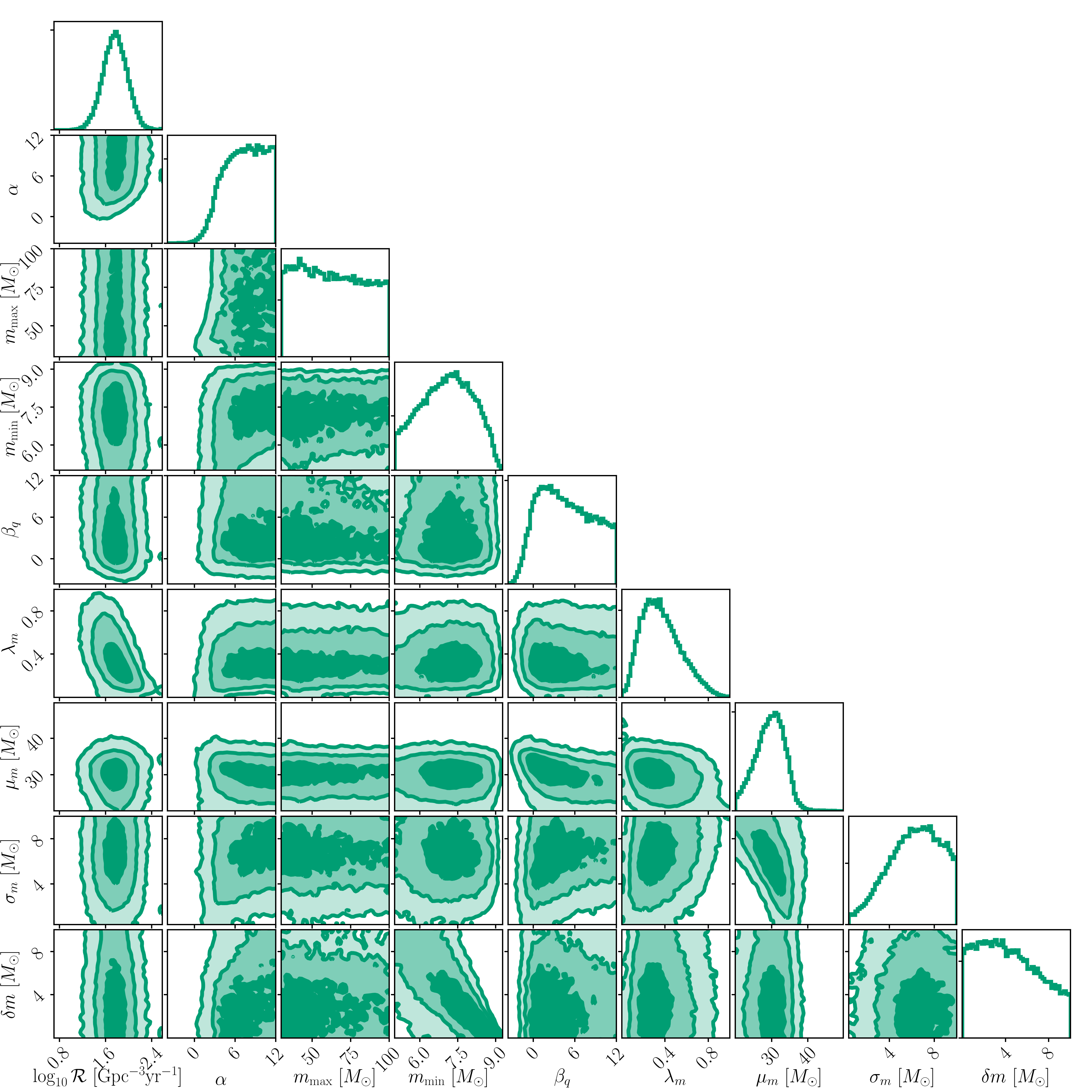}
\caption{
    One- and two-dimensional posterior distributions for the hyperparameters describing Model C.
    This model consists of the power-law distribution in Model B with an additional Gaussian component at high mass.
    The parameters $\alpha$, $\beta$, $\mmax$, and $\mmin$ describe the power-law component.
    The Gaussian has mean $\mu_{m}$ and standard deviation $\sigma_{m}$.
    The fraction of black holes in the Gaussian component is $\lambda_{m}$.
    This model also allows for a gradual turn-on at low masses over a mass range $\delta_{m}$.
}
\label{fig:dndm:hyperparameters_c}
\end{figure*}

Figure~\ref{fig:dndm} shows our updated inference for the compact binary primary mass $m_1$ and mass ratio $q$ distributions for several increasingly general population models.
In addition to inferring the mass distribution, all of these calculations self-consistently marginalize over the parameterized spin distribution presented in Section \ref{sec:spin_distr} and the merger rate.
Figures \ref{fig:dndm:hyperparameters_ab} and  \ref{fig:dndm:hyperparameters_c} show the posterior distribution on selected model hyperparameters.

If we  assume the black hole masses follow a power-law distributed and fix the minimum black hole mass to be $\mmin=5M_\odot$ (Model A), we find $\alpha = \ALPHAMimrpNoSpinVTQuadApm$ and $\mmax = \MMAXimrpNoSpinVTQuadApmU$.
In Model B we infer the power-law index of the primary mass to be $\alpha = \ALPHAMimrpNoSpinVTQuadBpm$ with corresponding limits $\mmin = \MMINimrpNoSpinVTQuadBpmU$ and $\mmax = \MMAXimrpNoSpinVTQuadBpmU$.

Figure \ref{fig:dndm:hyperparameters_c}, shows the posterior over the population parameters present in A and B, as well as a second, Gaussian population parameterized with $\mmax$ and $\sigma_m$.
$\lambda_m$ is the mixing fraction of binaries in the Gaussian population versus the power law, with $\lambda_m=0$ indicating only the power law component.
The Gaussian component is centered at $\mu_m = \MUMimrpNoSpinVTQuadCpmU$, has a width $\sigma_m = \SIGMAMimrpNoSpinVTQuadCpmU$, and is consistent with the parameters of the seven highest mass events in our sample as seen in Figure \ref{fig:mrem_mzams_gaps}.
Also as a consequence of this mixture, \changes{the second component can account for many of the high mass events. Without needing to accommodate higher mass events, the inferred power-law is much steeper $\alpha = \ALPHAMimrpNoSpinVTQuadCpm$ than Models A or B, however, the posterior distribution for Model C is less informative for $\alpha \gtrsim 4$.}
This in turn means that we cannot constrain the parameter $\mmax$ in Model C since the power-law component has negligible support above $\sim 45 \Msol$ (see the upper panel of Figure~\ref{fig:dndm}).
In the intermediate regime, $\sim 15 M_{\odot} - 25 M_{\odot}$, Model C infers a smaller rate than Models A or B as a consequence of the steeper power-law behavior.
The low mass smoothing allowed in this model also weakens constraints we can place on the minimum black hole mass, in this model we find $\mmin = \MMINimrpNoSpinVTQuadCpmU$.
\changes{All three models produce consistent results for the marginal merger rate distribution, as is further discussed in Section \ref{sec:rates}.}

All models feature a parameter, \mmax, which defines a cutoff of the power law.
However, the interpretation of that parameter within Model C is not a straightforward comparison with Models A and B, due to the presence of the Gaussian component at high mass and the large value of the power-law spectral index.
Instead, to compare those two features, we compute the 99th percentile of the mass distribution inferred from the model PPDs (see Equation~\ref{eqn:ppd}).
Model A obtains $\MPRIMARYimrpNoSpinVTQuadAboundninetynine$ $M_\odot$, Model B obtains $\MPRIMARYimrpNoSpinVTQuadBboundninetynine$ $M_\odot$, and Model C obtains $\MPRIMARYimrpNoSpinVTQuadCboundninetynine$ $M_\odot$.
Therefore, all models self-consistently infer a dearth of black holes above $\sim 45$ \Msol.
This is determined by the lower limit for the mass of the most massive black hole in the sample because \mmax{} can be \emph{no smaller} than this value.
Similarly, the models which allow $\mmin$ to vary (B and C) disfavor populations with $\mmin$ above $\simeq 9 M_\odot$.
This parameter is close to the largest allowed mass for the least massive black hole in the sample, for similar reasons.

The lower limits we place on $\mmin$ are dominated by our prior choices that
constrain $\mmin \in [5, 10] \, \Msol$ (see Table \ref{tab:mass_priors}).  For
example, in Figure \ref{fig:dndm:hyperparameters_ab}, the posterior on $\mmin$
becomes flat as $\mmin$ approaches the prior boundary at $5 \, \Msol$.  Given
current sensitivities, this is to be expected
\citep{2015ApJ...807L..24L,2015MNRAS.450L..85M}.  In the inspiral-dominated
regime, the sensitive time-volume scales as $VT \sim m^{15/6}$
\citep{1993PhRvD..47.2198F}; extending our inferred mass distributions and
merger rates into the possible lower black hole mass gap from $3$--$5\, \Msol$
\citep{2010ApJ...725.1918O,2011ApJ...741..103F,2012ApJ...757...36K} yields an
expected number of detected BBH mergers $\lesssim 1$. Thus, we are unable to
place meaningful constraints on the presence or absence of a mass gap at low black hole
mass.

Models B and C also allow the distribution of mass ratios to vary according to $\beta_q$.
In these cases the inferred mass-ratio distribution favors comparable-mass binaries (i.e., distributions with most support near $q\simeq 1$), see panel two of Figure~\ref{fig:dndm}.
Within the context of our parameterization, we find $\beta_q = \BETAQimrpNoSpinVTQuadBpm$ for Model B and $\beta_q = \BETAQimrpNoSpinVTQuadCpm$ for Model C.
These values are consistent with each other and are bounded above zero at 95\% confidence, thus implying that the mass ratio distribution is nearly flat or declining with more extreme mass ratios.
The posterior on $\beta_q$ returns the prior for $\beta_q \gtrsim 4$.
Thus, we cannot say much about the relative likelihood of asymmetric binaries, beyond their overall rarity.

The distribution of the parameter controlling the fraction of the power law versus the Gaussian component in Model C is $\lambda_m = \LAMBDAMimrpNoSpinVTQuadCpm$, which peaks away from zero, implying that this model prefers a contribution to the mass distribution from the Gaussian population in addition to the power laws modeled in A and B.
To determine preference amongst the three models presented in this Section, we compute the Bayes factors comparing the mass models using a nested sampler~\citep{2004AIPC..735..395S}, \texttt{CPNest}~\citep{john_veitch_2017_835874}.
These are shown in Table~\ref{tab:mass_bayes_factors}.
Model B, which allows $\mmin$ and $\beta_q$ to vary is preferred over Model A ($\ln \mathrm{BF}^{A}_{B} = \LNBFMassModelABIMRPhenomP$).
To isolate the contributions of the Gaussian component and low mass smoothing in Model C, we compute the Savage-Dickey density ratio, $p(\theta=0)/p_{\rm prior}(\theta=0)$, equivalent to the Bayes factor comparing without and with the feature.
The model including a Gaussian component in addition to the power-law distribution is preferred over the pure power-law models ($\ln \mathrm{BF}^{\lambda=0}_C = \SDDRLambdavsCIMRPhenomP$); nevertheless, all models infer mass distributions that agree within their 90\% credible bounds (see Figure \ref{fig:dndm}). \changes{We caution that the mild preferences in Table~\ref{tab:mass_bayes_factors} are influenced by our choices of the range and shape of the priors we apply to the parameters, particularly for models where the number of parameters is comparable to the number of events. Moreover, the credible intervals on the distributions of the primary mass overlap, indicating that the model predictions agree to within the individual model uncertainties.}
We are unable to distinguish between a gradual or sharp cutoff at low mass ($\ln \mathrm{BF}^{\delta_{m}=0}_C = \SDDRDeltaMvsCIMRPhenomP$).
This is unsurprising, since we are less sensitive to structure in the mass distribution at low masses~\citep{2018ApJ...856..173T}.

\begin{table}[tb]
\centering
\begin{tabular}{|c|c|c|c|c|} \hline
Model & A & B & C, $\lambda_m = 0$ & C, $\delta_m = 0$ \\ \hline
$\ln \mathrm{BF}^{i}_{C}$ & \LNBFMassModelACIMRPhenomP & \LNBFMassModelBCIMRPhenomP & \SDDRLambdavsCIMRPhenomP & \SDDRDeltaMvsCIMRPhenomP \\ \hline
\end{tabular}
\caption{
The log Bayes factor comparing each of the models described in Table~\ref{tab:mass_priors} to the most complex model, Model C.
The evidence for the three mass models is computed using nested sampling, while the limits $\lambda_{m}=0$ and $\delta_{m}=0$ of Model C are computed using the Savage-Dickey density ratio.
}
\label{tab:mass_bayes_factors}
\end{table}

The analysis above includes all ten binary black hole detections, though not all events have the same statistical detection confidence~\citep{2019MNRAS.tmp..230G}. To assess the stability of our results against systematics in the estimated significance, we have repeated these analyses after omitting the least significant detection.
For our sample, the least significant detection, GW170729, is also the most massive binary.
Most features we derive from our observations remain unchanged, with one exception shown in Figure~\ref{fig:dndm:hyperparameters_ab}:
since we have omitted the most massive binary, the maximum black hole mass $\mmax$ reported in models A and B is decreased
by about 5 \Msol. Without GW170729, the \mmax{} distribution is $\MMAXimrpNoSpinVTQuadANoBigDetectionpmU$  for Model A and $\MMAXimrpNoSpinVTQuadBNoBigDetectionpmU$ for Model B. This is consistent with the difference between GW170729 and the next highest mass binary, GW170823, when comparing the less massive end of their primary mass posteriors.

\subsection{Comparison with Theoretical and Observational Models}

Previous modeling of the primary mass distribution with a power law distribution~\citep{2016PhRvX...6d1015A} was last updated with the discovery of GW170104~\citep{2017PhRvL.118v1101A}. This analysis measured spectral index of the the power law to be $\alpha = {2.3}^{+1.3}_{-1.4}$ at 90\% confidence assuming a minimum black hole mass of 5 \Msol\ and maximum total mass of 100 \Msol. None of our models directly emulate this one, but Model A is the closest analog. When allowing \mmax{} to vary, 100 \Msol{} is strongly disfavored. As a consequence of the lower \mmax{}, the power law index inferred is also \changes{shallower than previously obtained \citep{2017ApJ...851L..25F}, but remains consistent with the previous distribution.}

\begin{figure*}[htbp]
\includegraphics[width=\textwidth]{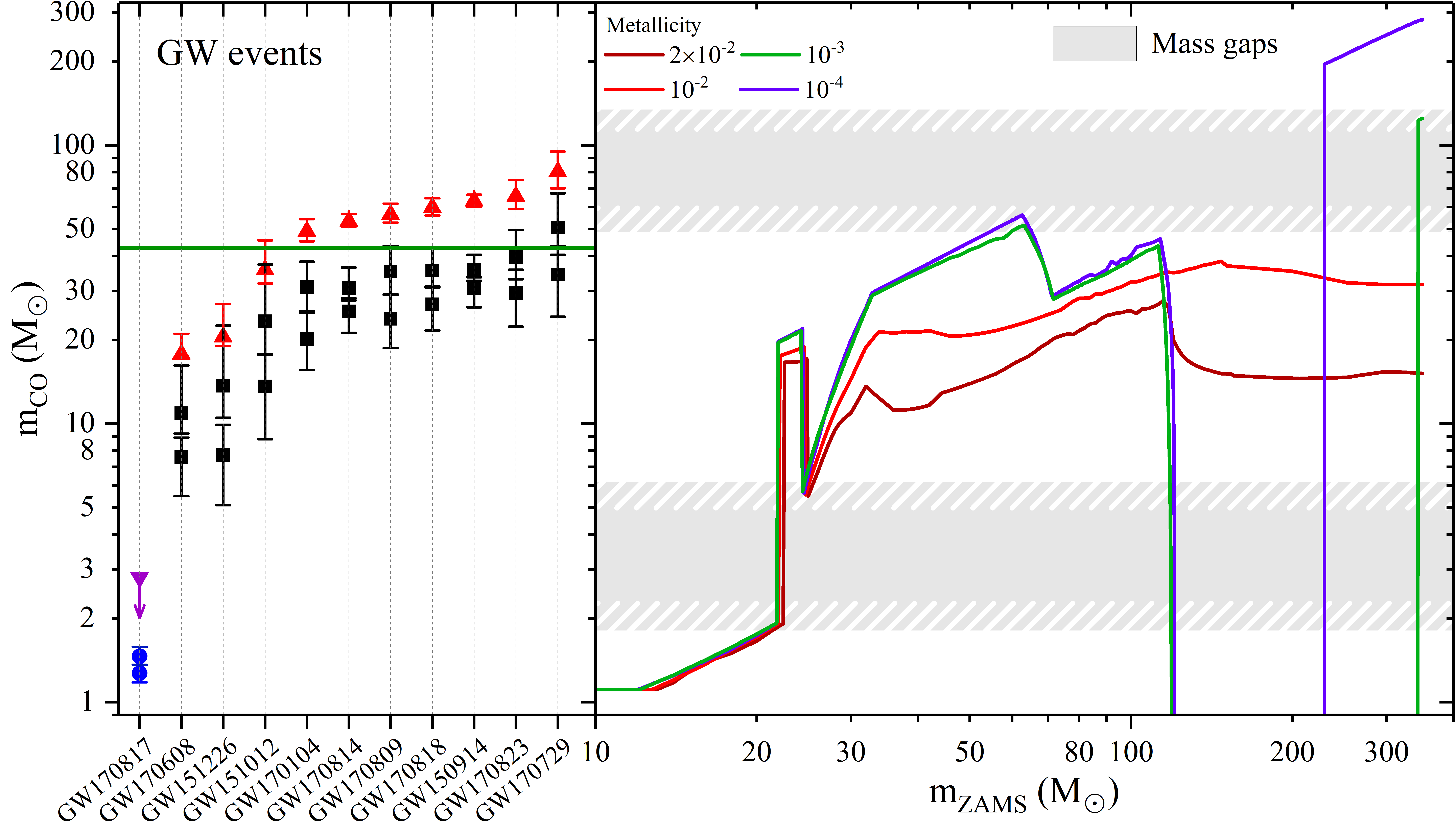}
\caption{The left-hand panel shows compact object masses ($m_{\rm CO}$) from GW detections in O1 and O2, with the black squares and error bars representing the component masses of the merging black holes and their uncertainties, and red triangles representing the mass and associated uncertainties of the merger products.
	The horizontal green line shows the 99th percentile of the mass distribution inferred from the Model B PPD.
	In the right-hand panel, the predicted compact-object mass is shown as a function of the zero-age main sequence mass of the progenitor star ($m_{\rm ZAMS}$) and for four different metallicities of the progenitor star (ranging from $Z=10^{-4}$ to $Z=2\times 10^{-2}$, \citealt{2017MNRAS.470.4739S}).
This model accounts for single stellar evolution from the {\sc PARSEC} stellar-evolution code~\citep{2012MNRAS.427..127B}, for core-collapse supernovae~\citep{2012ApJ...749...91F}, and for pulsational-pair instability and pair-instability supernovae~\citep{2017ApJ...836..244W}. The shaded areas represent the lower and upper mass gaps.
There is uncertainty as to the final product of GW170817. It is shown in the left-hand panel to emphasize that BNS mergers might fill the lower gap.}
\label{fig:mrem_mzams_gaps}
\end{figure*}

In Figure~\ref{fig:mrem_mzams_gaps}, we highlight the two mass gaps predicted by models of stellar evolution: the first gap between $\sim{}2$ and $\sim{}5$ M$_\odot$ and the second between $\sim{}50$ M$_\odot$ and $\sim{}150$ M$_\odot$, compared against the observed black holes. A set of tracks~\citep{2017MNRAS.470.4739S} relating the progenitor mass and compact object is also shown for reference purposes. The tracks are subject to many uncertainties in stellar and binary evolution, and only serve as representative examples. We discuss some of those uncertainties in the context of our results below.

The minimum mass of a black hole and the existence of a mass gap between neutron stars and black holes (lower gray shaded area, right panel of Figure \ref{fig:mrem_mzams_gaps}) are currently debated.
Claims~\citep{2010ApJ...725.1918O,2011ApJ...741..103F} of the existence of a mass gap between the heaviest neutron stars ($\sim{}2$ M$_\odot$) and the lightest black holes ($\sim{}5$ M$_\odot$) are based on the sample of about a dozen X-ray binaries with dynamical mass measurements.
However, \citet{2012ApJ...757...36K} suggested that the dearth of observed black hole masses in the gap could be due to a systematic offset in mass measurements.
\changes{Moreover, only a subset of theoretical models (e.g., the ``rapid'' model in~\cite{2012ApJ...749...91F}) reproduce this gap in stellar modelling.}
We can see in Figure~\ref{fig:mrem_mzams_gaps} that none of the observed binaries sit in this gap, but the sample is not sufficient to definitively confirm or refute the existence of this mass gap.

From the first six announced BBH detections, \citet{2017ApJ...851L..25F} argued that there is evidence for missing black holes with mass greater than $\gtrsim{}40$ M$_\odot$.
The existence of this second mass gap --- see the upper grey shaded area in the right panel of Figure \ref{fig:mrem_mzams_gaps} between $\sim{}50$ M$_\odot$ and $\sim{}150$ M$_\odot$ --- has been further explored by~\cite{2018ApJ...856..173T,2018arXiv180506442W,2018arXiv180204909B,2019MNRAS.tmp..231R}.
This gap might arise from the combined effect of pulsational pair instability \citep{1967PhRvL..18..379B,2003ApJ...591..288H,2007Natur.450..390W,2017ApJ...836..244W} and pair instability \citep{1964ApJS....9..201F,1983A&A...119...61O,1984ApJ...280..825B} supernovae. \changes{Uncertainties in stellar evolution models (e.g. stellar winds, rotation) and in the treatment of the final outcomes of (pulsational) pair instability lead to a range of possible low-mass edges for the upper mass gap as well as the shape and abundance in a putative build-up.}
Predictions for the maximum mass of black holes born after pulsational pair-instability supernovae are $\sim{}50 \Msol$~\citep{2016A&A...594A..97B,2017MNRAS.470.4739S}.
Our inferred maximum mass is consistent with these predictions.

\section{Merger Rates and Evolution with Redshift} \label{sec:rates}
As illustrated in previous work~\citep{2016ApJ...833L...1A,o2catalog,2017ApJ...851L..25F,2018arXiv180506442W,2018ApJ...863L..41F}, the inferred binary black hole
merger rate depends on and correlates with our assumptions about their intrinsic mass (and to a lesser extent, spin)
distribution.
In the most recent catalog of GW BBH events~\citep{o2catalog}, we infer the overall BBH merger rate for two fixed-parameter populations.
The first of these populations follows the power-law model given by Equation~\ref{eq:power-law:concrete} with $\alpha = 2.3$, $\beta_q = 0$, $\mmin = 5 \Msol$, and $\mmax = 50 \Msol$.
The second population follows a distribution in which both black hole masses are independently drawn from a flat-in-log distribution:
\begin{equation}
\label{eq:flat-in-log}
p(m_1, m_2) \propto \frac{1}{m_1 m_2},
\end{equation}
subject to the same mass cutoffs $5 \Msol < m_2 < m_1 < 50 \Msol$ as the fixed power-law population.
Both the power-law and flat-in-log populations assume an isotropic and uniform-magnitude spin distribution ($\alpha_a=\beta_a=1$).
These two fixed-parameter populations are used to estimate the population-averaged sensitive volume $\VT$ with a Monte-Carlo injection campaign as described in \cite{o2catalog}, with each population corresponding to a different $\VT$ because of the strong correlation between the mass spectrum and the sensitive volume.
Under the assumption of a constant-in-redshift rate density, these $\VT$ estimates yield two different estimates of the rate: \bbhPlawCombR \invstvol for the $\alpha = 2.3$ population, and \bbhUlmCombR \invstvol for the flat-in-log population (90\% credibility; combining the rate posteriors from the two analysis pipelines).

The two fixed-parameter distributions do not incorporate all information
about the mass, mass ratio, spin distribution, and redshift evolution suggested by our observations in O1 and O2.
In this section, rather than fixing the mass and spin distribution, we estimate the rate by marginalizing over the uncertainty in the underlying population, which we parameterize with the mass and spin models employed in Sections~\ref{sec:mass_distr} and \ref{sec:spin_distr}.
When carrying out these analyses, it is computationally infeasible to determine $VT(\xi)$ for each point in parameter space with the full Monte-Carlo injection campaign described in~\cite{o2catalog}, so we employ the semi-analytic methods described in Appendix~\ref{sec:systematics}.
Furthermore, while the rate calculations in~\cite{o2catalog} incorporate all triggers down to a very low threshold and fit the number of detections by modeling the signal and background distributions in the detection pipelines~\citep{2015PhRvD..91b3005F,2016ApJ...833L...1A},
in this work we fix a high detection threshold~\cite{o2catalog}, which sets the number of detections to $N_\mathrm{obs} = 10$.
In principle, our results are sensitive to the choice of threshold, but this effect has been shown to be much smaller than the statistical uncertainties~\citep{2019MNRAS.tmp..230G}.
The choice of detection threshold is further discussed in Appendix~\ref{sec:systematics}.
The full set of models used in this section are enumerated in Table~\ref{tab:rate_priors}.

\begin{table*}[htbp]
\centering
\begin{tabular}{|c|c|c|cc|cc|cc|} \hline
 & Mass Model & Rate Parameters & \multicolumn{4}{c|}{Spin Parameters} \\
Model & & $\lambda$ & $\alpha_a$ & $\beta_a$ & $\mathbb{E}[a]$ & $\mathrm{Var}[a]$ \\ \hline
\multirow{2}*{Fixed Parameter (power-law)} & A, with $\alpha = 2.3$, & \multirow{2}*{\textbf{0}} & \multirow{2}*{\textbf{1}} & \multirow{2}*{\textbf{1}} & \multirow{2}*{N/A} & \multirow{2}*{N/A}  \\
 & $\mmax = 50 \Msol$ & & & & & \\
Fixed Parameter (flat-in-log) & Equation~\ref{eq:flat-in-log} & \textbf{0} & \textbf{1} & \textbf{1} & N/A & N/A \\ \hline
Non-Evolving & A, B, C & \textbf{0} & N/A & N/A  & [0,1] & [0, 0.25] \\
Evolving\footnote{This model assumes the black holes have zero spin.} & A & [-25, 25] & N/A & N/A & \textbf{0} & \textbf{0} \\ \hline
\end{tabular}
\caption{Summary of models in Section \ref{sec:rates}, with prior ranges for the population parameters determining the
rate models. The fixed parameter models are drawn from~\citet{o2catalog}. The fixed parameters are in bold. Each of these distributions is uniform over the stated range; as
previously, we require $\alpha_a,\beta_a\ge 1$. Details of the mass models listed here are described in Table~\ref{tab:mass_priors}.}
\label{tab:rate_priors}
\end{table*}

In these calculations, we first maintain the assumption in~\cite{o2catalog} that the merger rate is uniform in comoving volume and source-frame time, as discussed in Section~\ref{sec:notation}.
We then relax this assumption and consider a merger rate that evolves in redshift according to Equation~\ref{eq:Rz}, fitting the mass distribution jointly with the rate density as a function of redshift.

\subsection{Non-Evolving Merger Rate}
\label{subsec:rate}
We first consider the case of a uniform in volume merger rate, and examine the effects of fitting the rate jointly with the distribution of masses and spins.
The first column in Figures~\ref{fig:dndm:hyperparameters_ab} and~\ref{fig:dndm:hyperparameters_c} shows the results of self-consistently determining the rate using
the models for the mass and spin distribution described in the previous two sections.

\begin{table}[tb]
\centering
\begin{tabular}{|c|c|c|c|} \hline
Model & A & B & C \\ \hline
$R_0$ (Gpc$^{-3}$ yr$^{-1}$) & \RATEimrpNoSpinVTQuadApm & \RATEimrpNoSpinVTQuadBpm & \RATEimrpNoSpinVTQuadCpm \\ \hline
\end{tabular}
\caption{This table lists the BBH merger rate intervals for each of the mass models tested. These rates assume no evolution in redshift, but otherwise marginalize over all other population parameters.
}
\label{tab:model_rates}
\end{table}

Table~\ref{tab:model_rates} contains the intervals on the distribution of $R_0$ for all three models.
For Models B and C we deduce a
merger rate between $R_0 = \RATEimrpNoSpinVTQuadBboundU$.
Adopting Model A for the mass distribution yields a slightly higher rate estimate, $R_0 = \RATEimrpNoSpinVTQuadAboundU$, as this model fixes $\mmin = 5 \Msol$, whereas Models B and C favor a higher minimum mass and therefore larger population-averaged sensitive volumes.
The rate estimates are consistent between all mass models considered, including the results presented for the fixed-parameter power-law model in~\cite{o2catalog}.
However, the fixed-parameter models in~\cite{o2catalog} are disfavored by our full fit to the mass distribution, particularly with respect to the maximum mass.
Our results favor maximum masses $\lesssim 45 \Msol$, rather than $50 \Msol$ as used in \cite{o2catalog}, and power-law slopes closer to $\alpha \sim 1$.
For this reason, although we infer a mass distribution slope that is similar to the flat-in-log population from~\cite{o2catalog}, we infer a rate that is closer to the rate inferred for the fixed-parameter power-law model\footnote{The flat-in-log population (Equation~\ref{eq:flat-in-log}) cannot be parameterized by the mass models A, B and C used in this work, because the mass ratio distribution takes a different form. However, it is very close to Model A with $\alpha = 1$.}.
While $\VT$ gets larger (implying a smaller rate estimate) as $\alpha$ is decreased, decreasing $\mmax$ has the opposite effect, and so the $\VT$ for the fixed-parameter power-law model is similar to the $\VT$s for our best-fit mass distributions, which favor smaller $\alpha$ and smaller $\mmax$.

We note that while our analysis differs from the rate calculations in~\cite{o2catalog} by the choice of prior on the rate parameter (log-uniform in this work compared to a Jeffreys prior $p(R_0) \propto R_0^{-0.5}$ in~\cite{o2catalog}), adopting a Jeffreys prior has a negligible effect on our rate posteriors.
For example, under a log-uniform prior, we recover a rate for Model A of \RATEimrpSpinVTQuadApmU, whereas under a Jeffreys prior this shifts by only $\sim 10\%$ to  $\RATEimrpSpinVTQuadAJeffriespmU$.

\subsection{Evolution of the Merger Rate with Redshift}

As discussed in the introduction, most formation channels predict some evolution of the merger rate with redshift, due to factors including the
star-formation rate, time-delay distribution, metallicity evolution, and globular cluster formation rate~\citep{2013ApJ...779...72D,2016Natur.534..512B,2016MNRAS.458.2634M,2018ApJ...866L...5R}.
Therefore, in this section, we allow the merger rate to evolve with redshift, and infer the redshift evolution jointly with the mass distribution.
For simplicity, we adopt the two-parameter Model A for the mass distribution and fix spins to zero for this analysis.
As discussed in Section~\ref{sec:mass_distr}, the additional mass and spin degrees of freedom have only a weak effect on the inferred merger rate.
We assume the redshift evolution model given by Equation~\ref{eq:Rz}.
Because massive binaries are detectable at higher redshifts, the observed redshift evolution correlates with the observed mass distribution of the population, and so we must fit them simultaneously.
However, as in~\cite{2018ApJ...863L..41F}, we assume that the underlying mass distribution does not vary with redshift.
We therefore fit the joint mass-redshift distribution according to the model:
\begin{equation}
\frac{d R}{dm_1dm_2} \left( z \right) = R_0 p(m_1, m_2 \mid \alpha, \mmax) (1 + z)^\lambda
\end{equation}
\changes{Note that this model assumes that the merger rate density increases or decreases monotonically with redshift over the sensitive range $z < 1$.
If the merger rate follows the star formation rate, we expect the rate to peak around $z \sim 2$, which is currently far beyond the horizon redshift for BBH detections.}

\begin{figure}[b!]
\includegraphics[width=\columnwidth]{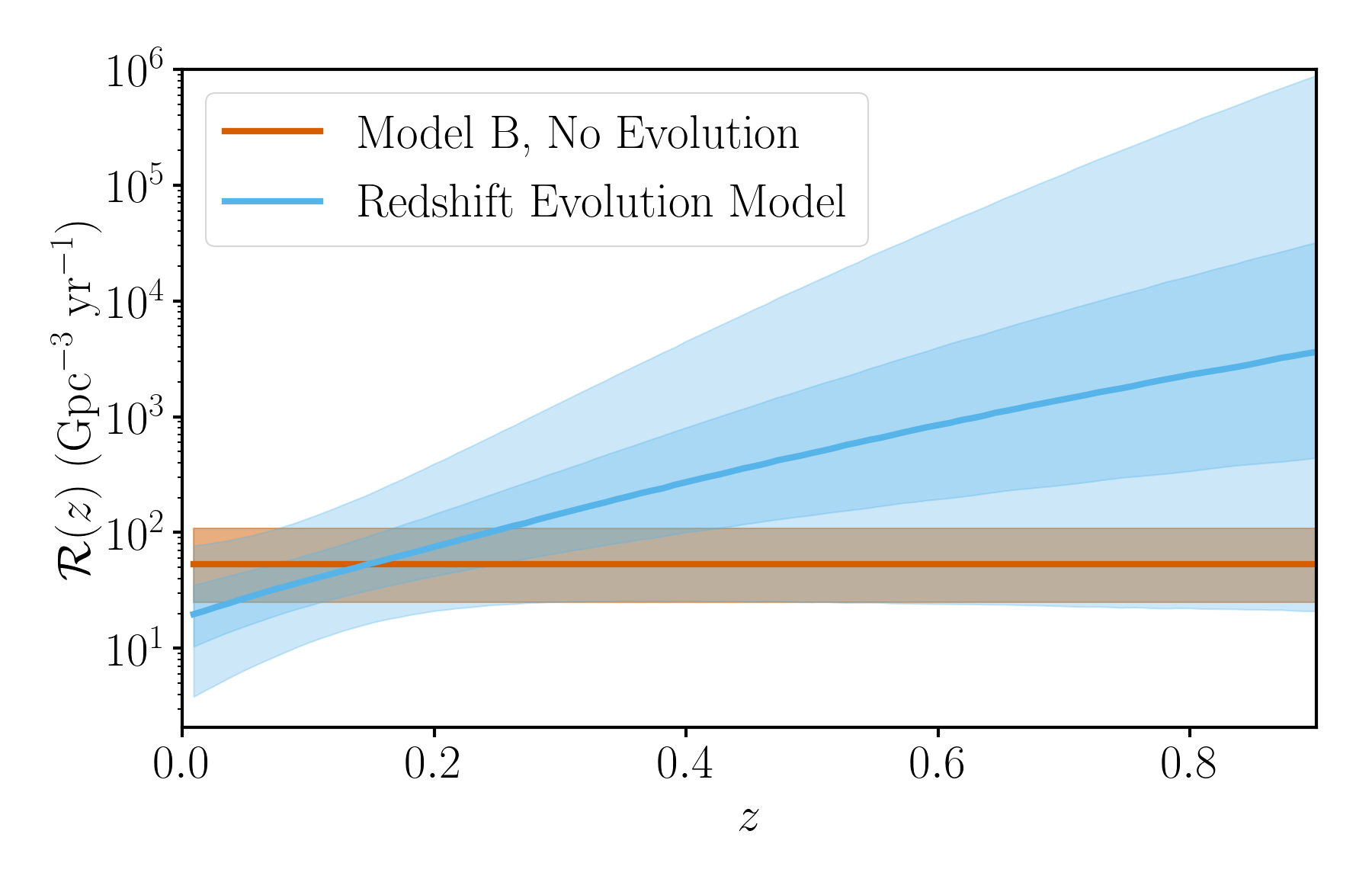}
\caption{\label{fig:dNdz} Constraints on evolution of the BBH merger rate density as a function of redshift.
Including the 10 BBHs from O1 and O2 in our analysis, we find a preference for a merger rate that increases with increasing redshift.
The solid blue line gives the posterior median merger rate density and dark and light bands give 50\% and 90\% credible intervals.
\changes{In orange, the solid line and shaded region shows the median and 90\% credible interval of the rate inferred for Model B as discussed in subsection~\ref{subsec:rate}, assuming a non-evolving merger rate.}}
\end{figure}

\begin{figure}[htbp]
\includegraphics[width=\columnwidth]{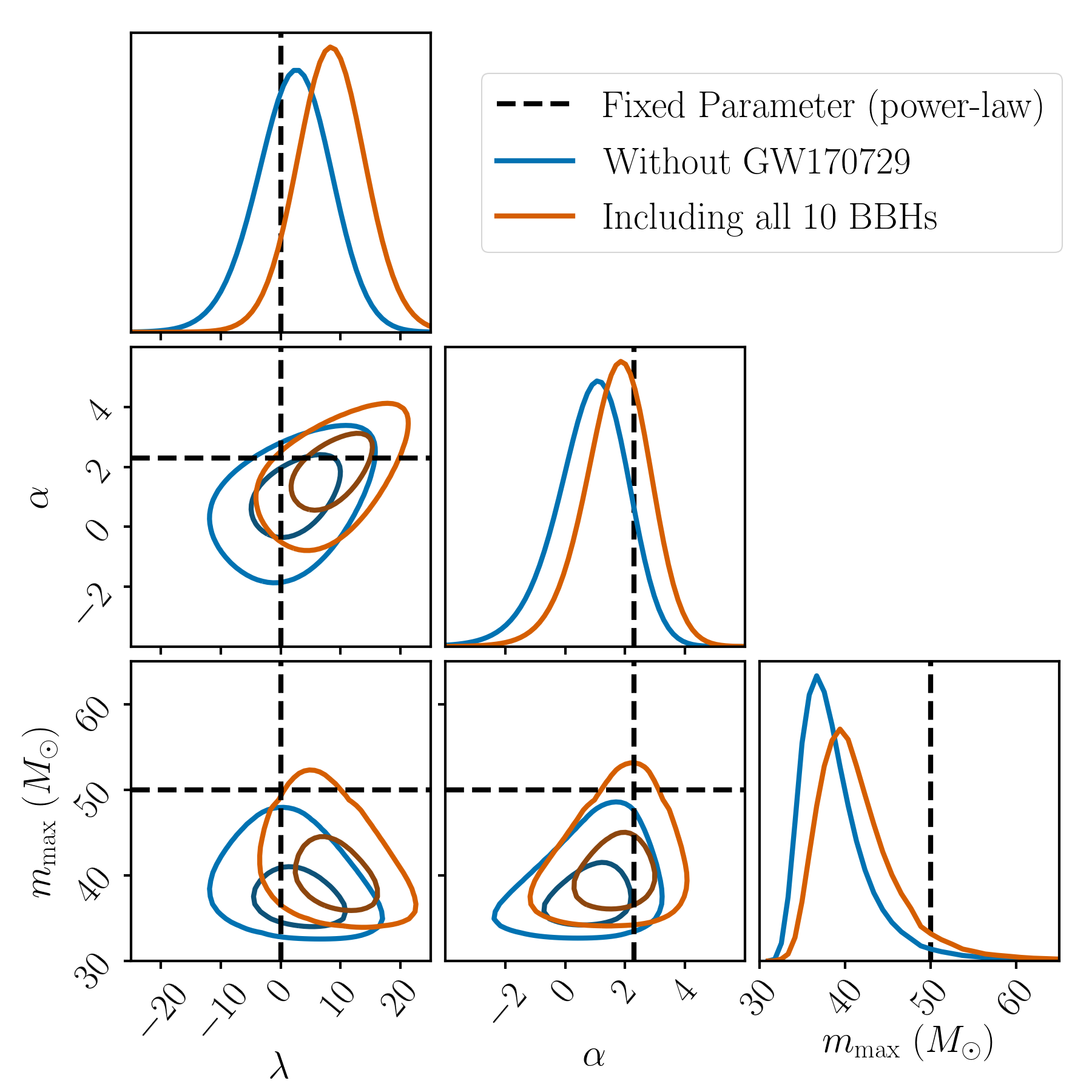}
\caption{\label{fig:p-lambda-alpha} The posterior PDF on the redshift evolution
parameter $\lambda$, mass power-law slope $\alpha$, and maximum mass $\mmax$,
marginalized over the local rate parameter $R_0$, and assuming a flat
prior on $\lambda$, $\alpha$, and $\mmax$ and a flat-in-log prior on
$R_0$. In order to analyze the stability of the model against
outliers, we repeat the analysis once with the sample of 10 BBHs (results shown
in blue), and once excluding the most distant and massive event in our sample,
GW170729 (results shown in red). The contours show 50\% and 90\% credible
intervals. The dashed black lines show the values of hyper-parameters assumed
for the fixed-parameter power-law model. We infer a redshift evolution that is
consistent with a flat in comoving volume and source-frame time merger rate
($\lambda = 0$) with a preference for $\lambda \geq 0$ at $\FRACLambdaGTZero{}$
credibility when considering all 10 events. This preference becomes less
significant with the exclusion of GW170729 from the analysis. The inferred
power-law slope and maximum mass is consistent with the values inferred in
Section~\ref{sec:mass_distr}. This analysis recovers a broader posterior on the
mass power-law slope because of the correlation with the redshift evolution
parameter, but the maximum mass remains well-constrained at $\lesssim 45
\Msol$.}
\end{figure}

\changes{Figure~\ref{fig:dNdz} shows the merger rate density as a function of redshift (blue band), compared to the rate inferred in subsection~\ref{subsec:rate} for the non-evolving Model B (orange band).}
The joint posterior PDF on $\lambda$, $\alpha$, and $\mmax$, marginalized over the local rate parameter $R_0$, is shown in Figure~\ref{fig:p-lambda-alpha}.
There is a strong correlation between the mass power-law slope and the redshift evolution parameter.
\changes{This is due to the fact that higher mass BBHs are detectable at higher redshifts, and so, for the same underlying mass distribution, an increasing rate evolution with redshift implies that a greater fraction of detected BBHs will be massive. This effect is hard to disentangle from a shallower mass distribution, which will also produce comparatively more massive BBH detections.
Note that the constraints on $\alpha$ and $\mmax$ in Section~\ref{sec:mass_distr} are correlated by the same effect.}
Compared to the constraints on $\alpha$ and $\mmax$ discussed in Section~\ref{sec:mass_distr}, which assume a constant-in-redshift merger rate density,
allowing for additional freedom in the redshift distribution of BBHs relaxes the constraints on the mass distribution parameters, especially the power-law slope $\alpha$ \changes{($\mmax$ is sufficiently well-measured that the correlation with $\lambda$ is not as noticeable)}.
Under the assumption of a constant merger rate density, Model A in Section~\ref{sec:mass_distr} finds $\alpha = \ALPHAMimrpNoSpinVTQuadApm$, $\mmax = \MMAXimrpNoSpinVTQuadApmU$, whereas allowing for redshift evolution yields $\alpha = \alphaintervalIMRP$, $\mmax = \mmaxintervalIMRP$ \Msol~when analyzing the sample of 10 BBHs from O1 and O2.
As in Section~\ref{sec:mass_distr}, we carry out a leave-one-out analysis, excluding the most massive and distant BBH, GW170729, from the sample (red curves in Figure~\ref{fig:p-lambda-alpha}).
Without GW170729, the marginalized mass-distribution posteriors become $\alpha = \alphaintervalExcIMRP$, $\mmax = \mmaxintervalExcIMRP$ \Msol.

Marginalizing over the two mass distribution parameters and the redshift-evolution parameter, the merger rate density is consistent with being constant in redshift ($\lambda = 0$), and in particular, it is consistent with the rate estimates recovered under the different mass distribution models in subsection~\ref{subsec:rate} above. 
However, we find a preference for a merger rate density that increases at higher redshift ($\lambda \geq 0$) with probability~$\FRACLambdaGTZero$.
\changes{This implies that models that predict a constant, or slightly decreasing merger rate with redshift, such as certain models of primordial black holes~\citep{2016PhRvL.117t1102M}, are disfavored.}
This preference for a merger rate that increases with increasing redshift becomes less significant when GW170729 is excluded from the analysis, because this event likely merged at redshift $z \gtrsim 0.5$, close to the O1-O2 detection horizon.
Although GW170729 shifts the posterior towards larger values of $\lambda$, implying a stronger redshift evolution of the merger rate, the posterior remains well within the uncertainties inferred from the remaining nine BBHs.
When including GW170729 in the analysis, we find $\lambda = \lambdaintervalIMRP$ at 90\% credibility, compared to $\lambda = \lambdaintervalExcIMRP$ when excluding GW170729 from the analysis.
With only 10 BBH detections so far, the wide range of possible values for $\lambda$ is consistent with most astrophysical formation channels.
The precision of this measurement will improve as we accumulate more detections in future observing runs and may enable us to discriminate between different formation rate histories or time-delay distributions~\citep{2012CQGra..29l4013S,2014JPhCS.484a2008V,2018ApJ...863L..41F}.

\section{The Spin Distribution} \label{sec:spin_distr}
The GW signal depends on spins in a complicated way, but at leading order, and in the regime we are interested in here, some combinations of parameters have more impact on our inferences than others, and thus are measurable.  One such parameter is \chieff.
 For binaries which are near equal mass, we can see from Equation~\ref{eq:chieff} that only when black hole spins are high and aligned with the orbital angular momentum \chieff{} will be measurably greater than zero. Figure \CatalogFigureQChieffAll{} in~\cite{o2catalog} illustrates the inferred \chieff{} spin distributions for all of the BBHs identified in our GW surveys in O1 and O2. Only GW170729 and GW151226 show significant evidence for positive $\chieff$; the rest of the posteriors cluster around $\chieff=0$.

Despite these degeneracies, several tests have been proposed to use spins to constrain BBH formation channels~\citep{2017CQGra..34cLT01V,2017Natur.548..426F,2018ApJ...854L...9F,2017MNRAS.471.2801S,2017PhRvD..96b3012T,2017PhRvD..95l4046G,2018arXiv180506442W,2018PhRvD..98h4036G}. Drawing upon these methods, we now seek to estimate the black hole spin magnitude and misalignment distributions, under different assumptions regarding isotropy or alignment.

\subsection{Spin Magnitude and Tilt Distributions}\label{sec:spin_param_infer}

We examine here the individual spin magnitudes and tilt distributions.
Throughout this section, when referring to the parametric models, we also allow the merger rate and population parameters describing the most general mass model to vary (Model C, see
Table~\ref{tab:mass_priors}). Changing the parameterization of the mass model does not significantly change our inferences about the spin distribution. However, to account for degeneracies between mass and spin that grow increasingly significant for longer, low-mass signals \citep{2013PhRvD..87b4035B}, we must consistently model the mass and spin distributions together. See Table~\ref{tab:spin_priors} for a summary of the models and priors used in this Section.

\begin{table*}[htbp]
\centering
\begin{tabular}{|c|c|ccc|cc|} \hline
 & Mass Model & \multicolumn{5}{c|}{Spin Parameters} \\
Model & & $\mathbb{E}[a]$ & $\mathrm{Var}[a]$ &$\alpha_a, \beta_a$  & $\zeta$ & $\sigma_i$ \\ \hline
Gaussian (G) & C & [0, 1] & [0, 0.25] & $\ge  1$ & \textbf{1} & [0, 4] \\
Mixture (M) & C & [0, 1] & [0, 0.25] & $\ge 1$ & [0, 1] & [0, 4] \\ \hline
\end{tabular}
\caption{Summary of spin distribution models examined in Section \ref{sec:spin_param_infer}, with prior ranges for the
population parameters determining the spin models. The fixed parameters are in bold. Each of these distributions is
uniform over the stated range, with boundary conditions such that the inferred parameters $\alpha_a,\beta_a$ must be
$\ge1$.  Details of the mass model listed here is described in Table~\ref{tab:mass_priors}.}
\label{tab:spin_priors}
\end{table*}

\begin{figure}[htbp]
  \centering
  \includegraphics[width=\columnwidth]{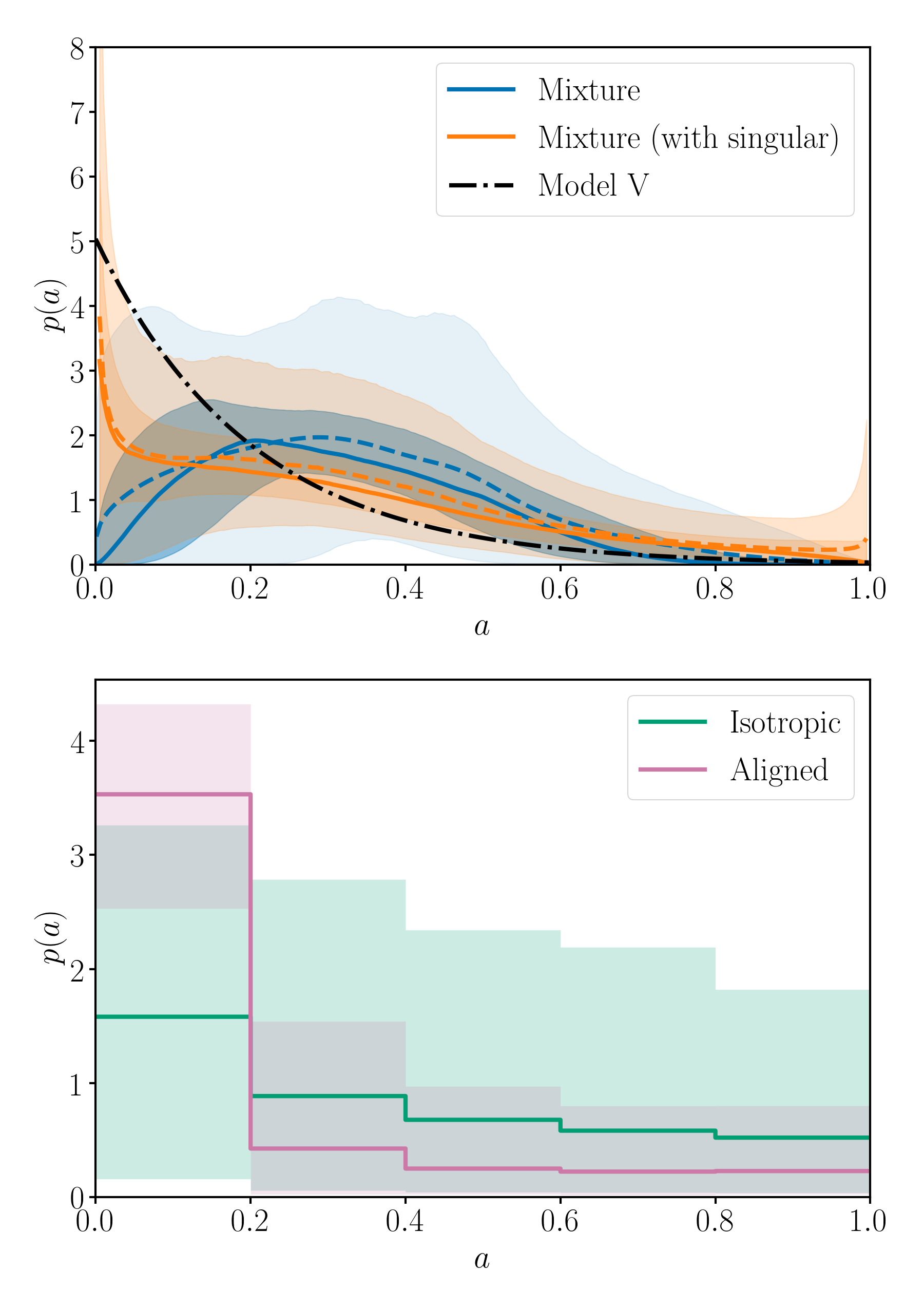}
  \caption{
Inferred distribution of spin magnitude for a parametric (top) and non-parametric binned model (bottom). \changes{Both component magnitudes are included in these distributions.}
The solid lines show the median and the dashed line shows the PPD.
The shaded regions denote the 50\% and 90\% symmetric intervals.
\changes{In the upper panel, the parametric model is presented with both singular (orange) and non-singular (blue) model configurations. For comparison purposes, the V (very low spin magnitude) model is plotted in a dash-dotted black line.}
In the bottom panel, the distribution of spin magnitude is inferred over five bins, assuming either perfectly aligned (pink) or isotropic (green) population. The solid lines denote the median, and the shaded regions denote the central 90\% posterior credible bounds. In both cases, the magnitude is consistent within the uncertainties with the parametric \changes{(singular and non-singular)} results. \changes{The number of bins in the model were chosen to balance resolution with the amount of information in the data; analyses with more bins do not indicate any additional features in the spin distributions.}
 \label{fig:parametric_spin}
}
\end{figure}

The inferred distributions of spin magnitude are shown in Figure~\ref{fig:parametric_spin}. The top panel shows the PPD
as well as the median and associated uncertainties on the spin magnitude inferred from the parametric Mixture model defined in Section~\ref{sec:spin_models} and using prior distributions shown in~\ref{tab:spin_priors}. It
marginalizes over all other parameters, including the mass parameters in Model C, and the spin mixture
fraction. We observe that spin distributions which decline with increasing magnitude are preferred. In
terms of our Beta function parameterization --- $\mathbb{E}[a]$ and $\mathrm{Var}[a]$, defined in Equation~\ref{eq:beta_distr_params} --- these have mean spin $\mathbb{E}[a]<1/2$ or equivalently have
$\beta_a>\alpha_a$, at posterior probability $\FRACBETAGTALPHAimrpNoSpinVTQuadB$. We
find that 90\% of black hole spins in BBHs are less than $a \le \APPDimrpNoSpinVTQuadCboundninety$ from the PPD, and 50\% of black hole spins are less than $a \le \APPDimrpNoSpinVTQuadCboundfifty$.
We find similar conclusions if both black hole spins are drawn from different distributions (i.e., 90\% of black hole spins on the more
massive black hole are less than $\APRIMARYimrpNoSpinVTQuadBboundninety$).
\changes{When avoiding singular values in the spin magnitude model distribution, the distribution exhibits a peak structure, i.e., $p(a=0)=p(a=1)=0$. If allowed to capture the full range of model parameters including ``singular'' configurations, the support for small values of $a$ is more pronounced. However, this scenario forces a small --- and otherwise observationally unsupported --- uptick of probability mass at $a$ near maximal spins. In both cases, the recovered spin distribution in the top panel of Figure~\ref{fig:parametric_spin} is driven by favoring declining spin distributions, which are more compatible with the observed population. This conclusion is also consistent with the preference in Appendix~\ref{sec:alt_spins} for the very low spin magnitude model.}

We also compute the posterior distribution for the magnitude of black hole spins from $\chieff$ measurements by modeling the distribution of black hole spin magnitudes non-parametrically with five bins, assuming either an isotropic or perfectly aligned population following~\cite{2018ApJ...854L...9F}. We show in the bottom panel of Figure~\ref{fig:parametric_spin} that under the perfectly aligned scenario there is preference for small black hole spin, inferring 90\% of black holes to have spin magnitudes below $0.6^{+0.24}_{-0.28}$. However, when spins are assumed to be isotropic the distribution is relatively flat, with 90\% of black hole spin magnitudes below $0.8^{+0.15}_{-0.24}$.
Thus, the non-parametric analysis produces conclusions consistent with our parametric analyses described above. These conclusions are also reinforced by computing the Bayes factor for a set of fixed parameter models of spin magnitude and orientation in Appendix~\ref{sec:alt_spins}. \changes{There we find that the very low spin magnitude model is preferred by a log Bayes factor of 1 or greater in most mass and spin orientation configurations tested (see Figure~\ref{fig:chieff_toymodels} and Table~\ref{tab:odds_ratio} for details).}

\begin{figure}[b!]
  \centering
  \includegraphics[width=\columnwidth]{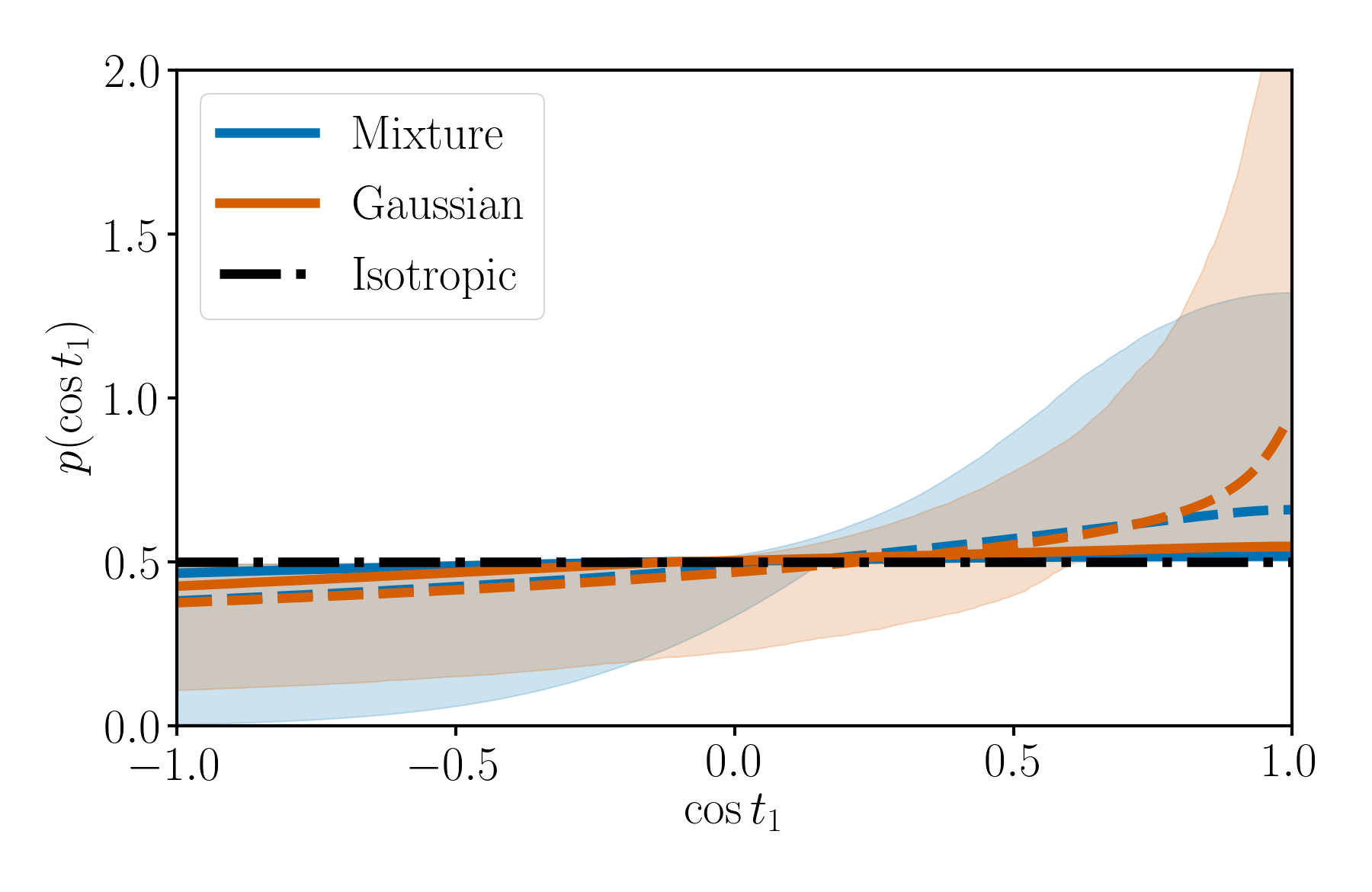}
  \caption{
Inferred distribution of cosine spin tilt for the more massive black hole for two choices of prior (see Section~\ref{sec:spin_models}).
The dash-dotted line denotes a completely isotropic distribution (see Appendix~\ref{sec:alt_spins}).
The solid lines show the median.
The shaded regions \changes{denote the 90\% symmetric intervals} and the dashed line denotes the PPD.
  }
  \label{fig:parametric_spin_tilts}
\end{figure}

Figure~\ref{fig:parametric_spin_tilts} shows the inferred distribution of the primary spin tilt for the more massive black hole.
These results were obtained without including the effects of component spins on the detection probability: see Appendix~\ref{sec:systematics} for further discussion.
In the Gaussian model ($\zeta=1$), all black hole spin orientations are drawn from spin tilt distributions which are preferentially aligned and parameterized with $\sigma_i$. In that model, the $\sigma_i$ distributions do not differ appreciably from the their flat priors. As such, the inferred spin tilt distribution are influenced by large $\sigma_i$ and the result resembles an isotropic distribution. The Mixture distribution does not return a decisive measurement of the mixture fraction, obtaining $\zeta = \XIimrpNoSpinVTQuadCpm$. Since the Gaussian model is a subset of the Mixture model, we can compare preferences via the Savage-Dickey ratio. The log Bayes factor for $\zeta=1$ is $\ln \textrm{BF} = \SDDRGaussianvsMixtureIMRPhenomP$, indicating virtually no preference for any particular orientation distribution.
While we allow both black holes to have different typical misalignment, the inference on the second tilt is less informative than the primary.
The inferred distribution for $\cos t_2$ is similar to $\cos t_1$, but also closer to the prior.

The mixture fraction distribution is also modelled with the fixed parameter models in Appendix~\ref{sec:alt_spins}.
The fixed magnitude distributions considered in Appendix~\ref{sec:alt_spins} prefer isotropic to aligned, but the preference is weakened for distributions concentrated at lower spins. A few exceptions occur for the very low spin fixed mass ratio models, with aligned models being slightly preferred.

In general, we are not able to place strong constraints on the distribution of spin orientations. We elaborate in Appendix~\ref{sec:three_bin} on how our black hole spin measurements are not yet informative enough to discern between isotropic and aligned orientation distribution via \chieff.

\subsection{Interpretation of Spin Distributions} \label{sec:spin_systematics}

The spins of black holes are affected by a number of uncertain processes which occur during the evolution of the binary. As a consequence, the magnitude distribution is difficult to predict from theoretical models of these processes alone.
While the spin of a black hole should be related to the rotation of the core of its progenitor star, the amount of spin which is lost during the final stages of the progenitor's life is still highly uncertain. \changes{While we have modeled the spins independently, correlations from binary evolution and stellar collapse are possible~\citep{2017arXiv170607053B,2018PhRvD..98h4036G,2018A&A...616A..28Q,2019MNRAS.483.3288P,2019MNRAS.482.2991A}.}
The core rotational angular momentum before the supernova can be changed from the birth spin of the progenitor by several processes~\citep{2012ARA&A..50..107L,2013ApJ...764..166D,2016MNRAS.458.3075A}.
Examples include mass transfer~\citep{1981ARA&A..19..277S,1981A&A...102...17P}, or tidal interactions~\citep{2005A&A...435.1013P}, as well as internal mixing of the stellar layers across the core-envelope boundary via magnetic torquing~\citep{2002A&A...381..923S,2003A&A...411..543M} and gravity waves~\citep{2005A&A...440..981T,2008A&A...482..597T,2015ApJ...810..101F}.
In principle, an off-center supernova explosion could also impart significant angular momentum and tilt the spin of the remnant into the collapsing star~\citep{2011ApJ...742...81F}.

Once a black hole is formed, however, changing the spin magnitude is more difficult due to limitations on mass accretion rates affecting how much a black hole can be spun up~\citep{1974ApJ...191..507T,2010Natur.468...77V,2012ApJ...747..111W,2019ApJ...870L..18Q}.
Once the binary black hole system is formed, the spin magnitudes do not change appreciably over the inspiral~\citep{2014PhRvD..90b4014F}.

No BBH detected to date has a component with confidently high and aligned spin magnitude.
The results in the previous section imply that black holes tend to be born with spin less than our PPD bound of $\APPDimrpNoSpinVTQuadCboundninety$, or that another process (e.g., supernova kicks or dynamical processes involved in binary formation) induces tilts such that \chieff{} is small.

The possibility of a spin magnitude distribution that peaks at low spins incurs a degeneracy between models that is not easily overcome: when the spin magnitudes are small enough models produce features which cannot be distinguished within observational uncertainties.

\section{Discussion and Conclusions} \label{sec:conclude}
We have presented a variety of estimates for  the mass, spin, and redshift distributions of BBH, based on the observed
sample of 10 BBH \changes{and generic phenomenological population models motivated by electromagnetic observations and theory}.  Some model independent features are evident
from the observations.   Notably, no binary black holes more massive than GW170729 have been observed to date,
but several binaries have component masses likely between $20-40 M_\odot$.   No
highly asymmetric (small $q$) system has been observed.  Only two systems (GW151226 and GW170729) produce a \chieff{}
distribution which is confidently different from zero; conversely,  most BH binaries are consistent with \chieff{}
near zero.  These features drive our inferences about the mass and spin distribution.

Despite exploring a wide range of mass and spin distributions, we find the  BBH merger rate density is $R = \RATEimrpNoSpinVTQuadApm\,\mbox{Gpc}^{-3}\,\mbox{yr}^{-1}$ for Model A and is within $R
= \RATEimrpNoSpinVTQuadBpm\,\mbox{Gpc}^{-3}\,\mbox{yr}^{-1}$ for Models B and C.   This result is consistent with the fixed model assumptions reported in the  combined O1 and O2 observational periods~\citep{o2catalog}.
We find a significant reduction in the merger rate for binary black holes with primary masses larger than $\sim
45 M_\odot$.
We do not have enough sensitivity to binaries with a black hole mass less than 5 \Msol{} to be able to place meaningful constraints on the minimum mass of black holes.
We find mild evidence that the mass distribution of coalescing black holes may not be a pure power law, instead being
slightly better fit by a model including a broad gaussian distribution at high mass.
We find the best-fitting models preferentially produce comparable-mass binaries (i.e., $\beta_q>0$ is preferred).

The mass models in this work supersede results from an older model from O1 which inferred only the power law index~\citep{2016PhRvX...6d1015A,2017PhRvL.118v1101A}. That model found systematically larger values of $\alpha$ than its nearest counterpart in this work, Model A, because the older model used a fixed value for the minimum and maximum mass of 5 and 100 \Msol, respectively. This extreme \mmax{} is highly disfavored by our current results, and so the older model is also disfavored.
Moreover, volumetric sensitivity grows as a strong function of mass. The lack of detections near the older \mmax{} drives a preference for a much smaller maximum BH mass in the new models~\citep{2017ApJ...851L..25F}.  A reduced maximum mass is associated with a shallower power-law fit.

Inferring the redshift distribution is difficult with only a small sample of local events~\citep{2018ApJ...863L..41F}. We have constrained models with extreme variation over redshift, favoring instead those which are uniform in the comoving volume or have increasing merger rates with higher redshift. Many potential formation channels in the literature~\citep{2014ApJ...789..120B,2016ApJ...824L...8R,2016ApJ...831..187A,2016MNRAS.458.2634M,2016MNRAS.461.2722I,2017MNRAS.472.2422M,2017ApJ...835..165B,2018MNRAS.481.1908K} produce event rates which are compatible with those from the previous observing runs~\citep{o2catalog} and this work. It is, of course, plausible that several are contributing simultaneously, and no combination of mass, rate, or redshift dependence explored here rules out any of the channels proposed to date. The next generation of interferometers will allow for an exquisite probe into this dependence at large redshifts~\citep{2012CQGra..29l4013S,2014JPhCS.484a2008V,2018arXiv180800901V}.

We have modeled the spin distribution in several ways, forming inferences on the spin magnitude and tilt
distributions.
In all of our analysis,
 the evidence disfavors
distributions with large spin components aligned (or nearly aligned) with the orbital angular momentum; specifically, we find that 90\% of the spin magnitude PPD is smaller than $\APPDimrpNoSpinVTQuadCboundninety$. We cannot
significantly constrain the degree of spin-orbit misalignment in the population.
However, regardless of the mass or assumed spin tilt distribution, there is a preference (demonstrated in
Figure~\ref{fig:parametric_spin} and Appendix~\ref{sec:alt_spins}) for distributions which emphasize lower spin magnitudes.  Our inferences suggest  90\% of coalescing black hole binaries are formed with  $\chieff < \CHIEFFimrpNoSpinVTQuadAboundninety$.
Low spins argue against so-called second generation mergers, where at least one of the components of the binary is a black hole formed from a previous merger~\citep{2007PhRvL..98i1101G,2007PhRvD..76f4034B} and possesses spins near 0.7~\citep{2017ApJ...840L..24F}.

GW170729 is notable in several ways: it is the most massive, largest \chieff, and most distant redshift event detected so far. To quantify the impact it has on our results, where possible, we have presented model posteriors which reflect its presence in or exclusion from the event set. Many of our predictions are robust despite its extreme values --- by far, and not unexpectedly, its influence is most significant in the distribution of \mmax. It also impacts our conclusions about redshift evolution, where its absence flattens the inferred redshift evolution.

Recent modelling using only the first six released events~\citep{2018arXiv180506442W,2019MNRAS.tmp..231R} have come to similar conclusions about low spin magnitudes and the shape of the power law distribution.  The presence of an apparent upper limit to the merging BBH mass distribution was also observed after the first six released events \citep{2017ApJ...851L..25F}.  An enhancement which will benefit these types of analyses in the future is a simultaneous fit of the astrophysical model and its parameters and noise background model~\citep{2019MNRAS.tmp..230G}.

Several studies have noted that population
features~\citep{2010CQGra..27k4007M,2015ApJ...810...58S,2017ApJ...851L..25F,2017MNRAS.471.2801S,2017ApJ...846...82Z,2017PhRvD..95j3010K,2017Natur.548..426F,2017PhRvD..96b3012T,2017ApJ...840L..24F,2017PhRvD..95l4046G,2018ApJ...856..173T,2018ApJ...854L...9F,2018MNRAS.477.4685B,2018arXiv180506442W,2018PhRvD..98h4036G}
and complementary physics~\citep{2016PhRvL.116m1102A,2017ApJ...846...82Z,2017MNRAS.471.2801S,2018Natur.562..545C} will
be increasingly accessible as observations accumulate. \changes{Additional events will also permit the enhancement of the simple phenomenological models used in this work and comparison with modeling of astrophysical processes.}
Given the event merger rates estimated here and anticipated
improvements in sensitivity~\citep{2018LRR....21....3A}, hundreds of BBHs and tens of binary neutron stars are expected
to be collected in the operational lifetime of second generation GW instruments.
 Thus, the inventory of BBH in the coming
years will enable inquiries into astrophysics which were previously unobtainable.

\acknowledgements
The authors gratefully acknowledge the support of the United States
National Science Foundation (NSF) for the construction and operation of the
LIGO Laboratory and Advanced LIGO as well as the Science and Technology Facilities Council (STFC) of the
United Kingdom, the Max-Planck-Society (MPS), and the State of
Niedersachsen/Germany for support of the construction of Advanced LIGO 
and construction and operation of the GEO600 detector. 
Additional support for Advanced LIGO was provided by the Australian Research Council.
The authors gratefully acknowledge the Italian Istituto Nazionale di Fisica Nucleare (INFN),  
the French Centre National de la Recherche Scientifique (CNRS) and
the Foundation for Fundamental Research on Matter supported by the Netherlands Organisation for Scientific Research, 
for the construction and operation of the Virgo detector
and the creation and support  of the EGO consortium. 
The authors also gratefully acknowledge research support from these agencies as well as by 
the Council of Scientific and Industrial Research of India, 
the Department of Science and Technology, India,
the Science \& Engineering Research Board (SERB), India,
the Ministry of Human Resource Development, India,
the Spanish  Agencia Estatal de Investigaci\'on,
the Vicepresid\`encia i Conselleria d'Innovaci\'o, Recerca i Turisme and the Conselleria d'Educaci\'o i Universitat del Govern de les Illes Balears,
the Conselleria d'Educaci\'o, Investigaci\'o, Cultura i Esport de la Generalitat Valenciana,
the National Science Centre of Poland,
the Swiss National Science Foundation (SNSF),
the Russian Foundation for Basic Research, 
the Russian Science Foundation,
the European Commission,
the European Regional Development Funds (ERDF),
the Royal Society, 
the Scottish Funding Council, 
the Scottish Universities Physics Alliance, 
the Hungarian Scientific Research Fund (OTKA),
the Lyon Institute of Origins (LIO),
the Paris \^{I}le-de-France Region, 
the National Research, Development and Innovation Office Hungary (NKFIH), 
the National Research Foundation of Korea,
Industry Canada and the Province of Ontario through the Ministry of Economic Development and Innovation, 
the Natural Science and Engineering Research Council Canada,
the Canadian Institute for Advanced Research,
the Brazilian Ministry of Science, Technology, Innovations, and Communications,
the International Center for Theoretical Physics South American Institute for Fundamental Research (ICTP-SAIFR), 
the Research Grants Council of Hong Kong,
the National Natural Science Foundation of China (NSFC),
the Leverhulme Trust, 
the Research Corporation, 
the Ministry of Science and Technology (MOST), Taiwan
and
the Kavli Foundation.
The authors gratefully acknowledge the support of the NSF, STFC, MPS, INFN, CNRS and the
State of Niedersachsen/Germany for provision of computational resources.

\bibliography{references}

\begin{thebibliography}{}
\expandafter\ifx\csname natexlab\endcsname\relax\def\natexlab#1{#1}\fi
\providecommand{\url}[1]{\href{#1}{#1}}
\providecommand{\dodoi}[1]{doi:~\href{http://doi.org/#1}{\nolinkurl{#1}}}
\providecommand{\doeprint}[1]{\href{http://ascl.net/#1}{\nolinkurl{http://ascl.net/#1}}}
\providecommand{\doarXiv}[1]{\href{https://arxiv.org/abs/#1}{\nolinkurl{https://arxiv.org/abs/#1}}}

\bibitem[{ase(2018)}]{asensio_ramos_arregui_2018}
 2018, Bayesian Astrophysics, Canary Islands Winter School of Astrophysics
  (Cambridge University Press), \dodoi{10.1017/9781316182406}.
\newblock \url{https://doi.org/10.1017/9781316182406}

\bibitem[{{Abbott} {et~al.}(2016{\natexlab{a}}){Abbott}, {Abbott}, {Abbott},
  {Abernathy}, {Acernese}, {Ackley}, {Adams}, {Adams}, {Addesso}, {Adhikari},
  {Adya}, {Affeldt}, {Agathos}, {Agatsuma}, {Aggarwal}, {Aguiar}, {Aiello},
  {Ain}, {Ajith}, {Allen}, {Allocca}, {Altin}, {Anderson}, {Anderson}, {Arai},
  {Araya}, {Arceneaux}, {Areeda}, {Arnaud}, {Arun}, {Ascenzi}, {Ashton}, {Ast},
  {Aston}, {Astone}, {Aufmuth}, {Aulbert}, {Babak}, {Bacon}, {Bader}, {Baker},
  {Baldaccini}, {Ballardin}, {Ballmer}, {Barayoga}, {Barclay}, {Barish},
  {Barker}, {Barone}, {Barr}, {Barsotti}, {Barsuglia}, {Barta}, {Bartlett},
  {Bartos}, {Bassiri}, {Basti}, {Batch}, {Baune}, {Bavigadda}, {Bazzan},
  {Behnke}, {Bejger}, {Bell}, {Bell}, {Berger}, {Bergman}, {Bergmann}, {Berry},
  {Bersanetti}, {Bertolini}, {Betzwieser}, {Bhagwat}, {Bhandare}, {Bilenko},
  {Billingsley}, {Birch}, {Birney}, {Biscans}, {Bisht}, {Bitossi}, {Biwer},
  {Bizouard}, {Blackburn}, {Blair}, {Blair}, {Blair}, {Bloemen}, {Bock},
  {Bodiya}, {Boer}, {Bogaert}, {Bogan}, {Bohe}, {Bojtos}, {Bond}, {Bondu},
  {Bonnand}, {Boom}, {Bork}, {Boschi}, {Bose}, {Bouffanais}, {Bozzi},
  {Bradaschia}, {Brady}, {Braginsky}, {Branchesi}, {Brau}, {Briant}, {Brillet},
  {Brinkmann}, {Brisson}, {Brockill}, {Brooks}, {Brown}, {Brown}, {Brown},
  {Buchanan}, {Buikema}, {Bulik}, {Bulten}, {Buonanno}, {Buskulic}, {Buy},
  {Byer}, {Cadonati}, {Cagnoli}, {Cahillane}, {Calder{\'o}n Bustillo},
  {Callister}, {Calloni}, {Camp}, {Cannon}, {Cao}, {Capano}, {Capocasa},
  {Carbognani}, {Caride}, {Casanueva Diaz}, {Casentini}, {Caudill},
  {Cavagli{\`a}}, {Cavalier}, {Cavalieri}, {Cella}, {Cepeda}, {Cerboni
  Baiardi}, {Cerretani}, {Cesarini}, {Chakraborty}, {Chalermsongsak},
  {Chamberlin}, {Chan}, {Chao}, {Charlton}, {Chassande-Mottin}, {Chen}, {Chen},
  {Cheng}, {Chincarini}, {Chiummo}, {Cho}, {Cho}, {Chow}, {Christensen}, {Chu},
  {Chua}, {Chung}, {Ciani}, {Clara}, {Clark}, {Cleva}, {Coccia}, {Cohadon},
  {Colla}, {Collette}, {Cominsky}, {Constancio}, {Conte}, {Conti}, {Cook},
  {Corbitt}, {Cornish}, {Corsi}, {Cortese}, {Costa}, {Coughlin}, {Coughlin},
  {Coulon}, {Countryman}, {Couvares}, {Cowan}, {Coward}, {Cowart}, {Coyne},
  {Coyne}, {Craig}, {Creighton}, {Cripe}, {Crowder}, {Cumming}, {Cunningham},
  {Cuoco}, {Dal Canton}, {Danilishin}, {D'Antonio}, {Danzmann}, {Darman},
  {Dattilo}, {Dave}, {Daveloza}, {Davier}, {Davies}, {Daw}, {Day}, {DeBra},
  {Debreczeni}, {Degallaix}, {De Laurentis}, {Del{\'e}glise}, {Del Pozzo},
  {Denker}, {Dent}, {Dereli}, {Dergachev}, {DeRosa}, {De Rosa}, {DeSalvo},
  {Dhurandhar}, {D{\'\i}az}, {Di Fiore}, {Di Giovanni}, {Di Lieto}, {Di Pace},
  {Di Palma}, {Di Virgilio}, {Dojcinoski}, {Dolique}, {Donovan}, {Dooley},
  {Doravari}, {Douglas}, {Downes}, {Drago}, {Drever}, {Driggers}, {Du},
  {Ducrot}, {Dwyer}, {Edo}, {Edwards}, {Effler}, {Eggenstein}, {Ehrens},
  {Eichholz}, {Eikenberry}, {Engels}, {Essick}, {Etzel}, {Evans}, {Evans},
  {Everett}, {Factourovich}, {Fafone}, {Fair}, {Fairhurst}, {Fan}, {Fang},
  {Farinon}, {Farr}, {Farr}, {Favata}, {Fays}, {Fehrmann}, {Fejer}, {Ferrante},
  {Ferreira}, {Ferrini}, {Fidecaro}, {Fiori}, {Fiorucci}, {Fisher}, {Flaminio},
  {Fletcher}, {Fournier}, {Franco}, {Frasca}, {Frasconi}, {Frei}, {Freise},
  {Frey}, {Frey}, {Fricke}, {Fritschel}, {Frolov}, {Fulda}, {Fyffe}, {Gabbard},
  {Gair}, {Gammaitoni}, {Gaonkar}, {Garufi}, {Gatto}, {Gaur}, {Gehrels},
  {Gemme}, {Gendre}, {Genin}, {Gennai}, {George}, {Gergely}, {Germain},
  {Ghosh}, {Ghosh}, {Giaime}, {Giardina}, {Giazotto}, {Gill}, {Glaefke},
  {Goetz}, {Goetz}, {Gondan}, {Gonz{\'a}lez}, {Gonzalez Castro}, {Gopakumar},
  {Gordon}, {Gorodetsky}, {Gossan}, {Gosselin}, {Gouaty}, {Graef}, {Graff},
  {Granata}, {Grant}, {Gras}, {Gray}, {Greco}, {Green}, {Groot}, {Grote},
  {Grunewald}, {Guidi}, {Guo}, {Gupta}, {Gupta}, {Gushwa}, {Gustafson},
  {Gustafson}, {Hacker}, {Hall}, {Hall}, {Hammond}, {Haney}, {Hanke}, {Hanks},
  {Hanna}, {Hannam}, {Hanson}, {Hardwick}, {Haris}, {Harms}, {Harry}, {Harry},
  {Hart}, {Hartman}, {Haster}, {Haughian}, {Heidmann}, {Heintze}, {Heitmann},
  {Hello}, {Hemming}, {Hendry}, {Heng}, {Hennig}, {Heptonstall}, {Heurs},
  {Hild}, {Hoak}, {Hodge}, {Hofman}, {Hollitt}, {Holt}, {Holz}, {Hopkins},
  {Hosken}, {Hough}, {Houston}, {Howell}, {Hu}, {Huang}, {Huerta}, {Huet},
  {Hughey}, {Husa}, {Huttner}, {Huynh-Dinh}, {Idrisy}, {Indik}, {Ingram},
  {Inta}, {Isa}, {Isac}, {Isi}, {Islas}, {Isogai}, {Iyer}, {Izumi}, {Jacqmin},
  {Jang}, {Jani}, {Jaranowski}, {Jawahar}, {Jim{\'e}nez-Forteza}, {Johnson},
  {Jones}, {Jones}, {Jonker}, {Ju}, {Kalaghatgi}, {Kalogera}, {Kandhasamy},
  {Kang}, {Kanner}, {Karki}, {Kasprzack}, {Katsavounidis}, {Katzman}, {Kaufer},
  {Kaur}, {Kawabe}, {Kawazoe}, {K{\'e}f{\'e}lian}, {Kehl}, {Keitel}, {Kelley},
  {Kells}, {Kennedy}, {Key}, {Khalaidovski}, {Khalili}, {Khan}, {Khan}, {Khan},
  {Khazanov}, {Kijbunchoo}, {Kim}, {Kim}, {Kim}, {Kim}, {Kim}, {Kim}, {King},
  {King}, {Kinzel}, {Kissel}, {Kleybolte}, {Klimenko}, {Koehlenbeck},
  {Kokeyama}, {Koley}, {Kondrashov}, {Kontos}, {Korobko}, {Korth}, {Kowalska},
  {Kozak}, {Kringel}, {Kr{\'o}lak}, {Krueger}, {Kuehn}, {Kumar}, {Kuo},
  {Kutynia}, {Lackey}, {Landry}, {Lange}, {Lantz}, {Lasky}, {Lazzarini},
  {Lazzaro}, {Leaci}, {Leavey}, {Lebigot}, {Lee}, {Lee}, {Lee}, {Lee}, {Lenon},
  {Leonardi}, {Leong}, {Leroy}, {Letendre}, {Levin}, {Levine}, {Li}, {Libson},
  {Littenberg}, {Lockerbie}, {Logue}, {Lombardi}, {Lord}, {Lorenzini},
  {Loriette}, {Lormand}, {Losurdo}, {Lough}, {L{\"u}ck}, {Lundgren}, {Luo},
  {Lynch}, {Ma}, {MacDonald}, {Machenschalk}, {MacInnis}, {Macleod},
  {Maga{\\textasciitilde n}a-Sandoval}, {Magee}, {Mageswaran}, {Majorana},
  {Maksimovic}, {Malvezzi}, {Man}, {Mandel}, {Mandic}, {Mangano}, {Mansell},
  {Manske}, {Mantovani}, {Marchesoni}, {Marion}, {M{\'a}rka}, {M{\'a}rka},
  {Markosyan}, {Maros}, {Martelli}, {Martellini}, {Martin}, {Martin},
  {Martynov}, {Marx}, {Mason}, {Masserot}, {Massinger}, {Masso-Reid},
  {Matichard}, {Matone}, {Mavalvala}, {Mazumder}, {Mazzolo}, {McCarthy},
  {McClelland}, {McCormick}, {McGuire}, {McIntyre}, {McIver}, {McManus},
  {McWilliams}, {Meacher}, {Meadors}, {Meidam}, {Melatos}, {Mendell}, {Mendoza-
  Gandara}, {Mercer}, {Merilh}, {Merzougui}, {Meshkov}, {Messenger}, {Messick},
  {Meyers}, {Mezzani}, {Miao}, {Michel}, {Middleton}, {Mikhailov}, {Milano},
  {Miller}, {Millhouse}, {Minenkov}, {Ming}, {Mirshekari}, {Mishra}, {Mitra},
  {Mitrofanov}, {Mitselmakher}, {Mittleman}, {Moggi}, {Mohan}, {Mohapatra},
  {Montani}, {Moore}, {Moore}, {Moraru}, {Moreno}, {Morriss}, {Mossavi},
  {Mours}, {Mow-Lowry}, {Mueller}, {Mueller}, {Muir}, {Mukherjee}, {Mukherjee},
  {Mukherjee}, {Mukund}, {Mullavey}, {Munch}, {Murphy}, {Murray}, {Mytidis},
  {Nardecchia}, {Naticchioni}, {Nayak}, {Necula}, {Nedkova}, {Nelemans},
  {Neri}, {Neunzert}, {Newton}, {Nguyen}, {Nielsen}, {Nissanke}, {Nitz},
  {Nocera}, {Nolting}, {Normandin}, {Nuttall}, {Oberling}, {Ochsner}, {O'Dell},
  {Oelker}, {Ogin}, {Oh}, {Oh}, {Ohme}, {Oliver}, {Oppermann}, {Oram},
  {O'Reilly}, {O'Shaughnessy}, {Ottaway}, {Ottens}, {Overmier}, {Owen}, {Pai},
  {Pai}, {Palamos}, {Palashov}, {Palomba}, {Pal-Singh}, {Pan}, {Pankow},
  {Pannarale}, {Pant}, {Paoletti}, {Paoli}, {Papa}, {Paris}, {Parker},
  {Pascucci}, {Pasqualetti}, {Passaquieti}, {Passuello}, {Patricelli},
  {Patrick}, {Pearlstone}, {Pedraza}, {Pedurand}, {Pekowsky}, {Pele}, {Penn},
  {Perreca}, {Phelps}, {Piccinni}, {Pichot}, {Piergiovanni}, {Pierro},
  {Pillant}, {Pinard}, {Pinto}, {Pitkin}, {Poggiani}, {Popolizio}, {Post},
  {Powell}, {Prasad}, {Predoi}, {Premachandra}, {Prestegard}, {Price},
  {Prijatelj}, {Principe}, {Privitera}, {Prodi}, {Prokhorov}, {Puncken},
  {Punturo}, {Puppo}, {P{\"u}rrer}, {Qi}, {Qin}, {Quetschke}, {Quintero},
  {Quitzow-James}, {Raab}, {Rabeling}, {Radkins}, {Raffai}, {Raja},
  {Rakhmanov}, {Rapagnani}, {Raymond}, {Razzano}, {Re}, {Read}, {Reed},
  {Regimbau}, {Rei}, {Reid}, {Reitze}, {Rew}, {Reyes}, {Ricci}, {Riles},
  {Robertson}, {Robie}, {Robinet}, {Rocchi}, {Rolland}, {Rollins}, {Roma},
  {Romano}, {Romanov}, {Romie}, {Rosi{\'n}ska}, {Rowan}, {R{\"u}diger},
  {Ruggi}, {Ryan}, {Sachdev}, {Sadecki}, {Sadeghian}, {Salconi}, {Saleem},
  {Salemi}, {Samajdar}, {Sammut}, {Sanchez}, {Sandberg}, {Sandeen}, {Sanders},
  {Sassolas}, {Sathyaprakash}, {Saulson}, {Sauter}, {Savage}, {Sawadsky},
  {Schale}, {Schilling}, {Schmidt}, {Schmidt}, {Schnabel}, {Schofield},
  {Sch{\"o}nbeck}, {Schreiber}, {Schuette}, {Schutz}, {Scott}, {Scott},
  {Sellers}, {Sengupta}, {Sentenac}, {Sequino}, {Sergeev}, {Serna},
  {Setyawati}, {Sevigny}, {Shaddock}, {Shah}, {Shahriar}, {Shaltev}, {Shao},
  {Shapiro}, {Shawhan}, {Sheperd}, {Shoemaker}, {Shoemaker}, {Siellez},
  {Siemens}, {Sigg}, {Silva}, {Simakov}, {Singer}, {Singer}, {Singh}, {Singh},
  {Singhal}, {Sintes}, {Slagmolen}, {Smith}, {Smith}, {Smith}, {Son}, {Sorazu},
  {Sorrentino}, {Souradeep}, {Srivastava}, {Staley}, {Steinke}, {Steinlechner},
  {Steinlechner}, {Steinmeyer}, {Stephens}, {Stone}, {Strain}, {Straniero},
  {Stratta}, {Strauss}, {Strigin}, {Sturani}, {Stuver}, {Summerscales}, {Sun},
  {Sutton}, {Swinkels}, {Szczepa{\'n}czyk}, {Tacca}, {Talukder}, {Tanner},
  {T{\'a}pai}, {Tarabrin}, {Taracchini}, {Taylor}, {Theeg},
  {Thirugnanasambandam}, {Thomas}, {Thomas}, {Thomas}, {Thorne}, {Thorne},
  {Thrane}, {Tiwari}, {Tiwari}, {Tokmakov}, {Tomlinson}, {Tonelli}, {Torres},
  {Torrie}, {T{\"o}yr{\"a}}, {Travasso}, {Traylor}, {Trifir{\`o}}, {Tringali},
  {Trozzo}, {Tse}, {Turconi}, {Tuyenbayev}, {Ugolini}, {Unnikrishnan}, {Urban},
  {Usman}, {Vahlbruch}, {Vajente}, {Valdes}, {van Bakel}, {van Beuzekom}, {van
  den Brand}, {Van Den Broeck}, {Vander-Hyde}, {van der Schaaf}, {van
  Heijningen}, {van Veggel}, {Vardaro}, {Vass}, {Vas{\'u}th}, {Vaulin},
  {Vecchio}, {Vedovato}, {Veitch}, {Veitch}, {Venkateswara}, {Verkindt},
  {Vetrano}, {Vicer{\'e}}, {Vinciguerra}, {Vine}, {Vinet}, {Vitale}, {Vo},
  {Vocca}, {Vorvick}, {Voss}, {Vousden}, {Vyatchanin}, {Wade}, {Wade}, {Wade},
  {Walker}, {Wallace}, {Walsh}, {Wang}, {Wang}, {Wang}, {Wang}, {Wang}, {Ward},
  {Warner}, {Was}, {Weaver}, {Wei}, {Weinert}, {Weinstein}, {Weiss}, {Welborn},
  {Wen}, {We{\ss}els}, {Westphal}, {Wette}, {Whelan}, {Whitcomb}, {White},
  {Whiting}, {Williams}, {Williamson}, {Willis}, {Willke}, {Wimmer}, {Winkler},
  {Wipf}, {Wittel}, {Woan}, {Worden}, {Wright}, {Wu}, {Yablon}, {Yam},
  {Yamamoto}, {Yancey}, {Yap}, {Yu}, {Yvert}, {Zadro{\.Z}ny}, {Zangrando},
  {Zanolin}, {Zendri}, {Zevin}, {Zhang}, {Zhang}, {Zhang}, {Zhang}, {Zhao},
  {Zhou}, {Zhou}, {Zhu}, {Zucker}, {Zuraw}, {Zweizig}, {LIGO Scientific
  Collaboration}, \& {Virgo Collaboration}}]{2016PhRvL.116m1103A}
{Abbott}, B.~P., {Abbott}, R., {Abbott}, T.~D., {et~al.} 2016{\natexlab{a}},
  \prl, 116, 131103, \dodoi{10.1103/PhysRevLett.116.131103}

\bibitem[{{Abbott} {et~al.}(2016{\natexlab{b}}){Abbott}, {Abbott}, {Abbott},
  {Abernathy}, {Acernese}, {Ackley}, {Adams}, {Adams}, {Addesso}, {Adhikari},
  {Adya}, {Affeldt}, {Agathos}, {Agatsuma}, {Aggarwal}, {Aguiar}, {Aiello},
  {Ain}, {Ajith}, {Allen}, {Allocca}, {Altin}, {Anderson}, {Anderson}, {Arai},
  {Araya}, {Arceneaux}, {Areeda}, {Arnaud}, {Arun}, {Ascenzi}, {Ashton}, {Ast},
  {Aston}, {Astone}, {Aufmuth}, {Aulbert}, {Babak}, {Bacon}, {Bader}, {Baker},
  {Baldaccini}, {Ballardin}, {Ballmer}, {Barayoga}, {Barclay}, {Barish},
  {Barker}, {Barone}, {Barr}, {Barsotti}, {Barsuglia}, {Barta}, {Bartlett},
  {Bartos}, {Bassiri}, {Basti}, {Batch}, {Baune}, {Bavigadda}, {Bazzan},
  {Bejger}, {Bell}, {Berger}, {Bergmann}, {Berry}, {Bersanetti}, {Bertolini},
  {Betzwieser}, {Bhagwat}, {Bhandare}, {Bilenko}, {Billingsley}, {Birch},
  {Birney}, {Birnholtz}, {Biscans}, {Bisht}, {Bitossi}, {Biwer}, {Bizouard},
  {Blackburn}, {Blair}, {Blair}, {Blair}, {Bloemen}, {Bock}, {Boer}, {Bogaert},
  {Bogan}, {Bohe}, {Bond}, {Bondu}, {Bonnand}, {Boom}, {Bork}, {Boschi},
  {Bose}, {Bouffanais}, {Bozzi}, {Bradaschia}, {Brady}, {Braginsky},
  {Branchesi}, {Brau}, {Briant}, {Brillet}, {Brinkmann}, {Brisson}, {Brockill},
  {Broida}, {Brooks}, {Brown}, {Brown}, {Brown}, {Brunett}, {Buchanan},
  {Buikema}, {Bulik}, {Bulten}, {Buonanno}, {Buskulic}, {Buy}, {Byer},
  {Cabero}, {Cadonati}, {Cagnoli}, {Cahillane}, {Calder{\'o}n Bustillo},
  {Callister}, {Calloni}, {Camp}, {Cannon}, {Cao}, {Capano}, {Capocasa},
  {Carbognani}, {Caride}, {Casanueva Diaz}, {Casentini}, {Caudill},
  {Cavagli{\`a}}, {Cavalier}, {Cavalieri}, {Cella}, {Cepeda}, {Cerboni
  Baiardi}, {Cerretani}, {Cesarini}, {Chamberlin}, {Chan}, {Chao}, {Charlton},
  {Chassande- Mottin}, {Cheeseboro}, {Chen}, {Chen}, {Cheng}, {Chincarini},
  {Chiummo}, {Cho}, {Cho}, {Chow}, {Christensen}, {Chu}, {Chua}, {Chung},
  {Ciani}, {Clara}, {Clark}, {Cleva}, {Coccia}, {Cohadon}, {Colla}, {Collette},
  {Cominsky}, {Constancio}, {Conte}, {Conti}, {Cook}, {Corbitt}, {Cornish},
  {Corsi}, {Cortese}, {Costa}, {Coughlin}, {Coughlin}, {Coulon}, {Countryman},
  {Couvares}, {Cowan}, {Coward}, {Cowart}, {Coyne}, {Coyne}, {Craig},
  {Creighton}, {Cripe}, {Crowder}, {Cumming}, {Cunningham}, {Cuoco}, {Dal
  Canton}, {Danilishin}, {D'Antonio}, {Danzmann}, {Darman}, {Dasgupta}, {Da
  Silva Costa}, {Dattilo}, {Dave}, {Davier}, {Davies}, {Daw}, {Day}, {De},
  {DeBra}, {Debreczeni}, {Degallaix}, {De Laurentis}, {Del{\'e}glise}, {Del
  Pozzo}, {Denker}, {Dent}, {Dergachev}, {De Rosa}, {DeRosa}, {DeSalvo},
  {Devine}, {Dhurandhar}, {D{\'\i}az}, {Di Fiore}, {Di Giovanni}, {Di
  Girolamo}, {Di Lieto}, {Di Pace}, {Di Palma}, {Di Virgilio}, {Dolique},
  {Donovan}, {Dooley}, {Doravari}, {Douglas}, {Downes}, {Drago}, {Drever},
  {Driggers}, {Ducrot}, {Dwyer}, {Edo}, {Edwards}, {Effler}, {Eggenstein},
  {Ehrens}, {Eichholz}, {Eikenberry}, {Engels}, {Essick}, {Etzel}, {Evans},
  {Evans}, {Everett}, {Factourovich}, {Fafone}, {Fair}, {Fairhurst}, {Fan},
  {Fang}, {Farinon}, {Farr}, {Farr}, {Favata}, {Fays}, {Fehrmann}, {Fejer},
  {Fenyvesi}, {Ferrante}, {Ferreira}, {Ferrini}, {Fidecaro}, {Fiori},
  {Fiorucci}, {Fisher}, {Flaminio}, {Fletcher}, {Fong}, {Fournier}, {Frasca},
  {Frasconi}, {Frei}, {Freise}, {Frey}, {Frey}, {Fritschel}, {Frolov}, {Fulda},
  {Fyffe}, {Gabbard}, {Gair}, {Gammaitoni}, {Gaonkar}, {Garufi}, {Gaur},
  {Gehrels}, {Gemme}, {Geng}, {Genin}, {Gennai}, {George}, {Gergely},
  {Germain}, {Ghosh}, {Ghosh}, {Ghosh}, {Giaime}, {Giardina}, {Giazotto},
  {Gill}, {Glaefke}, {Goetz}, {Goetz}, {Gondan}, {Gonz{\'a}lez}, {Gonzalez
  Castro}, {Gopakumar}, {Gordon}, {Gorodetsky}, {Gossan}, {Gosselin}, {Gouaty},
  {Grado}, {Graef}, {Graff}, {Granata}, {Grant}, {Gras}, {Gray}, {Greco},
  {Green}, {Groot}, {Grote}, {Grunewald}, {Guidi}, {Guo}, {Gupta}, {Gupta},
  {Gushwa}, {Gustafson}, {Gustafson}, {Hacker}, {Hall}, {Hall}, {Hamilton},
  {Hammond}, {Haney}, {Hanke}, {Hanks}, {Hanna}, {Hannam}, {Hanson},
  {Hardwick}, {Harms}, {Harry}, {Harry}, {Hart}, {Hartman}, {Haster},
  {Haughian}, {Healy}, {Heidmann}, {Heintze}, {Heitmann}, {Hello}, {Hemming},
  {Hendry}, {Heng}, {Hennig}, {Henry}, {Heptonstall}, {Heurs}, {Hild}, {Hoak},
  {Hofman}, {Holt}, {Holz}, {Hopkins}, {Hough}, {Houston}, {Howell}, {Hu},
  {Huang}, {Huerta}, {Huet}, {Hughey}, {Husa}, {Huttner}, {Huynh-Dinh},
  {Indik}, {Ingram}, {Inta}, {Isa}, {Isac}, {Isi}, {Isogai}, {Iyer}, {Izumi},
  {Jacqmin}, {Jang}, {Jani}, {Jaranowski}, {Jawahar}, {Jian}, {Jim{\'e
  }nez-Forteza}, {Johnson}, {Johnson-McDaniel}, {Jones}, {Jones}, {Jonker},
  {Ju}, {K}, {Kalaghatgi}, {Kalogera}, {Kandhasamy}, {Kang}, {Kanner},
  {Kapadia}, {Karki}, {Karvinen}, {Kasprzack}, {Katsavounidis}, {Katzman},
  {Kaufer}, {Kaur}, {Kawabe}, {K{\'e}f{\'e}lian}, {Kehl}, {Keitel}, {Kelley},
  {Kells}, {Kennedy}, {Key}, {Khalili}, {Khan}, {Khan}, {Khan}, {Khazanov},
  {Kijbunchoo}, {Kim}, {Kim}, {Kim}, {Kim}, {Kim}, {Kim}, {Kim}, {Kimbrell},
  {King}, {King}, {Kissel}, {Klein}, {Kleybolte}, {Klimenko}, {Koehlenbeck},
  {Koley}, {Kondrashov}, {Kontos}, {Korobko}, {Korth}, {Kowalska}, {Kozak},
  {Kringel}, {Krishnan}, {Kr{\'o}lak}, {Krueger}, {Kuehn}, {Kumar}, {Kumar},
  {Kuo}, {Kutynia}, {Lackey}, {Landry}, {Lange}, {Lantz}, {Lasky}, {Laxen},
  {Lazzarini}, {Lazzaro}, {Leaci}, {Leavey}, {Lebigot}, {Lee}, {Lee}, {Lee},
  {Lee}, {Lenon}, {Leonardi}, {Leong}, {Leroy}, {Letendre}, {Levin}, {Lewis},
  {Li}, {Libson}, {Littenberg}, {Lockerbie}, {Lombardi}, {London}, {Lord},
  {Lorenzini}, {Loriette}, {Lormand}, {Losurdo}, {Lough}, {Lousto}, {L{\"u}ck},
  {Lundgren}, {Lynch}, {Ma}, {Machenschalk}, {MacInnis}, {Macleod},
  {Maga{\\textasciitilde n}a-Sandoval}, {Maga{\\textasciitilde n}a Zertuche},
  {Magee}, {Majorana}, {Maksimovic}, {Malvezzi}, {Man}, {Mandel}, {Mandic},
  {Mangano}, {Mansell}, {Manske}, {Mantovani}, {Marchesoni}, {Marion},
  {M{\'a}rka}, {M{\'a}rka}, {Markosyan}, {Maros}, {Martelli}, {Martellini},
  {Martin}, {Martynov}, {Marx}, {Mason}, {Masserot}, {Massinger}, {Masso-Reid},
  {Mastrogiovanni}, {Matichard}, {Matone}, {Mavalvala}, {Mazumder}, {McCarthy},
  {McClelland}, {McCormick}, {McGuire}, {McIntyre}, {McIver}, {McManus},
  {McRae}, {McWilliams}, {Meacher}, {Meadors}, {Meidam}, {Melatos}, {Mendell},
  {Mercer}, {Merilh}, {Merzougui}, {Meshkov}, {Messenger}, {Messick},
  {Metzdorff}, {Meyers}, {Mezzani}, {Miao}, {Michel}, {Middleton}, {Mikhailov},
  {Milano}, {Miller}, {Miller}, {Miller}, {Miller}, {Millhouse}, {Minenkov},
  {Ming}, {Mirshekari}, {Mishra}, {Mitra}, {Mitrofanov}, {Mitselmakher},
  {Mittleman}, {Moggi}, {Mohan}, {Mohapatra}, {Montani}, {Moore}, {Moore},
  {Moraru}, {Moreno}, {Morriss}, {Mossavi}, {Mours}, {Mow-Lowry}, {Mueller},
  {Muir}, {Mukherjee}, {Mukherjee}, {Mukherjee}, {Mukund}, {Mullavey}, {Munch},
  {Murphy}, {Murray}, {Mytidis}, {Nardecchia}, {Naticchioni}, {Nayak},
  {Nedkova}, {Nelemans}, {Nelson}, {Neri}, {Neunzert}, {Newton}, {Nguyen},
  {Nielsen}, {Nissanke}, {Nitz}, {Nocera}, {Nolting}, {Normandin}, {Nuttall},
  {Oberling}, {Ochsner}, {O'Dell}, {Oelker}, {Ogin}, {Oh}, {Oh}, {Ohme},
  {Oliver}, {Oppermann}, {Oram}, {O'Reilly}, {O'Shaughnessy}, {Ottaway},
  {Overmier}, {Owen}, {Pai}, {Pai}, {Palamos}, {Palashov}, {Palomba},
  {Pal-Singh}, {Pan}, {Pankow}, {Pannarale}, {Pant}, {Paoletti}, {Paoli},
  {Papa}, {Paris}, {Parker}, {Pascucci}, {Pasqualetti}, {Passaquieti},
  {Passuello}, {Patricelli}, {Patrick}, {Pearlstone}, {Pedraza}, {Pedurand},
  {Pekowsky}, {Pele}, {Penn}, {Perreca}, {Perri}, {Pfeiffer}, {Phelps},
  {Piccinni}, {Pichot}, {Piergiovanni}, {Pierro}, {Pillant}, {Pinard}, {Pinto},
  {Pitkin}, {Poe}, {Poggiani}, {Popolizio}, {Post}, {Powell}, {Prasad},
  {Predoi}, {Prestegard}, {Price}, {Prijatelj}, {Principe}, {Privitera},
  {Prix}, {Prodi}, {Prokhorov}, {Puncken}, {Punturo}, {Puppo}, {P{\"u}rrer},
  {Qi}, {Qin}, {Qiu}, {Quetschke}, {Quintero}, {Quitzow-James}, {Raab},
  {Rabeling}, {Radkins}, {Raffai}, {Raja}, {Rajan}, {Rakhmanov}, {Rapagnani},
  {Raymond}, {Razzano}, {Re}, {Read}, {Reed}, {Regimbau}, {Rei}, {Reid},
  {Reitze}, {Rew}, {Reyes}, {Ricci}, {Riles}, {Rizzo}, {Robertson}, {Robie},
  {Robinet}, {Rocchi}, {Rolland}, {Rollins}, {Roma}, {Romano}, {Romano},
  {Romanov}, {Romie}, {Rosi{\'n}ska}, {Rowan}, {R{\"u}diger}, {Ruggi}, {Ryan},
  {Sachdev}, {Sadecki}, {Sadeghian}, {Sakellariadou}, {Salconi}, {Saleem},
  {Salemi}, {Samajdar}, {Sammut}, {Sanchez}, {Sandberg}, {Sandeen}, {Sanders},
  {Sassolas}, {Sathyaprakash}, {Saulson}, {Sauter}, {Savage}, {Sawadsky},
  {Schale}, {Schilling}, {Schmidt}, {Schmidt}, {Schnabel}, {Schofield},
  {Sch{\"o}nbeck}, {Schreiber}, {Schuette}, {Schutz}, {Scott}, {Scott},
  {Sellers}, {Sengupta}, {Sentenac}, {Sequino}, {Sergeev}, {Setyawati},
  {Shaddock}, {Shaffer}, {Shahriar}, {Shaltev}, {Shapiro}, {Shawhan},
  {Sheperd}, {Shoemaker}, {Shoemaker}, {Siellez}, {Siemens}, {Sieniawska},
  {Sigg}, {Silva}, {Singer}, {Singer}, {Singh}, {Singh}, {Singhal}, {Sintes},
  {Slagmolen}, {Smith}, {Smith}, {Smith}, {Son}, {Sorazu}, {Sorrentino},
  {Souradeep}, {Srivastava}, {Staley}, {Steinke}, {Steinlechner},
  {Steinlechner}, {Steinmeyer}, {Stephens}, {Stevenson}, {Stone}, {Strain},
  {Straniero}, {Stratta}, {Strauss}, {Strigin}, {Sturani}, {Stuver},
  {Summerscales}, {Sun}, {Sunil}, {Sutton}, {Swinkels}, {Szczepa{\'n}czyk},
  {Tacca}, {Talukder}, {Tanner}, {T{\'a}pai}, {Tarabrin}, {Taracchini},
  {Taylor}, {Theeg}, {Thirugnanasambandam}, {Thomas}, {Thomas}, {Thomas},
  {Thorne}, {Thrane}, {Tiwari}, {Tiwari}, {Tokmakov}, {Toland}, {Tomlinson},
  {Tonelli}, {Tornasi}, {Torres}, {Torrie}, {T{\"o}yr{\"a}}, {Travasso},
  {Traylor}, {Trifir{\`o}}, {Tringali}, {Trozzo}, {Tse}, {Turconi},
  {Tuyenbayev}, {Ugolini}, {Unnikrishnan}, {Urban}, {Usman}, {Vahlbruch},
  {Vajente}, {Valdes}, {Vallisneri}, {van Bakel}, {van Beuzekom}, {van den
  Brand}, {Van Den Broeck}, {Vander- Hyde}, {van der Schaaf}, {van Heijningen},
  {van Veggel}, {Vardaro}, {Vass}, {Vas{\'u}th}, {Vaulin}, {Vecchio},
  {Vedovato}, {Veitch}, {Veitch}, {Venkateswara}, {Verkindt}, {Vetrano},
  {Vicer{\'e}}, {Vinciguerra}, {Vine}, {Vinet}, {Vitale}, {Vo}, {Vocca},
  {Vorvick}, {Voss}, {Vousden}, {Vyatchanin}, {Wade}, {Wade}, {Wade}, {Walker},
  {Wallace}, {Walsh}, {Wang}, {Wang}, {Wang}, {Wang}, {Wang}, {Ward}, {Warner},
  {Was}, {Weaver}, {Wei}, {Weinert}, {Weinstein}, {Weiss}, {Wen}, {We{\ss}els},
  {Westphal}, {Wette}, {Whelan}, {Whiting}, {Williams}, {Williamson}, {Willis},
  {Willke}, {Wimmer}, {Winkler}, {Wipf}, {Wittel}, {Woan}, {Woehler}, {Worden},
  {Wright}, {Wu}, {Wu}, {Yablon}, {Yam}, {Yamamoto}, {Yancey}, {Yu}, {Yvert},
  {Zadro{\.z}ny}, {Zangrando}, {Zanolin}, {Zendri}, {Zevin}, {Zhang}, {Zhang},
  {Zhang}, {Zhao}, {Zhou}, {Zhou}, {Zhu}, {Zucker}, {Zuraw}, {Zweizig},
  {Boyle}, {Hemberger}, {Kidder}, {Lovelace}, {Ossokine}, {Scheel}, {Szilagyi},
  {Teukolsky}, \& {LIGO Scientific Collaboration}}]{2016PhRvL.116x1103A}
---. 2016{\natexlab{b}}, \prl, 116, 241103,
  \dodoi{10.1103/PhysRevLett.116.241103}

\bibitem[{{Abbott} {et~al.}(2016{\natexlab{c}}){Abbott}, {Abbott}, {Abbott},
  {Abernathy}, {Acernese}, {Ackley}, {Adams}, {Adams}, {Addesso}, {Adhikari},
  {Adya}, {Affeldt}, {Agathos}, {Agatsuma}, {Aggarwal}, {Aguiar}, {Aiello},
  {Ain}, {Ajith}, {Allen}, {Allocca}, {Altin}, {Anderson}, {Anderson}, {Arai},
  {Araya}, {Arceneaux}, {Areeda}, {Arnaud}, {Arun}, {Ascenzi}, {Ashton}, {Ast},
  {Aston}, {Astone}, {Aufmuth}, {Aulbert}, {Babak}, {Bacon}, {Bader}, {Baker},
  {Baldaccini}, {Ballardin}, {Ballmer}, {Barayoga}, {Barclay}, {Barish},
  {Barker}, {Barone}, {Barr}, {Barsotti}, {Barsuglia}, {Barta}, {Bartlett},
  {Bartos}, {Bassiri}, {Basti}, {Batch}, {Baune}, {Bavigadda}, {Bazzan},
  {Bejger}, {Bell}, {Berger}, {Bergmann}, {Berry}, {Bersanetti}, {Bertolini},
  {Betzwieser}, {Bhagwat}, {Bhandare}, {Bilenko}, {Billingsley}, {Birch},
  {Birney}, {Birnholtz}, {Biscans}, {Bisht}, {Bitossi}, {Biwer}, {Bizouard},
  {Blackburn}, {Blair}, {Blair}, {Blair}, {Bloemen}, {Bock}, {Boer}, {Bogaert},
  {Bogan}, {Bohe}, {Bond}, {Bondu}, {Bonnand}, {Boom}, {Bork}, {Boschi},
  {Bose}, {Bouffanais}, {Bozzi}, {Bradaschia}, {Brady}, {Braginsky},
  {Branchesi}, {Brau}, {Briant}, {Brillet}, {Brinkmann}, {Brisson}, {Brockill},
  {Broida}, {Brooks}, {Brown}, {Brown}, {Brown}, {Brunett}, {Buchanan},
  {Buikema}, {Bulik}, {Bulten}, {Buonanno}, {Buskulic}, {Buy}, {Byer},
  {Cabero}, {Cadonati}, {Cagnoli}, {Cahillane}, {Calder{\'o}n Bustillo},
  {Callister}, {Calloni}, {Camp}, {Cannon}, {Cao}, {Capano}, {Capocasa},
  {Carbognani}, {Caride}, {Casanueva Diaz}, {Casentini}, {Caudill},
  {Cavagli{\`a}}, {Cavalier}, {Cavalieri}, {Cella}, {Cepeda}, {Cerboni
  Baiardi}, {Cerretani}, {Cesarini}, {Chamberlin}, {Chan}, {Chao}, {Charlton},
  {Chassande- Mottin}, {Cheeseboro}, {Chen}, {Chen}, {Cheng}, {Chincarini},
  {Chiummo}, {Cho}, {Cho}, {Chow}, {Christensen}, {Chu}, {Chua}, {Chung},
  {Ciani}, {Clara}, {Clark}, {Cleva}, {Coccia}, {Cohadon}, {Colla}, {Collette},
  {Cominsky}, {Constancio}, {Conte}, {Conti}, {Cook}, {Corbitt}, {Cornish},
  {Corsi}, {Cortese}, {Costa}, {Coughlin}, {Coughlin}, {Coulon}, {Countryman},
  {Couvares}, {Cowan}, {Coward}, {Cowart}, {Coyne}, {Coyne}, {Craig},
  {Creighton}, {Cripe}, {Crowder}, {Cumming}, {Cunningham}, {Cuoco}, {Dal
  Canton}, {Danilishin}, {D'Antonio}, {Danzmann}, {Darman}, {Dasgupta}, {Da
  Silva Costa}, {Dattilo}, {Dave}, {Davier}, {Davies}, {Daw}, {Day}, {De},
  {DeBra}, {Debreczeni}, {Degallaix}, {De Laurentis}, {Del{\'e}glise}, {Del
  Pozzo}, {Denker}, {Dent}, {Dergachev}, {De Rosa}, {DeRosa}, {DeSalvo},
  {Devine}, {Dhurandhar}, {D{\'\i}az}, {Di Fiore}, {Di Giovanni}, {Di
  Girolamo}, {Di Lieto}, {Di Pace}, {Di Palma}, {Di Virgilio}, {Dolique},
  {Donovan}, {Dooley}, {Doravari}, {Douglas}, {Downes}, {Drago}, {Drever},
  {Driggers}, {Ducrot}, {Dwyer}, {Edo}, {Edwards}, {Effler}, {Eggenstein},
  {Ehrens}, {Eichholz}, {Eikenberry}, {Engels}, {Essick}, {Etzel}, {Evans},
  {Evans}, {Everett}, {Factourovich}, {Fafone}, {Fair}, {Fairhurst}, {Fan},
  {Fang}, {Farinon}, {Farr}, {Farr}, {Favata}, {Fays}, {Fehrmann}, {Fejer},
  {Fenyvesi}, {Ferrante}, {Ferreira}, {Ferrini}, {Fidecaro}, {Fiori},
  {Fiorucci}, {Fisher}, {Flaminio}, {Fletcher}, {Fong}, {Fournier}, {Frasca},
  {Frasconi}, {Frei}, {Freise}, {Frey}, {Frey}, {Fritschel}, {Frolov}, {Fulda},
  {Fyffe}, {Gabbard}, {Gaebel}, {Gair}, {Gammaitoni}, {Gaonkar}, {Garufi},
  {Gaur}, {Gehrels}, {Gemme}, {Geng}, {Genin}, {Gennai}, {George}, {Gergely},
  {Germain}, {Ghosh}, {Ghosh}, {Ghosh}, {Giaime}, {Giardina}, {Giazotto},
  {Gill}, {Glaefke}, {Goetz}, {Goetz}, {Gondan}, {Gonz{\'a}lez}, {Gonzalez
  Castro}, {Gopakumar}, {Gordon}, {Gorodetsky}, {Gossan}, {Gosselin}, {Gouaty},
  {Grado}, {Graef}, {Graff}, {Granata}, {Grant}, {Gras}, {Gray}, {Greco},
  {Green}, {Groot}, {Grote}, {Grunewald}, {Guidi}, {Guo}, {Gupta}, {Gupta},
  {Gushwa}, {Gustafson}, {Gustafson}, {Hacker}, {Hall}, {Hall}, {Hamilton},
  {Hammond}, {Haney}, {Hanke}, {Hanks}, {Hanna}, {Hannam}, {Hanson},
  {Hardwick}, {Harms}, {Harry}, {Harry}, {Hart}, {Hartman}, {Haster},
  {Haughian}, {Healy}, {Heidmann}, {Heintze}, {Heitmann}, {Hello}, {Hemming},
  {Hendry}, {Heng}, {Hennig}, {Henry}, {Heptonstall}, {Heurs}, {Hild}, {Hoak},
  {Hofman}, {Holt}, {Holz}, {Hopkins}, {Hough}, {Houston}, {Howell}, {Hu},
  {Huang}, {Huerta}, {Huet}, {Hughey}, {Husa}, {Huttner}, {Huynh-Dinh},
  {Indik}, {Ingram}, {Inta}, {Isa}, {Isac}, {Isi}, {Isogai}, {Iyer}, {Izumi},
  {Jacqmin}, {Jang}, {Jani}, {Jaranowski}, {Jawahar}, {Jian},
  {Jim{\'e}nez-Forteza}, {Johnson}, {Johnson-McDaniel}, {Jones}, {Jones},
  {Jonker}, {Ju}, {K}, {Kalaghatgi}, {Kalogera}, {Kandhasamy}, {Kang},
  {Kanner}, {Kapadia}, {Karki}, {Karvinen}, {Kasprzack}, {Katsavounidis},
  {Katzman}, {Kaufer}, {Kaur}, {Kawabe}, {K{\'e}f{\'e}lian}, {Kehl}, {Keitel},
  {Kelley}, {Kells}, {Kennedy}, {Key}, {Khalili}, {Khan}, {Khan}, {Khan},
  {Khazanov}, {Kijbunchoo}, {Kim}, {Kim}, {Kim}, {Kim}, {Kim}, {Kim}, {Kim},
  {Kimbrell}, {King}, {King}, {Kissel}, {Klein}, {Kleybolte}, {Klimenko},
  {Koehlenbeck}, {Koley}, {Kondrashov}, {Kontos}, {Korobko}, {Korth},
  {Kowalska}, {Kozak}, {Kringel}, {Krishnan}, {Kr{\'o}lak}, {Krueger}, {Kuehn},
  {Kumar}, {Kumar}, {Kuo}, {Kutynia}, {Lackey}, {Landry}, {Lange}, {Lantz},
  {Lasky}, {Laxen}, {Lazzarini}, {Lazzaro}, {Leaci}, {Leavey}, {Lebigot},
  {Lee}, {Lee}, {Lee}, {Lee}, {Lenon}, {Leonardi}, {Leong}, {Leroy},
  {Letendre}, {Levin}, {Lewis}, {Li}, {Libson}, {Littenberg}, {Lockerbie},
  {Lombardi}, {London}, {Lord}, {Lorenzini}, {Loriette}, {Lormand}, {Losurdo},
  {Lough}, {Lousto}, {L{\"u}ck}, {Lundgren}, {Lynch}, {Ma}, {Machenschalk},
  {MacInnis}, {Macleod}, {Maga{\\textasciitilde n}a-Sandoval},
  {Maga{\\textasciitilde n}a Zertuche}, {Magee}, {Majorana}, {Maksimovic},
  {Malvezzi}, {Man}, {Mandel}, {Mandic}, {Mangano}, {Mansell}, {Manske},
  {Mantovani}, {Marchesoni}, {Marion}, {M{\'a}rka}, {M{\'a}rka}, {Markosyan},
  {Maros}, {Martelli}, {Martellini}, {Martin}, {Martynov}, {Marx}, {Mason},
  {Masserot}, {Massinger}, {Masso-Reid}, {Mastrogiovanni}, {Matichard},
  {Matone}, {Mavalvala}, {Mazumder}, {McCarthy}, {McClelland}, {McCormick},
  {McGuire}, {McIntyre}, {McIver}, {McManus}, {McRae}, {McWilliams}, {Meacher},
  {Meadors}, {Meidam}, {Melatos}, {Mendell}, {Mercer}, {Merilh}, {Merzougui},
  {Meshkov}, {Messenger}, {Messick}, {Metzdorff}, {Meyers}, {Mezzani}, {Miao},
  {Michel}, {Middleton}, {Mikhailov}, {Milano}, {Miller}, {Miller}, {Miller},
  {Miller}, {Millhouse}, {Minenkov}, {Ming}, {Mirshekari}, {Mishra}, {Mitra},
  {Mitrofanov}, {Mitselmakher}, {Mittleman}, {Moggi}, {Mohan}, {Mohapatra},
  {Montani}, {Moore}, {Moore}, {Moraru}, {Moreno}, {Morriss}, {Mossavi},
  {Mours}, {Mow-Lowry}, {Mueller}, {Muir}, {Mukherjee}, {Mukherjee},
  {Mukherjee}, {Mukund}, {Mullavey}, {Munch}, {Murphy}, {Murray}, {Mytidis},
  {Nardecchia}, {Naticchioni}, {Nayak}, {Nedkova}, {Nelemans}, {Nelson},
  {Neri}, {Neunzert}, {Newton}, {Nguyen}, {Nielsen}, {Nissanke}, {Nitz},
  {Nocera}, {Nolting}, {Normandin}, {Nuttall}, {Oberling}, {Ochsner}, {O'Dell},
  {Oelker}, {Ogin}, {Oh}, {Oh}, {Ohme}, {Oliver}, {Oppermann}, {Oram},
  {O'Reilly}, {O'Shaughnessy}, {Ottaway}, {Overmier}, {Owen}, {Pai}, {Pai},
  {Palamos}, {Palashov}, {Palomba}, {Pal-Singh}, {Pan}, {Pan}, {Pankow},
  {Pannarale}, {Pant}, {Paoletti}, {Paoli}, {Papa}, {Paris}, {Parker},
  {Pascucci}, {Pasqualetti}, {Passaquieti}, {Passuello}, {Patricelli},
  {Patrick}, {Pearlstone}, {Pedraza}, {Pedurand}, {Pekowsky}, {Pele}, {Penn},
  {Perreca}, {Perri}, {Pfeiffer}, {Phelps}, {Piccinni}, {Pichot},
  {Piergiovanni}, {Pierro}, {Pillant}, {Pinard}, {Pinto}, {Pitkin}, {Poe},
  {Poggiani}, {Popolizio}, {Porter}, {Post}, {Powell}, {Prasad}, {Predoi},
  {Prestegard}, {Price}, {Prijatelj}, {Principe}, {Privitera}, {Prix}, {Prodi},
  {Prokhorov}, {Puncken}, {Punturo}, {Puppo}, {P{\"u}rrer}, {Qi}, {Qin}, {Qiu},
  {Quetschke}, {Quintero}, {Quitzow-James}, {Raab}, {Rabeling}, {Radkins},
  {Raffai}, {Raja}, {Rajan}, {Rakhmanov}, {Rapagnani}, {Raymond}, {Razzano},
  {Re}, {Read}, {Reed}, {Regimbau}, {Rei}, {Reid}, {Reitze}, {Rew}, {Reyes},
  {Ricci}, {Riles}, {Rizzo}, {Robertson}, {Robie}, {Robinet}, {Rocchi},
  {Rolland}, {Rollins}, {Roma}, {Romano}, {Romano}, {Romanov}, {Romie},
  {Rosi{\'n}ska}, {Rowan}, {R{\"u}diger}, {Ruggi}, {Ryan}, {Sachdev},
  {Sadecki}, {Sadeghian}, {Sakellariadou}, {Salconi}, {Saleem}, {Salemi},
  {Samajdar}, {Sammut}, {Sanchez}, {Sandberg}, {Sandeen}, {Sanders},
  {Sassolas}, {Sathyaprakash}, {Saulson}, {Sauter}, {Savage}, {Sawadsky},
  {Schale}, {Schilling}, {Schmidt}, {Schmidt}, {Schnabel}, {Schofield},
  {Sch{\"o}nbeck}, {Schreiber}, {Schuette}, {Schutz}, {Scott}, {Scott},
  {Sellers}, {Sengupta}, {Sentenac}, {Sequino}, {Sergeev}, {Setyawati},
  {Shaddock}, {Shaffer}, {Shahriar}, {Shaltev}, {Shapiro}, {Shawhan},
  {Sheperd}, {Shoemaker}, {Shoemaker}, {Siellez}, {Siemens}, {Sieniawska},
  {Sigg}, {Silva}, {Singer}, {Singer}, {Singh}, {Singh}, {Singhal}, {Sintes},
  {Slagmolen}, {Smith}, {Smith}, {Smith}, {Son}, {Sorazu}, {Sorrentino},
  {Souradeep}, {Srivastava}, {Staley}, {Steinke}, {Steinlechner},
  {Steinlechner}, {Steinmeyer}, {Stephens}, {Stevenson}, {Stone}, {Strain},
  {Straniero}, {Stratta}, {Strauss}, {Strigin}, {Sturani}, {Stuver},
  {Summerscales}, {Sun}, {Sunil}, {Sutton}, {Swinkels}, {Szczepa{\'n}czyk},
  {Tacca}, {Talukder}, {Tanner}, {T{\'a}pai}, {Tarabrin}, {Taracchini},
  {Taylor}, {Theeg}, {Thirugnanasambandam}, {Thomas}, {Thomas}, {Thomas},
  {Thorne}, {Thrane}, {Tiwari}, {Tiwari}, {Tokmakov}, {Toland}, {Tomlinson},
  {Tonelli}, {Tornasi}, {Torres}, {Torrie}, {T{\"o}yr{\"a}}, {Travasso},
  {Traylor}, {Trifir{\`o}}, {Tringali}, {Trozzo}, {Tse}, {Turconi},
  {Tuyenbayev}, {Ugolini}, {Unnikrishnan}, {Urban}, {Usman}, {Vahlbruch},
  {Vajente}, {Valdes}, {Vallisneri}, {van Bakel}, {van Beuzekom}, {van den
  Brand}, {Van Den Broeck}, {Vander-Hyde}, {van der Schaaf}, {van Heijningen},
  {van Veggel}, {Vardaro}, {Vass}, {Vas{\'u}th}, {Vaulin}, {Vecchio},
  {Vedovato}, {Veitch}, {Veitch}, {Venkateswara}, {Verkindt}, {Vetrano},
  {Vicer{\'e}}, {Vinciguerra}, {Vine}, {Vinet}, {Vitale}, {Vo}, {Vocca},
  {Vorvick}, {Voss}, {Vousden}, {Vyatchanin}, {Wade}, {Wade}, {Wade}, {Walker},
  {Wallace}, {Walsh}, {Wang}, {Wang}, {Wang}, {Wang}, {Wang}, {Ward}, {Warner},
  {Was}, {Weaver}, {Wei}, {Weinert}, {Weinstein}, {Weiss}, {Wen}, {We{\ss}els},
  {Westphal}, {Wette}, {Whelan}, {Whitcomb}, {Whiting}, {Williams},
  {Williamson}, {Willis}, {Willke}, {Wimmer}, {Winkler}, {Wipf}, {Wittel},
  {Woan}, {Woehler}, {Worden}, {Wright}, {Wu}, {Wu}, {Yablon}, {Yam},
  {Yamamoto}, {Yancey}, {Yu}, {Yvert}, {Zadro{\.Z}ny}, {Zangrando}, {Zanolin},
  {Zendri}, {Zevin}, {Zhang}, {Zhang}, {Zhang}, {Zhao}, {Zhou}, {Zhou}, {Zhu},
  {Zucker}, {Zuraw}, {Zweizig}, {LIGO Scientific Collaboration}, \& {Virgo
  Collaboration}}]{2016PhRvX...6d1015A}
---. 2016{\natexlab{c}}, Physical Review X, 6, 041015,
  \dodoi{10.1103/PhysRevX.6.041015}

\bibitem[{{Abbott} {et~al.}(2016{\natexlab{d}}){Abbott}, {Abbott}, {Abbott},
  {Abernathy}, {Acernese}, {Ackley}, {Adams}, {Adams}, {Addesso}, {Adhikari},
  {Adya}, {Affeldt}, {Agathos}, {Agatsuma}, {Aggarwal}, {Aguiar}, {Aiello},
  {Ain}, {Ajith}, {Allen}, {Allocca}, {Altin}, {Anderson}, {Anderson}, {Arai},
  {Araya}, {Arceneaux}, {Areeda}, {Arnaud}, {Arun}, {Ascenzi}, {Ashton}, {Ast},
  {Aston}, {Astone}, {Aufmuth}, {Aulbert}, {Babak}, {Bacon}, {Bader}, {Baker},
  {Baldaccini}, {Ballardin}, {Ballmer}, {Barayoga}, {Barclay}, {Barish},
  {Barker}, {Barone}, {Barr}, {Barsotti}, {Barsuglia}, {Barta}, {Bartlett},
  {Bartos}, {Bassiri}, {Basti}, {Batch}, {Baune}, {Bavigadda}, {Bazzan},
  {Behnke}, {Bejger}, {Belczynski}, {Bell}, {Bell}, {Berger}, {Bergman},
  {Bergmann}, {Berry}, {Bersanetti}, {Bertolini}, {Betzwieser}, {Bhagwat},
  {Bhandare}, {Bilenko}, {Billingsley}, {Birch}, {Birney}, {Biscans}, {Bisht},
  {Bitossi}, {Biwer}, {Bizouard}, {Blackburn}, {Blair}, {Blair}, {Blair},
  {Bloemen}, {Bock}, {Bodiya}, {Boer}, {Bogaert}, {Bogan}, {Bohe}, {Bojtos},
  {Bond}, {Bondu}, {Bonnand}, {Boom}, {Bork}, {Boschi}, {Bose}, {Bouffanais},
  {Bozzi}, {Bradaschia}, {Brady}, {Braginsky}, {Branchesi}, {Brau}, {Briant},
  {Brillet}, {Brinkmann}, {Brisson}, {Brockill}, {Brooks}, {Brown}, {Brown},
  {Brown}, {Buchanan}, {Buikema}, {Bulik}, {Bulten}, {Buonanno}, {Buskulic},
  {Buy}, {Byer}, {Cadonati}, {Cagnoli}, {Cahillane}, {Calder{\'o}n Bustillo},
  {Callister}, {Calloni}, {Camp}, {Cannon}, {Cao}, {Capano}, {Capocasa},
  {Carbognani}, {Caride}, {Casanueva Diaz}, {Casentini}, {Caudill},
  {Cavagli{\`a}}, {Cavalier}, {Cavalieri}, {Cella}, {Cepeda}, {Cerboni
  Baiardi}, {Cerretani}, {Cesarini}, {Chakraborty}, {Chalermsongsak},
  {Chamberlin}, {Chan}, {Chao}, {Charlton}, {Chassande-Mottin}, {Chen}, {Chen},
  {Cheng}, {Chincarini}, {Chiummo}, {Cho}, {Cho}, {Chow}, {Christensen}, {Chu},
  {Chua}, {Chung}, {Ciani}, {Clara}, {Clark}, {Cleva}, {Coccia}, {Cohadon},
  {Colla}, {Collette}, {Cominsky}, {Constancio}, {Conte}, {Conti}, {Cook},
  {Corbitt}, {Cornish}, {Corsi}, {Cortese}, {Costa}, {Coughlin}, {Coughlin},
  {Coulon}, {Countryman}, {Couvares}, {Cowan}, {Coward}, {Cowart}, {Coyne},
  {Coyne}, {Craig}, {Creighton}, {Cripe}, {Crowder}, {Cumming}, {Cunningham},
  {Cuoco}, {Dal Canton}, {Danilishin}, {D'Antonio}, {Danzmann}, {Darman},
  {Dattilo}, {Dave}, {Daveloza}, {Davier}, {Davies}, {Daw}, {Day}, {DeBra},
  {Debreczeni}, {Degallaix}, {De Laurentis}, {Del{\'e}glise}, {Del Pozzo},
  {Denker}, {Dent}, {Dereli}, {Dergachev}, {DeRosa}, {DeRosa}, {DeSalvo},
  {Dhurandhar}, {D{\'\i}az}, {Di Fiore}, {Di Giovanni}, {Di Lieto}, {Di Pace},
  {Di Palma}, {Di Virgilio}, {Dojcinoski}, {Dolique}, {Donovan}, {Dooley},
  {Doravari}, {Douglas}, {Downes}, {Drago}, {Drever}, {Driggers}, {Du},
  {Ducrot}, {Dwyer}, {Edo}, {Edwards}, {Effler}, {Eggenstein}, {Ehrens},
  {Eichholz}, {Eikenberry}, {Engels}, {Essick}, {Etzel}, {Evans}, {Evans},
  {Everett}, {Factourovich}, {Fafone}, {Fair}, {Fairhurst}, {Fan}, {Fang},
  {Farinon}, {Farr}, {Farr}, {Favata}, {Fays}, {Fehrmann}, {Fejer}, {Ferrante},
  {Ferreira}, {Ferrini}, {Fidecaro}, {Fiori}, {Fiorucci}, {Fisher}, {Flaminio},
  {Fletcher}, {Fournier}, {Franco}, {Frasca}, {Frasconi}, {Frei}, {Freise},
  {Frey}, {Frey}, {Fricke}, {Fritschel}, {Frolov}, {Fulda}, {Fyffe}, {Gabbard},
  {Gair}, {Gammaitoni}, {Gaonkar}, {Garufi}, {Gatto}, {Gaur}, {Gehrels},
  {Gemme}, {Gendre}, {Genin}, {Gennai}, {George}, {Gergely}, {Germain},
  {Ghosh}, {Ghosh}, {Giaime}, {Giardina}, {Giazotto}, {Gill}, {Glaefke},
  {Goetz}, {Goetz}, {Gondan}, {Gonz{\'a}lez}, {Gonzalez Castro}, {Gopakumar},
  {Gordon}, {Gorodetsky}, {Gossan}, {Gosselin}, {Gouaty}, {Graef}, {Graff},
  {Granata}, {Grant}, {Gras}, {Gray}, {Greco}, {Green}, {Groot}, {Grote},
  {Grunewald}, {Guidi}, {Guo}, {Gupta}, {Gupta}, {Gushwa}, {Gustafson},
  {Gustafson}, {Hacker}, {Hall}, {Hall}, {Hammond}, {Haney}, {Hanke}, {Hanks},
  {Hanna}, {Hannam}, {Hanson}, {Hardwick}, {Harms}, {Harry}, {Harry}, {Hart},
  {Hartman}, {Haster}, {Haughian}, {Heidmann}, {Heintze}, {Heitmann}, {Hello},
  {Hemming}, {Hendry}, {Heng}, {Hennig}, {Heptonstall}, {Heurs}, {Hild},
  {Hoak}, {Hodge}, {Hofman}, {Hollitt}, {Holt}, {Holz}, {Hopkins}, {Hosken},
  {Hough}, {Houston}, {Howell}, {Hu}, {Huang}, {Huerta}, {Huet}, {Hughey},
  {Husa}, {Huttner}, {Huynh-Dinh}, {Idrisy}, {Indik}, {Ingram}, {Inta}, {Isa},
  {Isac}, {Isi}, {Islas}, {Isogai}, {Iyer}, {Izumi}, {Jacqmin}, {Jang}, {Jani},
  {Jaranowski}, {Jawahar}, {Jim{\'e}nez-Forteza}, {Johnson}, {Jones}, {Jones},
  {Jonker}, {Ju}, {K}, {Kalaghatgi}, {Kalogera}, {Kandhasamy}, {Kang},
  {Kanner}, {Karki}, {Kasprzack}, {Katsavounidis}, {Katzman}, {Kaufer}, {Kaur},
  {Kawabe}, {Kawazoe}, {K{\'e}f{\'e}lian}, {Kehl}, {Keitel}, {Kelley}, {Kells},
  {Kennedy}, {Key}, {Khalaidovski}, {Khalili}, {Khan}, {Khan}, {Khan},
  {Khazanov}, {Kijbunchoo}, {Kim}, {Kim}, {Kim}, {Kim}, {Kim}, {Kim}, {King},
  {King}, {Kinzel}, {Kissel}, {Kleybolte}, {Klimenko}, {Koehlenbeck},
  {Kokeyama}, {Koley}, {Kondrashov}, {Kontos}, {Korobko}, {Korth}, {Kowalska},
  {Kozak}, {Kringel}, {Krishnan}, {Kr{\'o}lak}, {Krueger}, {Kuehn}, {Kumar},
  {Kuo}, {Kutynia}, {Lackey}, {Landry}, {Lange}, {Lantz}, {Lasky}, {Lazzarini},
  {Lazzaro}, {Leaci}, {Leavey}, {Lebigot}, {Lee}, {Lee}, {Lee}, {Lee}, {Lenon},
  {Leonardi}, {Leong}, {Leroy}, {Letendre}, {Levin}, {Levine}, {Li}, {Libson},
  {Littenberg}, {Lockerbie}, {Logue}, {Lombardi}, {Lord}, {Lorenzini},
  {Loriette}, {Lormand}, {Losurdo}, {Lough}, {L{\"u}ck}, {Lundgren}, {Luo},
  {Lynch}, {Ma}, {MacDonald}, {Machenschalk}, {MacInnis}, {Macleod},
  {Maga{\\textasciitilde n}a-Sandoval}, {Magee}, {Mageswaran}, {Majorana},
  {Maksimovic}, {Malvezzi}, {Man}, {Mandel}, {Mandic}, {Mangano}, {Mansell},
  {Manske}, {Mantovani}, {Marchesoni}, {Marion}, {M{\'a}rka}, {M{\'a}rka},
  {Markosyan}, {Maros}, {Martelli}, {Martellini}, {Martin}, {Martin},
  {Martynov}, {Marx}, {Mason}, {Masserot}, {Massinger}, {Masso-Reid},
  {Matichard}, {Matone}, {Mavalvala}, {Mazumder}, {Mazzolo}, {McCarthy},
  {McClelland}, {McCormick}, {McGuire}, {McIntyre}, {McIver}, {McManus},
  {McWilliams}, {Meacher}, {Meadors}, {Meidam}, {Melatos}, {Mendell}, {Mendoza-
  Gandara}, {Mercer}, {Merilh}, {Merzougui}, {Meshkov}, {Messenger}, {Messick},
  {Meyers}, {Mezzani}, {Miao}, {Michel}, {Middleton}, {Mikhailov}, {Milano},
  {Miller}, {Millhouse}, {Minenkov}, {Ming}, {Mirshekari}, {Mishra}, {Mitra},
  {Mitrofanov}, {Mitselmakher}, {Mittleman}, {Moggi}, {Mohan}, {Mohapatra},
  {Montani}, {Moore}, {Moore}, {Moraru}, {Moreno}, {Morriss}, {Mossavi},
  {Mours}, {Mow-Lowry}, {Mueller}, {Mueller}, {Muir}, {Mukherjee}, {Mukherjee},
  {Mukherjee}, {Mukund}, {Mullavey}, {Munch}, {Murphy}, {Murray}, {Mytidis},
  {Nardecchia}, {Naticchioni}, {Nayak}, {Necula}, {Nedkova}, {Nelemans},
  {Neri}, {Neunzert}, {Newton}, {Nguyen}, {Nielsen}, {Nissanke}, {Nitz},
  {Nocera}, {Nolting}, {Normandin}, {Nuttall}, {Oberling}, {Ochsner}, {O'Dell},
  {Oelker}, {Ogin}, {Oh}, {Oh}, {Ohme}, {Oliver}, {Oppermann}, {Oram},
  {O'Reilly}, {O'Shaughnessy}, {Ottaway}, {Ottens}, {Overmier}, {Owen}, {Pai},
  {Pai}, {Palamos}, {Palashov}, {Palomba}, {Pal-Singh}, {Pan}, {Pankow},
  {Pannarale}, {Pant}, {Paoletti}, {Paoli}, {Papa}, {Paris}, {Parker},
  {Pascucci}, {Pasqualetti}, {Passaquieti}, {Passuello}, {Patricelli},
  {Patrick}, {Pearlstone}, {Pedraza}, {Pedurand}, {Pekowsky}, {Pele}, {Penn},
  {Perreca}, {Phelps}, {Piccinni}, {Pichot}, {Piergiovanni}, {Pierro},
  {Pillant}, {Pinard}, {Pinto}, {Pitkin}, {Poggiani}, {Popolizio}, {Post},
  {Powell}, {Prasad}, {Predoi}, {Premachandra}, {Prestegard}, {Price},
  {Prijatelj}, {Principe}, {Privitera}, {Prix}, {Prodi}, {Prokhorov},
  {Puncken}, {Punturo}, {Puppo}, {P{\"u}rrer}, {Qi}, {Qin}, {Quetschke},
  {Quintero}, {Quitzow-James}, {Raab}, {Rabeling}, {Radkins}, {Raffai}, {Raja},
  {Rakhmanov}, {Rapagnani}, {Raymond}, {Razzano}, {Re}, {Read}, {Reed},
  {Regimbau}, {Rei}, {Reid}, {Reitze}, {Rew}, {Reyes}, {Ricci}, {Riles},
  {Robertson}, {Robie}, {Robinet}, {Rocchi}, {Rolland}, {Rollins}, {Roma},
  {Romano}, {Romano}, {Romanov}, {Romie}, {Rosi{\'n}ska}, {Rowan},
  {R{\"u}diger}, {Ruggi}, {Ryan}, {Sachdev}, {Sadecki}, {Sadeghian}, {Salconi},
  {Saleem}, {Salemi}, {Samajdar}, {Sammut}, {Sanchez}, {Sandberg}, {Sandeen},
  {Sanders}, {Sassolas}, {Sathyaprakash}, {Saulson}, {Sauter}, {Savage},
  {Sawadsky}, {Schale}, {Schilling}, {Schmidt}, {Schmidt}, {Schnabel},
  {Schofield}, {Sch{\"o}nbeck}, {Schreiber}, {Schuette}, {Schutz}, {Scott},
  {Scott}, {Sellers}, {Sentenac}, {Sequino}, {Sergeev}, {Serna}, {Setyawati},
  {Sevigny}, {Shaddock}, {Shah}, {Shahriar}, {Shaltev}, {Shao}, {Shapiro},
  {Shawhan}, {Sheperd}, {Shoemaker}, {Shoemaker}, {Siellez}, {Siemens}, {Sigg},
  {Silva}, {Simakov}, {Singer}, {Singer}, {Singh}, {Singh}, {Singhal},
  {Sintes}, {Slagmolen}, {Smith}, {Smith}, {Smith}, {Son}, {Sorazu},
  {Sorrentino}, {Souradeep}, {Srivastava}, {Staley}, {Steinke}, {Steinlechner},
  {Steinlechner}, {Steinmeyer}, {Stephens}, {Stevenson}, {Stone}, {Strain},
  {Straniero}, {Stratta}, {Strauss}, {Strigin}, {Sturani}, {Stuver},
  {Summerscales}, {Sun}, {Sutton}, {Swinkels}, {Szczepa{\'n}czyk}, {Tacca},
  {Talukder}, {Tanner}, {T{\'a}pai}, {Tarabrin}, {Taracchini}, {Taylor},
  {Theeg}, {Thirugnanasambandam}, {Thomas}, {Thomas}, {Thomas}, {Thorne},
  {Thorne}, {Thrane}, {Tiwari}, {Tiwari}, {Tokmakov}, {Tomlinson}, {Tonelli},
  {Torres}, {Torrie}, {T{\"o}yr{\"a}}, {Travasso}, {Traylor}, {Trifir{\`o}},
  {Tringali}, {Trozzo}, {Tse}, {Turconi}, {Tuyenbayev}, {Ugolini},
  {Unnikrishnan}, {Urban}, {Usman}, {Vahlbruch}, {Vajente}, {Valdes}, {van
  Bakel}, {van Beuzekom}, {van den Brand}, {van den Broeck}, {Vander-Hyde},
  {van der Schaaf}, {van Heijningen}, {van Veggel}, {Vardaro}, {Vass},
  {Vas{\'u}th}, {Vaulin}, {Vecchio}, {Vedovato}, {Veitch}, {Veitch},
  {Venkateswara}, {Verkindt}, {Vetrano}, {Vicer{\'e}}, {Vinciguerra}, {Vine},
  {Vinet}, {Vitale}, {Vo}, {Vocca}, {Vorvick}, {Voss}, {Vousden}, {Vyatchanin},
  {Wade}, {Wade}, {Wade}, {Walker}, {Wallace}, {Walsh}, {Wang}, {Wang}, {Wang},
  {Wang}, {Wang}, {Ward}, {Warner}, {Was}, {Weaver}, {Wei}, {Weinert},
  {Weinstein}, {Weiss}, {Welborn}, {Wen}, {We{\ss}els}, {Westphal}, {Wette},
  {Whelan}, {White}, {Whiting}, {Williams}, {Williamson}, {Willis}, {Willke},
  {Wimmer}, {Winkler}, {Wipf}, {Wittel}, {Woan}, {Worden}, {Wright}, {Wu},
  {Yablon}, {Yam}, {Yamamoto}, {Yancey}, {Yap}, {Yu}, {Yvert}, {Zadro{\.z}ny},
  {Zangrando}, {Zanolin}, {Zendri}, {Zevin}, {Zhang}, {Zhang}, {Zhang},
  {Zhang}, {Zhao}, {Zhou}, {Zhou}, {Zhu}, {Zucker}, {Zuraw}, {and}, {Zweizig},
  {LIGO Scientific Collaboration}, \& {Virgo
  Collaboration}}]{2016ApJ...818L..22A}
---. 2016{\natexlab{d}}, \apj, 818, L22, \dodoi{10.3847/2041-8205/818/2/L22}

\bibitem[{{Abbott} {et~al.}(2016{\natexlab{e}}){Abbott}, {Abbott}, {Abbott},
  {Abernathy}, {Acernese}, {Ackley}, {Adams}, {Adams}, {Addesso}, {Adhikari},
  {Adya}, {Affeldt}, {Agathos}, {Agatsuma}, {Aggarwal}, {Aguiar}, {Aiello},
  {Ain}, {Ajith}, {Allen}, {Allocca}, {Altin}, {Anderson}, {Anderson}, {Arai},
  {Araya}, {Arceneaux}, {Areeda}, {Arnaud}, {Arun}, {Ascenzi}, {Ashton}, {Ast},
  {Aston}, {Astone}, {Aufmuth}, {Aulbert}, {Babak}, {Bacon}, {Bader}, {Baker},
  {Baldaccini}, {Ballardin}, {Ballmer}, {Barayoga}, {Barclay}, {Barish},
  {Barker}, {Barone}, {Barr}, {Barsotti}, {Barsuglia}, {Barta}, {Bartlett},
  {Bartos}, {Bassiri}, {Basti}, {Batch}, {Baune}, {Bavigadda}, {Bazzan},
  {Behnke}, {Bejger}, {Bell}, {Bell}, {Berger}, {Bergman}, {Bergmann}, {Berry},
  {Bersanetti}, {Bertolini}, {Betzwieser}, {Bhagwat}, {Bhandare}, {Bilenko},
  {Billingsley}, {Birch}, {Birney}, {Biscans}, {Bisht}, {Bitossi}, {Biwer},
  {Bizouard}, {Blackburn}, {Blair}, {Blair}, {Blair}, {Bloemen}, {Bock},
  {Bodiya}, {Boer}, {Bogaert}, {Bogan}, {Bohe}, {Bojtos}, {Bond}, {Bondu},
  {Bonnand}, {Boom}, {Bork}, {Boschi}, {Bose}, {Bouffanais}, {Bozzi},
  {Bradaschia}, {Brady}, {Braginsky}, {Branchesi}, {Brau}, {Briant}, {Brillet},
  {Brinkmann}, {Brisson}, {Brockill}, {Brooks}, {Brown}, {Brown}, {Brown},
  {Buchanan}, {Buikema}, {Bulik}, {Bulten}, {Buonanno}, {Buskulic}, {Buy},
  {Byer}, {Cadonati}, {Cagnoli}, {Cahillane}, {Calder{\'o}n Bustillo},
  {Callister}, {Calloni}, {Camp}, {Cannon}, {Cao}, {Capano}, {Capocasa},
  {Carbognani}, {Caride}, {Casanueva Diaz}, {Casentini}, {Caudill},
  {Cavagli{\`a}}, {Cavalier}, {Cavalieri}, {Cella}, {Cepeda}, {Cerboni
  Baiardi}, {Cerretani}, {Cesarini}, {Chakraborty}, {Chalermsongsak},
  {Chamberlin}, {Chan}, {Chao}, {Charlton}, {Chassande-Mottin}, {Chen}, {Chen},
  {Cheng}, {Chincarini}, {Chiummo}, {Cho}, {Cho}, {Chow}, {Christensen}, {Chu},
  {Chua}, {Chung}, {Ciani}, {Clara}, {Clark}, {Cleva}, {Coccia}, {Cohadon},
  {Colla}, {Collette}, {Cominsky}, {Constancio}, {Conte}, {Conti}, {Cook},
  {Corbitt}, {Cornish}, {Corsi}, {Cortese}, {Costa}, {Coughlin}, {Coughlin},
  {Coulon}, {Countryman}, {Couvares}, {Cowan}, {Coward}, {Cowart}, {Coyne},
  {Coyne}, {Craig}, {Creighton}, {Cripe}, {Crowder}, {Cumming}, {Cunningham},
  {Cuoco}, {Dal Canton}, {Danilishin}, {D'Antonio}, {Danzmann}, {Darman},
  {Dattilo}, {Dave}, {Daveloza}, {Davier}, {Davies}, {Daw}, {Day}, {De},
  {DeBra}, {Debreczeni}, {Degallaix}, {De Laurentis}, {Del{\'e}glise}, {Del
  Pozzo}, {Denker}, {Dent}, {Dereli}, {Dergachev}, {De Rosa}, {DeRosa},
  {DeSalvo}, {Dhurandhar}, {D{\'\i}az}, {Di Fiore}, {Di Giovanni}, {Di Lieto},
  {Di Pace}, {Di Palma}, {Di Virgilio}, {Dojcinoski}, {Dolique}, {Donovan},
  {Dooley}, {Doravari}, {Douglas}, {Downes}, {Drago}, {Drever}, {Driggers},
  {Du}, {Ducrot}, {Dwyer}, {Edo}, {Edwards}, {Effler}, {Eggenstein}, {Ehrens},
  {Eichholz}, {Eikenberry}, {Engels}, {Essick}, {Etzel}, {Evans}, {Evans},
  {Everett}, {Factourovich}, {Fafone}, {Fair}, {Fairhurst}, {Fan}, {Fang},
  {Farinon}, {Farr}, {Farr}, {Favata}, {Fays}, {Fehrmann}, {Fejer}, {Ferrante},
  {Ferreira}, {Ferrini}, {Fidecaro}, {Fiori}, {Fiorucci}, {Fisher}, {Flaminio},
  {Fletcher}, {Fong}, {Fournier}, {Franco}, {Frasca}, {Frasconi}, {Frei},
  {Freise}, {Frey}, {Frey}, {Fricke}, {Fritschel}, {Frolov}, {Fulda}, {Fyffe},
  {Gabbard}, {Gair}, {Gammaitoni}, {Gaonkar}, {Garufi}, {Gatto}, {Gaur},
  {Gehrels}, {Gemme}, {Gendre}, {Genin}, {Gennai}, {George}, {Gergely},
  {Germain}, {Ghosh}, {Ghosh}, {Giaime}, {Giardina}, {Giazotto}, {Gill},
  {Glaefke}, {Goetz}, {Goetz}, {Gondan}, {Gonz{\'a}lez}, {Gonzalez Castro},
  {Gopakumar}, {Gordon}, {Gorodetsky}, {Gossan}, {Gosselin}, {Gouaty}, {Graef},
  {Graff}, {Granata}, {Grant}, {Gras}, {Gray}, {Greco}, {Green}, {Groot},
  {Grote}, {Grunewald}, {Guidi}, {Guo}, {Gupta}, {Gupta}, {Gushwa},
  {Gustafson}, {Gustafson}, {Hacker}, {Hall}, {Hall}, {Hammond}, {Haney},
  {Hanke}, {Hanks}, {Hanna}, {Hannam}, {Hanson}, {Hardwick}, {Harms}, {Harry},
  {Harry}, {Hart}, {Hartman}, {Haster}, {Haughian}, {Heidmann}, {Heintze},
  {Heitmann}, {Hello}, {Hemming}, {Hendry}, {Heng}, {Hennig}, {Heptonstall},
  {Heurs}, {Hild}, {Hoak}, {Hodge}, {Hofman}, {Hollitt}, {Holt}, {Holz},
  {Hopkins}, {Hosken}, {Hough}, {Houston}, {Howell}, {Hu}, {Huang}, {Huerta},
  {Huet}, {Hughey}, {Husa}, {Huttner}, {Huynh-Dinh}, {Idrisy}, {Indik},
  {Ingram}, {Inta}, {Isa}, {Isac}, {Isi}, {Islas}, {Isogai}, {Iyer}, {Izumi},
  {Jacqmin}, {Jang}, {Jani}, {Jaranowski}, {Jawahar}, {Jim{\'e}nez-Forteza},
  {Johnson}, {Jones}, {Jones}, {Jonker}, {Ju}, {K}, {Kalaghatgi}, {Kalogera},
  {Kandhasamy}, {Kang}, {Kanner}, {Karki}, {Kasprzack}, {Katsavounidis},
  {Katzman}, {Kaufer}, {Kaur}, {Kawabe}, {Kawazoe}, {K{\'e}f{\'e}lian}, {Kehl},
  {Keitel}, {Kelley}, {Kells}, {Kennedy}, {Key}, {Khalaidovski}, {Khalili},
  {Khan}, {Khan}, {Khan}, {Khazanov}, {Kijbunchoo}, {Kim}, {Kim}, {Kim}, {Kim},
  {Kim}, {Kim}, {King}, {King}, {Kinzel}, {Kissel}, {Kleybolte}, {Klimenko},
  {Koehlenbeck}, {Kokeyama}, {Koley}, {Kondrashov}, {Kontos}, {Korobko},
  {Korth}, {Kowalska}, {Kozak}, {Kringel}, {Krishnan}, {Kr{\'o}lak}, {Krueger},
  {Kuehn}, {Kumar}, {Kuo}, {Kutynia}, {Lackey}, {Landry}, {Lange}, {Lantz},
  {Lasky}, {Lazzarini}, {Lazzaro}, {Leaci}, {Leavey}, {Lebigot}, {Lee}, {Lee},
  {Lee}, {Lee}, {Lenon}, {Leonardi}, {Leong}, {Leroy}, {Letendre}, {Levin},
  {Levine}, {Li}, {Libson}, {Littenberg}, {Lockerbie}, {Logue}, {Lombardi},
  {Lord}, {Lorenzini}, {Loriette}, {Lormand}, {Losurdo}, {Lough}, {L{\"u}ck},
  {Lundgren}, {Luo}, {Lynch}, {Ma}, {MacDonald}, {Machenschalk}, {MacInnis},
  {Macleod}, {Maga{\\textasciitilde n}a-Sandoval}, {Magee}, {Mageswaran},
  {Majorana}, {Maksimovic}, {Malvezzi}, {Man}, {Mandel}, {Mandic}, {Mangano},
  {Mansell}, {Manske}, {Mantovani}, {Marchesoni}, {Marion}, {M{\'a}rka},
  {M{\'a}rka}, {Markosyan}, {Maros}, {Martelli}, {Martellini}, {Martin},
  {Martin}, {Martynov}, {Marx}, {Mason}, {Masserot}, {Massinger}, {Masso-Reid},
  {Matichard}, {Matone}, {Mavalvala}, {Mazumder}, {Mazzolo}, {McCarthy},
  {McClelland}, {McCormick}, {McGuire}, {McIntyre}, {McIver}, {McManus},
  {McWilliams}, {Meacher}, {Meadors}, {Meidam}, {Melatos}, {Mendell}, {Mendoza-
  Gandara}, {Mercer}, {Merilh}, {Merzougui}, {Meshkov}, {Messenger}, {Messick},
  {Meyers}, {Mezzani}, {Miao}, {Michel}, {Middleton}, {Mikhailov}, {Milano},
  {Miller}, {Millhouse}, {Minenkov}, {Ming}, {Mirshekari}, {Mishra}, {Mitra},
  {Mitrofanov}, {Mitselmakher}, {Mittleman}, {Moggi}, {Mohan}, {Mohapatra},
  {Montani}, {Moore}, {Moore}, {Moraru}, {Moreno}, {Morriss}, {Mossavi},
  {Mours}, {Mow-Lowry}, {Mueller}, {Mueller}, {Muir}, {Mukherjee}, {Mukherjee},
  {Mukherjee}, {Mukund}, {Mullavey}, {Munch}, {Murphy}, {Murray}, {Mytidis},
  {Nardecchia}, {Naticchioni}, {Nayak}, {Necula}, {Nedkova}, {Nelemans},
  {Neri}, {Neunzert}, {Newton}, {Nguyen}, {Nielsen}, {Nissanke}, {Nitz},
  {Nocera}, {Nolting}, {Normandin}, {Nuttall}, {Oberling}, {Ochsner}, {O'Dell},
  {Oelker}, {Ogin}, {Oh}, {Oh}, {Ohme}, {Oliver}, {Oppermann}, {Oram},
  {O'Reilly}, {O'Shaughnessy}, {Ottaway}, {Ottens}, {Overmier}, {Owen}, {Pai},
  {Pai}, {Palamos}, {Palashov}, {Palomba}, {Pal-Singh}, {Pan}, {Pankow},
  {Pannarale}, {Pant}, {Paoletti}, {Paoli}, {Papa}, {Paris}, {Parker},
  {Pascucci}, {Pasqualetti}, {Passaquieti}, {Passuello}, {Patricelli},
  {Patrick}, {Pearlstone}, {Pedraza}, {Pedurand}, {Pekowsky}, {Pele}, {Penn},
  {Perreca}, {Phelps}, {Piccinni}, {Pichot}, {Piergiovanni}, {Pierro},
  {Pillant}, {Pinard}, {Pinto}, {Pitkin}, {Poggiani}, {Popolizio}, {Porter},
  {Post}, {Powell}, {Prasad}, {Predoi}, {Premachandra}, {Prestegard}, {Price},
  {Prijatelj}, {Principe}, {Privitera}, {Prodi}, {Prokhorov}, {Puncken},
  {Punturo}, {Puppo}, {P{\"u}rrer}, {Qi}, {Qin}, {Quetschke}, {Quintero},
  {Quitzow-James}, {Raab}, {Rabeling}, {Radkins}, {Raffai}, {Raja},
  {Rakhmanov}, {Rapagnani}, {Raymond}, {Razzano}, {Re}, {Read}, {Reed},
  {Regimbau}, {Rei}, {Reid}, {Reitze}, {Rew}, {Reyes}, {Ricci}, {Riles},
  {Robertson}, {Robie}, {Robinet}, {Rocchi}, {Rolland}, {Rollins}, {Roma},
  {Romano}, {Romanov}, {Romie}, {Rosi{\'n}ska}, {Rowan}, {R{\"u}diger},
  {Ruggi}, {Ryan}, {Sachdev}, {Sadecki}, {Sadeghian}, {Salconi}, {Saleem},
  {Salemi}, {Samajdar}, {Sammut}, {Sampson}, {Sanchez}, {Sandberg}, {Sandeen},
  {Sanders}, {Sassolas}, {Sathyaprakash}, {Saulson}, {Sauter}, {Savage},
  {Sawadsky}, {Schale}, {Schilling}, {Schmidt}, {Schmidt}, {Schnabel},
  {Schofield}, {Sch{\"o}nbeck}, {Schreiber}, {Schuette}, {Schutz}, {Scott},
  {Scott}, {Sellers}, {Sengupta}, {Sentenac}, {Sequino}, {Sergeev}, {Serna},
  {Setyawati}, {Sevigny}, {Shaddock}, {Shah}, {Shahriar}, {Shaltev}, {Shao},
  {Shapiro}, {Shawhan}, {Sheperd}, {Shoemaker}, {Shoemaker}, {Siellez},
  {Siemens}, {Sigg}, {Silva}, {Simakov}, {Singer}, {Singer}, {Singh}, {Singh},
  {Singhal}, {Sintes}, {Slagmolen}, {Smith}, {Smith}, {Smith}, {Son}, {Sorazu},
  {Sorrentino}, {Souradeep}, {Srivastava}, {Staley}, {Steinke}, {Steinlechner},
  {Steinlechner}, {Steinmeyer}, {Stephens}, {Stevenson}, {Stone}, {Strain},
  {Straniero}, {Stratta}, {Strauss}, {Strigin}, {Sturani}, {Stuver},
  {Summerscales}, {Sun}, {Sutton}, {Swinkels}, {Szczepa{\'n}czyk}, {Tacca},
  {Talukder}, {Tanner}, {T{\'a}pai}, {Tarabrin}, {Taracchini}, {Taylor},
  {Theeg}, {Thirugnanasambandam}, {Thomas}, {Thomas}, {Thomas}, {Thorne},
  {Thorne}, {Thrane}, {Tiwari}, {Tiwari}, {Tokmakov}, {Tomlinson}, {Tonelli},
  {Torres}, {Torrie}, {T{\"o}yr{\"a}}, {Travasso}, {Traylor}, {Trifir{\`o}},
  {Tringali}, {Trozzo}, {Tse}, {Turconi}, {Tuyenbayev}, {Ugolini},
  {Unnikrishnan}, {Urban}, {Usman}, {Vahlbruch}, {Vajente}, {Valdes},
  {Vallisneri}, {van Bakel}, {van Beuzekom}, {van den Brand}, {Van Den Broeck},
  {Vander-Hyde}, {van der Schaaf}, {van Heijningen}, {van Veggel}, {Vardaro},
  {Vass}, {Vas{\'u}th}, {Vaulin}, {Vecchio}, {Vedovato}, {Veitch}, {Veitch},
  {Venkateswara}, {Verkindt}, {Vetrano}, {Vicer{\'e}}, {Vinciguerra}, {Vine},
  {Vinet}, {Vitale}, {Vo}, {Vocca}, {Vorvick}, {Voss}, {Vousden}, {Vyatchanin},
  {Wade}, {Wade}, {Wade}, {Walker}, {Wallace}, {Walsh}, {Wang}, {Wang}, {Wang},
  {Wang}, {Wang}, {Ward}, {Warner}, {Was}, {Weaver}, {Wei}, {Weinert},
  {Weinstein}, {Weiss}, {Welborn}, {Wen}, {We{\ss}els}, {Westphal}, {Wette},
  {Whelan}, {White}, {Whiting}, {Williams}, {Williamson}, {Willis}, {Willke},
  {Wimmer}, {Winkler}, {Wipf}, {Wittel}, {Woan}, {Worden}, {Wright}, {Wu},
  {Yablon}, {Yam}, {Yamamoto}, {Yancey}, {Yap}, {Yu}, {Yvert}, {Zadro{\.z}ny},
  {Zangrando}, {Zanolin}, {Zendri}, {Zevin}, {Zhang}, {Zhang}, {Zhang},
  {Zhang}, {Zhao}, {Zhou}, {Zhou}, {Zhu}, {Zucker}, {Zuraw}, {Zweizig}, {LIGO
  Scientific Collaboration}, \& {Virgo Collaboration}}]{2016ApJ...833L...1A}
---. 2016{\natexlab{e}}, \apj, 833, L1, \dodoi{10.3847/2041-8205/833/1/L1}

\bibitem[{{Abbott} {et~al.}(2016{\natexlab{f}}){Abbott}, {Abbott}, {Abbott},
  {Abernathy}, {Acernese}, {Ackley}, {Adams}, {Adams}, {Addesso}, {Adhikari},
  {Adya}, {Affeldt}, {Agathos}, {Agatsuma}, {Aggarwal}, {Aguiar}, {Aiello},
  {Ain}, {Ajith}, {Allen}, {Allocca}, {Altin}, {Anderson}, {Anderson}, {Arai},
  {Araya}, {Arceneaux}, {Areeda}, {Arnaud}, {Arun}, {Ascenzi}, {Ashton}, {Ast},
  {Aston}, {Astone}, {Aufmuth}, {Aulbert}, {Babak}, {Bacon}, {Bader}, {Baker},
  {Baldaccini}, {Ballardin}, {Ballmer}, {Barayoga}, {Barclay}, {Barish},
  {Barker}, {Barone}, {Barr}, {Barsotti}, {Barsuglia}, {Barta}, {Bartlett},
  {Bartos}, {Bassiri}, {Basti}, {Batch}, {Baune}, {Bavigadda}, {Bazzan},
  {Behnke}, {Bejger}, {Bell}, {Bell}, {Berger}, {Bergman}, {Bergmann}, {Berry},
  {Bersanetti}, {Bertolini}, {Betzwieser}, {Bhagwat}, {Bhandare}, {Bilenko},
  {Billingsley}, {Birch}, {Birney}, {Biscans}, {Bisht}, {Bitossi}, {Biwer},
  {Bizouard}, {Blackburn}, {Blair}, {Blair}, {Blair}, {Bloemen}, {Bock},
  {Bodiya}, {Boer}, {Bogaert}, {Bogan}, {Bohe}, {Bojtos}, {Bond}, {Bondu},
  {Bonnand}, {Boom}, {Bork}, {Boschi}, {Bose}, {Bouffanais}, {Bozzi},
  {Bradaschia}, {Brady}, {Braginsky}, {Branchesi}, {Brau}, {Briant}, {Brillet},
  {Brinkmann}, {Brisson}, {Brockill}, {Brooks}, {Brown}, {Brown}, {Buchanan},
  {Buikema}, {Bulik}, {Bulten}, {Buonanno}, {Buskulic}, {Buy}, {Byer},
  {Cadonati}, {Cagnoli}, {Cahillane}, {Bustillo}, {Callister}, {Calloni},
  {Camp}, {Cannon}, {Cao}, {Capano}, {Capocasa}, {Carbognani}, {Caride},
  {Diaz}, {Casentini}, {Caudill}, {Cavagli{\`a}}, {Cavalier}, {Cavalieri},
  {Cella}, {Cepeda}, {Baiardi}, {Cerretani}, {Cesarini}, {Chakraborty},
  {Chalermsongsak}, {Chamberlin}, {Chan}, {Chao}, {Charlton},
  {Chassande-Mottin}, {Chen}, {Chen}, {Cheng}, {Chincarini}, {Chiummo}, {Cho},
  {Cho}, {Chow}, {Christensen}, {Chu}, {Chua}, {Chung}, {Ciani}, {Clara},
  {Clark}, {Cleva}, {Coccia}, {Cohadon}, {Colla}, {Collette}, {Cominsky},
  {Constancio}, {Conte}, {Conti}, {Cook}, {Corbitt}, {Cornish}, {Corsi},
  {Cortese}, {Costa}, {Coughlin}, {Coughlin}, {Coulon}, {Countryman},
  {Couvares}, {Cowan}, {Coward}, {Cowart}, {Coyne}, {Coyne}, {Craig},
  {Creighton}, {Cripe}, {Crowder}, {Cumming}, {Cunningham}, {Cuoco}, {Dal
  Canton}, {Danilishin}, {D'Antonio}, {Danzmann}, {Darman}, {Dattilo}, {Dave},
  {Daveloza}, {Davier}, {Davies}, {Daw}, {Day}, {DeBra}, {Debreczeni},
  {Degallaix}, {De Laurentis}, {Del{\'e}glise}, {Del Pozzo}, {Denker}, {Dent},
  {Dereli}, {Dergachev}, {DeRosa}, {De Rosa}, {DeSalvo}, {Dhurandhar},
  {D{\'\i}az}, {Di Fiore}, {Di Giovanni}, {Di Lieto}, {Di Pace}, {Di Palma},
  {Di Virgilio}, {Dojcinoski}, {Dolique}, {Donovan}, {Dooley}, {Doravari},
  {Douglas}, {Downes}, {Drago}, {Drever}, {Driggers}, {Du}, {Ducrot}, {Dwyer},
  {Edo}, {Edwards}, {Effler}, {Eggenstein}, {Ehrens}, {Eichholz}, {Eikenberry},
  {Engels}, {Essick}, {Etzel}, {Evans}, {Evans}, {Everett}, {Factourovich},
  {Fafone}, {Fair}, {Fairhurst}, {Fan}, {Fang}, {Farinon}, {Farr}, {Farr},
  {Favata}, {Fays}, {Fehrmann}, {Fejer}, {Ferrante}, {Ferreira}, {Ferrini},
  {Fidecaro}, {Fiori}, {Fiorucci}, {Fisher}, {Flaminio}, {Fletcher},
  {Fournier}, {Franco}, {Frasca}, {Frasconi}, {Frei}, {Freise}, {Frey}, {Frey},
  {Fricke}, {Fritschel}, {Frolov}, {Fulda}, {Fyffe}, {Gabbard}, {Gair},
  {Gammaitoni}, {Gaonkar}, {Garufi}, {Gatto}, {Gaur}, {Gehrels}, {Gemme},
  {Gendre}, {Genin}, {Gennai}, {George}, {Gergely}, {Germain}, {Ghosh},
  {Ghosh}, {Giaime}, {Giardina}, {Giazotto}, {Gill}, {Glaefke}, {Goetz},
  {Goetz}, {Gondan}, {Gonz{\'a}lez}, {Castro}, {Gopakumar}, {Gordon},
  {Gorodetsky}, {Gossan}, {Gosselin}, {Gouaty}, {Graef}, {Graff}, {Granata},
  {Grant}, {Gras}, {Gray}, {Greco}, {Green}, {Groot}, {Grote}, {Grunewald},
  {Guidi}, {Guo}, {Gupta}, {Gupta}, {Gushwa}, {Gustafson}, {Gustafson},
  {Hacker}, {Hall}, {Hall}, {Hammond}, {Haney}, {Hanke}, {Hanks}, {Hanna},
  {Hannam}, {Hanson}, {Hardwick}, {Haris}, {Harms}, {Harry}, {Harry}, {Hart},
  {Hartman}, {Haster}, {Haughian}, {Heidmann}, {Heintze}, {Heitmann}, {Hello},
  {Hemming}, {Hendry}, {Heng}, {Hennig}, {Heptonstall}, {Heurs}, {Hild},
  {Hoak}, {Hodge}, {Hofman}, {Hollitt}, {Holt}, {Holz}, {Hopkins}, {Hosken},
  {Hough}, {Houston}, {Howell}, {Hu}, {Huang}, {Huerta}, {Huet}, {Hughey},
  {Husa}, {Huttner}, {Huynh-Dinh}, {Idrisy}, {Indik}, {Ingram}, {Inta}, {Isa},
  {Isac}, {Isi}, {Islas}, {Isogai}, {Iyer}, {Izumi}, {Jacqmin}, {Jang}, {Jani},
  {Jaranowski}, {Jawahar}, {Jim{\'e}nez-Forteza}, {Johnson}, {Jones}, {Jones},
  {Jonker}, {Ju}, {Kalaghatgi}, {Kalogera}, {Kandhasamy}, {Kang}, {Kanner},
  {Karki}, {Kasprzack}, {Katsavounidis}, {Katzman}, {Kaufer}, {Kaur}, {Kawabe},
  {Kawazoe}, {K{\'e}f{\'e}lian}, {Kehl}, {Keitel}, {Kelley}, {Kells},
  {Kennedy}, {Key}, {Khalaidovski}, {Khalili}, {Khan}, {Khan}, {Khan},
  {Khazanov}, {Kijbunchoo}, {Kim}, {Kim}, {Kim}, {Kim}, {Kim}, {Kim}, {King},
  {King}, {Kinzel}, {Kissel}, {Kleybolte}, {Klimenko}, {Koehlenbeck},
  {Kokeyama}, {Koley}, {Kondrashov}, {Kontos}, {Korobko}, {Korth}, {Kowalska},
  {Kozak}, {Kringel}, {Kr{\'o}lak}, {Krueger}, {Kuehn}, {Kumar}, {Kuo},
  {Kutynia}, {Lackey}, {Landry}, {Lange}, {Lantz}, {Lasky}, {Lazzarini},
  {Lazzaro}, {Leaci}, {Leavey}, {Lebigot}, {Lee}, {Lee}, {Lee}, {Lee}, {Lenon},
  {Leonardi}, {Leong}, {Leroy}, {Letendre}, {Levin}, {Levine}, {Li}, {Libson},
  {Littenberg}, {Lockerbie}, {Logue}, {Lombardi}, {Lord}, {Lorenzini},
  {Loriette}, {Lormand}, {Losurdo}, {Lough}, {L{\"u}ck}, {Lundgren}, {Luo},
  {Lynch}, {Ma}, {MacDonald}, {Machenschalk}, {MacInnis}, {Macleod},
  {Maga{\\textasciitilde n}a-Sandoval}, {Magee}, {Mageswaran}, {Majorana},
  {Maksimovic}, {Malvezzi}, {Man}, {Mandel}, {Mandic}, {Mangano}, {Mansell},
  {Manske}, {Mantovani}, {Marchesoni}, {Marion}, {M{\'a}rka}, {M{\'a}rka},
  {Markosyan}, {Maros}, {Martelli}, {Martellini}, {Martin}, {Martin},
  {Martynov}, {Marx}, {Mason}, {Masserot}, {Massinger}, {Masso-Reid},
  {Matichard}, {Matone}, {Mavalvala}, {Mazumder}, {Mazzolo}, {McCarthy},
  {McClelland}, {McCormick}, {McGuire}, {McIntyre}, {McIver}, {McManus},
  {McWilliams}, {Meacher}, {Meadors}, {Meidam}, {Melatos}, {Mendell}, {Mendoza-
  Gandara}, {Mercer}, {Merilh}, {Merzougui}, {Meshkov}, {Messenger}, {Messick},
  {Meyers}, {Mezzani}, {Miao}, {Michel}, {Middleton}, {Mikhailov}, {Milano},
  {Miller}, {Millhouse}, {Minenkov}, {Ming}, {Mirshekari}, {Mishra}, {Mitra},
  {Mitrofanov}, {Mitselmakher}, {Mittleman}, {Moggi}, {Mohan}, {Mohapatra},
  {Montani}, {Moore}, {Moore}, {Moraru}, {Moreno}, {Morriss}, {Mossavi},
  {Mours}, {Mow-Lowry}, {Mueller}, {Mueller}, {Muir}, {Mukherjee}, {Mukherjee},
  {Mukherjee}, {Mukund}, {Mullavey}, {Munch}, {Murphy}, {Murray}, {Mytidis},
  {Nardecchia}, {Naticchioni}, {Nayak}, {Necula}, {Nedkova}, {Nelemans},
  {Neri}, {Neunzert}, {Newton}, {Nguyen}, {Nielsen}, {Nissanke}, {Nitz},
  {Nocera}, {Nolting}, {Normandin}, {Nuttall}, {Oberling}, {Ochsner}, {O'Dell},
  {Oelker}, {Ogin}, {Oh}, {Oh}, {Ohme}, {Oliver}, {Oppermann}, {Oram},
  {O'Reilly}, {O'Shaughnessy}, {Ottaway}, {Ottens}, {Overmier}, {Owen}, {Pai},
  {Pai}, {Palamos}, {Palashov}, {Palomba}, {Pal-Singh}, {Pan}, {Pankow},
  {Pannarale}, {Pant}, {Paoletti}, {Paoli}, {Papa}, {Paris}, {Parker},
  {Pascucci}, {Pasqualetti}, {Passaquieti}, {Passuello}, {Patricelli},
  {Patrick}, {Pearlstone}, {Pedraza}, {Pedurand}, {Pekowsky}, {Pele}, {Penn},
  {Perreca}, {Phelps}, {Piccinni}, {Pichot}, {Piergiovanni}, {Pierro},
  {Pillant}, {Pinard}, {Pinto}, {Pitkin}, {Poggiani}, {Popolizio}, {Post},
  {Powell}, {Prasad}, {Predoi}, {Premachandra}, {Prestegard}, {Price},
  {Prijatelj}, {Principe}, {Privitera}, {Prodi}, {Prokhorov}, {Puncken},
  {Punturo}, {Puppo}, {P{\"u}rrer}, {Qi}, {Qin}, {Quetschke}, {Quintero},
  {Quitzow-James}, {Raab}, {Rabeling}, {Radkins}, {Raffai}, {Raja},
  {Rakhmanov}, {Rapagnani}, {Raymond}, {Razzano}, {Re}, {Read}, {Reed},
  {Regimbau}, {Rei}, {Reid}, {Reitze}, {Rew}, {Reyes}, {Ricci}, {Riles},
  {Robertson}, {Robie}, {Robinet}, {Rocchi}, {Rolland}, {Rollins}, {Roma},
  {Romano}, {Romano}, {Romanov}, {Romie}, {Rosi{\'n}ska}, {Rowan},
  {R{\"u}diger}, {Ruggi}, {Ryan}, {Sachdev}, {Sadecki}, {Sadeghian}, {Salconi},
  {Saleem}, {Salemi}, {Samajdar}, {Sammut}, {Sanchez}, {Sandberg}, {Sandeen},
  {Sanders}, {Sassolas}, {Sathyaprakash}, {Saulson}, {Sauter}, {Savage},
  {Sawadsky}, {Schale}, {Schilling}, {Schmidt}, {Schmidt}, {Schnabel},
  {Schofield}, {Sch{\"o}nbeck}, {Schreiber}, {Schuette}, {Schutz}, {Scott},
  {Scott}, {Sellers}, {Sentenac}, {Sequino}, {Sergeev}, {Serna}, {Setyawati},
  {Sevigny}, {Shaddock}, {Shah}, {Shahriar}, {Shaltev}, {Shao}, {Shapiro},
  {Shawhan}, {Sheperd}, {Shoemaker}, {Shoemaker}, {Siellez}, {Siemens}, {Sigg},
  {Silva}, {Simakov}, {Singer}, {Singer}, {Singh}, {Singh}, {Singhal},
  {Sintes}, {Slagmolen}, {Smith}, {Smith}, {Smith}, {Son}, {Sorazu},
  {Sorrentino}, {Souradeep}, {Srivastava}, {Staley}, {Steinke}, {Steinlechner},
  {Steinlechner}, {Steinmeyer}, {Stephens}, {Stone}, {Strain}, {Straniero},
  {Stratta}, {Strauss}, {Strigin}, {Sturani}, {Stuver}, {Summerscales}, {Sun},
  {Sutton}, {Swinkels}, {Szczepa{\'n}czyk}, {Tacca}, {Talukder}, {Tanner},
  {T{\'a}pai}, {Tarabrin}, {Taracchini}, {Taylor}, {Theeg},
  {Thirugnanasambandam}, {Thomas}, {Thomas}, {Thomas}, {Thorne}, {Thorne},
  {Thrane}, {Tiwari}, {Tiwari}, {Tokmakov}, {Tomlinson}, {Tonelli}, {Torres},
  {Torrie}, {T{\"o}yr{\"a}}, {Travasso}, {Traylor}, {Trifir{\`o}}, {Tringali},
  {Trozzo}, {Tse}, {Turconi}, {Tuyenbayev}, {Ugolini}, {Unnikrishnan}, {Urban},
  {Usman}, {Vahlbruch}, {Vajente}, {Valdes}, {van Bakel}, {van Beuzekom}, {van
  den Brand}, {Van Den Broeck}, {Vander- Hyde}, {van der Schaaf}, {van
  Heijningen}, {van Veggel}, {Vardaro}, {Vass}, {Vas{\'u}th}, {Vaulin},
  {Vecchio}, {Vedovato}, {Veitch}, {Veitch}, {Venkateswara}, {Verkindt},
  {Vetrano}, {Vicer{\'e}}, {Vinciguerra}, {Vine}, {Vinet}, {Vitale}, {Vo},
  {Vocca}, {Vorvick}, {Voss}, {Vousden}, {Vyatchanin}, {Wade}, {Wade}, {Wade},
  {Walker}, {Wallace}, {Walsh}, {Wang}, {Wang}, {Wang}, {Wang}, {Wang}, {Ward},
  {Warner}, {Was}, {Weaver}, {Wei}, {Weinert}, {Weinstein}, {Weiss}, {Welborn},
  {Wen}, {We{\ss}els}, {Westphal}, {Wette}, {Whelan}, {White}, {Whiting},
  {Williams}, {Williamson}, {Willis}, {Willke}, {Wimmer}, {Winkler}, {Wipf},
  {Wittel}, {Woan}, {Worden}, {Wright}, {Wu}, {Yablon}, {Yam}, {Yamamoto},
  {Yancey}, {Yap}, {Yu}, {Yvert}, {Zadro{\.Z}ny}, {Zangrando}, {Zanolin},
  {Zendri}, {Zevin}, {Zhang}, {Zhang}, {Zhang}, {Zhang}, {Zhao}, {Zhou},
  {Zhou}, {Zhu}, {Zucker}, {Zuraw}, {Zweizig}, {LIGO Scientific Collaboration},
  \& {Virgo Collaboration}}]{2016PhRvL.116m1102A}
---. 2016{\natexlab{f}}, \prl, 116, 131102,
  \dodoi{10.1103/PhysRevLett.116.131102}

\bibitem[{{Abbott} {et~al.}(2017{\natexlab{a}}){Abbott}, {Abbott}, {Abbott},
  {Acernese}, {Ackley}, {Adams}, {Adams}, {Addesso}, {Adhikari}, {Adya},
  {Affeldt}, {Afrough}, {Agarwal}, {Agathos}, {Agatsuma}, {Aggarwal}, {Aguiar},
  {Aiello}, {Ain}, {Ajith}, {Allen}, {Allen}, {Allocca}, {Altin}, {Amato},
  {Ananyeva}, {Anderson}, {Anderson}, {Angelova}, {Antier}, {Appert}, {Arai},
  {Araya}, {Areeda}, {Arnaud}, {Arun}, {Ascenzi}, {Ashton}, {Ast}, {Aston},
  {Astone}, {Atallah}, {Aufmuth}, {Aulbert}, {AultONeal}, {Austin},
  {Avila-Alvarez}, {Babak}, {Bacon}, {Bader}, {Bae}, {Baker}, {Baldaccini},
  {Ballardin}, {Ballmer}, {Banagiri}, {Barayoga}, {Barclay}, {Barish},
  {Barker}, {Barkett}, {Barone}, {Barr}, {Barsotti}, {Barsuglia}, {Barta},
  {Bartlett}, {Bartos}, {Bassiri}, {Basti}, {Batch}, {Bawaj}, {Bayley},
  {Bazzan}, {B{\'e}csy}, {Beer}, {Bejger}, {Belahcene}, {Bell}, {Berger},
  {Bergmann}, {Bero}, {Berry}, {Bersanetti}, {Bertolini}, {Betzwieser},
  {Bhagwat}, {Bhandare}, {Bilenko}, {Billingsley}, {Billman}, {Birch},
  {Birney}, {Birnholtz}, {Biscans}, {Biscoveanu}, {Bisht}, {Bitossi}, {Biwer},
  {Bizouard}, {Blackburn}, {Blackman}, {Blair}, {Blair}, {Blair}, {Bloemen},
  {Bock}, {Bode}, {Boer}, {Bogaert}, {Bohe}, {Bondu}, {Bonilla}, {Bonnand},
  {Boom}, {Bork}, {Boschi}, {Bose}, {Bossie}, {Bouffanais}, {Bozzi},
  {Bradaschia}, {Brady}, {Branchesi}, {Brau}, {Briant}, {Brillet}, {Brinkmann},
  {Brisson}, {Brockill}, {Broida}, {Brooks}, {Brown}, {Brown}, {Brunett},
  {Buchanan}, {Buikema}, {Bulik}, {Bulten}, {Buonanno}, {Buskulic}, {Buy},
  {Byer}, {Cabero}, {Cadonati}, {Cagnoli}, {Cahillane}, {Calder{\'o}n
  Bustillo}, {Callister}, {Calloni}, {Camp}, {Canepa}, {Canizares}, {Cannon},
  {Cao}, {Cao}, {Capano}, {Capocasa}, {Carbognani}, {Caride}, {Carney},
  {Casanueva Diaz}, {Casentini}, {Caudill}, {Cavagli{\`a}}, {Cavalier},
  {Cavalieri}, {Cella}, {Cepeda}, {Cerd{\'a}-Dur{\'a}n}, {Cerretani},
  {Cesarini}, {Chamberlin}, {Chan}, {Chao}, {Charlton}, {Chase}, {Chassande-
  Mottin}, {Chatterjee}, {Chatziioannou}, {Cheeseboro}, {Chen}, {Chen}, {Chen},
  {Cheng}, {Chia}, {Chincarini}, {Chiummo}, {Chmiel}, {Cho}, {Cho}, {Chow},
  {Christensen}, {Chu}, {Chua}, {Chua}, {Chung}, {Chung}, {Ciani}, {Ciolfi},
  {Cirelli}, {Cirone}, {Clara}, {Clark}, {Clearwater}, {Cleva}, {Cocchieri},
  {Coccia}, {Cohadon}, {Cohen}, {Colla}, {Collette}, {Cominsky}, {Constancio},
  {Conti}, {Cooper}, {Corban}, {Corbitt}, {Cordero-Carri{\'o}n}, {Corley},
  {Cornish}, {Corsi}, {Cortese}, {Costa}, {Coughlin}, {Coughlin}, {Coulon},
  {Countryman}, {Couvares}, {Covas}, {Cowan}, {Coward}, {Cowart}, {Coyne},
  {Coyne}, {Creighton}, {Creighton}, {Cripe}, {Crowder}, {Cullen}, {Cumming},
  {Cunningham}, {Cuoco}, {Dal Canton}, {D{\'a}lya}, {Danilishin}, {D'Antonio},
  {Danzmann}, {Dasgupta}, {Da Silva Costa}, {Dattilo}, {Dave}, {Davier},
  {Davis}, {Daw}, {Day}, {De}, {DeBra}, {Degallaix}, {De Laurentis},
  {Del{\'e}glise}, {Del Pozzo}, {Demos}, {Denker}, {Dent}, {De Pietri},
  {Dergachev}, {De Rosa}, {DeRosa}, {De Rossi}, {DeSalvo}, {de Varona},
  {Devenson}, {Dhurandhar}, {D{\'\i}az}, {Di Fiore}, {Di Giovanni}, {Di
  Girolamo}, {Di Lieto}, {Di Pace}, {Di Palma}, {Di Renzo}, {Doctor},
  {Dolique}, {Donovan}, {Dooley}, {Doravari}, {Dorrington}, {Douglas}, {Dovale
  {\'A}lvarez}, {Downes}, {Drago}, {Dreissigacker}, {Driggers}, {Du}, {Ducrot},
  {Dupej}, {Dwyer}, {Edo}, {Edwards}, {Effler}, {Eggenstein}, {Ehrens},
  {Eichholz}, {Eikenberry}, {Eisenstein}, {Essick}, {Estevez}, {Etienne},
  {Etzel}, {Evans}, {Evans}, {Factourovich}, {Fafone}, {Fair}, {Fairhurst},
  {Fan}, {Farinon}, {Farr}, {Farr}, {Fauchon-Jones}, {Favata}, {Fays}, {Fee},
  {Fehrmann}, {Feicht}, {Fejer}, {Fernandez-Galiana}, {Ferrante}, {Ferreira},
  {Ferrini}, {Fidecaro}, {Finstad}, {Fiori}, {Fiorucci}, {Fishbach}, {Fisher},
  {Fitz-Axen}, {Flaminio}, {Fletcher}, {Fong}, {Font}, {Forsyth}, {Forsyth},
  {Fournier}, {Frasca}, {Frasconi}, {Frei}, {Freise}, {Frey}, {Frey}, {Fries},
  {Fritschel}, {Frolov}, {Fulda}, {Fyffe}, {Gabbard}, {Gadre}, {Gaebel},
  {Gair}, {Gammaitoni}, {Ganija}, {Gaonkar}, {Garcia-Quiros}, {Garufi},
  {Gateley}, {Gaudio}, {Gaur}, {Gayathri}, {Gehrels}, {Gemme}, {Genin},
  {Gennai}, {George}, {George}, {Gergely}, {Germain}, {Ghonge}, {Ghosh},
  {Ghosh}, {Ghosh}, {Giaime}, {Giardina}, {Giazotto}, {Gill}, {Glover},
  {Goetz}, {Goetz}, {Gomes}, {Goncharov}, {Gonz{\'a}lez}, {Gonzalez Castro},
  {Gopakumar}, {Gorodetsky}, {Gossan}, {Gosselin}, {Gouaty}, {Grado}, {Graef},
  {Granata}, {Grant}, {Gras}, {Gray}, {Greco}, {Green}, {Gretarsson}, {Groot},
  {Grote}, {Grunewald}, {Gruning}, {Guidi}, {Guo}, {Gupta}, {Gupta}, {Gushwa},
  {Gustafson}, {Gustafson}, {Halim}, {Hall}, {Hall}, {Hamilton}, {Hammond},
  {Haney}, {Hanke}, {Hanks}, {Hanna}, {Hannam}, {Hannuksela}, {Hanson},
  {Hardwick}, {Harms}, {Harry}, {Harry}, {Hart}, {Haster}, {Haughian}, {Healy},
  {Heidmann}, {Heintze}, {Heitmann}, {Hello}, {Hemming}, {Hendry}, {Heng},
  {Hennig}, {Heptonstall}, {Heurs}, {Hild}, {Hinderer}, {Hoak}, {Hofman},
  {Holt}, {Holz}, {Hopkins}, {Horst}, {Hough}, {Houston}, {Howell}, {Hreibi},
  {Hu}, {Huerta}, {Huet}, {Hughey}, {Husa}, {Huttner}, {Huynh-Dinh}, {Indik},
  {Inta}, {Intini}, {Isa}, {Isac}, {Isi}, {Iyer}, {Izumi}, {Jacqmin}, {Jani},
  {Jaranowski}, {Jawahar}, {Jim{\'e}nez-Forteza}, {Johnson},
  {Johnson-McDaniel}, {Jones}, {Jones}, {Jonker}, {Ju}, {Junker}, {Kalaghatgi},
  {Kalogera}, {Kamai}, {Kandhasamy}, {Kang}, {Kanner}, {Kapadia}, {Karki},
  {Karvinen}, {Kasprzack}, {Katolik}, {Katsavounidis}, {Katzman}, {Kaufer},
  {Kawabe}, {K{\'e}f{\'e}lian}, {Keitel}, {Kemball}, {Kennedy}, {Kent}, {Key},
  {Khalili}, {Khan}, {Khan}, {Khan}, {Khazanov}, {Kijbunchoo}, {Kim}, {Kim},
  {Kim}, {Kim}, {Kim}, {Kim}, {Kimbrell}, {King}, {King}, {Kinley-Hanlon},
  {Kirchhoff}, {Kissel}, {Kleybolte}, {Klimenko}, {Knowles}, {Koch},
  {Koehlenbeck}, {Koley}, {Kondrashov}, {Kontos}, {Korobko}, {Korth},
  {Kowalska}, {Kozak}, {Kr{\"a}mer}, {Kringel}, {Krishnan}, {Kr{\'o}lak},
  {Kuehn}, {Kumar}, {Kumar}, {Kumar}, {Kuo}, {Kutynia}, {Kwang}, {Lackey},
  {Lai}, {Landry}, {Lang}, {Lange}, {Lantz}, {Lanza}, {Lartaux-Vollard},
  {Lasky}, {Laxen}, {Lazzarini}, {Lazzaro}, {Leaci}, {Leavey}, {Lee}, {Lee},
  {Lee}, {Lee}, {Lee}, {Lehmann}, {Lenon}, {Leonardi}, {Leroy}, {Letendre},
  {Levin}, {Li}, {Linker}, {Littenberg}, {Liu}, {Lo}, {Lockerbie}, {London},
  {Lord}, {Lorenzini}, {Loriette}, {Lormand}, {Losurdo}, {Lough}, {Lousto},
  {Lovelace}, {L{\"u}ck}, {Lumaca}, {Lundgren}, {Lynch}, {Ma}, {Macas},
  {Macfoy}, {Machenschalk}, {MacInnis}, {Macleod}, {Maga{\\textasciitilde n}a
  Hernandez}, {Maga{\\textasciitilde n}a-Sandoval}, {Maga{\\textasciitilde n}a
  Zertuche}, {Magee}, {Majorana}, {Maksimovic}, {Man}, {Mandic}, {Mangano},
  {Mansell}, {Manske}, {Mantovani}, {Marchesoni}, {Marion}, {M{\'a}rka},
  {M{\'a}rka}, {Markakis}, {Markosyan}, {Markowitz}, {Maros}, {Marquina},
  {Martelli}, {Martellini}, {Martin}, {Martin}, {Martynov}, {Mason}, {Massera},
  {Masserot}, {Massinger}, {Masso-Reid}, {Mastrogiovanni}, {Matas},
  {Matichard}, {Matone}, {Mavalvala}, {Mazumder}, {McCarthy}, {McClelland},
  {McCormick}, {McCuller}, {McGuire}, {McIntyre}, {McIver}, {McManus},
  {McNeill}, {McRae}, {McWilliams}, {Meacher}, {Meadors}, {Mehmet}, {Meidam},
  {Mejuto-Villa}, {Melatos}, {Mendell}, {Mercer}, {Merilh}, {Merzougui},
  {Meshkov}, {Messenger}, {Messick}, {Metzdorff}, {Meyers}, {Miao}, {Michel},
  {Middleton}, {Mikhailov}, {Milano}, {Miller}, {Miller}, {Miller},
  {Millhouse}, {Milovich-Goff}, {Minazzoli}, {Minenkov}, {Ming}, {Mishra},
  {Mitra}, {Mitrofanov}, {Mitselmakher}, {Mittleman}, {Moffa}, {Moggi},
  {Mogushi}, {Mohan}, {Mohapatra}, {Montani}, {Moore}, {Moraru}, {Moreno},
  {Morriss}, {Mours}, {Mow-Lowry}, {Mueller}, {Muir}, {Mukherjee}, {Mukherjee},
  {Mukherjee}, {Mukund}, {Mullavey}, {Munch}, {Mu{\\textasciitilde n}iz},
  {Muratore}, {Murray}, {Napier}, {Nardecchia}, {Naticchioni}, {Nayak},
  {Neilson}, {Nelemans}, {Nelson}, {Nery}, {Neunzert}, {Nevin}, {Newport},
  {Newton}, {Ng}, {Nguyen}, {Nichols}, {Nielsen}, {Nissanke}, {Nitz}, {Noack},
  {Nocera}, {Nolting}, {North}, {Nuttall}, {Oberling}, {O'Dea}, {Ogin}, {Oh},
  {Oh}, {Ohme}, {Okada}, {Oliver}, {Oppermann}, {Oram}, {O'Reilly}, {Ormiston},
  {Ortega}, {O'Shaughnessy}, {Ossokine}, {Ottaway}, {Overmier}, {Owen}, {Pace},
  {Page}, {Page}, {Pai}, {Pai}, {Palamos}, {Palashov}, {Palomba}, {Pal- Singh},
  {Pan}, {Pan}, {Pang}, {Pang}, {Pankow}, {Pannarale}, {Pant}, {Paoletti},
  {Paoli}, {Papa}, {Parida}, {Parker}, {Pascucci}, {Pasqualetti},
  {Passaquieti}, {Passuello}, {Patil}, {Patricelli}, {Pearlstone}, {Pedraza},
  {Pedurand}, {Pekowsky}, {Pele}, {Penn}, {Perez}, {Perreca}, {Perri},
  {Pfeiffer}, {Phelps}, {Piccinni}, {Pichot}, {Piergiovanni}, {Pierro},
  {Pillant}, {Pinard}, {Pinto}, {Pirello}, {Pitkin}, {Poe}, {Poggiani},
  {Popolizio}, {Porter}, {Post}, {Powell}, {Prasad}, {Pratt}, {Pratten},
  {Predoi}, {Prestegard}, {Prijatelj}, {Principe}, {Privitera}, {Prodi},
  {Prokhorov}, {Puncken}, {Punturo}, {Puppo}, {P{\"u}rrer}, {Qi}, {Quetschke},
  {Quintero}, {Quitzow-James}, {Raab}, {Rabeling}, {Radkins}, {Raffai}, {Raja},
  {Rajan}, {Rajbhandari}, {Rakhmanov}, {Ramirez}, {Ramos-Buades}, {Rapagnani},
  {Raymond}, {Razzano}, {Read}, {Regimbau}, {Rei}, {Reid}, {Reitze}, {Ren},
  {Reyes}, {Ricci}, {Ricker}, {Rieger}, {Riles}, {Rizzo}, {Robertson}, {Robie},
  {Robinet}, {Rocchi}, {Rolland}, {Rollins}, {Roma}, {Romano}, {Romel},
  {Romie}, {Rosi{\'n}ska}, {Ross}, {Rowan}, {R{\"u}diger}, {Ruggi}, {Rutins},
  {Ryan}, {Sachdev}, {Sadecki}, {Sadeghian}, {Sakellariadou}, {Salconi},
  {Saleem}, {Salemi}, {Samajdar}, {Sammut}, {Sampson}, {Sanchez}, {Sanchez},
  {Sanchis-Gual}, {Sandberg}, {Sanders}, {Sassolas}, {Sathyaprakash},
  {Saulson}, {Sauter}, {Savage}, {Sawadsky}, {Schale}, {Scheel}, {Scheuer},
  {Schmidt}, {Schmidt}, {Schnabel}, {Schofield}, {Sch{\"o}nbeck}, {Schreiber},
  {Schuette}, {Schulte}, {Schutz}, {Schwalbe}, {Scott}, {Scott}, {Seidel},
  {Sellers}, {Sengupta}, {Sentenac}, {Sequino}, {Sergeev}, {Shaddock},
  {Shaffer}, {Shah}, {Shahriar}, {Shaner}, {Shao}, {Shapiro}, {Shawhan},
  {Sheperd}, {Shoemaker}, {Shoemaker}, {Siellez}, {Siemens}, {Sieniawska},
  {Sigg}, {Silva}, {Singer}, {Singh}, {Singhal}, {Sintes}, {Slagmolen},
  {Smith}, {Smith}, {Smith}, {Somala}, {Son}, {Sonnenberg}, {Sorazu},
  {Sorrentino}, {Souradeep}, {Spencer}, {Srivastava}, {Staats}, {Staley},
  {Steinke}, {Steinlechner}, {Steinlechner}, {Steinmeyer}, {Stevenson},
  {Stone}, {Stops}, {Strain}, {Stratta}, {Strigin}, {Strunk}, {Sturani},
  {Stuver}, {Summerscales}, {Sun}, {Sunil}, {Suresh}, {Sutton}, {Swinkels},
  {Szczepa{\'n}czyk}, {Tacca}, {Tait}, {Talbot}, {Talukder}, {Tanner},
  {T{\'a}pai}, {Taracchini}, {Tasson}, {Taylor}, {Taylor}, {Tewari}, {Theeg},
  {Thies}, {Thomas}, {Thomas}, {Thomas}, {Thorne}, {Thrane}, {Tiwari},
  {Tiwari}, {Tokmakov}, {Toland}, {Tonelli}, {Tornasi}, {Torres- Forn{\'e}},
  {Torrie}, {T{\"o}yr{\"a}}, {Travasso}, {Traylor}, \&
  {Trinastic}}]{2017ApJ...851L..35A}
---. 2017{\natexlab{a}}, \apj, 851, L35, \dodoi{10.3847/2041-8213/aa9f0c}

\bibitem[{{Abbott} {et~al.}(2017{\natexlab{b}}){Abbott}, {Abbott}, {Abbott},
  {Acernese}, {Ackley}, {Adams}, {Adams}, {Addesso}, {Adhikari}, {Adya},
  {Affeldt}, {Afrough}, {Agarwal}, {Agathos}, {Agatsuma}, {Aggarwal}, {Aguiar},
  {Aiello}, {Ain}, {Ajith}, {Allen}, {Allen}, {Allocca}, {Altin}, {Amato},
  {Ananyeva}, {Anderson}, {Anderson}, {Angelova}, {Antier}, {Appert}, {Arai},
  {Araya}, {Areeda}, {Arnaud}, {Arun}, {Ascenzi}, {Ashton}, {Ast}, {Aston},
  {Astone}, {Atallah}, {Aufmuth}, {Aulbert}, {AultONeal}, {Austin},
  {Avila-Alvarez}, {Babak}, {Bacon}, {Bader}, {Bae}, {Bailes}, {Baker},
  {Baldaccini}, {Ballardin}, {Ballmer}, {Banagiri}, {Barayoga}, {Barclay},
  {Barish}, {Barker}, {Barkett}, {Barone}, {Barr}, {Barsotti}, {Barsuglia},
  {Barta}, {Barthelmy}, {Bartlett}, {Bartos}, {Bassiri}, {Basti}, {Batch},
  {Bawaj}, {Bayley}, {Bazzan}, {B{\'e}csy}, {Beer}, {Bejger}, {Belahcene},
  {Bell}, {Berger}, {Bergmann}, {Bernuzzi}, {Bero}, {Berry}, {Bersanetti},
  {Bertolini}, {Betzwieser}, {Bhagwat}, {Bhandare}, {Bilenko}, {Billingsley},
  {Billman}, {Birch}, {Birney}, {Birnholtz}, {Biscans}, {Biscoveanu}, {Bisht},
  {Bitossi}, {Biwer}, {Bizouard}, {Blackburn}, {Blackman}, {Blair}, {Blair},
  {Blair}, {Bloemen}, {Bock}, {Bode}, {Boer}, {Bogaert}, {Bohe}, {Bondu},
  {Bonilla}, {Bonnand}, {Boom}, {Bork}, {Boschi}, {Bose}, {Bossie},
  {Bouffanais}, {Bozzi}, {Bradaschia}, {Brady}, {Branchesi}, {Brau}, {Briant},
  {Brillet}, {Brinkmann}, {Brisson}, {Brockill}, {Broida}, {Brooks}, {Brown},
  {Brown}, {Brunett}, {Buchanan}, {Buikema}, {Bulik}, {Bulten}, {Buonanno},
  {Buskulic}, {Buy}, {Byer}, {Cabero}, {Cadonati}, {Cagnoli}, {Cahillane},
  {Calder{\'o}n Bustillo}, {Callister}, {Calloni}, {Camp}, {Canepa},
  {Canizares}, {Cannon}, {Cao}, {Cao}, {Capano}, {Capocasa}, {Carbognani},
  {Caride}, {Carney}, {Carullo}, {Casanueva Diaz}, {Casentini}, {Caudill},
  {Cavagli{\`a}}, {Cavalier}, {Cavalieri}, {Cella}, {Cepeda},
  {Cerd{\'a}-Dur{\'a}n}, {Cerretani}, {Cesarini}, {Chamberlin}, {Chan}, {Chao},
  {Charlton}, {Chase}, {Chassande-Mottin}, {Chatterjee}, {Chatziioannou},
  {Cheeseboro}, {Chen}, {Chen}, {Chen}, {Cheng}, {Chia}, {Chincarini},
  {Chiummo}, {Chmiel}, {Cho}, {Cho}, {Chow}, {Christensen}, {Chu}, {Chua},
  {Chua}, {Chung}, {Chung}, {Ciani}, {Ciolfi}, {Cirelli}, {Cirone}, {Clara},
  {Clark}, {Clearwater}, {Cleva}, {Cocchieri}, {Coccia}, {Cohadon}, {Cohen},
  {Colla}, {Collette}, {Cominsky}, {Constancio}, {Conti}, {Cooper}, {Corban},
  {Corbitt}, {Cordero-Carri{\'o}n}, {Corley}, {Cornish}, {Corsi}, {Cortese},
  {Costa}, {Coughlin}, {Coughlin}, {Coulon}, {Countryman}, {Couvares}, {Covas},
  {Cowan}, {Coward}, {Cowart}, {Coyne}, {Coyne}, {Creighton}, {Creighton},
  {Cripe}, {Crowder}, {Cullen}, {Cumming}, {Cunningham}, {Cuoco}, {Dal Canton},
  {D{\'a}lya}, {Danilishin}, {D'Antonio}, {Danzmann}, {Dasgupta}, {Da Silva
  Costa}, {Dattilo}, {Dave}, {Davier}, {Davis}, {Daw}, {Day}, {De}, {DeBra},
  {Degallaix}, {De Laurentis}, {Del{\'e}glise}, {Del Pozzo}, {Demos}, {Denker},
  {Dent}, {De Pietri}, {Dergachev}, {De Rosa}, {DeRosa}, {De Rossi}, {DeSalvo},
  {de Varona}, {Devenson}, {Dhurandhar}, {D{\'\i}az}, {Dietrich}, {Di Fiore},
  {Di Giovanni}, {Di Girolamo}, {Di Lieto}, {Di Pace}, {Di Palma}, {Di Renzo},
  {Doctor}, {Dolique}, {Donovan}, {Dooley}, {Doravari}, {Dorrington},
  {Douglas}, {Dovale {\'A}lvarez}, {Downes}, {Drago}, {Dreissigacker},
  {Driggers}, {Du}, {Ducrot}, {Dudi}, {Dupej}, {Dwyer}, {Edo}, {Edwards},
  {Effler}, {Eggenstein}, {Ehrens}, {Eichholz}, {Eikenberry}, {Eisenstein},
  {Essick}, {Estevez}, {Etienne}, {Etzel}, {Evans}, {Evans}, {Factourovich},
  {Fafone}, {Fair}, {Fairhurst}, {Fan}, {Farinon}, {Farr}, {Farr},
  {Fauchon-Jones}, {Favata}, {Fays}, {Fee}, {Fehrmann}, {Feicht}, {Fejer},
  {Fernandez-Galiana}, {Ferrante}, {Ferreira}, {Ferrini}, {Fidecaro},
  {Finstad}, {Fiori}, {Fiorucci}, {Fishbach}, {Fisher}, {Fitz-Axen},
  {Flaminio}, {Fletcher}, {Fong}, {Font}, {Forsyth}, {Forsyth}, {Fournier},
  {Frasca}, {Frasconi}, {Frei}, {Freise}, {Frey}, {Frey}, {Fries}, {Fritschel},
  {Frolov}, {Fulda}, {Fyffe}, {Gabbard}, {Gadre}, {Gaebel}, {Gair},
  {Gammaitoni}, {Ganija}, {Gaonkar}, {Garcia-Quiros}, {Garufi}, {Gateley},
  {Gaudio}, {Gaur}, {Gayathri}, {Gehrels}, {Gemme}, {Genin}, {Gennai},
  {George}, {George}, {Gergely}, {Germain}, {Ghonge}, {Ghosh}, {Ghosh},
  {Ghosh}, {Giaime}, {Giardina}, {Giazotto}, {Gill}, {Glover}, {Goetz},
  {Goetz}, {Gomes}, {Goncharov}, {Gonz{\'a}lez}, {Gonzalez Castro},
  {Gopakumar}, {Gorodetsky}, {Gossan}, {Gosselin}, {Gouaty}, {Grado}, {Graef},
  {Granata}, {Grant}, {Gras}, {Gray}, {Greco}, {Green}, {Gretarsson}, {Groot},
  {Grote}, {Grunewald}, {Gruning}, {Guidi}, {Guo}, {Gupta}, {Gupta}, {Gushwa},
  {Gustafson}, {Gustafson}, {Halim}, {Hall}, {Hall}, {Hamilton}, {Hammond},
  {Haney}, {Hanke}, {Hanks}, {Hanna}, {Hannam}, {Hannuksela}, {Hanson},
  {Hardwick}, {Harms}, {Harry}, {Harry}, {Hart}, {Haster}, {Haughian}, {Healy},
  {Heidmann}, {Heintze}, {Heitmann}, {Hello}, {Hemming}, {Hendry}, {Heng},
  {Hennig}, {Heptonstall}, {Heurs}, {Hild}, {Hinderer}, {Ho}, {Hoak}, {Hofman},
  {Holt}, {Holz}, {Hopkins}, {Horst}, {Hough}, {Houston}, {Howell}, {Hreibi},
  {Hu}, {Huerta}, {Huet}, {Hughey}, {Husa}, {Huttner}, {Huynh-Dinh}, {Indik},
  {Inta}, {Intini}, {Isa}, {Isac}, {Isi}, {Iyer}, {Izumi}, {Jacqmin}, {Jani},
  {Jaranowski}, {Jawahar}, {Jim{\'e}nez- Forteza}, {Johnson},
  {Johnson-McDaniel}, {Jones}, {Jones}, {Jonker}, {Ju}, {Junker}, {Kalaghatgi},
  {Kalogera}, {Kamai}, {Kandhasamy}, {Kang}, {Kanner}, {Kapadia}, {Karki},
  {Karvinen}, {Kasprzack}, {Kastaun}, {Katolik}, {Katsavounidis}, {Katzman},
  {Kaufer}, {Kawabe}, {K{\'e}f{\'e}lian}, {Keitel}, {Kemball}, {Kennedy},
  {Kent}, {Key}, {Khalili}, {Khan}, {Khan}, {Khan}, {Khazanov}, {Kijbunchoo},
  {Kim}, {Kim}, {Kim}, {Kim}, {Kim}, {Kim}, {Kimbrell}, {King}, {King},
  {Kinley-Hanlon}, {Kirchhoff}, {Kissel}, {Kleybolte}, {Klimenko}, {Knowles},
  {Koch}, {Koehlenbeck}, {Koley}, {Kondrashov}, {Kontos}, {Korobko}, {Korth},
  {Kowalska}, {Kozak}, {Kr{\"a}mer}, {Kringel}, {Krishnan}, {Kr{\'o}lak},
  {Kuehn}, {Kumar}, {Kumar}, {Kumar}, {Kuo}, {Kutynia}, {Kwang}, {Lackey},
  {Lai}, {Landry}, {Lang}, {Lange}, {Lantz}, {Lanza}, {Larson},
  {Lartaux-Vollard}, {Lasky}, {Laxen}, {Lazzarini}, {Lazzaro}, {Leaci},
  {Leavey}, {Lee}, {Lee}, {Lee}, {Lee}, {Lee}, {Lehmann}, {Lenon}, {Leon},
  {Leonardi}, {Leroy}, {Letendre}, {Levin}, {Li}, {Linker}, {Littenberg},
  {Liu}, {Liu}, {Lo}, {Lockerbie}, {London}, {Lord}, {Lorenzini}, {Loriette},
  {Lormand}, {Losurdo}, {Lough}, {Lousto}, {Lovelace}, {L{\"u}ck}, {Lumaca},
  {Lundgren}, {Lynch}, {Ma}, {Macas}, {Macfoy}, {Machenschalk}, {MacInnis},
  {Macleod}, {Maga{\\textasciitilde n}a Hernandez}, {Maga{\\textasciitilde
  n}a-Sandoval}, {Maga{\\textasciitilde n}a Zertuche}, {Magee}, {Majorana},
  {Maksimovic}, {Man}, {Mandic}, {Mangano}, {Mansell}, {Manske}, {Mantovani},
  {Marchesoni}, {Marion}, {M{\'a}rka}, {M{\'a}rka}, {Markakis}, {Markosyan},
  {Markowitz}, {Maros}, {Marquina}, {Marsh}, {Martelli}, {Martellini},
  {Martin}, {Martin}, {Martynov}, {Marx}, {Mason}, {Massera}, {Masserot},
  {Massinger}, {Masso-Reid}, {Mastrogiovanni}, {Matas}, {Matichard}, {Matone},
  {Mavalvala}, {Mazumder}, {McCarthy}, {McClelland}, {McCormick}, {McCuller},
  {McGuire}, {McIntyre}, {McIver}, {McManus}, {McNeill}, {McRae}, {McWilliams},
  {Meacher}, {Meadors}, {Mehmet}, {Meidam}, {Mejuto-Villa}, {Melatos},
  {Mendell}, {Mercer}, {Merilh}, {Merzougui}, {Meshkov}, {Messenger},
  {Messick}, {Metzdorff}, {Meyers}, {Miao}, {Michel}, {Middleton}, {Mikhailov},
  {Milano}, {Miller}, {Miller}, {Miller}, {Millhouse}, {Milovich-Goff},
  {Minazzoli}, {Minenkov}, {Ming}, {Mishra}, {Mitra}, {Mitrofanov},
  {Mitselmakher}, {Mittleman}, {Moffa}, {Moggi}, {Mogushi}, {Mohan},
  {Mohapatra}, {Molina}, {Montani}, {Moore}, {Moraru}, {Moreno}, {Morisaki},
  {Morriss}, {Mours}, {Mow-Lowry}, {Mueller}, {Muir}, {Mukherjee}, {Mukherjee},
  {Mukherjee}, {Mukund}, {Mullavey}, {Munch}, {Mu{\\textasciitilde n}iz},
  {Muratore}, {Murray}, {Nagar}, {Napier}, {Nardecchia}, {Naticchioni},
  {Nayak}, {Neilson}, {Nelemans}, {Nelson}, {Nery}, {Neunzert}, {Nevin},
  {Newport}, {Newton}, {Ng}, {Nguyen}, {Nguyen}, {Nichols}, {Nielsen},
  {Nissanke}, {Nitz}, {Noack}, {Nocera}, {Nolting}, {North}, {Nuttall},
  {Oberling}, {O'Dea}, {Ogin}, {Oh}, {Oh}, {Ohme}, {Okada}, {Oliver},
  {Oppermann}, {Oram}, {O'Reilly}, {Ormiston}, {Ortega}, {O'Shaughnessy},
  {Ossokine}, {Ottaway}, {Overmier}, {Owen}, {Pace}, {Page}, {Page}, {Pai},
  {Pai}, {Palamos}, {Palashov}, {Palomba}, {Pal- Singh}, {Pan}, {Pan}, {Pang},
  {Pang}, {Pankow}, {Pannarale}, {Pant}, {Paoletti}, {Paoli}, {Papa}, {Parida},
  {Parker}, {Pascucci}, {Pasqualetti}, {Passaquieti}, {Passuello}, {Patil},
  {Patricelli}, {Pearlstone}, {Pedraza}, {Pedurand}, {Pekowsky}, {Pele},
  {Penn}, {Perez}, {Perreca}, {Perri}, {Pfeiffer}, {Phelps}, {Piccinni},
  {Pichot}, {Piergiovanni}, {Pierro}, {Pillant}, {Pinard}, {Pinto}, {Pirello},
  {Pitkin}, {Poe}, {Poggiani}, {Popolizio}, {Porter}, {Post}, {Powell},
  {Prasad}, {Pratt}, {Pratten}, {Predoi}, {Prestegard}, {Prijatelj},
  {Principe}, {Privitera}, {Prix}, {Prodi}, {Prokhorov}, {Puncken}, {Punturo},
  {Puppo}, {P{\"u}rrer}, {Qi}, {Quetschke}, {Quintero}, {Quitzow-James},
  {Raab}, {Rabeling}, {Radkins}, {Raffai}, {Raja}, {Rajan}, {Rajbhandari},
  {Rakhmanov}, {Ramirez}, {Ramos-Buades}, {Rapagnani}, {Raymond}, {Razzano},
  {Read}, {Regimbau}, {Rei}, {Reid}, {Reitze}, {Ren}, {Reyes}, {Ricci},
  {Ricker}, {Rieger}, {Riles}, {Rizzo}, {Robertson}, {Robie}, {Robinet},
  {Rocchi}, {Rolland}, {Rollins}, {Roma}, {Romano}, {Romano}, {Romel}, {Romie},
  {Rosi{\'n}ska}, {Ross}, {Rowan}, {R{\"u}diger}, {Ruggi}, {Rutins}, {Ryan},
  {Sachdev}, {Sadecki}, {Sadeghian}, {Sakellariadou}, {Salconi}, {Saleem},
  {Salemi}, {Samajdar}, {Sammut}, {Sampson}, {Sanchez}, {Sanchez},
  {Sanchis-Gual}, {Sandberg}, {Sanders}, {Sassolas}, {Sathyaprakash},
  {Saulson}, {Sauter}, {Savage}, {Sawadsky}, {Schale}, {Scheel}, {Scheuer},
  {Schmidt}, {Schmidt}, {Schnabel}, {Schofield}, {Sch{\"o}nbeck}, {Schreiber},
  {Schuette}, {Schulte}, {Schutz}, {Schwalbe}, {Scott}, {Scott}, {Seidel},
  {Sellers}, {Sengupta}, {Sentenac}, {Sequino}, {Sergeev}, {Shaddock},
  {Shaffer}, {Shah}, {Shahriar}, {Shaner}, {Shao}, {Shapiro}, {Shawhan},
  {Sheperd}, {Shoemaker}, {Shoemaker}, {Siellez}, {Siemens}, {Sieniawska},
  {Sigg}, {Silva}, {Singer}, {Singh}, {Singhal}, {Sintes}, {Slagmolen},
  {Smith}, {Smith}, {Smith}, {Somala}, {Son}, {Sonnenberg}, {Sorazu},
  {Sorrentino}, {Souradeep}, {Spencer}, {Srivastava}, {Staats}, {Staley},
  {Steinke}, {Steinlechner}, {Steinlechner}, {Steinmeyer}, {Stevenson},
  {Stone}, {Stops}, {Strain}, {Stratta}, {Strigin}, {Strunk}, {Sturani},
  {Stuver}, {Summerscales}, {Sun}, {Sunil}, {Suresh}, {Sutton}, {Swinkels},
  {Szczepa{\'n}czyk}, {Tacca}, {Tait}, {Talbot}, {Talukder}, {Tanner},
  {T{\'a}pai}, {Taracchini}, {Tasson}, {Taylor}, {Taylor}, {Tewari}, \&
  {Theeg}}]{2017PhRvL.119p1101A}
---. 2017{\natexlab{b}}, \prl, 119, 161101,
  \dodoi{10.1103/PhysRevLett.119.161101}

\bibitem[{{Abbott} {et~al.}(2017{\natexlab{c}}){Abbott}, {Abbott}, {Abbott},
  {Acernese}, {Ackley}, {Adams}, {Adams}, {Addesso}, {Adhikari}, {Adya},
  {Affeldt}, {Afrough}, {Agarwal}, {Agathos}, {Agatsuma}, {Aggarwal}, {Aguiar},
  {Aiello}, {Ain}, {Ajith}, {Allen}, {Allen}, {Allocca}, {Aloy}, {Altin},
  {Amato}, {Ananyeva}, {Anderson}, {Anderson}, {Angelova}, {Antier}, {Appert},
  {Arai}, {Araya}, {Areeda}, {Arnaud}, {Arun}, {Ascenzi}, {Ashton}, {Ast},
  {Aston}, {Astone}, {Atallah}, {Aufmuth}, {Aulbert}, {AultONeal}, {Austin},
  {Avila-Alvarez}, {Babak}, {Bacon}, {Bader}, {Bae}, {Baker}, {Baldaccini},
  {Ballardin}, {Ballmer}, {Banagiri}, {Barayoga}, {Barclay}, {Barish},
  {Barker}, {Barkett}, {Barone}, {Barr}, {Barsotti}, {Barsuglia}, {Barta},
  {Bartlett}, {Bartos}, {Bassiri}, {Basti}, {Batch}, {Bawaj}, {Bayley},
  {Bazzan}, {B{\'e}csy}, {Beer}, {Bejger}, {Belahcene}, {Bell}, {Berger},
  {Bergmann}, {Bero}, {Berry}, {Bersanetti}, {Bertolini}, {Betzwieser},
  {Bhagwat}, {Bhandare}, {Bilenko}, {Billingsley}, {Billman}, {Birch},
  {Birney}, {Birnholtz}, {Biscans}, {Biscoveanu}, {Bisht}, {Bitossi}, {Biwer},
  {Bizouard}, {Blackburn}, {Blackman}, {Blair}, {Blair}, {Blair}, {Bloemen},
  {Bock}, {Bode}, {Boer}, {Bogaert}, {Bohe}, {Bondu}, {Bonilla}, {Bonnand},
  {Boom}, {Bork}, {Boschi}, {Bose}, {Bossie}, {Bouffanais}, {Bozzi},
  {Bradaschia}, {Brady}, {Branchesi}, {Brau}, {Briant}, {Brillet}, {Brinkmann},
  {Brisson}, {Brockill}, {Broida}, {Brooks}, {Brown}, {Brown}, {Brunett},
  {Buchanan}, {Buikema}, {Bulik}, {Bulten}, {Buonanno}, {Buskulic}, {Buy},
  {Byer}, {Cabero}, {Cadonati}, {Cagnoli}, {Cahillane}, {Calder{\'o}n
  Bustillo}, {Callister}, {Calloni}, {Camp}, {Canepa}, {Canizares}, {Cannon},
  {Cao}, {Cao}, {Capano}, {Capocasa}, {Carbognani}, {Caride}, {Carney},
  {Casanueva Diaz}, {Casentini}, {Caudill}, {Cavagli{\`a}}, {Cavalier},
  {Cavalieri}, {Cella}, {Cepeda}, {Cerd{\'a}-Dur{\'a}n}, {Cerretani},
  {Cesarini}, {Chamberlin}, {Chan}, {Chao}, {Charlton}, {Chase}, {Chassande-
  Mottin}, {Chatterjee}, {Chatziioannou}, {Cheeseboro}, {Chen}, {Chen}, {Chen},
  {Cheng}, {Chia}, {Chincarini}, {Chiummo}, {Chmiel}, {Cho}, {Cho}, {Chow},
  {Christensen}, {Chu}, {Chua}, {Chua}, {Chung}, {Chung}, {Ciani}, {Ciolfi},
  {Cirelli}, {Cirone}, {Clara}, {Clark}, {Clearwater}, {Cleva}, {Cocchieri},
  {Coccia}, {Cohadon}, {Cohen}, {Colla}, {Collette}, {Cominsky}, {Constancio},
  {Conti}, {Cooper}, {Corban}, {Corbitt}, {Cordero-Carri{\'o}n}, {Corley},
  {Cornish}, {Corsi}, {Cortese}, {Costa}, {Coughlin}, {Coughlin}, {Coulon},
  {Countryman}, {Couvares}, {Covas}, {Cowan}, {Coward}, {Cowart}, {Coyne},
  {Coyne}, {Creighton}, {Creighton}, {Cripe}, {Crowder}, {Cullen}, {Cumming},
  {Cunningham}, {Cuoco}, {Dal Canton}, {D{\'a}lya}, {Danilishin}, {D'Antonio},
  {Danzmann}, {Dasgupta}, {Da Silva Costa}, {Dattilo}, {Dave}, {Davier},
  {Davis}, {Daw}, {Day}, {De}, {DeBra}, {Degallaix}, {De Laurentis},
  {Del{\'e}glise}, {Del Pozzo}, {Demos}, {Denker}, {Dent}, {De Pietri},
  {Dergachev}, {De Rosa}, {DeRosa}, {De Rossi}, {DeSalvo}, {de Varona},
  {Devenson}, {Dhurandhar}, {D{\'\i}az}, {Di Fiore}, {Di Giovanni}, {Di
  Girolamo}, {Di Lieto}, {Di Pace}, {Di Palma}, {Di Renzo}, {Doctor},
  {Dolique}, {Donovan}, {Dooley}, {Doravari}, {Dorrington}, {Douglas}, {Dovale
  {\'A}lvarez}, {Downes}, {Drago}, {Dreissigacker}, {Driggers}, {Du}, {Ducrot},
  {Dupej}, {Dwyer}, {Edo}, {Edwards}, {Effler}, {Eggenstein}, {Ehrens},
  {Eichholz}, {Eikenberry}, {Eisenstein}, {Essick}, {Estevez}, {Etienne},
  {Etzel}, {Evans}, {Evans}, {Factourovich}, {Fafone}, {Fair}, {Fairhurst},
  {Fan}, {Farinon}, {Farr}, {Farr}, {Fauchon-Jones}, {Favata}, {Fays}, {Fee},
  {Fehrmann}, {Feicht}, {Fejer}, {Fernandez-Galiana}, {Ferrante}, {Ferreira},
  {Ferrini}, {Fidecaro}, {Finstad}, {Fiori}, {Fiorucci}, {Fishbach}, {Fisher},
  {Fitz-Axen}, {Flaminio}, {Fletcher}, {Fong}, {Font}, {Forsyth}, {Forsyth},
  {Fournier}, {Frasca}, {Frasconi}, {Frei}, {Freise}, {Frey}, {Frey}, {Fries},
  {Fritschel}, {Frolov}, {Fulda}, {Fyffe}, {Gabbard}, {Gadre}, {Gaebel},
  {Gair}, {Gammaitoni}, {Ganija}, {Gaonkar}, {Garcia-Quiros}, {Garufi},
  {Gateley}, {Gaudio}, {Gaur}, {Gayathri}, {Gehrels}, {Gemme}, {Genin},
  {Gennai}, {George}, {George}, {Gergely}, {Germain}, {Ghonge}, {Ghosh},
  {Ghosh}, {Ghosh}, {Giaime}, {Giardina}, {Giazotto}, {Gill}, {Glover},
  {Goetz}, {Goetz}, {Gomes}, {Goncharov}, {Gonz{\'a}lez}, {Gonzalez Castro},
  {Gopakumar}, {Gorodetsky}, {Gossan}, {Gosselin}, {Gouaty}, {Grado}, {Graef},
  {Granata}, {Grant}, {Gras}, {Gray}, {Greco}, {Green}, {Gretarsson}, {Groot},
  {Grote}, {Grunewald}, {Gruning}, {Guidi}, {Guo}, {Gupta}, {Gupta}, {Gushwa},
  {Gustafson}, {Gustafson}, {Halim}, {Hall}, {Hall}, {Hamilton}, {Hammond},
  {Haney}, {Hanke}, {Hanks}, {Hanna}, {Hannam}, {Hannuksela}, {Hanson},
  {Hardwick}, {Harms}, {Harry}, {Harry}, {Hart}, {Haster}, {Haughian}, {Healy},
  {Heidmann}, {Heintze}, {Heitmann}, {Hello}, {Hemming}, {Hendry}, {Heng},
  {Hennig}, {Heptonstall}, {Heurs}, {Hild}, {Hinderer}, {Hoak}, {Hofman},
  {Holt}, {Holz}, {Hopkins}, {Horst}, {Hough}, {Houston}, {Howell}, {Hreibi},
  {Hu}, {Huerta}, {Huet}, {Hughey}, {Husa}, {Huttner}, {Huynh-Dinh}, {Indik},
  {Inta}, {Intini}, {Isa}, {Isac}, {Isi}, {Iyer}, {Izumi}, {Jacqmin}, {Jani},
  {Jaranowski}, {Jawahar}, {Jim{\'e}nez-Forteza}, {Johnson},
  {Johnson-McDaniel}, {Jones}, {Jones}, {Jonker}, {Ju}, {Junker}, {Kalaghatgi},
  {Kalogera}, {Kamai}, {Kandhasamy}, {Kang}, {Kanner}, {Kapadia}, {Karki},
  {Karvinen}, {Kasprzack}, {Kastaun}, {Katolik}, {Katsavounidis}, {Katzman},
  {Kaufer}, {Kawabe}, {K{\'e}f{\'e}lian}, {Keitel}, {Kemball}, {Kennedy},
  {Kent}, {Key}, {Khalili}, {Khan}, {Khan}, {Khan}, {Khazanov}, {Kijbunchoo},
  {Kim}, {Kim}, {Kim}, {Kim}, {Kim}, {Kim}, {Kimbrell}, {King}, {King},
  {Kinley-Hanlon}, {Kirchhoff}, {Kissel}, {Kleybolte}, {Klimenko}, {Knowles},
  {Koch}, {Koehlenbeck}, {Koley}, {Kondrashov}, {Kontos}, {Korobko}, {Korth},
  {Kowalska}, {Kozak}, {Kr{\"a}mer}, {Kringel}, {Krishnan}, {Kr{\'o}lak},
  {Kuehn}, {Kumar}, {Kumar}, {Kumar}, {Kuo}, {Kutynia}, {Kwang}, {Lackey},
  {Lai}, {Landry}, {Lang}, {Lange}, {Lantz}, {Lanza}, {Lartaux-Vollard},
  {Lasky}, {Laxen}, {Lazzarini}, {Lazzaro}, {Leaci}, {Leavey}, {Lee}, {Lee},
  {Lee}, {Lee}, {Lee}, {Lehmann}, {Lenon}, {Leonardi}, {Leroy}, {Letendre},
  {Levin}, {Li}, {Linker}, {Littenberg}, {Liu}, {Lo}, {Lockerbie}, {London},
  {Lord}, {Lorenzini}, {Loriette}, {Lormand}, {Losurdo}, {Lough}, {Lousto},
  {Lovelace}, {L{\"u}ck}, {Lumaca}, {Lundgren}, {Lynch}, {Ma}, {Macas},
  {Macfoy}, {Machenschalk}, {MacInnis}, {Macleod}, {Maga{\\textasciitilde n}a
  Hernandez}, {Maga{\\textasciitilde n}a-Sandoval}, {Maga{\\textasciitilde n}a
  Zertuche}, {Magee}, {Majorana}, {Maksimovic}, {Man}, {Mandic}, {Mangano},
  {Mansell}, {Manske}, {Mantovani}, {Marchesoni}, {Marion}, {M{\'a}rka},
  {M{\'a}rka}, {Markakis}, {Markosyan}, {Markowitz}, {Maros}, {Marquina},
  {Martelli}, {Martellini}, {Martin}, {Martin}, {Martynov}, {Mason}, {Massera},
  {Masserot}, {Massinger}, {Masso-Reid}, {Mastrogiovanni}, {Matas},
  {Matichard}, {Matone}, {Mavalvala}, {Mazumder}, {McCarthy}, {McClelland},
  {McCormick}, {McCuller}, {McGuire}, {McIntyre}, {McIver}, {McManus},
  {McNeill}, {McRae}, {McWilliams}, {Meacher}, {Meadors}, {Mehmet}, {Meidam},
  {Mejuto-Villa}, {Melatos}, {Mendell}, {Mercer}, {Merilh}, {Merzougui},
  {Meshkov}, {Messenger}, {Messick}, {Metzdorff}, {Meyers}, {Miao}, {Michel},
  {Middleton}, {Mikhailov}, {Milano}, {Miller}, {Miller}, {Miller},
  {Millhouse}, {Milovich-Goff}, {Minazzoli}, {Minenkov}, {Ming}, {Mishra},
  {Mitra}, {Mitrofanov}, {Mitselmakher}, {Mittleman}, {Moffa}, {Moggi},
  {Mogushi}, {Mohan}, {Mohapatra}, {Montani}, {Moore}, {Moraru}, {Moreno},
  {Morriss}, {Mours}, {Mow-Lowry}, {Mueller}, {Muir}, {Mukherjee}, {Mukherjee},
  {Mukherjee}, {Mukund}, {Mullavey}, {Munch}, {Mu{\\textasciitilde n}iz},
  {Muratore}, {Murray}, {Napier}, {Nardecchia}, {Naticchioni}, {Nayak},
  {Neilson}, {Nelemans}, {Nelson}, {Nery}, {Neunzert}, {Nevin}, {Newport},
  {Newton}, {Ng}, {Nguyen}, {Nichols}, {Nielsen}, {Nissanke}, {Nitz}, {Noack},
  {Nocera}, {Nolting}, {North}, {Nuttall}, {Oberling}, {O'Dea}, {Ogin}, {Oh},
  {Oh}, {Ohme}, {Okada}, {Oliver}, {Oppermann}, {Oram}, {O'Reilly}, {Ormiston},
  {Ortega}, {O'Shaughnessy}, {Ossokine}, {Ottaway}, {Overmier}, {Owen}, {Pace},
  {Page}, {Page}, {Pai}, {Pai}, {Palamos}, {Palashov}, {Palomba}, {Pal- Singh},
  {Pan}, {Pan}, {Pang}, {Pang}, {Pankow}, {Pannarale}, {Pant}, {Paoletti},
  {Paoli}, {Papa}, {Parida}, {Parker}, {Pascucci}, {Pasqualetti},
  {Passaquieti}, {Passuello}, {Patil}, {Patricelli}, {Pearlstone}, {Pedraza},
  {Pedurand}, {Pekowsky}, {Pele}, {Penn}, {Perez}, {Perreca}, {Perri},
  {Pfeiffer}, {Phelps}, {Piccinni}, {Pichot}, {Piergiovanni}, {Pierro},
  {Pillant}, {Pinard}, {Pinto}, {Pirello}, {Pitkin}, {Poe}, {Poggiani},
  {Popolizio}, {Porter}, {Post}, {Powell}, {Prasad}, {Pratt}, {Pratten},
  {Predoi}, {Prestegard}, {Prijatelj}, {Principe}, {Privitera}, {Prodi},
  {Prokhorov}, {Puncken}, {Punturo}, {Puppo}, {P{\"u}rrer}, {Qi}, {Quetschke},
  {Quintero}, {Quitzow-James}, {Raab}, {Rabeling}, {Radkins}, {Raffai}, {Raja},
  {Rajan}, {Rajbhandari}, {Rakhmanov}, {Ramirez}, {Ramos-Buades}, {Rapagnani},
  {Raymond}, {Razzano}, {Read}, {Regimbau}, {Rei}, {Reid}, {Reitze}, {Ren},
  {Reyes}, {Ricci}, {Ricker}, {Rieger}, {Riles}, {Rizzo}, {Robertson}, {Robie},
  {Robinet}, {Rocchi}, {Rolland}, {Rollins}, {Roma}, {Romano}, {Romel},
  {Romie}, {Rosi{\'n}ska}, {Ross}, {Rowan}, {R{\"u}diger}, {Ruggi}, {Rutins},
  {Ryan}, {Sachdev}, {Sadecki}, {Sadeghian}, {Sakellariadou}, {Salconi},
  {Saleem}, {Salemi}, {Samajdar}, {Sammut}, {Sampson}, {Sanchez}, {Sanchez},
  {Sanchis-Gual}, {Sandberg}, {Sanders}, {Sassolas}, {Sathyaprakash},
  {Saulson}, {Sauter}, {Savage}, {Sawadsky}, {Schale}, {Scheel}, {Scheuer},
  {Schmidt}, {Schmidt}, {Schnabel}, {Schofield}, {Sch{\"o}nbeck}, {Schreiber},
  {Schuette}, {Schulte}, {Schutz}, {Schwalbe}, {Scott}, {Scott}, {Seidel},
  {Sellers}, {Sengupta}, {Sentenac}, {Sequino}, {Sergeev}, {Shaddock},
  {Shaffer}, {Shah}, {Shahriar}, {Shaner}, {Shao}, {Shapiro}, {Shawhan},
  {Sheperd}, {Shoemaker}, {Shoemaker}, {Siellez}, {Siemens}, {Sieniawska},
  {Sigg}, {Silva}, {Singer}, {Singh}, {Singhal}, {Sintes}, {Slagmolen},
  {Smith}, {Smith}, {Smith}, {Somala}, {Son}, {Sonnenberg}, {Sorazu},
  {Sorrentino}, {Souradeep}, {Spencer}, {Srivastava}, {Staats}, {Staley},
  {Steinke}, {Steinlechner}, {Steinlechner}, {Steinmeyer}, {Stevenson},
  {Stone}, {Stops}, {Strain}, {Stratta}, {Strigin}, {Strunk}, {Sturani},
  {Stuver}, {Summerscales}, {Sun}, {Sunil}, {Suresh}, {Sutton}, {Swinkels},
  {Szczepa{\'n}czyk}, {Tacca}, {Tait}, {Talbot}, {Talukder}, {Tanner},
  {T{\'a}pai}, {Taracchini}, {Tasson}, {Taylor}, {Taylor}, {Tewari}, {Theeg},
  {Thies}, {Thomas}, {Thomas}, {Thomas}, {Thorne}, {Thorne}, {Thrane},
  {Tiwari}, {Tiwari}, {Tokmakov}, {Toland}, {Tonelli}, {Tornasi},
  {Torres-Forn{\'e}}, {Torrie}, \& {T{\"o}yr{\"a}}}]{2017ApJ...848L..13A}
---. 2017{\natexlab{c}}, \apj, 848, L13, \dodoi{10.3847/2041-8213/aa920c}

\bibitem[{{Abbott} {et~al.}(2017{\natexlab{d}}){Abbott}, {Abbott}, {Abbott},
  {Abernathy}, {Acernese}, {Ackley}, {Adams}, {Adams}, {Addesso}, {Adhikari},
  {Adya}, {Affeldt}, {Agathos}, {Agatsuma}, {Aggarwal}, {Aguiar}, {Aiello},
  {Ain}, {Ajith}, {Allen}, {Allocca}, {Altin}, {Ananyeva}, {Anderson},
  {Anderson}, {Appert}, {Arai}, {Araya}, {Areeda}, {Arnaud}, {Arun}, {Ascenzi},
  {Ashton}, {Ast}, {Aston}, {Astone}, {Aufmuth}, {Aulbert}, {Avila- Alvarez},
  {Babak}, {Bacon}, {Bader}, {Baker}, {Baldaccini}, {Ballardin}, {Ballmer},
  {Barayoga}, {E Barclay}, {Barish}, {Barker}, {Barone}, {Barr}, {Barsotti},
  {Barsuglia}, {Barta}, {Bartlett}, {Bartos}, {Bassiri}, {Basti}, {Batch},
  {Baune}, {Bavigadda}, {Bazzan}, {Beer}, {Bejger}, {Belahcene}, {Belgin},
  {Bell}, {Berger}, {Bergmann}, {Berry}, {Bersanetti}, {Bertolini},
  {Betzwieser}, {Bhagwat}, {Bhandare}, {Bilenko}, {Billingsley}, {Billman},
  {Birch}, {Birney}, {Birnholtz}, {Biscans}, {Bisht}, {Bitossi}, {Biwer},
  {Bizouard}, {Blackburn}, {Blackman}, {Blair}, {Blair}, {Blair}, {Bloemen},
  {Bock}, {Boer}, {Bogaert}, {Bohe}, {Bondu}, {Bonnand}, {Boom}, {Bork},
  {Boschi}, {Bose}, {Bouffanais}, {Bozzi}, {Bradaschia}, {Brady}, {Braginsky},
  {Branchesi}, {E Brau}, {Briant}, {Brillet}, {Brinkmann}, {Brisson},
  {Brockill}, {E Broida}, {Brooks}, {Brown}, {Brown}, {Brown}, {Brunett},
  {Buchanan}, {Buikema}, {Bulik}, {Bulten}, {Buonanno}, {Buskulic}, {Buy},
  {Byer}, {Cabero}, {Cadonati}, {Cagnoli}, {Cahillane}, {Calder{\'o}n
  Bustillo}, {Callister}, {Calloni}, {Camp}, {Cannon}, {Cao}, {Cao}, {Capano},
  {Capocasa}, {Carbognani}, {Caride}, {Casanueva Diaz}, {Casentini}, {Caudill},
  {Cavagli{\`a}}, {Cavalier}, {Cavalieri}, {Cella}, {Cepeda}, {Cerboni
  Baiardi}, {Cerretani}, {Cesarini}, {Chamberlin}, {Chan}, {Chao}, {Charlton},
  {Chassande-Mottin}, {Cheeseboro}, {Chen}, {Chen}, {Cheng}, {Chincarini},
  {Chiummo}, {Chmiel}, {Cho}, {Cho}, {Chow}, {Christensen}, {Chu}, {Chua},
  {Chua}, {Chung}, {Ciani}, {Clara}, {Clark}, {Cleva}, {Cocchieri}, {Coccia},
  {Cohadon}, {Colla}, {Collette}, {Cominsky}, {Constancio}, {Conti}, {Cooper},
  {Corbitt}, {Cornish}, {Corsi}, {Cortese}, {Costa}, {Coughlin}, {Coughlin},
  {Coulon}, {Countryman}, {Couvares}, {Covas}, {E Cowan}, {Coward}, {Cowart},
  {Coyne}, {Coyne}, {E Creighton}, {Creighton}, {Cripe}, {Crowder}, {Cullen},
  {Cumming}, {Cunningham}, {Cuoco}, {Dal Canton}, {Danilishin}, {D'Antonio},
  {Danzmann}, {Dasgupta}, {Da Silva Costa}, {Dattilo}, {Dave}, {Davier},
  {Davies}, {Davis}, {Daw}, {Day}, {Day}, {De}, {DeBra}, {Debreczeni},
  {Degallaix}, {De Laurentis}, {Del{\'e}glise}, {Del Pozzo}, {Denker}, {Dent},
  {Dergachev}, {De Rosa}, {DeRosa}, {DeSalvo}, {Devenson}, {Devine},
  {Dhurandhar}, {D{\'\i}az}, {Di Fiore}, {Di Giovanni}, {Di Girolamo}, {Di
  Lieto}, {Di Pace}, {Di Palma}, {Di Virgilio}, {Doctor}, {Dolique}, {Donovan},
  {Dooley}, {Doravari}, {Dorrington}, {Douglas}, {Dovale {\'A}lvarez},
  {Downes}, {Drago}, {Drever}, {Driggers}, {Du}, {Ducrot}, {E Dwyer}, {Edo},
  {Edwards}, {Effler}, {Eggenstein}, {Ehrens}, {Eichholz}, {Eikenberry},
  {Eisenstein}, {Essick}, {Etienne}, {Etzel}, {Evans}, {Evans}, {Everett},
  {Factourovich}, {Fafone}, {Fair}, {Fairhurst}, {Fan}, {Farinon}, {Farr},
  {Farr}, {Fauchon-Jones}, {Favata}, {Fays}, {Fehrmann}, {Fejer},
  {Fern{\'a}ndez Galiana}, {Ferrante}, {Ferreira}, {Ferrini}, {Fidecaro},
  {Fiori}, {Fiorucci}, {Fisher}, {Flaminio}, {Fletcher}, {Fong}, {Forsyth},
  {Fournier}, {Frasca}, {Frasconi}, {Frei}, {Freise}, {Frey}, {Frey}, {Fries},
  {Fritschel}, {Frolov}, {Fulda}, {Fyffe}, {Gabbard}, {Gadre}, {Gaebel},
  {Gair}, {Gammaitoni}, {Gaonkar}, {Garufi}, {Gaur}, {Gayathri}, {Gehrels},
  {Gemme}, {Genin}, {Gennai}, {George}, {Gergely}, {Germain}, {Ghonge},
  {Ghosh}, {Ghosh}, {Ghosh}, {Giaime}, {Giardina}, {Giazotto}, {Gill},
  {Glaefke}, {Goetz}, {Goetz}, {Gondan}, {Gonz{\'a}lez}, {Gonzalez Castro},
  {Gopakumar}, {Gorodetsky}, {E Gossan}, {Gosselin}, {Gouaty}, {Grado},
  {Graef}, {Granata}, {Grant}, {Gras}, {Gray}, {Greco}, {Green}, {Groot},
  {Grote}, {Grunewald}, {Guidi}, {Guo}, {Gupta}, {Gupta}, {E Gushwa},
  {Gustafson}, {Gustafson}, {Hacker}, {Hall}, {Hall}, {Hammond}, {Haney},
  {Hanke}, {Hanks}, {Hanna}, {Hannam}, {Hanson}, {Hardwick}, {Harms}, {Harry},
  {Harry}, {Hart}, {Hartman}, {Haster}, {Haughian}, {Healy}, {Heidmann},
  {Heintze}, {Heitmann}, {Hello}, {Hemming}, {Hendry}, {Heng}, {Hennig},
  {Henry}, {Heptonstall}, {Heurs}, {Hild}, {Hoak}, {Hofman}, {Holt}, {E Holz},
  {Hopkins}, {Hough}, {Houston}, {Howell}, {Hu}, {Huerta}, {Huet}, {Hughey},
  {Husa}, {Huttner}, {Huynh-Dinh}, {Indik}, {Ingram}, {Inta}, {Isa}, {Isac},
  {Isi}, {Isogai}, {Iyer}, {Izumi}, {Jacqmin}, {Jani}, {Jaranowski}, {Jawahar},
  {Jim{\'e}nez-Forteza}, {Johnson}, {Jones}, {Jones}, {Jonker}, {Ju}, {Junker},
  {Kalaghatgi}, {Kalogera}, {Kandhasamy}, {Kang}, {Kanner}, {Karki},
  {Karvinen}, {Kasprzack}, {Katsavounidis}, {Katzman}, {Kaufer}, {Kaur},
  {Kawabe}, {K{\'e}f{\'e}lian}, {Keitel}, {Kelley}, {Kennedy}, {Key},
  {Khalili}, {Khan}, {Khan}, {Khan}, {Khazanov}, {Kijbunchoo}, {Kim}, {Kim},
  {Kim}, {Kim}, {Kim}, {Kimbrell}, {King}, {King}, {Kirchhoff}, {Kissel},
  {Klein}, {Kleybolte}, {Klimenko}, {Koch}, {Koehlenbeck}, {Koley},
  {Kondrashov}, {Kontos}, {Korobko}, {Korth}, {Kowalska}, {Kozak},
  {Kr{\"a}mer}, {Kringel}, {Krishnan}, {Kr{\'o}lak}, {Kuehn}, {Kumar}, {Kumar},
  {Kuo}, {Kutynia}, {Lackey}, {Landry}, {Lang}, {Lange}, {Lantz}, {Lanza},
  {Lartaux-Vollard}, {Lasky}, {Laxen}, {Lazzarini}, {Lazzaro}, {Leaci},
  {Leavey}, {Lebigot}, {Lee}, {Lee}, {Lee}, {Lee}, {Lehmann}, {Lenon},
  {Leonardi}, {Leong}, {Leroy}, {Letendre}, {Levin}, {Li}, {Libson},
  {Littenberg}, {Liu}, {Lockerbie}, {Lombardi}, {London}, {E Lord},
  {Lorenzini}, {Loriette}, {Lormand}, {Losurdo}, {Lough}, {Lovelace},
  {L{\"u}ck}, {Lundgren}, {Lynch}, {Ma}, {Macfoy}, {Machenschalk}, {MacInnis},
  {Macleod}, {Maga{\\textasciitilde n}a-Sandoval}, {Majorana}, {Maksimovic},
  {Malvezzi}, {Man}, {Mandic}, {Mangano}, {Mansell}, {Manske}, {Mantovani},
  {Marchesoni}, {Marion}, {M{\'a}rka}, {M{\'a}rka}, {Markosyan}, {Maros},
  {Martelli}, {Martellini}, {Martin}, {Martynov}, {Mason}, {Masserot},
  {Massinger}, {Masso-Reid}, {Mastrogiovanni}, {Matichard}, {Matone},
  {Mavalvala}, {Mazumder}, {McCarthy}, {E McClelland}, {McCormick}, {McGrath},
  {McGuire}, {McIntyre}, {McIver}, {McManus}, {McRae}, {McWilliams}, {Meacher},
  {Meadors}, {Meidam}, {Melatos}, {Mendell}, {Mendoza- Gandara}, {Mercer},
  {Merilh}, {Merzougui}, {Meshkov}, {Messenger}, {Messick}, {Metzdorff},
  {Meyers}, {Mezzani}, {Miao}, {Michel}, {Middleton}, {E Mikhailov}, {Milano},
  {Miller}, {Miller}, {Miller}, {Miller}, {Millhouse}, {Minenkov}, {Ming},
  {Mirshekari}, {Mishra}, {Mitra}, {Mitrofanov}, {Mitselmakher}, {Mittleman},
  {Moggi}, {Mohan}, {Mohapatra}, {Montani}, {Moore}, {Moore}, {Moraru},
  {Moreno}, {Morriss}, {Mours}, {Mow-Lowry}, {Mueller}, {Muir}, {Mukherjee},
  {Mukherjee}, {Mukherjee}, {Mukund}, {Mullavey}, {Munch}, {Muniz}, {Murray},
  {Mytidis}, {Napier}, {Nardecchia}, {Naticchioni}, {Nelemans}, {Nelson},
  {Neri}, {Nery}, {Neunzert}, {Newport}, {Newton}, {Nguyen}, {Nielsen},
  {Nissanke}, {Nitz}, {Noack}, {Nocera}, {Nolting}, {Normandin}, {Nuttall},
  {Oberling}, {Ochsner}, {Oelker}, {Ogin}, {Oh}, {Oh}, {Ohme}, {Oliver},
  {Oppermann}, {Oram}, {O'Reilly}, {O'Shaughnessy}, {Ottaway}, {Overmier},
  {Owen}, {E Pace}, {Page}, {Pai}, {Pai}, {Palamos}, {Palashov}, {Palomba},
  {Pal-Singh}, {Pan}, {Pankow}, {Pannarale}, {Pant}, {Paoletti}, {Paoli},
  {Papa}, {Paris}, {Parker}, {Pascucci}, {Pasqualetti}, {Passaquieti},
  {Passuello}, {Patricelli}, {Pearlstone}, {Pedraza}, {Pedurand}, {Pekowsky},
  {Pele}, {Penn}, {Perez}, {Perreca}, {Perri}, {Pfeiffer}, {Phelps},
  {Piccinni}, {Pichot}, {Piergiovanni}, {Pierro}, {Pillant}, {Pinard}, {Pinto},
  {Pitkin}, {Poe}, {Poggiani}, {Popolizio}, {Post}, {Powell}, {Prasad},
  {Pratt}, {Predoi}, {Prestegard}, {Prijatelj}, {Principe}, {Privitera},
  {Prodi}, {Prokhorov}, {Puncken}, {Punturo}, {Puppo}, {P{\"u}rrer}, {Qi},
  {Qin}, {Qiu}, {Quetschke}, {Quintero}, {Quitzow-James}, {Raab}, {Rabeling},
  {Radkins}, {Raffai}, {Raja}, {Rajan}, {Rakhmanov}, {Rapagnani}, {Raymond},
  {Razzano}, {Re}, {Read}, {Regimbau}, {Rei}, {Reid}, {Reitze}, {Rew}, {Reyes},
  {Rhoades}, {Ricci}, {Riles}, {Rizzo}, {Robertson}, {Robie}, {Robinet},
  {Rocchi}, {Rolland}, {Rollins}, {Roma}, {Romano}, {Romano}, {Romie},
  {Rosi{\'n}ska}, {Rowan}, {R{\"u}diger}, {Ruggi}, {Ryan}, {Sachdev},
  {Sadecki}, {Sadeghian}, {Sakellariadou}, {Salconi}, {Saleem}, {Salemi},
  {Samajdar}, {Sammut}, {Sampson}, {Sanchez}, {Sandberg}, {Sanders},
  {Sassolas}, {Sathyaprakash}, {Saulson}, {Sauter}, {Savage}, {Sawadsky},
  {Schale}, {Scheuer}, {Schmidt}, {Schmidt}, {Schmidt}, {Schnabel},
  {Schofield}, {Sch{\"o}nbeck}, {Schreiber}, {Schuette}, {Schutz}, {Schwalbe},
  {Scott}, {Scott}, {Sellers}, {Sengupta}, {Sentenac}, {Sequino}, {Sergeev},
  {Setyawati}, {Shaddock}, {Shaffer}, {Shahriar}, {Shapiro}, {Shawhan},
  {Sheperd}, {Shoemaker}, {Shoemaker}, {Siellez}, {Siemens}, {Sieniawska},
  {Sigg}, {Silva}, {Singer}, {Singer}, {Singh}, {Singh}, {Singhal}, {Sintes},
  {Slagmolen}, {Smith}, {Smith}, {E Smith}, {Son}, {Sorazu}, {Sorrentino},
  {Souradeep}, {Spencer}, {Srivastava}, {Staley}, {Steinke}, {Steinlechner},
  {Steinlechner}, {Steinmeyer}, {Stephens}, {Stevenson}, {Stone}, {Strain},
  {Straniero}, {Stratta}, {E Strigin}, {Sturani}, {Stuver}, {Summerscales},
  {Sun}, {Sunil}, {Sutton}, {Swinkels}, {Szczepa{\'n}czyk}, {Tacca},
  {Talukder}, {Tanner}, {T{\'a}pai}, {Taracchini}, {Taylor}, {Theeg}, {Thomas},
  {Thomas}, {Thomas}, {Thorne}, {Thrane}, {Tippens}, {Tiwari}, {Tiwari},
  {Tokmakov}, {Toland}, {Tomlinson}, {Tonelli}, {Tornasi}, {Torrie},
  {T{\"o}yr{\"a}}, {Travasso}, {Traylor}, {Trifir{\`o}}, {Trinastic},
  {Tringali}, {Trozzo}, {Tse}, {Tso}, {Turconi}, {Tuyenbayev}, {Ugolini},
  {Unnikrishnan}, {Urban}, {Usman}, {Vahlbruch}, {Vajente}, {Valdes}, {van
  Bakel}, {van Beuzekom}, {van den Brand}, {Van Den Broeck}, {Vander-Hyde},
  {van der Schaaf}, {van Heijningen}, {van Veggel}, {Vardaro}, {Varma}, {Vass},
  {Vas{\'u}th}, {Vecchio}, {Vedovato}, {Veitch}, {Veitch}, {Venkateswara},
  {Venugopalan}, {Verkindt}, {Vetrano}, {Vicer{\'e}}, {Viets}, {Vinciguerra},
  {Vine}, {Vinet}, {Vitale}, {Vo}, {Vocca}, {Vorvick}, {Voss}, {Vousden},
  {Vyatchanin}, {Wade}, {E Wade}, {Wade}, {Walker}, {Wallace}, {Walsh}, {Wang},
  {Wang}, {Wang}, {Wang}, {Ward}, {Warner}, {Was}, {Watchi}, {Weaver}, {Wei},
  {Weinert}, {Weinstein}, {Weiss}, {Wen}, {We{\ss}els}, {Westphal}, {Wette},
  {Whelan}, {Whiting}, {Whittle}, {Williams}, {Williams}, {Williamson},
  {Willis}, {Willke}, {Wimmer}, {Winkler}, {Wipf}, {Wittel}, {Woan}, {Woehler},
  {Worden}, {Wright}, {Wu}, {Wu}, {Yam}, {Yamamoto}, {Yancey}, {Yap}, {Yu},
  {Yu}, {Yvert}, {Zadro{\.z}ny}, {Zangrando}, {Zanolin}, {Zendri}, {Zevin}, \&
  {Zhang}}]{2017CQGra..34j4002A}
---. 2017{\natexlab{d}}, Classical and Quantum Gravity, 34, 104002,
  \dodoi{10.1088/1361-6382/aa6854}

\bibitem[{{Abbott} {et~al.}(2017{\natexlab{e}}){Abbott}, {Abbott}, {Abbott},
  {Acernese}, {Ackley}, {Adams}, {Adams}, {Addesso}, {Adhikari}, {Adya},
  {Affeldt}, {Afrough}, {Agarwal}, {Agathos}, {Agatsuma}, {Aggarwal}, {Aguiar},
  {Aiello}, {Ain}, {Ajith}, {Allen}, {Allen}, {Allocca}, {Altin}, {Amato},
  {Ananyeva}, {Anderson}, {Anderson}, {Antier}, {Appert}, {Arai}, {Araya},
  {Areeda}, {Arnaud}, {Arun}, {Ascenzi}, {Ashton}, {Ast}, {Aston}, {Astone},
  {Aufmuth}, {Aulbert}, {AultONeal}, {Avila-Alvarez}, {Babak}, {Bacon},
  {Bader}, {Bae}, {Baker}, {Baldaccini}, {Ballardin}, {Ballmer}, {Banagiri},
  {Barayoga}, {Barclay}, {Barish}, {Barker}, {Barone}, {Barr}, {Barsotti},
  {Barsuglia}, {Barta}, {Bartlett}, {Bartos}, {Bassiri}, {Basti}, {Batch},
  {Baune}, {Bawaj}, {Bazzan}, {B{\'e}csy}, {Beer}, {Bejger}, {Belahcene},
  {Bell}, {Berger}, {Bergmann}, {Berry}, {Bersanetti}, {Bertolini},
  {Betzwieser}, {Bhagwat}, {Bhandare}, {Bilenko}, {Billingsley}, {Billman},
  {Birch}, {Birney}, {Birnholtz}, {Biscans}, {Bisht}, {Bitossi}, {Biwer},
  {Bizouard}, {Blackburn}, {Blackman}, {Blair}, {Blair}, {Blair}, {Bloemen},
  {Bock}, {Bode}, {Boer}, {Bogaert}, {Bohe}, {Bondu}, {Bonnand}, {Boom},
  {Bork}, {Boschi}, {Bose}, {Bouffanais}, {Bozzi}, {Bradaschia}, {Brady},
  {Braginsky}, {Branchesi}, {Brau}, {Briant}, {Brillet}, {Brinkmann},
  {Brisson}, {Brockill}, {Broida}, {Brooks}, {Brown}, {Brown}, {Brown},
  {Brunett}, {Buchanan}, {Buikema}, {Bulik}, {Bulten}, {Buonanno}, {Buskulic},
  {Buy}, {Byer}, {Cabero}, {Cadonati}, {Cagnoli}, {Cahillane}, {Calder{\'o}n
  Bustillo}, {Callister}, {Calloni}, {Camp}, {Canepa}, {Canizares}, {Cannon},
  {Cao}, {Cao}, {Capano}, {Capocasa}, {Carbognani}, {Caride}, {Carney},
  {Casanueva Diaz}, {Casentini}, {Caudill}, {Cavagli{\`a}}, {Cavalier},
  {Cavalieri}, {Cella}, {Cepeda}, {Cerboni Baiardi}, {Cerretani}, {Cesarini},
  {Chamberlin}, {Chan}, {Chao}, {Charlton}, {Chassande-Mottin}, {Chatterjee},
  {Chatziioannou}, {Cheeseboro}, {Chen}, {Chen}, {Cheng}, {Chincarini},
  {Chiummo}, {Chmiel}, {Cho}, {Cho}, {Chow}, {Christensen}, {Chu}, {Chua},
  {Chua}, {Chung}, {Chung}, {Ciani}, {Ciolfi}, {Cirelli}, {Cirone}, {Clara},
  {Clark}, {Cleva}, {Cocchieri}, {Coccia}, {Cohadon}, {Colla}, {Collette},
  {Cominsky}, {Constancio}, {Conti}, {Cooper}, {Corban}, {Corbitt}, {Corley},
  {Cornish}, {Corsi}, {Cortese}, {Costa}, {Coughlin}, {Coughlin}, {Coulon},
  {Countryman}, {Couvares}, {Covas}, {Cowan}, {Coward}, {Cowart}, {Coyne},
  {Coyne}, {Creighton}, {Creighton}, {Cripe}, {Crowder}, {Cullen}, {Cumming},
  {Cunningham}, {Cuoco}, {Dal Canton}, {Danilishin}, {D'Antonio}, {Danzmann},
  {Dasgupta}, {Da Silva Costa}, {Dattilo}, {Dave}, {Davier}, {Davis}, {Daw},
  {Day}, {De}, {DeBra}, {Deelman}, {Degallaix}, {De Laurentis},
  {Del{\'e}glise}, {Del Pozzo}, {Denker}, {Dent}, {Dergachev}, {De Rosa},
  {DeRosa}, {DeSalvo}, {Devenson}, {Devine}, {Dhurandhar}, {D{\'\i}az}, {Di
  Fiore}, {Di Giovanni}, {Di Girolamo}, {Di Lieto}, {Di Pace}, {Di Palma}, {Di
  Renzo}, {Doctor}, {Dolique}, {Donovan}, {Dooley}, {Doravari}, {Dorrington},
  {Douglas}, {Dovale {\'A}lvarez}, {Downes}, {Drago}, {Drever}, {Driggers},
  {Du}, {Ducrot}, {Duncan}, {Dwyer}, {Edo}, {Edwards}, {Effler}, {Eggenstein},
  {Ehrens}, {Eichholz}, {Eikenberry}, {Eisenstein}, {Essick}, {Etienne},
  {Etzel}, {Evans}, {Evans}, {Factourovich}, {Fafone}, {Fair}, {Fairhurst},
  {Fan}, {Farinon}, {Farr}, {Farr}, {Fauchon-Jones}, {Favata}, {Fays},
  {Fehrmann}, {Feicht}, {Fejer}, {Fernandez-Galiana}, {Ferrante}, {Ferreira},
  {Ferrini}, {Fidecaro}, {Fiori}, {Fiorucci}, {Fisher}, {Flaminio}, {Fletcher},
  {Fong}, {Forsyth}, {Forsyth}, {Fournier}, {Frasca}, {Frasconi}, {Frei},
  {Freise}, {Frey}, {Frey}, {Fries}, {Fritschel}, {Frolov}, {Fulda}, {Fyffe},
  {Gabbard}, {Gabel}, {Gadre}, {Gaebel}, {Gair}, {Gammaitoni}, {Ganija},
  {Gaonkar}, {Garufi}, {Gaudio}, {Gaur}, {Gayathri}, {Gehrels}, {Gemme},
  {Genin}, {Gennai}, {George}, {George}, {Gergely}, {Germain}, {Ghonge},
  {Ghosh}, {Ghosh}, {Ghosh}, {Giaime}, {Giardina}, {Giazotto}, {Gill},
  {Glover}, {Goetz}, {Goetz}, {Gomes}, {Gonz{\'a}lez}, {Gonzalez Castro},
  {Gopakumar}, {Gorodetsky}, {Gossan}, {Gosselin}, {Gouaty}, {Grado}, {Graef},
  {Granata}, {Grant}, {Gras}, {Gray}, {Greco}, {Green}, {Groot}, {Grote},
  {Grunewald}, {Gruning}, {Guidi}, {Guo}, {Gupta}, {Gupta}, {Gushwa},
  {Gustafson}, {Gustafson}, {Hall}, {Hall}, {Hammond}, {Haney}, {Hanke},
  {Hanks}, {Hanna}, {Hannam}, {Hannuksela}, {Hanson}, {Hardwick}, {Harms},
  {Harry}, {Harry}, {Hart}, {Haster}, {Haughian}, {Healy}, {Heidmann},
  {Heintze}, {Heitmann}, {Hello}, {Hemming}, {Hendry}, {Heng}, {Hennig},
  {Henry}, {Heptonstall}, {Heurs}, {Hild}, {Hoak}, {Hofman}, {Holt}, {Holz},
  {Hopkins}, {Horst}, {Hough}, {Houston}, {Howell}, {Hu}, {Huerta}, {Huet},
  {Hughey}, {Husa}, {Huttner}, {Huynh-Dinh}, {Indik}, {Ingram}, {Inta},
  {Intini}, {Isa}, {Isac}, {Isi}, {Iyer}, {Izumi}, {Jacqmin}, {Jani},
  {Jaranowski}, {Jawahar}, {Jim{\'e}nez-Forteza}, {Johnson},
  {Johnson-McDaniel}, {Jones}, {Jones}, {Jonker}, {Ju}, {Junker}, {Kalaghatgi},
  {Kalogera}, {Kandhasamy}, {Kang}, {Kanner}, {Karki}, {Karvinen}, {Kasprzack},
  {Katolik}, {Katsavounidis}, {Katzman}, {Kaufer}, {Kawabe},
  {K{\'e}f{\'e}lian}, {Keitel}, {Kemball}, {Kennedy}, {Kent}, {Key}, {Khalili},
  {Khan}, {Khan}, {Khan}, {Khazanov}, {Kijbunchoo}, {Kim}, {Kim}, {Kim}, {Kim},
  {Kim}, {Kimbrell}, {King}, {King}, {Kirchhoff}, {Kissel}, {Kleybolte},
  {Klimenko}, {Koch}, {Koehlenbeck}, {Koley}, {Kondrashov}, {Kontos},
  {Korobko}, {Korth}, {Kowalska}, {Kozak}, {Kr{\"a}mer}, {Kringel}, {Krishnan},
  {Kr{\'o}lak}, {Kuehn}, {Kumar}, {Kumar}, {Kumar}, {Kuo}, {Kutynia}, {Kwang},
  {Lackey}, {Lai}, {Landry}, {Lang}, {Lange}, {Lantz}, {Lanza},
  {Lartaux-Vollard}, {Lasky}, {Laxen}, {Lazzarini}, {Lazzaro}, {Leaci},
  {Leavey}, {Lee}, {Lee}, {Lee}, {Lee}, {Lee}, {Lehmann}, {Lenon}, {Leonardi},
  {Leroy}, {Letendre}, {Levin}, {Li}, {Libson}, {Littenberg}, {Liu}, {Lo},
  {Lockerbie}, {London}, {Lord}, {Lorenzini}, {Loriette}, {Lormand}, {Losurdo},
  {Lough}, {Lovelace}, {L{\"u}ck}, {Lumaca}, {Lundgren}, {Lynch}, {Ma},
  {Macfoy}, {Machenschalk}, {MacInnis}, {Macleod}, {Maga{\\textasciitilde n}a
  Hernandez}, {Maga{\\textasciitilde n}a-Sandoval}, {Maga{\\textasciitilde n}a
  Zertuche}, {Magee}, {Majorana}, {Maksimovic}, {Man}, {Mandic}, {Mangano},
  {Mansell}, {Manske}, {Mantovani}, {Marchesoni}, {Marion}, {M{\'a}rka},
  {M{\'a}rka}, {Markakis}, {Markosyan}, {Maros}, {Martelli}, {Martellini},
  {Martin}, {Martynov}, {Marx}, {Mason}, {Masserot}, {Massinger}, {Masso-Reid},
  {Mastrogiovanni}, {Matas}, {Matichard}, {Matone}, {Mavalvala}, {Mayani},
  {Mazumder}, {McCarthy}, {McClelland}, {McCormick}, {McCuller}, {McGuire},
  {McIntyre}, {McIver}, {McManus}, {McRae}, {McWilliams}, {Meacher}, {Meadors},
  {Meidam}, {Mejuto-Villa}, {Melatos}, {Mendell}, {Mercer}, {Merilh},
  {Merzougui}, {Meshkov}, {Messenger}, {Messick}, {Metzdorff}, {Meyers},
  {Mezzani}, {Miao}, {Michel}, {Middleton}, {Mikhailov}, {Milano}, {Miller},
  {Miller}, {Miller}, {Miller}, {Millhouse}, {Minazzoli}, {Minenkov}, {Ming},
  {Mishra}, {Mitra}, {Mitrofanov}, {Mitselmakher}, {Mittleman}, {Moggi},
  {Mohan}, {Mohapatra}, {Montani}, {Moore}, {Moore}, {Moraru}, {Moreno},
  {Morriss}, {Mours}, {Mow-Lowry}, {Mueller}, {Muir}, {Mukherjee}, {Mukherjee},
  {Mukherjee}, {Mukund}, {Mullavey}, {Munch}, {Muniz}, {Murray}, {Napier},
  {Nardecchia}, {Naticchioni}, {Nayak}, {Nelemans}, {Nelson}, {Neri}, {Nery},
  {Neunzert}, {Newport}, {Newton}, {Ng}, {Nguyen}, {Nichols}, {Nielsen},
  {Nissanke}, {Nitz}, {Noack}, {Nocera}, {Nolting}, {Normandin}, {Nuttall},
  {Oberling}, {Ochsner}, {Oelker}, {Ogin}, {Oh}, {Oh}, {Ohme}, {Oliver},
  {Oppermann}, {Oram}, {O'Reilly}, {Ormiston}, {Ortega}, {O'Shaughnessy},
  {Ottaway}, {Overmier}, {Owen}, {Pace}, {Page}, {Page}, {Pai}, {Pai},
  {Palamos}, {Palashov}, {Palomba}, {Pal-Singh}, {Pan}, {Pang}, {Pang},
  {Pankow}, {Pannarale}, {Pant}, {Paoletti}, {Paoli}, {Papa}, {Paris},
  {Parker}, {Pascucci}, {Pasqualetti}, {Passaquieti}, {Passuello},
  {Patricelli}, {Pearlstone}, {Pedraza}, {Pedurand}, {Pekowsky}, {Pele},
  {Penn}, {Perez}, {Perreca}, {Perri}, {Pfeiffer}, {Phelps}, {Piccinni},
  {Pichot}, {Piergiovanni}, {Pierro}, {Pillant}, {Pinard}, {Pinto}, {Pitkin},
  {Poggiani}, {Popolizio}, {Porter}, {Post}, {Powell}, {Prasad}, {Pratt},
  {Predoi}, {Prestegard}, {Prijatelj}, {Principe}, {Privitera}, {Prodi},
  {Prokhorov}, {Puncken}, {Punturo}, {Puppo}, {P{\"u}rrer}, {Qi}, {Qin}, {Qiu},
  {Quetschke}, {Quintero}, {Quitzow-James}, {Raab}, {Rabeling}, {Radkins},
  {Raffai}, {Raja}, {Rajan}, {Rakhmanov}, {Ramirez}, {Rapagnani}, {Raymond},
  {Razzano}, {Read}, {Regimbau}, {Rei}, {Reid}, {Reitze}, {Rew}, {Reyes},
  {Ricci}, {Ricker}, {Rieger}, {Riles}, {Rizzo}, {Robertson}, {Robie},
  {Robinet}, {Rocchi}, {Rolland}, {Rollins}, {Roma}, {Romano}, {Romano},
  {Romel}, {Romie}, {Rosi{\'n}ska}, {Ross}, {Rowan}, {R{\"u}diger}, {Ruggi},
  {Ryan}, {Rynge}, {Sachdev}, {Sadecki}, {Sadeghian}, {Sakellariadou},
  {Salconi}, {Saleem}, {Salemi}, {Samajdar}, {Sammut}, {Sampson}, {Sanchez},
  {Sandberg}, {Sandeen}, {Sanders}, {Sassolas}, {Sathyaprakash}, {Saulson},
  {Sauter}, {Savage}, {Sawadsky}, {Schale}, {Scheuer}, {Schmidt}, {Schmidt},
  {Schmidt}, {Schnabel}, {Schofield}, {Sch{\"o}nbeck}, {Schreiber}, {Schuette},
  {Schulte}, {Schutz}, {Schwalbe}, {Scott}, {Scott}, {Seidel}, {Sellers},
  {Sengupta}, {Sentenac}, {Sequino}, {Sergeev}, {Shaddock}, {Shaffer}, {Shah},
  {Shahriar}, {Shao}, {Shapiro}, {Shawhan}, {Sheperd}, {Shoemaker},
  {Shoemaker}, {Siellez}, {Siemens}, {Sieniawska}, {Sigg}, {Silva}, {Singer},
  {Singer}, {Singh}, {Singh}, {Singhal}, {Sintes}, {Slagmolen}, {Smith},
  {Smith}, {Smith}, {Son}, {Sonnenberg}, {Sorazu}, {Sorrentino}, {Souradeep},
  {Spencer}, {Srivastava}, {Staley}, {Steinke}, {Steinlechner}, {Steinlechner},
  {Steinmeyer}, {Stephens}, {Stevenson}, {Stone}, {Strain}, {Stratta},
  {Strigin}, {Sturani}, {Stuver}, {Summerscales}, {Sun}, {Sunil}, {Sutton},
  {Swinkels}, {Szczepa{\'n}czyk}, {Tacca}, {Talukder}, {Tanner}, {T{\'a}pai},
  {Taracchini}, {Taylor}, {Taylor}, {Theeg}, {Thomas}, {Thomas}, {Thomas},
  {Thorne}, {Thorne}, {Thrane}, {Tiwari}, {Tiwari}, {Tokmakov}, {Toland},
  {Tonelli}, {Tornasi}, {Torrie}, {T{\"o}yr{\"a}}, {Travasso}, {Traylor},
  {Trifir{\`o}}, {Trinastic}, {Tringali}, {Trozzo}, {Tsang}, {Tse}, {Tso},
  {Tuyenbayev}, {Ueno}, {Ugolini}, {Unnikrishnan}, {Urban}, {Usman}, {Vahi},
  {Vahlbruch}, {Vajente}, {Valdes}, {Vallisneri}, {van Bakel}, {van Beuzekom},
  {van den Brand}, {Van Den Broeck}, {Vander-Hyde}, {van der Schaaf}, {van
  Heijningen}, {van Veggel}, {Vardaro}, {Varma}, {Vass}, {Vas{\'u}th},
  {Vecchio}, {Vedovato}, {Veitch}, {Veitch}, {Venkateswara}, {Venugopalan},
  {Verkindt}, {Vetrano}, {Vicer{\'e}}, {Viets}, {Vinciguerra}, {Vine}, {Vinet},
  {Vitale}, {Vo}, {Vocca}, {Vorvick}, {Voss}, {Vousden}, {Vyatchanin}, \&
  {Wade}}]{2017PhRvL.118v1101A}
---. 2017{\natexlab{e}}, \prl, 118, 221101,
  \dodoi{10.1103/PhysRevLett.118.221101}

\bibitem[{Abbott {et~al.}(2018)}]{o2catalog}
Abbott, B.~P., {et~al.} 2018, To be submitted

\bibitem[{{Abbott} {et~al.}(2018){Abbott}, {Abbott}, {Abbott}, {Abernathy},
  {Acernese}, {Ackley}, {Adams}, {Adams}, {Addesso}, {Adhikari}, {Adya},
  {Affeldt}, {Agathos}, {Agatsuma}, {Aggarwal}, {Aguiar}, {Aiello}, {Ain},
  {Ajith}, {Akutsu}, {Allen}, {Allocca}, {Altin}, {Ananyeva}, {Anderson},
  {Anderson}, {Ando}, {Appert}, {Arai}, {Araya}, {Araya}, {Areeda}, {Arnaud},
  {Arun}, {Asada}, {Ascenzi}, {Ashton}, {Aso}, {Ast}, {Aston}, {Astone},
  {Atsuta}, {Aufmuth}, {Aulbert}, {Avila-Alvarez}, {Awai}, {Babak}, {Bacon},
  {Bader}, {Baiotti}, {Baker}, {Baldaccini}, {Ballardin}, {Ballmer},
  {Barayoga}, {Barclay}, {Barish}, {Barker}, {Barone}, {Barr}, {Barsotti},
  {Barsuglia}, {Barta}, {Bartlett}, {Barton}, {Bartos}, {Bassiri}, {Basti},
  {Batch}, {Baune}, {Bavigadda}, {Bazzan}, {B{\'e}csy}, {Beer}, {Bejger},
  {Belahcene}, {Belgin}, {Bell}, {Berger}, {Bergmann}, {Berry}, {Bersanetti},
  {Bertolini}, {Betzwieser}, {Bhagwat}, {Bhandare}, {Bilenko}, {Billingsley},
  {Billman}, {Birch}, {Birney}, {Birnholtz}, {Biscans}, {Bisht}, {Bitossi},
  {Biwer}, {Bizouard}, {Blackburn}, {Blackman}, {Blair}, {Blair}, {Blair},
  {Bloemen}, {Bock}, {Boer}, {Bogaert}, {Bohe}, {Bondu}, {Bonnand}, {Boom},
  {Bork}, {Boschi}, {Bose}, {Bouffanais}, {Bozzi}, {Bradaschia}, {Brady},
  {Braginsky}, {Branchesi}, {Brau}, {Briant}, {Brillet}, {Brinkmann},
  {Brisson}, {Brockill}, {Broida}, {Brooks}, {Brown}, {Brown}, {Brown},
  {Brunett}, {Buchanan}, {Buikema}, {Bulik}, {Bulten}, {Buonanno}, {Buskulic},
  {Buy}, {Byer}, {Cabero}, {Cadonati}, {Cagnoli}, {Cahillane}, {Calder{\'o}n
  Bustillo}, {Callister}, {Calloni}, {Camp}, {Cannon}, {Cao}, {Cao}, {Capano},
  {Capocasa}, {Carbognani}, {Caride}, {Casanueva Diaz}, {Casentini}, {Caudill},
  {Cavagli{\`a}}, {Cavalier}, {Cavalieri}, {Cella}, {Cepeda}, {Cerboni
  Baiardi}, {Cerretani}, {Cesarini}, {Chamberlin}, {Chan}, {Chao}, {Charlton},
  {Chassande- Mottin}, {Cheeseboro}, {Chen}, {Chen}, {Cheng}, {Chincarini},
  {Chiummo}, {Chmiel}, {Cho}, {Cho}, {Chow}, {Christensen}, {Chu}, {Chua},
  {Chua}, {Chung}, {Ciani}, {Clara}, {Clark}, {Cleva}, {Cocchieri}, {Coccia},
  {Cohadon}, {Colla}, {Collette}, {Cominsky}, {Constancio}, {Conti}, {Cooper},
  {Corbitt}, {Cornish}, {Corsi}, {Cortese}, {Costa}, {Coughlin}, {Coughlin},
  {Coulon}, {Countryman}, {Couvares}, {Covas}, {Cowan}, {Coward}, {Cowart},
  {Coyne}, {Coyne}, {Creighton}, {Creighton}, {Cripe}, {Crowder}, {Cullen},
  {Cumming}, {Cunningham}, {Cuoco}, {Dal Canton}, {Danilishin}, {D'Antonio},
  {Danzmann}, {Dasgupta}, {Da Silva Costa}, {Dattilo}, {Dave}, {Davier},
  {Davies}, {Davis}, {Daw}, {Day}, {Day}, {De}, {DeBra}, {Debreczeni},
  {Degallaix}, {De Laurentis}, {Del{\'e}glise}, {Del Pozzo}, {Denker}, {Dent},
  {Dergachev}, {De Rosa}, {DeRosa}, {DeSalvo}, {Devine}, {Dhurandhar},
  {D{\'\i}az}, {Fiore}, {Giovanni}, {Girolamo}, {Lieto}, {Pace}, {Palma},
  {Virgilio}, {Doctor}, {Doi}, {Dolique}, {Donovan}, {Dooley}, {Doravari},
  {Dorrington}, {Douglas}, {Dovale {\'A}lvarez}, {Downes}, {Drago}, {Drever},
  {Driggers}, {Du}, {Ducrot}, {Dwyer}, {Eda}, {Edo}, {Edwards}, {Effler},
  {Eggenstein}, {Ehrens}, {Eichholz}, {Eikenberry}, {Eisenstein}, {Essick},
  {Etienne}, {Etzel}, {Evans}, {Evans}, {Everett}, {Factourovich}, {Fafone},
  {Fair}, {Fairhurst}, {Fan}, {Farinon}, {Farr}, {Farr}, {Fauchon-Jones},
  {Favata}, {Fays}, {Fehrmann}, {Fejer}, {Fern{\'a}ndez Galiana}, {Ferrante},
  {Ferreira}, {Ferrini}, {Fidecaro}, {Fiori}, {Fiorucci}, {Fisher}, {Flaminio},
  {Fletcher}, {Fong}, {Forsyth}, {Fournier}, {Frasca}, {Frasconi}, {Frei},
  {Freise}, {Frey}, {Frey}, {Fries}, {Fritschel}, {Frolov}, {Fujii},
  {Fujimoto}, {Fulda}, {Fyffe}, {Gabbard}, {Gadre}, {Gaebel}, {Gair},
  {Gammaitoni}, {Gaonkar}, {Garufi}, {Gaur}, {Gayathri}, {Gehrels}, {Gemme},
  {Genin}, {Gennai}, {George}, {Gergely}, {Germain}, {Ghonge}, {Ghosh},
  {Ghosh}, {Ghosh}, {Giaime}, {Giardina}, {Giazotto}, {Gill}, {Glaefke},
  {Goetz}, {Goetz}, {Gondan}, {Gonz{\'a}lez}, {Gonzalez Castro}, {Gopakumar},
  {Gorodetsky}, {Gossan}, {Gosselin}, {Gouaty}, {Grado}, {Graef}, {Granata},
  {Grant}, {Gras}, {Gray}, {Greco}, {Green}, {Groot}, {Grote}, {Grunewald},
  {Guidi}, {Guo}, {Gupta}, {Gupta}, {Gushwa}, {Gustafson}, {Gustafson},
  {Hacker}, {Hagiwara}, {Hall}, {Hall}, {Hammond}, {Haney}, {Hanke}, {Hanks},
  {Hanna}, {Hannam}, {Hanson}, {Hardwick}, {Harms}, {Harry}, {Harry}, {Hart},
  {Hartman}, {Haster}, {Haughian}, {Hayama}, {Healy}, {Heidmann}, {Heintze},
  {Heitmann}, {Hello}, {Hemming}, {Hendry}, {Heng}, {Hennig}, {Henry},
  {Heptonstall}, {Heurs}, {Hild}, {Hirose}, {Hoak}, {Hofman}, {Holt}, {Holz},
  {Hopkins}, {Hough}, {Houston}, {Howell}, {Hu}, {Huerta}, {Huet}, {Hughey},
  {Husa}, {Huttner}, {Huynh-Dinh}, {Indik}, {Ingram}, {Inta}, {Ioka}, {Isa},
  {Isac}, {Isi}, {Isogai}, {Itoh}, {Iyer}, {Izumi}, {Jacqmin}, {Jani},
  {Jaranowski}, {Jawahar}, {Jim{\'e}nez-Forteza}, {Johnson}, {Jones}, {Jones},
  {Jonker}, {Ju}, {Junker}, {Kagawa}, {Kajita}, {Kakizaki}, {Kalaghatgi},
  {Kalogera}, {Kamiizumi}, {Kanda}, {Kandhasamy}, {Kanemura}, {Kaneyama},
  {Kang}, {Kanner}, {Karki}, {Karvinen}, {Kasprzack}, {Kataoka},
  {Katsavounidis}, {Katzman}, {Kaufer}, {Kaur}, {Kawabe}, {Kawai}, {Kawamura},
  {K{\'e}f{\'e}lian}, {Keitel}, {Kelley}, {Kennedy}, {Key}, {Khalili}, {Khan},
  {Khan}, {Khan}, {Khazanov}, {Kijbunchoo}, {Kim}, {Kim}, {Kim}, {Kim}, {Kim},
  {Kim}, {Kimbrell}, {Kimura}, {King}, {King}, {Kirchhoff}, {Kissel}, {Klein},
  {Kleybolte}, {Klimenko}, {Koch}, {Koehlenbeck}, {Kojima}, {Kokeyama},
  {Koley}, {Komori}, {Kondrashov}, {Kontos}, {Korobko}, {Korth}, {Kotake},
  {Kowalska}, {Kozak}, {Kr{\"a}mer}, {Kringel}, {Krishnan}, {Kr{\'o}lak},
  {Kuehn}, {Kumar}, {Kumar}, {Kumar}, {Kuo}, {Kuroda}, {Kutynia}, {Kuwahara},
  {Lackey}, {Landry}, {Lang}, {Lange}, {Lantz}, {Lanza}, {Lartaux-Vollard},
  {Lasky}, {Laxen}, {Lazzarini}, {Lazzaro}, {Leaci}, {Leavey}, {Lebigot},
  {Lee}, {Lee}, {Lee}, {Lee}, {Lee}, {Lehmann}, {Lenon}, {Leonardi}, {Leong},
  {Leroy}, {Letendre}, {Levin}, {Li}, {Libson}, {Littenberg}, {Liu},
  {Lockerbie}, {Lombardi}, {London}, {Lord}, {Lorenzini}, {Loriette},
  {Lormand}, {Losurdo}, {Lough}, {Lousto}, {Lovelace}, {L{\"u}ck}, {Lundgren},
  {Lynch}, {Ma}, {Macfoy}, {Machenschalk}, {MacInnis}, {Macleod},
  {Maga{\\textasciitilde n}a-Sandoval}, {Majorana}, {Maksimovic}, {Malvezzi},
  {Man}, {Mandic}, {Mangano}, {Mano}, {Mansell}, {Manske}, {Mantovani},
  {Marchesoni}, {Marchio}, {Marion}, {M{\'a}rka}, {M{\'a}rka}, {Markosyan},
  {Maros}, {Martelli}, {Martellini}, {Martin}, {Martynov}, {Mason}, {Masserot},
  {Massinger}, {Masso-Reid}, {Mastrogiovanni}, {Matichard}, {Matone},
  {Matsumoto}, {Matsushima}, {Mavalvala}, {Mazumder}, {McCarthy}, {McClelland},
  {McCormick}, {McGrath}, {McGuire}, {McIntyre}, {McIver}, {McManus}, {McRae},
  {McWilliams}, {Meacher}, {Meadors}, {Meidam}, {Melatos}, {Mendell},
  {Mendoza-Gandara}, {Mercer}, {Merilh}, {Merzougui}, {Meshkov}, {Messenger},
  {Messick}, {Metzdorff}, {Meyers}, {Mezzani}, {Miao}, {Michel}, {Michimura},
  {Middleton}, {Mikhailov}, {Milano}, {Miller}, {Miller}, {Miller}, {Miller},
  {Millhouse}, {Minenkov}, {Ming}, {Mirshekari}, {Mishra}, {Mitrofanov},
  {Mitselmakher}, {Mittleman}, {Miyakawa}, {Miyamoto}, {Miyamoto}, {Miyoki},
  {Moggi}, {Mohan}, {Mohapatra}, {Montani}, {Moore}, {Moore}, {Moraru},
  {Moreno}, {Morii}, {Morisaki}, {Moriwaki}, {Morriss}, {Mours}, {Mow-Lowry},
  {Mueller}, {Muir}, {Mukherjee}, {Mukherjee}, {Mukherjee}, {Mukund},
  {Mullavey}, {Munch}, {Muniz}, {Murray}, {Mytidis}, {Nagano}, {Nakamura},
  {Nakamura}, {Nakano}, {Nakano}, {Nakano}, {Nakao}, {Napier}, {Nardecchia},
  {Narikawa}, {Naticchioni}, {Nelemans}, {Nelson}, {Neri}, {Nery}, {Neunzert},
  {Newport}, {Newton}, {Nguyen}, {Ni}, {Nielsen}, {Nissanke}, {Nitz}, {Noack},
  {Nocera}, {Nolting}, {Normandin}, {Nuttall}, {Oberling}, {Ochsner}, {Oelker},
  {Ogin}, {Oh}, {Oh}, {Ohashi}, {Ohishi}, {Ohkawa}, {Ohme}, {Okutomi},
  {Oliver}, {Ono}, {Ono}, {Oohara}, {Oppermann}, {Oram}, {O'Reilly},
  {O'Shaughnessy}, {Ottaway}, {Overmier}, {Owen}, {Pace}, {Page}, {Pai}, {Pai},
  {Palamos}, {Palashov}, {Palomba}, {Pal-Singh}, {Pan}, {Pankow}, {Pannarale},
  {Pant}, {Paoletti}, {Paoli}, {Papa}, {Paris}, {Parker}, {Pascucci},
  {Pasqualetti}, {Passaquieti}, {Passuello}, {Patricelli}, {Pearlstone},
  {Pedraza}, {Pedurand}, {Pekowsky}, {Pele}, {Pe{\\textasciitilde n}a
  Arellano}, {Penn}, {Perez}, {Perreca}, {Perri}, {Pfeiffer}, {Phelps},
  {Piccinni}, {Pichot}, {Piergiovanni}, {Pierro}, {Pillant}, {Pinard}, {Pinto},
  {Pitkin}, {Poe}, {Poggiani}, {Popolizio}, {Post}, {Powell}, {Prasad},
  {Pratt}, {Predoi}, {Prestegard}, {Prijatelj}, {Principe}, {Privitera},
  {Prodi}, {Prokhorov}, {Puncken}, {Punturo}, {Puppo}, {P{\"u}rrer}, {Qi},
  {Qin}, {Qiu}, {Quetschke}, {Quintero}, {Quitzow-James}, {Raab}, {Rabeling},
  {Radkins}, {Raffai}, {Raja}, {Rajan}, {Rakhmanov}, {Rapagnani}, {Raymond},
  {Razzano}, {Re}, {Read}, {Regimbau}, {Rei}, {Reid}, {Reitze}, {Rew}, {Reyes},
  {Rhoades}, {Ricci}, {Riles}, {Rizzo}, {Robertson}, {Robie}, {Robinet},
  {Rocchi}, {Rolland}, {Rollins}, {Roma}, {Romano}, {Romie}, {Rosi{\'n}ska},
  {Rowan}, {R{\"u}diger}, {Ruggi}, {Ryan}, {Sachdev}, {Sadecki}, {Sadeghian},
  {Sago}, {Saijo}, {Saito}, {Sakai}, {Sakellariadou}, {Salconi}, {Saleem},
  {Salemi}, {Samajdar}, {Sammut}, {Sampson}, {Sanchez}, {Sandberg}, {Sanders},
  {Sasaki}, {Sassolas}, {Sathyaprakash}, {Sato}, {Sato}, {Saulson}, {Sauter},
  {Savage}, {Sawadsky}, {Schale}, {Scheuer}, {Schmidt}, {Schmidt}, {Schmidt},
  {Schnabel}, {Schofield}, {Sch{\"o}nbeck}, {Schreiber}, {Schuette}, {Schutz},
  {Schwalbe}, {Scott}, {Scott}, {Sekiguchi}, {Sekiguchi}, {Sellers},
  {Sengupta}, {Sentenac}, {Sequino}, {Sergeev}, {Setyawati}, {Shaddock},
  {Shaffer}, {Shahriar}, {Shapiro}, {Shawhan}, {Sheperd}, {Shibata}, {Shikano},
  {Shimoda}, {Shoda}, {Shoemaker}, {Shoemaker}, {Siellez}, {Siemens},
  {Sieniawska}, {Sigg}, {Silva}, {Singer}, {Singer}, {Singh}, {Singh},
  {Singhal}, {Sintes}, {Slagmolen}, {Smith}, {Smith}, {Smith}, {Somiya}, {Son},
  {Sorazu}, {Sorrentino}, {Souradeep}, {Spencer}, {Srivastava}, {Staley},
  {Steinke}, {Steinlechner}, {Steinlechner}, {Steinmeyer}, {Stephens},
  {Stevenson}, {Stone}, {Strain}, {Straniero}, {Stratta}, {Strigin}, {Sturani},
  {Stuver}, {Sugimoto}, {Summerscales}, {Sun}, {Sunil}, {Sutton}, {Suzuki},
  {Swinkels}, {Szczepa{\'n}czyk}, {Tacca}, {Tagoshi}, {Takada}, {Takahashi},
  {Takahashi}, {Takamori}, {Talukder}, {Tanaka}, {Tanaka}, {Tanaka}, {Tanner},
  {T{\'a}pai}, {Taracchini}, {Tatsumi}, {Taylor}, {Telada}, {Theeg}, {Thomas},
  {Thomas}, {Thomas}, {Thorne}, {Thrane}, {Tippens}, {Tiwari}, {Tiwari},
  {Tokmakov}, {Toland}, {Tomaru}, {Tomlinson}, {Tonelli}, {Tornasi}, {Torrie},
  {T{\"o}yr{\"a}}, {Travasso}, {Traylor}, {Trifir{\`o}}, {Trinastic},
  {Tringali}, {Trozzo}, {Tse}, {Tso}, {Tsubono}, {Tsuzuki}, {Turconi}, \&
  {Tuyenbayev}}]{2018LRR....21....3A}
{Abbott}, B.~P., {Abbott}, R., {Abbott}, T.~D., {et~al.} 2018, Living Reviews
  in Relativity, 21, 3, \dodoi{10.1007/s41114-018-0012-9}

\bibitem[{{Acernese} {et~al.}(2015){Acernese}, {Agathos}, {Agatsuma}, {Aisa},
  {Allemandou}, {Allocca}, {Amarni}, {Astone}, {Balestri}, {Ballardin},
  {Barone}, {Baronick}, {Barsuglia}, {Basti}, {Basti}, {Bauer}, {Bavigadda},
  {Bejger}, {Beker}, {Belczynski}, {Bersanetti}, {Bertolini}, {Bitossi},
  {Bizouard}, {Bloemen}, {Blom}, {Boer}, {Bogaert}, {Bondi}, {Bondu},
  {Bonelli}, {Bonnand}, {Boschi}, {Bosi}, {Bouedo}, {Bradaschia}, {Branchesi},
  {Briant}, {Brillet}, {Brisson}, {Bulik}, {Bulten}, {Buskulic}, {Buy},
  {Cagnoli}, {Calloni}, {Campeggi}, {Canuel}, {Carbognani}, {Cavalier},
  {Cavalieri}, {Cella}, {Cesarini}, {Chassande-Mottin}, {Chincarini},
  {Chiummo}, {Chua}, {Cleva}, {Coccia}, {Cohadon}, {Colla}, {Colombini},
  {Conte}, {Coulon}, {Cuoco}, {Dalmaz}, {D'Antonio}, {Dattilo}, {Davier},
  {Day}, {Debreczeni}, {Degallaix}, {Del{\'e}glise}, {Del Pozzo}, {Dereli}, {De
  Rosa}, {Di Fiore}, {Di Lieto}, {Di Virgilio}, {Doets}, {Dolique}, {Drago},
  {Ducrot}, {Endr{\'{o}}czi}, {Fafone}, {Farinon}, {Ferrante}, {Ferrini},
  {Fidecaro}, {Fiori}, {Flaminio}, {Fournier}, {Franco}, {Frasca}, {Frasconi},
  {Gammaitoni}, {Garufi}, {Gaspard}, {Gatto}, {Gemme}, {Gendre}, {Genin},
  {Gennai}, {Ghosh}, {Giacobone}, {Giazotto}, {Gouaty}, {Granata}, {Greco},
  {Groot}, {Guidi}, {Harms}, {Heidmann}, {Heitmann}, {Hello}, {Hemming},
  {Hennes}, {Hofman}, {Jaranowski}, {Jonker}, {Kasprzack}, {K{\'e}f{\'e}lian},
  {Kowalska}, {Kraan}, {Kr{\'o}lak}, {Kutynia}, {Lazzaro}, {Leonardi}, {Leroy},
  {Letendre}, {Li}, {Lieunard}, {Lorenzini}, {Loriette}, {Losurdo},
  {Magazz{\`u}}, {Majorana}, {Maksimovic}, {Malvezzi}, {Man}, {Mangano},
  {Mantovani}, {Marchesoni}, {Marion}, {Marque}, {Martelli}, {Martellini},
  {Masserot}, {Meacher}, {Meidam}, {Mezzani}, {Michel}, {Milano}, {Minenkov},
  {Moggi}, {Mohan}, {Montani}, {Morgado}, {Mours}, {Mul}, {Nagy}, {Nardecchia},
  {Naticchioni}, {Nelemans}, {Neri}, {Neri}, {Nocera}, {Pacaud}, {Palomba},
  {Paoletti}, {Paoli}, {Pasqualetti}, {Passaquieti}, {Passuello}, {Perciballi},
  {Petit}, {Pichot}, {Piergiovanni}, {Pillant}, {Piluso}, {Pinard}, {Poggiani},
  {Prijatelj}, {Prodi}, {Punturo}, {Puppo}, {Rabeling}, {R{\'a}cz},
  {Rapagnani}, {Razzano}, {Re}, {Regimbau}, {Ricci}, {Robinet}, {Rocchi},
  {Rolland}, {Romano}, {Rosi{\'n}ska}, {Ruggi}, {Saracco}, {Sassolas},
  {Schimmel}, {Sentenac}, {Sequino}, {Shah}, {Siellez}, {Straniero},
  {Swinkels}, {Tacca}, {Tonelli}, {Travasso}, {Turconi}, {Vajente}, {van
  Bakel}, {van Beuzekom}, {van den Brand}, {Van Den Broeck}, {van der Sluys},
  {van Heijningen}, {Vas{\'u}th}, {Vedovato}, {Veitch}, {Verkindt}, {Vetrano},
  {Vicer{\'e}}, {Vinet}, {Visser}, {Vocca}, {Ward}, {Was}, {Wei}, {Yvert},
  {Zadro {\.z}ny}, \& {Zendri}}]{2015CQGra..32b4001A}
{Acernese}, F., {Agathos}, M., {Agatsuma}, K., {et~al.} 2015, Classical and
  Quantum Gravity, 32, 024001, \dodoi{10.1088/0264-9381/32/2/024001}

\bibitem[{{Ajith} {et~al.}(2011){Ajith}, {Hannam}, {Husa}, {Chen},
  {Br{\"u}gmann}, {Dorband}, {M{\"u}ller}, {Ohme}, {Pollney}, {Reisswig},
  {Santamar{\'\i}a}, \& {Seiler}}]{2011PhRvL.106x1101A}
{Ajith}, P., {Hannam}, M., {Husa}, S., {et~al.} 2011, \prl, 106, 241101,
  \dodoi{10.1103/PhysRevLett.106.241101}

\bibitem[{{Alexander} {et~al.}(2017){Alexander}, {Berger}, {Fong}, {Williams},
  {Guidorzi}, {Margutti}, {Metzger}, {Annis}, {Blanchard}, {Brout}, {Brown},
  {Chen}, {Chornock}, {Cowperthwaite}, {Drout}, {Eftekhari}, {Frieman}, {Holz},
  {Nicholl}, {Rest}, {Sako}, {Soares-Santos}, \&
  {Villar}}]{2017ApJ...848L..21A}
{Alexander}, K.~D., {Berger}, E., {Fong}, W., {et~al.} 2017, \apj, 848, L21,
  \dodoi{10.3847/2041-8213/aa905d}

\bibitem[{{Ali-Ha{\"\i}moud} {et~al.}(2017){Ali-Ha{\"\i}moud}, {Kovetz}, \&
  {Kamionkowski}}]{2017PhRvD..96l3523A}
{Ali-Ha{\"\i}moud}, Y., {Kovetz}, E.~D., \& {Kamionkowski}, M. 2017, \prd, 96,
  123523, \dodoi{10.1103/PhysRevD.96.123523}

\bibitem[{{Amaro-Seoane} \& {Chen}(2016)}]{2016MNRAS.458.3075A}
{Amaro-Seoane}, P., \& {Chen}, X. 2016, \mnras, 458, 3075,
  \dodoi{10.1093/mnras/stw503}

\bibitem[{{Ando} {et~al.}(2018){Ando}, {Inomata}, {Kawasaki}, {Mukaida}, \&
  {Yanagida}}]{2018PhRvD..97l3512A}
{Ando}, K., {Inomata}, K., {Kawasaki}, M., {Mukaida}, K., \& {Yanagida}, T.~T.
  2018, \prd, 97, 123512, \dodoi{10.1103/PhysRevD.97.123512}

\bibitem[{{Antonini} {et~al.}(2014){Antonini}, {Murray}, \&
  {Mikkola}}]{2014ApJ...781...45A}
{Antonini}, F., {Murray}, N., \& {Mikkola}, S. 2014, \apj, 781, 45,
  \dodoi{10.1088/0004-637X/781/1/45}

\bibitem[{{Antonini} \& {Perets}(2012)}]{2012ApJ...757...27A}
{Antonini}, F., \& {Perets}, H.~B. 2012, \apj, 757, 27,
  \dodoi{10.1088/0004-637X/757/1/27}

\bibitem[{{Antonini} \& {Rasio}(2016)}]{2016ApJ...831..187A}
{Antonini}, F., \& {Rasio}, F.~A. 2016, \apj, 831, 187,
  \dodoi{10.3847/0004-637X/831/2/187}

\bibitem[{{Antonini} {et~al.}(2017){Antonini}, {Toonen}, \&
  {Hamers}}]{2017ApJ...841...77A}
{Antonini}, F., {Toonen}, S., \& {Hamers}, A.~S. 2017, \apj, 841, 77,
  \dodoi{10.3847/1538-4357/aa6f5e}

\bibitem[{{Arca Sedda} \& {Benacquista}(2019)}]{2019MNRAS.482.2991A}
{Arca Sedda}, M., \& {Benacquista}, M. 2019, \mnras, 482, 2991,
  \dodoi{10.1093/mnras/sty2764}

\bibitem[{{Askar} {et~al.}(2017){Askar}, {Szkudlarek}, {Gondek-Rosi{\'n}ska},
  {Giersz}, \& {Bulik}}]{2017MNRAS.464L..36A}
{Askar}, A., {Szkudlarek}, M., {Gondek-Rosi{\'n}ska}, D., {Giersz}, M., \&
  {Bulik}, T. 2017, \mnras, 464, L36, \dodoi{10.1093/mnrasl/slw177}

\bibitem[{{Babak} {et~al.}(2017){Babak}, {Taracchini}, \&
  {Buonanno}}]{2017PhRvD..95b4010B}
{Babak}, S., {Taracchini}, A., \& {Buonanno}, A. 2017, \prd, 95, 024010,
  \dodoi{10.1103/PhysRevD.95.024010}

\bibitem[{{Bai} {et~al.}(2018){Bai}, {Barger}, \& {Lu}}]{2018arXiv180204909B}
{Bai}, Y., {Barger}, V., \& {Lu}, S. 2018, arXiv e-prints, arXiv:1802.04909.
\newblock \doarXiv{1802.04909}

\bibitem[{{Bailyn} {et~al.}(1998){Bailyn}, {Jain}, {Coppi}, \&
  {Orosz}}]{1998ApJ...499..367B}
{Bailyn}, C.~D., {Jain}, R.~K., {Coppi}, P., \& {Orosz}, J.~A. 1998, \apj, 499,
  367, \dodoi{10.1086/305614}

\bibitem[{{Baird} {et~al.}(2013){Baird}, {Fairhurst}, {Hannam}, \&
  {Murphy}}]{2013PhRvD..87b4035B}
{Baird}, E., {Fairhurst}, S., {Hannam}, M., \& {Murphy}, P. 2013, \prd, 87,
  024035, \dodoi{10.1103/PhysRevD.87.024035}

\bibitem[{{Banerjee}(2017)}]{2017MNRAS.467..524B}
{Banerjee}, S. 2017, \mnras, 467, 524, \dodoi{10.1093/mnras/stw3392}

\bibitem[{{Barkat} {et~al.}(1967){Barkat}, {Rakavy}, \&
  {Sack}}]{1967PhRvL..18..379B}
{Barkat}, Z., {Rakavy}, G., \& {Sack}, N. 1967, \prl, 18, 379,
  \dodoi{10.1103/PhysRevLett.18.379}

\bibitem[{{Barrett} {et~al.}(2018){Barrett}, {Gaebel}, {Neijssel},
  {Vigna-G{\'o}mez}, {Stevenson}, {Berry}, {Farr}, \&
  {Mandel}}]{2018MNRAS.477.4685B}
{Barrett}, J.~W., {Gaebel}, S.~M., {Neijssel}, C.~J., {et~al.} 2018, \mnras,
  477, 4685, \dodoi{10.1093/mnras/sty908}

\bibitem[{{Bartos} {et~al.}(2017){Bartos}, {Kocsis}, {Haiman}, \&
  {M{\'a}rka}}]{2017ApJ...835..165B}
{Bartos}, I., {Kocsis}, B., {Haiman}, Z., \& {M{\'a}rka}, S. 2017, \apj, 835,
  165, \dodoi{10.3847/1538-4357/835/2/165}

\bibitem[{{Belczynski} {et~al.}(2010){Belczynski}, {Bulik}, {Fryer}, {Ruiter},
  {Valsecchi}, {Vink}, \& {Hurley}}]{2010ApJ...714.1217B}
{Belczynski}, K., {Bulik}, T., {Fryer}, C.~L., {et~al.} 2010, \apj, 714, 1217,
  \dodoi{10.1088/0004-637X/714/2/1217}

\bibitem[{{Belczynski} {et~al.}(2014){Belczynski}, {Buonanno}, {Cantiello},
  {Fryer}, {Holz}, {Mandel}, {Miller}, \& {Walczak}}]{2014ApJ...789..120B}
{Belczynski}, K., {Buonanno}, A., {Cantiello}, M., {et~al.} 2014, \apj, 789,
  120, \dodoi{10.1088/0004-637X/789/2/120}

\bibitem[{{Belczynski} {et~al.}(2016{\natexlab{a}}){Belczynski}, {Holz},
  {Bulik}, \& {O'Shaughnessy}}]{2016Natur.534..512B}
{Belczynski}, K., {Holz}, D.~E., {Bulik}, T., \& {O'Shaughnessy}, R.
  2016{\natexlab{a}}, \nat, 534, 512, \dodoi{10.1038/nature18322}

\bibitem[{{Belczynski} {et~al.}(2002){Belczynski}, {Kalogera}, \&
  {Bulik}}]{2002ApJ...572..407B}
{Belczynski}, K., {Kalogera}, V., \& {Bulik}, T. 2002, \apj, 572, 407,
  \dodoi{10.1086/340304}

\bibitem[{{Belczynski} {et~al.}(2008){Belczynski}, {Kalogera}, {Rasio}, {Taam},
  {Zezas}, {Bulik}, {Maccarone}, \& {Ivanova}}]{2008ApJS..174..223B}
{Belczynski}, K., {Kalogera}, V., {Rasio}, F.~A., {et~al.} 2008, The
  Astrophysical Journal Supplement Series, 174, 223, \dodoi{10.1086/521026}

\bibitem[{{Belczynski} {et~al.}(2007){Belczynski}, {Taam}, {Kalogera}, {Rasio},
  \& {Bulik}}]{2007ApJ...662..504B}
{Belczynski}, K., {Taam}, R.~E., {Kalogera}, V., {Rasio}, F.~A., \& {Bulik}, T.
  2007, \apj, 662, 504, \dodoi{10.1086/513562}

\bibitem[{{Belczynski} {et~al.}(2016{\natexlab{b}}){Belczynski}, {Heger},
  {Gladysz}, {Ruiter}, {Woosley}, {Wiktorowicz}, {Chen}, {Bulik},
  {O'Shaughnessy}, {Holz}, {Fryer}, \& {Berti}}]{2016A&A...594A..97B}
{Belczynski}, K., {Heger}, A., {Gladysz}, W., {et~al.} 2016{\natexlab{b}},
  \aap, 594, A97, \dodoi{10.1051/0004-6361/201628980}

\bibitem[{{Belczynski} {et~al.}(2017){Belczynski}, {Klencki}, {Meynet},
  {Fryer}, {Brown}, {Chruslinska}, {Gladysz}, {O'Shaughnessy}, {Bulik},
  {Berti}, {Holz}, {Gerosa}, {Giersz}, {Ekstrom}, {Georgy}, {Askar}, {Wysocki},
  \& {Lasota}}]{2017arXiv170607053B}
{Belczynski}, K., {Klencki}, J., {Meynet}, G., {et~al.} 2017, arXiv e-prints,
  arXiv:1706.07053.
\newblock \doarXiv{1706.07053}

\bibitem[{{Berti} {et~al.}(2007){Berti}, {Cardoso}, {Gonzalez}, {Sperhake},
  {Hannam}, {Husa}, \& {Br{\"u}gmann}}]{2007PhRvD..76f4034B}
{Berti}, E., {Cardoso}, V., {Gonzalez}, J.~A., {et~al.} 2007, \prd, 76, 064034,
  \dodoi{10.1103/PhysRevD.76.064034}

\bibitem[{{Bethe} \& {Brown}(1998)}]{1998ApJ...506..780B}
{Bethe}, H.~A., \& {Brown}, G.~E. 1998, \apj, 506, 780, \dodoi{10.1086/306265}

\bibitem[{{Bird} {et~al.}(2016){Bird}, {Cholis}, {Mu{\~n}oz}, {Ali-
  Ha{\"\i}moud}, {Kamionkowski}, {Kovetz}, {Raccanelli}, \&
  {Riess}}]{2016PhRvL.116t1301B}
{Bird}, S., {Cholis}, I., {Mu{\~n}oz}, J.~B., {et~al.} 2016, \prl, 116, 201301,
  \dodoi{10.1103/PhysRevLett.116.201301}

\bibitem[{{Bond} {et~al.}(1984){Bond}, {Arnett}, \&
  {Carr}}]{1984ApJ...280..825B}
{Bond}, J.~R., {Arnett}, W.~D., \& {Carr}, B.~J. 1984, \apj, 280, 825,
  \dodoi{10.1086/162057}

\bibitem[{{Bressan} {et~al.}(2012){Bressan}, {Marigo}, {Girardi}, {Salasnich},
  {Dal Cero}, {Rubele}, \& {Nanni}}]{2012MNRAS.427..127B}
{Bressan}, A., {Marigo}, P., {Girardi}, L., {et~al.} 2012, \mnras, 427, 127,
  \dodoi{10.1111/j.1365-2966.2012.21948.x}

\bibitem[{{Brott} {et~al.}(2011){Brott}, {de Mink}, {Cantiello}, {Langer}, {de
  Koter}, {Evans}, {Hunter}, {Trundle}, \& {Vink}}]{2011A&A...530A.115B}
{Brott}, I., {de Mink}, S.~E., {Cantiello}, M., {et~al.} 2011, \aap, 530, A115,
  \dodoi{10.1051/0004-6361/201016113}

\bibitem[{{Byrnes} {et~al.}(2018){Byrnes}, {Hindmarsh}, {Young}, \&
  {Hawkins}}]{2018JCAP...08..041B}
{Byrnes}, C.~T., {Hindmarsh}, M., {Young}, S., \& {Hawkins}, M. R.~S. 2018,
  Journal of Cosmology and Astro-Particle Physics, 2018, 041,
  \dodoi{10.1088/1475-7516/2018/08/041}

\bibitem[{{Carr} {et~al.}(2016){Carr}, {K{\"u}hnel}, \&
  {Sandstad}}]{2016PhRvD..94h3504C}
{Carr}, B., {K{\"u}hnel}, F., \& {Sandstad}, M. 2016, \prd, 94, 083504,
  \dodoi{10.1103/PhysRevD.94.083504}

\bibitem[{{Carr} \& {Hawking}(1974)}]{1974MNRAS.168..399C}
{Carr}, B.~J., \& {Hawking}, S.~W. 1974, \mnras, 168, 399,
  \dodoi{10.1093/mnras/168.2.399}

\bibitem[{{Chatterjee} {et~al.}(2017){Chatterjee}, {Rodriguez}, {Kalogera}, \&
  {Rasio}}]{2017ApJ...836L..26C}
{Chatterjee}, S., {Rodriguez}, C.~L., {Kalogera}, V., \& {Rasio}, F.~A. 2017,
  \apj, 836, L26, \dodoi{10.3847/2041-8213/aa5caa}

\bibitem[{{Chen} {et~al.}(2018){Chen}, {Fishbach}, \&
  {Holz}}]{2018Natur.562..545C}
{Chen}, H.-Y., {Fishbach}, M., \& {Holz}, D.~E. 2018, \nat, 562, 545,
  \dodoi{10.1038/s41586-018-0606-0}

\bibitem[{{Chen} \& {Huang}(2018)}]{2018ApJ...864...61C}
{Chen}, Z.-C., \& {Huang}, Q.-G. 2018, \apj, 864, 61,
  \dodoi{10.3847/1538-4357/aad6e2}

\bibitem[{{Chornock} {et~al.}(2017){Chornock}, {Berger}, {Kasen},
  {Cowperthwaite}, {Nicholl}, {Villar}, {Alexander}, {Blanchard}, {Eftekhari},
  {Fong}, {Margutti}, {Williams}, {Annis}, {Brout}, {Brown}, {Chen}, {Drout},
  {Farr}, {Foley}, {Frieman}, {Fryer}, {Herner}, {Holz}, {Kessler}, {Matheson},
  {Metzger}, {Quataert}, {Rest}, {Sako}, {Scolnic}, {Smith}, \&
  {Soares-Santos}}]{2017ApJ...848L..19C}
{Chornock}, R., {Berger}, E., {Kasen}, D., {et~al.} 2017, \apj, 848, L19,
  \dodoi{10.3847/2041-8213/aa905c}

\bibitem[{{Chruslinska} {et~al.}(2018){Chruslinska}, {Belczynski}, {Klencki},
  \& {Benacquista}}]{2018MNRAS.474.2937C}
{Chruslinska}, M., {Belczynski}, K., {Klencki}, J., \& {Benacquista}, M. 2018,
  \mnras, 474, 2937, \dodoi{10.1093/mnras/stx2923}

\bibitem[{{Clausen} {et~al.}(2013){Clausen}, {Sigurdsson}, \&
  {Chernoff}}]{2013MNRAS.428.3618C}
{Clausen}, D., {Sigurdsson}, S., \& {Chernoff}, D.~F. 2013, \mnras, 428, 3618,
  \dodoi{10.1093/mnras/sts295}

\bibitem[{{Clesse} \& {Garc{\'\i}a-Bellido}(2017)}]{2017PDU....15..142C}
{Clesse}, S., \& {Garc{\'\i}a-Bellido}, J. 2017, Physics of the Dark Universe,
  15, 142, \dodoi{10.1016/j.dark.2016.10.002}

\bibitem[{{Corsaro} {et~al.}(2017){Corsaro}, {Lee}, {Garc{\'\i}a},
  {Hennebelle}, {Mathur}, {Beck}, {Mathis}, {Stello}, \&
  {Bouvier}}]{2017NatAs...1E..64C}
{Corsaro}, E., {Lee}, Y.-N., {Garc{\'\i}a}, R.~A., {et~al.} 2017, Nature
  Astronomy, 1, 0064, \dodoi{10.1038/s41550-017-0064}

\bibitem[{{Coughlin} {et~al.}(2015){Coughlin}, {Meyers}, {Thrane}, {Luo}, \&
  {Christensen}}]{2015PhRvD..91f3004C}
{Coughlin}, M., {Meyers}, P., {Thrane}, E., {Luo}, J., \& {Christensen}, N.
  2015, \prd, 91, 063004, \dodoi{10.1103/PhysRevD.91.063004}

\bibitem[{{Coulter} {et~al.}(2017){Coulter}, {Foley}, {Kilpatrick}, {Drout},
  {Piro}, {Shappee}, {Siebert}, {Simon}, {Ulloa}, {Kasen}, {Madore},
  {Murguia-Berthier}, {Pan}, {Prochaska}, {Ramirez-Ruiz}, {Rest}, \&
  {Rojas-Bravo}}]{2017Sci...358.1556C}
{Coulter}, D.~A., {Foley}, R.~J., {Kilpatrick}, C.~D., {et~al.} 2017, Science,
  358, 1556, \dodoi{10.1126/science.aap9811}

\bibitem[{{Cowperthwaite} {et~al.}(2017){Cowperthwaite}, {Berger}, {Villar},
  {Metzger}, {Nicholl}, {Chornock}, {Blanchard}, {Fong}, {Margutti},
  {Soares-Santos}, {Alexander}, {Allam}, {Annis}, {Brout}, {Brown}, {Butler},
  {Chen}, {Diehl}, {Doctor}, {Drout}, {Eftekhari}, {Farr}, {Finley}, {Foley},
  {Frieman}, {Fryer}, {Garc{\'\i}a-Bellido}, {Gill}, {Guillochon}, {Herner},
  {Holz}, {Kasen}, {Kessler}, {Marriner}, {Matheson}, {Neilsen}, {Quataert},
  {Palmese}, {Rest}, {Sako}, {Scolnic}, {Smith}, {Tucker}, {Williams},
  {Balbinot}, {Carlin}, {Cook}, {Durret}, {Li}, {Lopes}, {Louren{\c{c}}o},
  {Marshall}, {Medina}, {Muir}, {Mu{\\textasciitilde n}oz}, {Sauseda},
  {Schlegel}, {Secco}, {Vivas}, {Wester}, {Zenteno}, {Zhang}, {Abbott},
  {Banerji}, {Bechtol}, {Benoit-L{\'e}vy}, {Bertin}, {Buckley-Geer}, {Burke},
  {Capozzi}, {Carnero Rosell}, {Carrasco Kind}, {Castander}, {Crocce}, {Cunha},
  {D'Andrea}, {da Costa}, {Davis}, {DePoy}, {Desai}, {Dietrich},
  {Drlica-Wagner}, {Eifler}, {Evrard}, {Fernandez}, {Flaugher}, {Fosalba},
  {Gaztanaga}, {Gerdes}, {Giannantonio}, {Goldstein}, {Gruen}, {Gruendl},
  {Gutierrez}, {Honscheid}, {Jain}, {James}, {Jeltema}, {Johnson}, {Johnson},
  {Kent}, {Krause}, {Kron}, {Kuehn}, {Nuropatkin}, {Lahav}, {Lima}, {Lin},
  {Maia}, {March}, {Martini}, {McMahon}, {Menanteau}, {Miller}, {Miquel},
  {Mohr}, {Neilsen}, {Nichol}, {Ogando}, {Plazas}, {Roe}, {Romer}, {Roodman},
  {Rykoff}, {Sanchez}, {Scarpine}, {Schindler}, {Schubnell}, {Sevilla-Noarbe},
  {Smith}, {Smith}, {Sobreira}, {Suchyta}, {Swanson}, {Tarle}, {Thomas},
  {Thomas}, {Troxel}, {Vikram}, {Walker}, {Wechsler}, {Weller}, {Yanny}, \&
  {Zuntz}}]{2017ApJ...848L..17C}
{Cowperthwaite}, P.~S., {Berger}, E., {Villar}, V.~A., {et~al.} 2017, \apj,
  848, L17, \dodoi{10.3847/2041-8213/aa8fc7}

\bibitem[{{Damour}(2001)}]{2001PhRvD..64l4013D}
{Damour}, T. 2001, \prd, 64, 124013, \dodoi{10.1103/PhysRevD.64.124013}

\bibitem[{{de Mink} {et~al.}(2009){de Mink}, {Cantiello}, {Langer}, {Pols},
  {Brott}, \& {Yoon}}]{2009A&A...497..243D}
{de Mink}, S.~E., {Cantiello}, M., {Langer}, N., {et~al.} 2009, \aap, 497, 243,
  \dodoi{10.1051/0004-6361/200811439}

\bibitem[{{de Mink} {et~al.}(2013){de Mink}, {Langer}, {Izzard}, {Sana}, \& {de
  Koter}}]{2013ApJ...764..166D}
{de Mink}, S.~E., {Langer}, N., {Izzard}, R.~G., {Sana}, H., \& {de Koter}, A.
  2013, \apj, 764, 166, \dodoi{10.1088/0004-637X/764/2/166}

\bibitem[{{de Mink} \& {Mandel}(2016)}]{2016MNRAS.460.3545D}
{de Mink}, S.~E., \& {Mandel}, I. 2016, \mnras, 460, 3545,
  \dodoi{10.1093/mnras/stw1219}

\bibitem[{{Dewi} {et~al.}(2006){Dewi}, {Podsiadlowski}, \&
  {Sena}}]{2006MNRAS.368.1742D}
{Dewi}, J. D.~M., {Podsiadlowski}, P., \& {Sena}, A. 2006, \mnras, 368, 1742,
  \dodoi{10.1111/j.1365-2966.2006.10233.x}

\bibitem[{{Dominik} {et~al.}(2012){Dominik}, {Belczynski}, {Fryer}, {Holz},
  {Berti}, {Bulik}, {Mandel}, \& {O'Shaughnessy}}]{2012ApJ...759...52D}
{Dominik}, M., {Belczynski}, K., {Fryer}, C., {et~al.} 2012, \apj, 759, 52,
  \dodoi{10.1088/0004-637X/759/1/52}

\bibitem[{{Dominik} {et~al.}(2013){Dominik}, {Belczynski}, {Fryer}, {Holz},
  {Berti}, {Bulik}, {Mandel}, \& {O'Shaughnessy}}]{2013ApJ...779...72D}
---. 2013, \apj, 779, 72, \dodoi{10.1088/0004-637X/779/1/72}

\bibitem[{{Downing} {et~al.}(2010){Downing}, {Benacquista}, {Giersz}, \&
  {Spurzem}}]{2010MNRAS.407.1946D}
{Downing}, J. M.~B., {Benacquista}, M.~J., {Giersz}, M., \& {Spurzem}, R. 2010,
  \mnras, 407, 1946, \dodoi{10.1111/j.1365-2966.2010.17040.x}

\bibitem[{{Downing} {et~al.}(2011){Downing}, {Benacquista}, {Giersz}, \&
  {Spurzem}}]{2011MNRAS.416..133D}
---. 2011, \mnras, 416, 133, \dodoi{10.1111/j.1365-2966.2011.19023.x}

\bibitem[{{Eldridge} \& {Stanway}(2016)}]{2016MNRAS.462.3302E}
{Eldridge}, J.~J., \& {Stanway}, E.~R. 2016, \mnras, 462, 3302,
  \dodoi{10.1093/mnras/stw1772}

\bibitem[{{Ertl} {et~al.}(2016){Ertl}, {Janka}, {Woosley}, {Sukhbold}, \&
  {Ugliano}}]{2016ApJ...818..124E}
{Ertl}, T., {Janka}, H.~T., {Woosley}, S.~E., {Sukhbold}, T., \& {Ugliano}, M.
  2016, \apj, 818, 124, \dodoi{10.3847/0004-637X/818/2/124}

\bibitem[{{Farr} {et~al.}(2018){Farr}, {Holz}, \& {Farr}}]{2018ApJ...854L...9F}
{Farr}, B., {Holz}, D.~E., \& {Farr}, W.~M. 2018, \apj, 854, L9,
  \dodoi{10.3847/2041-8213/aaaa64}

\bibitem[{{Farr} {et~al.}(2014){Farr}, {Kalogera}, \&
  {Luijten}}]{2014PhRvD..90b4014F}
{Farr}, B., {Kalogera}, V., \& {Luijten}, E. 2014, \prd, 90, 024014,
  \dodoi{10.1103/PhysRevD.90.024014}

\bibitem[{Farr {et~al.}(2015)Farr, Farr, \&
  Littenberg}]{SplineCalMarg-T1400682}
Farr, W.~M., Farr, B., \& Littenberg, T. 2015, Modelling Calibration Errors In
  CBC Waveforms, Tech. Rep. {LIGO}-T1400682, LIGO Scientific Collaboration and
  Virgo Collaboration

\bibitem[{{Farr} {et~al.}(2015){Farr}, {Gair}, {Mandel}, \&
  {Cutler}}]{2015PhRvD..91b3005F}
{Farr}, W.~M., {Gair}, J.~R., {Mandel}, I., \& {Cutler}, C. 2015, \prd, 91,
  023005, \dodoi{10.1103/PhysRevD.91.023005}

\bibitem[{{Farr} {et~al.}(2011{\natexlab{a}}){Farr}, {Kremer}, {Lyutikov}, \&
  {Kalogera}}]{2011ApJ...742...81F}
{Farr}, W.~M., {Kremer}, K., {Lyutikov}, M., \& {Kalogera}, V.
  2011{\natexlab{a}}, \apj, 742, 81, \dodoi{10.1088/0004-637X/742/2/81}

\bibitem[{{Farr} {et~al.}(2011{\natexlab{b}}){Farr}, {Sravan}, {Cantrell},
  {Kreidberg}, {Bailyn}, {Mandel}, \& {Kalogera}}]{2011ApJ...741..103F}
{Farr}, W.~M., {Sravan}, N., {Cantrell}, A., {et~al.} 2011{\natexlab{b}}, \apj,
  741, 103, \dodoi{10.1088/0004-637X/741/2/103}

\bibitem[{{Farr} {et~al.}(2017){Farr}, {Stevenson}, {Miller}, {Mandel}, {Farr},
  \& {Vecchio}}]{2017Natur.548..426F}
{Farr}, W.~M., {Stevenson}, S., {Miller}, M.~C., {et~al.} 2017, \nat, 548, 426,
  \dodoi{10.1038/nature23453}

\bibitem[{Ferrari \& Cribari-Neto(2004)}]{BetaStats}
Ferrari, S., \& Cribari-Neto, F. 2004, Journal of Applied Statistics, 31, 799,
  \dodoi{10.1080/0266476042000214501}

\bibitem[{{Finn} \& {Chernoff}(1993)}]{1993PhRvD..47.2198F}
{Finn}, L.~S., \& {Chernoff}, D.~F. 1993, \prd, 47, 2198,
  \dodoi{10.1103/PhysRevD.47.2198}

\bibitem[{{Fishbach} \& {Holz}(2017)}]{2017ApJ...851L..25F}
{Fishbach}, M., \& {Holz}, D.~E. 2017, \apj, 851, L25,
  \dodoi{10.3847/2041-8213/aa9bf6}

\bibitem[{{Fishbach} {et~al.}(2017){Fishbach}, {Holz}, \&
  {Farr}}]{2017ApJ...840L..24F}
{Fishbach}, M., {Holz}, D.~E., \& {Farr}, B. 2017, \apj, 840, L24,
  \dodoi{10.3847/2041-8213/aa7045}

\bibitem[{{Fishbach} {et~al.}(2018){Fishbach}, {Holz}, \&
  {Farr}}]{2018ApJ...863L..41F}
{Fishbach}, M., {Holz}, D.~E., \& {Farr}, W.~M. 2018, \apj, 863, L41,
  \dodoi{10.3847/2041-8213/aad800}

\bibitem[{{Foreman-Mackey} {et~al.}(2014){Foreman-Mackey}, {Hogg}, \&
  {Morton}}]{2014ApJ...795...64F}
{Foreman-Mackey}, D., {Hogg}, D.~W., \& {Morton}, T.~D. 2014, \apj, 795, 64,
  \dodoi{10.1088/0004-637X/795/1/64}

\bibitem[{{Fowler} \& {Hoyle}(1964)}]{1964ApJS....9..201F}
{Fowler}, W.~A., \& {Hoyle}, F. 1964, The Astrophysical Journal Supplement
  Series, 9, 201, \dodoi{10.1086/190103}

\bibitem[{{Fragione} {et~al.}(2018){Fragione}, {Ginsburg}, \&
  {Kocsis}}]{2018ApJ...856...92F}
{Fragione}, G., {Ginsburg}, I., \& {Kocsis}, B. 2018, \apj, 856, 92,
  \dodoi{10.3847/1538-4357/aab368}

\bibitem[{{Fregeau}(2004)}]{2004PhDT.......185F}
{Fregeau}, J.~M. 2004, PhD thesis, MASSACHUSETTS INSTITUTE OF TECHNOLOGY.
\newblock \url{http://hdl.handle.net/1721.1/29454}

\bibitem[{{Freire} {et~al.}(2008){Freire}, {Ransom}, {B{\'e}gin}, {Stairs},
  {Hessels}, {Frey}, \& {Camilo}}]{2008ApJ...675..670F}
{Freire}, P. C.~C., {Ransom}, S.~M., {B{\'e}gin}, S., {et~al.} 2008, \apj, 675,
  670, \dodoi{10.1086/526338}

\bibitem[{{Fryer} {et~al.}(2012){Fryer}, {Belczynski}, {Wiktorowicz},
  {Dominik}, {Kalogera}, \& {Holz}}]{2012ApJ...749...91F}
{Fryer}, C.~L., {Belczynski}, K., {Wiktorowicz}, G., {et~al.} 2012, \apj, 749,
  91, \dodoi{10.1088/0004-637X/749/1/91}

\bibitem[{{Fuller} {et~al.}(2015){Fuller}, {Cantiello}, {Lecoanet}, \&
  {Quataert}}]{2015ApJ...810..101F}
{Fuller}, J., {Cantiello}, M., {Lecoanet}, D., \& {Quataert}, E. 2015, \apj,
  810, 101, \dodoi{10.1088/0004-637X/810/2/101}

\bibitem[{{Gaebel} {et~al.}(2019){Gaebel}, {Veitch}, {Dent}, \&
  {Farr}}]{2019MNRAS.tmp..230G}
{Gaebel}, S.~M., {Veitch}, J., {Dent}, T., \& {Farr}, W.~M. 2019, \mnras, 230,
  \dodoi{10.1093/mnras/stz225}

\bibitem[{Gelman {et~al.}(2004)Gelman, Carlin, Stern, Dunson, Vehtari, \&
  Rubin}]{gelman2004bayesian}
Gelman, A., Carlin, J.~B., Stern, H.~S., {et~al.} 2004, Bayesian data analysis
  (Boca Raton, Fla: Chapman \& Hall/CRC).
\newblock \url{https://doi.org/10.1103/PhysRev.136.B1224}

\bibitem[{{Georg} \& {Watson}(2017)}]{2017JHEP...09..138G}
{Georg}, J., \& {Watson}, S. 2017, Journal of High Energy Physics, 2017, 138,
  \dodoi{10.1007/JHEP09(2017)138}

\bibitem[{{Gerosa} \& {Berti}(2017)}]{2017PhRvD..95l4046G}
{Gerosa}, D., \& {Berti}, E. 2017, \prd, 95, 124046,
  \dodoi{10.1103/PhysRevD.95.124046}

\bibitem[{{Gerosa} {et~al.}(2018){Gerosa}, {Berti}, {O'Shaughnessy},
  {Belczynski}, {Kesden}, {Wysocki}, \& {Gladysz}}]{2018PhRvD..98h4036G}
{Gerosa}, D., {Berti}, E., {O'Shaughnessy}, R., {et~al.} 2018, \prd, 98,
  084036, \dodoi{10.1103/PhysRevD.98.084036}

\bibitem[{{Gerosa} {et~al.}(2015){Gerosa}, {Kesden}, {Sperhake}, {Berti}, \&
  {O'Shaughnessy}}]{2015PhRvD..92f4016G}
{Gerosa}, D., {Kesden}, M., {Sperhake}, U., {Berti}, E., \& {O'Shaughnessy}, R.
  2015, \prd, 92, 064016, \dodoi{10.1103/PhysRevD.92.064016}

\bibitem[{{Giacobbo} \& {Mapelli}(2018)}]{2018MNRAS.480.2011G}
{Giacobbo}, N., \& {Mapelli}, M. 2018, \mnras, 480, 2011,
  \dodoi{10.1093/mnras/sty1999}

\bibitem[{{Giacobbo} {et~al.}(2018){Giacobbo}, {Mapelli}, \&
  {Spera}}]{2018MNRAS.474.2959G}
{Giacobbo}, N., {Mapelli}, M., \& {Spera}, M. 2018, \mnras, 474, 2959,
  \dodoi{10.1093/mnras/stx2933}

\bibitem[{{Goldstein} {et~al.}(2017){Goldstein}, {Veres}, {Burns}, {Briggs},
  {Hamburg}, {Kocevski}, {Wilson-Hodge}, {Preece}, {Poolakkil}, {Roberts},
  {Hui}, {Connaughton}, {Racusin}, {von Kienlin}, {Dal Canton}, {Christensen},
  {Littenberg}, {Siellez}, {Blackburn}, {Broida}, {Bissaldi}, {Cleveland},
  {Gibby}, {Giles}, {Kippen}, {McBreen}, {McEnery}, {Meegan}, {Paciesas}, \&
  {Stanbro}}]{2017ApJ...848L..14G}
{Goldstein}, A., {Veres}, P., {Burns}, E., {et~al.} 2017, \apj, 848, L14,
  \dodoi{10.3847/2041-8213/aa8f41}

\bibitem[{{Gonz{\'a}lez} {et~al.}(2007){Gonz{\'a}lez}, {Sperhake},
  {Br{\"u}gmann}, {Hannam}, \& {Husa}}]{2007PhRvL..98i1101G}
{Gonz{\'a}lez}, J.~A., {Sperhake}, U., {Br{\"u}gmann}, B., {Hannam}, M., \&
  {Husa}, S. 2007, \prl, 98, 091101, \dodoi{10.1103/PhysRevLett.98.091101}

\bibitem[{{Gr{\"a}fener} \& {Hamann}(2008)}]{2008A&A...482..945G}
{Gr{\"a}fener}, G., \& {Hamann}, W.~R. 2008, \aap, 482, 945,
  \dodoi{10.1051/0004-6361:20066176}

\bibitem[{{Grandcl{\'e}ment} {et~al.}(2004){Grandcl{\'e}ment}, {Ihm},
  {Kalogera}, \& {Belczynski}}]{2004PhRvD..69j2002G}
{Grandcl{\'e}ment}, P., {Ihm}, M., {Kalogera}, V., \& {Belczynski}, K. 2004,
  \prd, 69, 102002, \dodoi{10.1103/PhysRevD.69.102002}

\bibitem[{{Grindlay} {et~al.}(2006){Grindlay}, {Portegies Zwart}, \&
  {McMillan}}]{2006NatPh...2..116G}
{Grindlay}, J., {Portegies Zwart}, S., \& {McMillan}, S. 2006, Nature Physics,
  2, 116, \dodoi{10.1038/nphys214}

\bibitem[{{Hannam} {et~al.}(2014){Hannam}, {Schmidt}, {Boh{\'e}}, {Haegel},
  {Husa}, {Ohme}, {Pratten}, \& {P{\"u}rrer}}]{2014PhRvL.113o1101H}
{Hannam}, M., {Schmidt}, P., {Boh{\'e}}, A., {et~al.} 2014, \prl, 113, 151101,
  \dodoi{10.1103/PhysRevLett.113.151101}

\bibitem[{{Heger} {et~al.}(2003){Heger}, {Fryer}, {Woosley}, {Langer}, \&
  {Hartmann}}]{2003ApJ...591..288H}
{Heger}, A., {Fryer}, C.~L., {Woosley}, S.~E., {Langer}, N., \& {Hartmann},
  D.~H. 2003, \apj, 591, 288, \dodoi{10.1086/375341}

\bibitem[{{Heger} \& {Woosley}(2002)}]{2002ApJ...567..532H}
{Heger}, A., \& {Woosley}, S.~E. 2002, \apj, 567, 532, \dodoi{10.1086/338487}

\bibitem[{Hilbe {et~al.}(2017)Hilbe, de~Souza, \&
  Ishida}]{hilbe_de_souza_ishida_2017}
Hilbe, J.~M., de~Souza, R.~S., \& Ishida, E. E.~O. 2017, Bayesian Models for
  Astrophysical Data: Using R, JAGS, Python, and Stan (Cambridge University
  Press), \dodoi{10.1017/CBO9781316459515}.
\newblock \url{https://doi.org/10.1017/CBO9781316459515}

\bibitem[{{Hinder} {et~al.}(2018){Hinder}, {Kidder}, \&
  {Pfeiffer}}]{2018PhRvD..98d4015H}
{Hinder}, I., {Kidder}, L.~E., \& {Pfeiffer}, H.~P. 2018, \prd, 98, 044015,
  \dodoi{10.1103/PhysRevD.98.044015}

\bibitem[{{Hinder} {et~al.}(2008){Hinder}, {Vaishnav}, {Herrmann}, {Shoemaker},
  \& {Laguna}}]{2008PhRvD..77h1502H}
{Hinder}, I., {Vaishnav}, B., {Herrmann}, F., {Shoemaker}, D.~M., \& {Laguna},
  P. 2008, \prd, 77, 081502, \dodoi{10.1103/PhysRevD.77.081502}

\bibitem[{{Hogg}(1999)}]{1999astro.ph..5116H}
{Hogg}, D.~W. 1999, arXiv e-prints, astro.
\newblock \doarXiv{astro-ph/9905116}

\bibitem[{{Hogg} {et~al.}(2010){Hogg}, {Myers}, \&
  {Bovy}}]{2010ApJ...725.2166H}
{Hogg}, D.~W., {Myers}, A.~D., \& {Bovy}, J. 2010, \apj, 725, 2166,
  \dodoi{10.1088/0004-637X/725/2/2166}

\bibitem[{{Huerta} {et~al.}(2014){Huerta}, {Kumar}, {McWilliams},
  {O'Shaughnessy}, \& {Yunes}}]{2014PhRvD..90h4016H}
{Huerta}, E.~A., {Kumar}, P., {McWilliams}, S.~T., {O'Shaughnessy}, R., \&
  {Yunes}, N. 2014, \prd, 90, 084016, \dodoi{10.1103/PhysRevD.90.084016}

\bibitem[{{Huerta} {et~al.}(2017){Huerta}, {Kumar}, {Agarwal}, {George},
  {Schive}, {Pfeiffer}, {Haas}, {Ren}, {Chu}, {Boyle}, {Hemberger}, {Kidder},
  {Scheel}, \& {Szilagyi}}]{2017PhRvD..95b4038H}
{Huerta}, E.~A., {Kumar}, P., {Agarwal}, B., {et~al.} 2017, \prd, 95, 024038,
  \dodoi{10.1103/PhysRevD.95.024038}

\bibitem[{{Husa} {et~al.}(2016){Husa}, {Khan}, {Hannam}, {P{\"u}rrer}, {Ohme},
  {Forteza}, \& {Boh{\'e}}}]{2016PhRvD..93d4006H}
{Husa}, S., {Khan}, S., {Hannam}, M., {et~al.} 2016, \prd, 93, 044006,
  \dodoi{10.1103/PhysRevD.93.044006}

\bibitem[{{Inayoshi} {et~al.}(2016){Inayoshi}, {Kashiyama}, {Visbal}, \&
  {Haiman}}]{2016MNRAS.461.2722I}
{Inayoshi}, K., {Kashiyama}, K., {Visbal}, E., \& {Haiman}, Z. 2016, \mnras,
  461, 2722, \dodoi{10.1093/mnras/stw1431}

\bibitem[{{Inomata} {et~al.}(2017){Inomata}, {Kawasaki}, {Mukaida}, {Tada}, \&
  {Yanagida}}]{2017PhRvD..95l3510I}
{Inomata}, K., {Kawasaki}, M., {Mukaida}, K., {Tada}, Y., \& {Yanagida}, T.~T.
  2017, \prd, 95, 123510, \dodoi{10.1103/PhysRevD.95.123510}

\bibitem[{{Ivanova} {et~al.}(2008){Ivanova}, {Heinke}, {Rasio}, {Belczynski},
  \& {Fregeau}}]{2008MNRAS.386..553I}
{Ivanova}, N., {Heinke}, C.~O., {Rasio}, F.~A., {Belczynski}, K., \& {Fregeau},
  J.~M. 2008, \mnras, 386, 553, \dodoi{10.1111/j.1365-2966.2008.13064.x}

\bibitem[{{Janka}(2012)}]{2012ARNPS..62..407J}
{Janka}, H.-T. 2012, Annual Review of Nuclear and Particle Science, 62, 407,
  \dodoi{10.1146/annurev-nucl-102711-094901}

\bibitem[{{Kalogera}(2000)}]{2000ApJ...541..319K}
{Kalogera}, V. 2000, \apj, 541, 319, \dodoi{10.1086/309400}

\bibitem[{{Kalogera} {et~al.}(2007){Kalogera}, {Belczynski}, {Kim},
  {O'Shaughnessy}, \& {Willems}}]{2007PhR...442...75K}
{Kalogera}, V., {Belczynski}, K., {Kim}, C., {O'Shaughnessy}, R., \& {Willems},
  B. 2007, \physrep, 442, 75, \dodoi{10.1016/j.physrep.2007.02.008}

\bibitem[{{Khan} {et~al.}(2016){Khan}, {Husa}, {Hannam}, {Ohme}, {P{\"u}rrer},
  {Forteza}, \& {Boh{\'e}}}]{2016PhRvD..93d4007K}
{Khan}, S., {Husa}, S., {Hannam}, M., {et~al.} 2016, \prd, 93, 044007,
  \dodoi{10.1103/PhysRevD.93.044007}

\bibitem[{{Kimpson} {et~al.}(2016){Kimpson}, {Spera}, {Mapelli}, \&
  {Ziosi}}]{2016MNRAS.463.2443K}
{Kimpson}, T.~O., {Spera}, M., {Mapelli}, M., \& {Ziosi}, B.~M. 2016, \mnras,
  463, 2443, \dodoi{10.1093/mnras/stw2085}

\bibitem[{{Klein} {et~al.}(2018){Klein}, {Boetzel}, {Gopakumar}, {Jetzer}, \&
  {de Vittori}}]{2018PhRvD..98j4043K}
{Klein}, A., {Boetzel}, Y., {Gopakumar}, A., {Jetzer}, P., \& {de Vittori}, L.
  2018, \prd, 98, 104043, \dodoi{10.1103/PhysRevD.98.104043}

\bibitem[{{Kocsis} \& {Levin}(2012)}]{2012PhRvD..85l3005K}
{Kocsis}, B., \& {Levin}, J. 2012, \prd, 85, 123005,
  \dodoi{10.1103/PhysRevD.85.123005}

\bibitem[{{Kovetz} {et~al.}(2017){Kovetz}, {Cholis}, {Breysse}, \&
  {Kamionkowski}}]{2017PhRvD..95j3010K}
{Kovetz}, E.~D., {Cholis}, I., {Breysse}, P.~C., \& {Kamionkowski}, M. 2017,
  \prd, 95, 103010, \dodoi{10.1103/PhysRevD.95.103010}

\bibitem[{{Kreidberg} {et~al.}(2012){Kreidberg}, {Bailyn}, {Farr}, \&
  {Kalogera}}]{2012ApJ...757...36K}
{Kreidberg}, L., {Bailyn}, C.~D., {Farr}, W.~M., \& {Kalogera}, V. 2012, \apj,
  757, 36, \dodoi{10.1088/0004-637X/757/1/36}

\bibitem[{{Kremer} {et~al.}(2019){Kremer}, {Chatterjee}, {Ye}, {Rodriguez}, \&
  {Rasio}}]{2019ApJ...871...38K}
{Kremer}, K., {Chatterjee}, S., {Ye}, C.~S., {Rodriguez}, C.~L., \& {Rasio},
  F.~A. 2019, \apj, 871, 38, \dodoi{10.3847/1538-4357/aaf646}

\bibitem[{{Kruckow} {et~al.}(2018){Kruckow}, {Tauris}, {Langer}, {Kramer}, \&
  {Izzard}}]{2018MNRAS.481.1908K}
{Kruckow}, M.~U., {Tauris}, T.~M., {Langer}, N., {Kramer}, M., \& {Izzard},
  R.~G. 2018, \mnras, 481, 1908, \dodoi{10.1093/mnras/sty2190}

\bibitem[{{Kudritzki} \& {Puls}(2000)}]{2000ARA&A..38..613K}
{Kudritzki}, R.-P., \& {Puls}, J. 2000, Annual Review of Astronomy and
  Astrophysics, 38, 613, \dodoi{10.1146/annurev.astro.38.1.613}

\bibitem[{{Kulkarni} {et~al.}(1993){Kulkarni}, {Hut}, \&
  {McMillan}}]{1993Natur.364..421K}
{Kulkarni}, S.~R., {Hut}, P., \& {McMillan}, S. 1993, \nat, 364, 421,
  \dodoi{10.1038/364421a0}

\bibitem[{{Langer}(2012)}]{2012ARA&A..50..107L}
{Langer}, N. 2012, Annual Review of Astronomy and Astrophysics, 50, 107,
  \dodoi{10.1146/annurev-astro-081811-125534}

\bibitem[{{LIGO Scientific Collaboration} {et~al.}(2015){LIGO Scientific
  Collaboration}, {Aasi}, {Abbott}, {Abbott}, {Abbott}, {Abernathy}, {Ackley},
  {Adams}, {Adams}, {Addesso}, {Adhikari}, {Adya}, {Affeldt}, {Aggarwal},
  {Aguiar}, {Ain}, {Ajith}, {Alemic}, {Allen}, {Amariutei}, {Anderson},
  {Anderson}, {Arai}, {Araya}, {Arceneaux}, {Areeda}, {Ashton}, {Ast}, {Aston},
  {Aufmuth}, {Aulbert}, {Aylott}, {Babak}, {Baker}, {Ballmer}, {Barayoga},
  {Barbet}, {Barclay}, {Barish}, {Barker}, {Barr}, {Barsotti}, {Bartlett},
  {Barton}, {Bartos}, {Bassiri}, {Batch}, {Baune}, {Behnke}, {Bell}, {Bell},
  {Benacquista}, {Bergman}, {Bergmann}, {Berry}, {Betzwieser}, {Bhagwat},
  {Bhandare}, {Bilenko}, {Billingsley}, {Birch}, {Biscans}, {Biwer},
  {Blackburn}, {Blackburn}, {Blair}, {Blair}, {Bock}, {Bodiya}, {Bojtos},
  {Bond}, {Bork}, {Born}, {Bose}, {Brady}, {Braginsky}, {Brau}, {Bridges},
  {Brinkmann}, {Brooks}, {Brown}, {Brown}, {Brown}, {Buchman}, {Buikema},
  {Buonanno}, {Cadonati}, {Calder{\'o}n Bustillo}, {Camp}, {Cannon}, {Cao},
  {Capano}, {Caride}, {Caudill}, {Cavagli{\`a}}, {Cepeda}, {Chakraborty},
  {Chalermsongsak}, {Chamberlin}, {Chao}, {Charlton}, {Chen}, {Cho}, {Cho},
  {Chow}, {Christensen}, {Chu}, {Chung}, {Ciani}, {Clara}, {Clark}, {Collette},
  {Cominsky}, {Constancio}, {Cook}, {Corbitt}, {Cornish}, {Corsi}, {Costa},
  {Coughlin}, {Countryman}, {Couvares}, {Coward}, {Cowart}, {Coyne}, {Coyne},
  {Craig}, {Creighton}, {Creighton}, {Cripe}, {Crowder}, {Cumming},
  {Cunningham}, {Cutler}, {Dahl}, {Dal Canton}, {Damjanic}, {Danilishin},
  {Danzmann}, {Dartez}, {Dave}, {Daveloza}, {Davies}, {Daw}, {DeBra}, {Del
  Pozzo}, {Denker}, {Dent}, {Dergachev}, {DeRosa}, {DeSalvo}, {Dhurandhar},
  {D́{\i}az}, {Di Palma}, {Dojcinoski}, {Dominguez}, {Donovan}, {Dooley},
  {Doravari}, {Douglas}, {Downes}, {Driggers}, {Du}, {Dwyer}, {Eberle}, {Edo},
  {Edwards}, {Edwards}, {Effler}, {Eggenstein}, {Ehrens}, {Eichholz},
  {Eikenberry}, {Essick}, {Etzel}, {Evans}, {Evans}, {Factourovich},
  {Fairhurst}, {Fan}, {Fang}, {Farr}, {Farr}, {Favata}, {Fays}, {Fehrmann},
  {Fejer}, {Feldbaum}, {Ferreira}, {Fisher}, {Frei}, {Freise}, {Frey},
  {Fricke}, {Fritschel}, {Frolov}, {Fuentes-Tapia}, {Fulda}, {Fyffe}, {Gair},
  {Gaonkar}, {Gehrels}, {Gergely}, {Giaime}, {Giardina}, {Gleason}, {Goetz},
  {Goetz}, {Gondan}, {Gonz{\'a}lez}, {Gordon}, {Gorodetsky}, {Gossan},
  {Go{\ss}ler}, {Gr{\"a}f}, {Graff}, {Grant}, {Gras}, {Gray}, {Greenhalgh},
  {Gretarsson}, {Grote}, {Grunewald}, {Guido}, {Guo}, {Gushwa}, {Gustafson},
  {Gustafson}, {Hacker}, {Hall}, {Hammond}, {Hanke}, {Hanks}, {Hanna},
  {Hannam}, {Hanson}, {Hardwick}, {Harry}, {Harry}, {Hart}, {Hartman},
  {Haster}, {Haughian}, {Hee}, {Heintze}, {Heinzel}, {Hendry}, {Heng},
  {Heptonstall}, {Heurs}, {Hewitson}, {Hild}, {Hoak}, {Hodge}, {Hollitt},
  {Holt}, {Hopkins}, {Hosken}, {Hough}, {Houston}, {Howell}, {Hu}, {Huerta},
  {Hughey}, {Husa}, {Huttner}, {Huynh}, {Huynh-Dinh}, {Idrisy}, {Indik},
  {Ingram}, {Inta}, {Islas}, {Isler}, {Isogai}, {Iyer}, {Izumi}, {Jacobson},
  {Jang}, {Jawahar}, {Ji}, {Jim{\'e}nez-Forteza}, {Johnson}, {Jones}, {Jones},
  {Ju}, {Haris}, {Kalogera}, {Kandhasamy}, {Kang}, {Kanner}, {Katsavounidis},
  {Katzman}, {Kaufer}, {Kaufer}, {Kaur}, {Kawabe}, {Kawazoe}, {Keiser},
  {Keitel}, {Kelley}, {Kells}, {Keppel}, {Key}, {Khalaidovski}, {Khalili},
  {Khazanov}, {Kim}, {Kim}, {Kim}, {Kim}, {Kim}, {King}, {King}, {Kinzel},
  {Kissel}, {Klimenko}, {Kline}, {Koehlenbeck}, {Kokeyama}, {Kondrashov},
  {Korobko}, {Korth}, {Kozak}, {Kringel}, {Krishnan}, {Krueger}, {Kuehn},
  {Kumar}, {Kumar}, {Kuo}, {Landry}, {Lantz}, {Larson}, {Lasky}, {Lazzarini},
  {Lazzaro}, {Le}, {Leaci}, {Leavey}, {Lebigot}, {Lee}, {Lee}, {Lee}, {Leong},
  {Levin}, {Levine}, {Lewis}, {Li}, {Libbrecht}, {Libson}, {Lin}, {Littenberg},
  {Lockerbie}, {Lockett}, {Logue}, {Lombardi}, {Lormand}, {Lough}, {Lubinski},
  {L{\"u}ck}, {Lundgren}, {Lynch}, {Ma}, {Macarthur}, {MacDonald},
  {Machenschalk}, {MacInnis}, {Macleod}, {Maga{\\textasciitilde n}a-Sandoval},
  {Magee}, {Mageswaran}, {Maglione}, {Mailand}, {Mandel}, {Mandic}, {Mangano},
  {Mansell}, {M{\'a}rka}, {M{\'a}rka}, {Markosyan}, {Maros}, {Martin},
  {Martin}, {Martynov}, {Marx}, {Mason}, {Massinger}, {Matichard}, {Matone},
  {Mavalvala}, {Mazumder}, {Mazzolo}, {McCarthy}, {McClelland}, {McCormick},
  {McGuire}, {McIntyre}, {McIver}, {McLin}, {McWilliams}, {Meadors},
  {Meinders}, {Melatos}, {Mendell}, {Mercer}, {Meshkov}, {Messenger}, {Meyers},
  {Miao}, {Middleton}, {Mikhailov}, {Miller}, {Miller}, {Millhouse}, {Ming},
  {Mirshekari}, {Mishra}, {Mitra}, {Mitrofanov}, {Mitselmakher}, {Mittleman},
  {Moe}, {Mohanty}, {Mohapatra}, {Moore}, {Moraru}, {Moreno}, {Morriss},
  {Mossavi}, {Mow-Lowry}, {Mueller}, {Mueller}, {Mukherjee}, {Mullavey},
  {Munch}, {Murphy}, {Murray}, {Mytidis}, {Nash}, {Nayak}, {Necula}, {Nedkova},
  {Newton}, {Nguyen}, {Nielsen}, {Nissanke}, {Nitz}, {Nolting}, {Normandin},
  {Nuttall}, {Ochsner}, {O'Dell}, {Oelker}, {Ogin}, {Oh}, {Oh}, {Ohme},
  {Oppermann}, {Oram}, {O'Reilly}, {Ortega}, {O'Shaughnessy}, {Osthelder},
  {Ott}, {Ottaway}, {Ottens}, {Overmier}, {Owen}, {Padilla}, {Pai}, {Pai},
  {Palashov}, {Pal-Singh}, {Pan}, {Pankow}, {Pannarale}, {Pant}, {Papa},
  {Paris}, {Patrick}, {Pedraza}, {Pekowsky}, {Pele}, {Penn}, {Perreca},
  {Phelps}, {Pierro}, {Pinto}, {Pitkin}, {Poeld}, {Post}, {Poteomkin},
  {Powell}, {Prasad}, {Predoi}, {Premachandra}, {Prestegard}, {Price},
  {Principe}, {Privitera}, {Prix}, {Prokhorov}, {Puncken}, {P{\"u}rrer}, {Qin},
  {Quetschke}, {Quintero}, {Quiroga}, {Quitzow-James}, {Raab}, {Rabeling},
  {Radkins}, {Raffai}, {Raja}, {Rajalakshmi}, {Rakhmanov}, {Ramirez},
  {Raymond}, {Reed}, {Reid}, {Reitze}, {Reula}, {Riles}, {Robertson}, {Robie},
  {Rollins}, {Roma}, {Romano}, {Romanov}, {Romie}, {Rowan}, {R{\"u}diger},
  {Ryan}, {Sachdev}, {Sadecki}, {Sadeghian}, {Saleem}, {Salemi}, {Sammut},
  {Sandberg}, {Sanders}, {Sannibale}, {Santiago-Prieto}, {Sathyaprakash},
  {Saulson}, {Savage}, {Sawadsky}, {Scheuer}, {Schilling}, {Schmidt},
  {Schnabel}, {Schofield}, {Schreiber}, {Schuette}, {Schutz}, {Scott}, {Scott},
  {Sellers}, {Sengupta}, {Sergeev}, {Serna}, {Sevigny}, {Shaddock}, {Shahriar},
  {Shaltev}, {Shao}, {Shapiro}, {Shawhan}, {Shoemaker}, {Sidery}, {Siemens},
  {Sigg}, {Silva}, {Simakov}, {Singer}, {Singer}, {Singh}, {Sintes},
  {Slagmolen}, {Smith}, {Smith}, {Smith}, {Smith-Lefebvre}, {Son}, {Sorazu},
  {Souradeep}, {Staley}, {Stebbins}, {Steinke}, {Steinlechner}, {Steinlechner},
  {Steinmeyer}, {Stephens}, {Steplewski}, {Stevenson}, {Stone}, {Strain},
  {Strigin}, {Sturani}, {Stuver}, {Summerscales}, {Sutton}, {Szczepanczyk},
  {Szeifert}, {Talukder}, {Tanner}, {T{\'a}pai}, {Tarabrin}, {Taracchini},
  {Taylor}, {Tellez}, {Theeg}, {Thirugnanasambandam}, {Thomas}, {Thomas},
  {Thorne}, {Thorne}, {Thrane}, {Tiwari}, {Tomlinson}, {Torres}, {Torrie},
  {Traylor}, {Tse}, {Tshilumba}, {Ugolini}, {Unnikrishnan}, {Urban}, {Usman},
  {Vahlbruch}, {Vajente}, {Valdes}, {Vallisneri}, {van Veggel}, {Vass},
  {Vaulin}, {Vecchio}, {Veitch}, {Veitch}, {Venkateswara}, {Vincent- Finley},
  {Vitale}, {Vo}, {Vorvick}, {Vousden}, {Vyatchanin}, {Wade}, {Wade}, {Wade},
  {Walker}, {Wallace}, {Walsh}, {Wang}, {Wang}, {Wang}, {Ward}, {Warner},
  {Was}, {Weaver}, {Weinert}, {Weinstein}, {Weiss}, {Welborn}, {Wen},
  {Wessels}, {Westphal}, {Wette}, {Whelan}, {Whitcomb}, {White}, {Whiting},
  {Wilkinson}, {Williams}, {Williams}, {Williamson}, {Willis}, {Willke},
  {Wimmer}, {Winkler}, {Wipf}, {Wittel}, {Woan}, {Worden}, {Xie}, {Yablon},
  {Yakushin}, {Yam}, {Yamamoto}, {Yancey}, {Yang}, {Zanolin}, {Zhang}, {Zhang},
  {Zhang}, {Zhang}, {Zhao}, {Zhou}, {Zhu}, {Zucker}, {Zuraw}, \&
  {Zweizig}}]{2015CQGra..32g4001L}
{LIGO Scientific Collaboration}, {Aasi}, J., {Abbott}, B.~P., {et~al.} 2015,
  Classical and Quantum Gravity, 32, 074001,
  \dodoi{10.1088/0264-9381/32/7/074001}

\bibitem[{{Littenberg} {et~al.}(2015){Littenberg}, {Farr}, {Coughlin},
  {Kalogera}, \& {Holz}}]{2015ApJ...807L..24L}
{Littenberg}, T.~B., {Farr}, B., {Coughlin}, S., {Kalogera}, V., \& {Holz},
  D.~E. 2015, \apj, 807, L24, \dodoi{10.1088/2041-8205/807/2/L24}

\bibitem[{{Liu} \& {Lai}(2018)}]{2018ApJ...863...68L}
{Liu}, B., \& {Lai}, D. 2018, \apj, 863, 68, \dodoi{10.3847/1538-4357/aad09f}

\bibitem[{{Loredo}(2004)}]{2004AIPC..735..195L}
{Loredo}, T.~J. 2004, in American Institute of Physics Conference Series, Vol.
  735, American Institute of Physics Conference Series, ed. R.~{Fischer},
  R.~{Preuss}, \& U.~V. {Toussaint}, 195--206.
\newblock \url{https://doi.org/10.1063/1.1835214}

\bibitem[{{Lower} {et~al.}(2018){Lower}, {Thrane}, {Lasky}, \&
  {Smith}}]{2018PhRvD..98h3028L}
{Lower}, M.~E., {Thrane}, E., {Lasky}, P.~D., \& {Smith}, R. 2018, \prd, 98,
  083028, \dodoi{10.1103/PhysRevD.98.083028}

\bibitem[{{Madau} \& {Dickinson}(2014)}]{2014ARA&A..52..415M}
{Madau}, P., \& {Dickinson}, M. 2014, Annual Review of Astronomy and
  Astrophysics, 52, 415, \dodoi{10.1146/annurev-astro-081811-125615}

\bibitem[{{Maeder} \& {Meynet}(2003)}]{2003A&A...411..543M}
{Maeder}, A., \& {Meynet}, G. 2003, \aap, 411, 543,
  \dodoi{10.1051/0004-6361:20031491}

\bibitem[{{Mandel}(2010)}]{2010PhRvD..81h4029M}
{Mandel}, I. 2010, \prd, 81, 084029, \dodoi{10.1103/PhysRevD.81.084029}

\bibitem[{{Mandel} \& {de Mink}(2016)}]{2016MNRAS.458.2634M}
{Mandel}, I., \& {de Mink}, S.~E. 2016, \mnras, 458, 2634,
  \dodoi{10.1093/mnras/stw379}

\bibitem[{{Mandel} {et~al.}(2017){Mandel}, {Farr}, {Colonna}, {Stevenson},
  {Ti{\v{n}}o}, \& {Veitch}}]{2017MNRAS.465.3254M}
{Mandel}, I., {Farr}, W.~M., {Colonna}, A., {et~al.} 2017, \mnras, 465, 3254,
  \dodoi{10.1093/mnras/stw2883}

\bibitem[{{Mandel} {et~al.}(2018){Mandel}, {Farr}, \&
  {Gair}}]{2018arXiv180902063M}
{Mandel}, I., {Farr}, W.~M., \& {Gair}, J.~R. 2018, arXiv e-prints,
  arXiv:1809.02063.
\newblock \doarXiv{1809.02063}

\bibitem[{{Mandel} {et~al.}(2015){Mandel}, {Haster}, {Dominik}, \&
  {Belczynski}}]{2015MNRAS.450L..85M}
{Mandel}, I., {Haster}, C.-J., {Dominik}, M., \& {Belczynski}, K. 2015, \mnras,
  450, L85, \dodoi{10.1093/mnrasl/slv054}

\bibitem[{{Mandel} \& {O'Shaughnessy}(2010)}]{2010CQGra..27k4007M}
{Mandel}, I., \& {O'Shaughnessy}, R. 2010, Classical and Quantum Gravity, 27,
  114007, \dodoi{10.1088/0264-9381/27/11/114007}

\bibitem[{{Mandic} {et~al.}(2016){Mandic}, {Bird}, \&
  {Cholis}}]{2016PhRvL.117t1102M}
{Mandic}, V., {Bird}, S., \& {Cholis}, I. 2016, \prl, 117, 201102,
  \dodoi{10.1103/PhysRevLett.117.201102}

\bibitem[{{Mapelli}(2016)}]{2016MNRAS.459.3432M}
{Mapelli}, M. 2016, \mnras, 459, 3432, \dodoi{10.1093/mnras/stw869}

\bibitem[{{Mapelli} {et~al.}(2009){Mapelli}, {Colpi}, \&
  {Zampieri}}]{2009MNRAS.395L..71M}
{Mapelli}, M., {Colpi}, M., \& {Zampieri}, L. 2009, \mnras, 395, L71,
  \dodoi{10.1111/j.1745-3933.2009.00645.x}

\bibitem[{{Mapelli} \& {Giacobbo}(2018)}]{2018MNRAS.479.4391M}
{Mapelli}, M., \& {Giacobbo}, N. 2018, \mnras, 479, 4391,
  \dodoi{10.1093/mnras/sty1613}

\bibitem[{{Mapelli} {et~al.}(2017){Mapelli}, {Giacobbo}, {Ripamonti}, \&
  {Spera}}]{2017MNRAS.472.2422M}
{Mapelli}, M., {Giacobbo}, N., {Ripamonti}, E., \& {Spera}, M. 2017, \mnras,
  472, 2422, \dodoi{10.1093/mnras/stx2123}

\bibitem[{{Marchant} {et~al.}(2016){Marchant}, {Langer}, {Podsiadlowski},
  {Tauris}, \& {Moriya}}]{2016A&A...588A..50M}
{Marchant}, P., {Langer}, N., {Podsiadlowski}, P., {Tauris}, T.~M., \&
  {Moriya}, T.~J. 2016, \aap, 588, A50, \dodoi{10.1051/0004-6361/201628133}

\bibitem[{{Marchant} {et~al.}(2018){Marchant}, {Renzo}, {Farmer}, {Pappas},
  {Taam}, {de Mink}, \& {Kalogera}}]{2018arXiv181013412M}
{Marchant}, P., {Renzo}, M., {Farmer}, R., {et~al.} 2018, arXiv e-prints,
  arXiv:1810.13412.
\newblock \doarXiv{1810.13412}

\bibitem[{{Margalit} \& {Metzger}(2017)}]{2017ApJ...850L..19M}
{Margalit}, B., \& {Metzger}, B.~D. 2017, \apj, 850, L19,
  \dodoi{10.3847/2041-8213/aa991c}

\bibitem[{{Margutti} {et~al.}(2017){Margutti}, {Berger}, {Fong}, {Guidorzi},
  {Alexander}, {Metzger}, {Blanchard}, {Cowperthwaite}, {Chornock},
  {Eftekhari}, {Nicholl}, {Villar}, {Williams}, {Annis}, {Brown}, {Chen},
  {Doctor}, {Frieman}, {Holz}, {Sako}, \&
  {Soares-Santos}}]{2017ApJ...848L..20M}
{Margutti}, R., {Berger}, E., {Fong}, W., {et~al.} 2017, \apj, 848, L20,
  \dodoi{10.3847/2041-8213/aa9057}

\bibitem[{{McKernan} {et~al.}(2012){McKernan}, {Ford}, {Lyra}, \&
  {Perets}}]{2012MNRAS.425..460M}
{McKernan}, B., {Ford}, K. E.~S., {Lyra}, W., \& {Perets}, H.~B. 2012, \mnras,
  425, 460, \dodoi{10.1111/j.1365-2966.2012.21486.x}

\bibitem[{{Mennekens} \& {Vanbeveren}(2014)}]{2014A&A...564A.134M}
{Mennekens}, N., \& {Vanbeveren}, D. 2014, \aap, 564, A134,
  \dodoi{10.1051/0004-6361/201322198}

\bibitem[{{Messick} {et~al.}(2017){Messick}, {Blackburn}, {Brady}, {Brockill},
  {Cannon}, {Cariou}, {Caudill}, {Chamberlin}, {Creighton}, {Everett}, {Hanna},
  {Keppel}, {Lang}, {Li}, {Meacher}, {Nielsen}, {Pankow}, {Privitera}, {Qi},
  {Sachdev}, {Sadeghian}, {Singer}, {Thomas}, {Wade}, {Wade}, {Weinstein}, \&
  {Wiesner}}]{2017PhRvD..95d2001M}
{Messick}, C., {Blackburn}, K., {Brady}, P., {et~al.} 2017, \prd, 95, 042001,
  \dodoi{10.1103/PhysRevD.95.042001}

\bibitem[{Misner {et~al.}(1973)Misner, Thorne, \&
  Wheeler}]{Misner1973Gravitation}
Misner, C.~W., Thorne, K.~S., \& Wheeler, J.~A. 1973, Gravitation, first
  edition edn., Physics Series (San Francisco: W. H. Freeman).
\newblock \url{https://doi.org/10.1103/PhysRev.136.B1224}

\bibitem[{{Morscher} {et~al.}(2013){Morscher}, {Umbreit}, {Farr}, \&
  {Rasio}}]{2013ApJ...763L..15M}
{Morscher}, M., {Umbreit}, S., {Farr}, W.~M., \& {Rasio}, F.~A. 2013, \apj,
  763, L15, \dodoi{10.1088/2041-8205/763/1/L15}

\bibitem[{{Mortlock} {et~al.}(2018){Mortlock}, {Feeney}, {Peiris},
  {Williamson}, \& {Nissanke}}]{2018arXiv181111723M}
{Mortlock}, D.~J., {Feeney}, S.~M., {Peiris}, H.~V., {Williamson}, A.~R., \&
  {Nissanke}, S.~M. 2018, arXiv e-prints, arXiv:1811.11723.
\newblock \doarXiv{1811.11723}

\bibitem[{{Nicholl} {et~al.}(2017){Nicholl}, {Berger}, {Kasen}, {Metzger},
  {Elias}, {Brice{\\textasciitilde n}o}, {Alexander}, {Blanchard}, {Chornock},
  {Cowperthwaite}, {Eftekhari}, {Fong}, {Margutti}, {Villar}, {Williams},
  {Brown}, {Annis}, {Bahramian}, {Brout}, {Brown}, {Chen}, {Clemens},
  {Dennihy}, {Dunlap}, {Holz}, {Marchesini}, {Massaro}, {Moskowitz},
  {Pelisoli}, {Rest}, {Ricci}, {Sako}, {Soares-Santos}, \&
  {Strader}}]{2017ApJ...848L..18N}
{Nicholl}, M., {Berger}, E., {Kasen}, D., {et~al.} 2017, \apj, 848, L18,
  \dodoi{10.3847/2041-8213/aa9029}

\bibitem[{{Nitz} {et~al.}(2017){Nitz}, {Dent}, {Dal Canton}, {Fairhurst}, \&
  {Brown}}]{2017ApJ...849..118N}
{Nitz}, A.~H., {Dent}, T., {Dal Canton}, T., {Fairhurst}, S., \& {Brown}, D.~A.
  2017, \apj, 849, 118, \dodoi{10.3847/1538-4357/aa8f50}

\bibitem[{{Ober} {et~al.}(1983){Ober}, {El Eid}, \&
  {Fricke}}]{1983A&A...119...61O}
{Ober}, W.~W., {El Eid}, M.~F., \& {Fricke}, K.~J. 1983, \aap, 119, 61

\bibitem[{{O'Connor} \& {Ott}(2011)}]{2011ApJ...730...70O}
{O'Connor}, E., \& {Ott}, C.~D. 2011, \apj, 730, 70,
  \dodoi{10.1088/0004-637X/730/2/70}

\bibitem[{{O'Leary} {et~al.}(2016){O'Leary}, {Meiron}, \&
  {Kocsis}}]{2016ApJ...824L..12O}
{O'Leary}, R.~M., {Meiron}, Y., \& {Kocsis}, B. 2016, \apj, 824, L12,
  \dodoi{10.3847/2041-8205/824/1/L12}

\bibitem[{{O'Leary} {et~al.}(2006){O'Leary}, {Rasio}, {Fregeau}, {Ivanova}, \&
  {O'Shaughnessy}}]{2006ApJ...637..937O}
{O'Leary}, R.~M., {Rasio}, F.~A., {Fregeau}, J.~M., {Ivanova}, N., \&
  {O'Shaughnessy}, R. 2006, \apj, 637, 937, \dodoi{10.1086/498446}

\bibitem[{{O'Shaughnessy} {et~al.}(2017){O'Shaughnessy}, {Gerosa}, \&
  {Wysocki}}]{2017PhRvL.119a1101O}
{O'Shaughnessy}, R., {Gerosa}, D., \& {Wysocki}, D. 2017, \prl, 119, 011101,
  \dodoi{10.1103/PhysRevLett.119.011101}

\bibitem[{{O'Shaughnessy} {et~al.}(2010){O'Shaughnessy}, {Kalogera}, \&
  {Belczynski}}]{2010ApJ...716..615O}
{O'Shaughnessy}, R., {Kalogera}, V., \& {Belczynski}, K. 2010, \apj, 716, 615,
  \dodoi{10.1088/0004-637X/716/1/615}

\bibitem[{{{\"O}zel} \& {Freire}(2016)}]{2016ARA&A..54..401O}
{{\"O}zel}, F., \& {Freire}, P. 2016, Annual Review of Astronomy and
  Astrophysics, 54, 401, \dodoi{10.1146/annurev-astro-081915-023322}

\bibitem[{{{\"O}zel} {et~al.}(2010){{\"O}zel}, {Psaltis}, {Narayan}, \&
  {McClintock}}]{2010ApJ...725.1918O}
{{\"O}zel}, F., {Psaltis}, D., {Narayan}, R., \& {McClintock}, J.~E. 2010,
  \apj, 725, 1918, \dodoi{10.1088/0004-637X/725/2/1918}

\bibitem[{{{\"O}zel} {et~al.}(2012){{\"O}zel}, {Psaltis}, {Narayan}, \& {Santos
  Villarreal}}]{2012ApJ...757...55O}
{{\"O}zel}, F., {Psaltis}, D., {Narayan}, R., \& {Santos Villarreal}, A. 2012,
  \apj, 757, 55, \dodoi{10.1088/0004-637X/757/1/55}

\bibitem[{{Packet}(1981)}]{1981A&A...102...17P}
{Packet}, W. 1981, \aap, 102, 17

\bibitem[{{Pan} {et~al.}(2014){Pan}, {Buonanno}, {Taracchini}, {Kidder},
  {Mrou{\'e}}, {Pfeiffer}, {Scheel}, \& {Szil{\'a}gyi}}]{2014PhRvD..89h4006P}
{Pan}, Y., {Buonanno}, A., {Taracchini}, A., {et~al.} 2014, \prd, 89, 084006,
  \dodoi{10.1103/PhysRevD.89.084006}

\bibitem[{{Peters}(1964)}]{1964PhRv..136.1224P}
{Peters}, P.~C. 1964, Physical Review, 136, 1224,
  \dodoi{10.1103/PhysRev.136.B1224}

\bibitem[{{Petrovic} {et~al.}(2005){Petrovic}, {Langer}, \& {van der
  Hucht}}]{2005A&A...435.1013P}
{Petrovic}, J., {Langer}, N., \& {van der Hucht}, K.~A. 2005, \aap, 435, 1013,
  \dodoi{10.1051/0004-6361:20042368}

\bibitem[{{Petrovich} \& {Antonini}(2017)}]{2017ApJ...846..146P}
{Petrovich}, C., \& {Antonini}, F. 2017, \apj, 846, 146,
  \dodoi{10.3847/1538-4357/aa8628}

\bibitem[{{Pian} {et~al.}(2017){Pian}, {D'Avanzo}, {Benetti}, {Branchesi},
  {Brocato}, {Campana}, {Cappellaro}, {Covino}, {D'Elia}, {Fynbo}, {Getman},
  {Ghirlanda}, {Ghisellini}, {Grado}, {Greco}, {Hjorth}, {Kouveliotou},
  {Levan}, {Limatola}, {Malesani}, {Mazzali}, {Melandri}, {M{\o}ller},
  {Nicastro}, {Palazzi}, {Piranomonte}, {Rossi}, {Salafia}, {Selsing},
  {Stratta}, {Tanaka}, {Tanvir}, {Tomasella}, {Watson}, {Yang}, {Amati},
  {Antonelli}, {Ascenzi}, {Bernardini}, {Bo{\"e}r}, {Bufano}, {Bulgarelli},
  {Capaccioli}, {Casella}, {Castro-Tirado}, {Chassande-Mottin}, {Ciolfi},
  {Copperwheat}, {Dadina}, {De Cesare}, {di Paola}, {Fan}, {Gendre},
  {Giuffrida}, {Giunta}, {Hunt}, {Israel}, {Jin}, {Kasliwal}, {Klose}, {Lisi},
  {Longo}, {Maiorano}, {Mapelli}, {Masetti}, {Nava}, {Patricelli}, {Perley},
  {Pescalli}, {Piran}, {Possenti}, {Pulone}, {Razzano}, {Salvaterra},
  {Schipani}, {Spera}, {Stamerra}, {Stella}, {Tagliaferri}, {Testa}, {Troja},
  {Turatto}, {Vergani}, \& {Vergani}}]{2017Natur.551...67P}
{Pian}, E., {D'Avanzo}, P., {Benetti}, S., {et~al.} 2017, \nat, 551, 67,
  \dodoi{10.1038/nature24298}

\bibitem[{{Planck Collaboration} {et~al.}(2016){Planck Collaboration}, {Ade},
  {Aghanim}, {Arnaud}, {Ashdown}, {Aumont}, {Baccigalupi}, {Banday},
  {Barreiro}, {Bartlett}, {Bartolo}, {Battaner}, {Battye}, {Benabed},
  {Beno{\^\i}t}, {Benoit-L{\'e}vy}, {Bernard}, {Bersanelli}, {Bielewicz},
  {Bock}, {Bonaldi}, {Bonavera}, {Bond}, {Borrill}, {Bouchet}, {Boulanger},
  {Bucher}, {Burigana}, {Butler}, {Calabrese}, {Cardoso}, {Catalano},
  {Challinor}, {Chamballu}, {Chary}, {Chiang}, {Chluba}, {Christensen},
  {Church}, {Clements}, {Colombi}, {Colombo}, {Combet}, {Coulais}, {Crill},
  {Curto}, {Cuttaia}, {Danese}, {Davies}, {Davis}, {de Bernardis}, {de Rosa},
  {de Zotti}, {Delabrouille}, {D{\'e}sert}, {Di Valentino}, {Dickinson},
  {Diego}, {Dolag}, {Dole}, {Donzelli}, {Dor{\'e}}, {Douspis}, {Ducout},
  {Dunkley}, {Dupac}, {Efstathiou}, {Elsner}, {En{\ss}lin}, {Eriksen},
  {Farhang}, {Fergusson}, {Finelli}, {Forni}, {Frailis}, {Fraisse},
  {Franceschi}, {Frejsel}, {Galeotta}, {Galli}, {Ganga}, {Gauthier}, {Gerbino},
  {Ghosh}, {Giard}, {Giraud-H{\'e}raud}, {Giusarma}, {Gjerl{\o}w},
  {Gonz{\'a}lez-Nuevo}, {G{\'o}rski}, {Gratton}, {Gregorio}, {Gruppuso},
  {Gudmundsson}, {Hamann}, {Hansen}, {Hanson}, {Harrison}, {Helou}, {Henrot-
  Versill{\'e}}, {Hern{\'a}ndez-Monteagudo}, {Herranz}, {Hildebrandt}, {Hivon},
  {Hobson}, {Holmes}, {Hornstrup}, {Hovest}, {Huang}, {Huffenberger}, {Hurier},
  {Jaffe}, {Jaffe}, {Jones}, {Juvela}, {Keih{\"a}nen}, {Keskitalo}, {Kisner},
  {Kneissl}, {Knoche}, {Knox}, {Kunz}, {Kurki-Suonio}, {Lagache},
  {L{\"a}hteenm{\"a}ki}, {Lamarre}, {Lasenby}, {Lattanzi}, {Lawrence}, {Leahy},
  {Leonardi}, {Lesgourgues}, {Levrier}, {Lewis}, {Liguori}, {Lilje},
  {Linden-V{\o}rnle}, {L{\'o}pez-Caniego}, {Lubin}, {Mac{\'\i}as-P{\'e}rez},
  {Maggio}, {Maino}, {Mandolesi}, {Mangilli}, {Marchini}, {Maris}, {Martin},
  {Martinelli}, {Mart{\'\i}nez-Gonz{\'a}lez}, {Masi}, {Matarrese}, {McGehee},
  {Meinhold}, {Melchiorri}, {Melin}, {Mendes}, {Mennella}, {Migliaccio},
  {Millea}, {Mitra}, {Miville-Desch{\^e}nes}, {Moneti}, {Montier}, {Morgante},
  {Mortlock}, {Moss}, {Munshi}, {Murphy}, {Naselsky}, {Nati}, {Natoli},
  {Netterfield}, {N{\o}rgaard-Nielsen}, {Noviello}, {Novikov}, {Novikov},
  {Oxborrow}, {Paci}, {Pagano}, {Pajot}, {Paladini}, {Paoletti}, {Partridge},
  {Pasian}, {Patanchon}, {Pearson}, {Perdereau}, {Perotto}, {Perrotta},
  {Pettorino}, {Piacentini}, {Piat}, {Pierpaoli}, {Pietrobon}, {Plaszczynski},
  {Pointecouteau}, {Polenta}, {Popa}, {Pratt}, {Pr{\'e}zeau}, {Prunet},
  {Puget}, {Rachen}, {Reach}, {Rebolo}, {Reinecke}, {Remazeilles}, {Renault},
  {Renzi}, {Ristorcelli}, {Rocha}, {Rosset}, {Rossetti}, {Roudier},
  {Rouill{\'e} d'Orfeuil}, {Rowan-Robinson}, {Rubi{\\textasciitilde
  n}o-Mart{\'\i}n}, {Rusholme}, {Said}, {Salvatelli}, {Salvati}, {Sandri},
  {Santos}, {Savelainen}, {Savini}, {Scott}, {Seiffert}, {Serra}, {Shellard},
  {Spencer}, {Spinelli}, {Stolyarov}, {Stompor}, {Sudiwala}, {Sunyaev},
  {Sutton}, {Suur-Uski}, {Sygnet}, {Tauber}, {Terenzi}, {Toffolatti}, {Tomasi},
  {Tristram}, {Trombetti}, {Tucci}, {Tuovinen}, {T{\"u}rler}, {Umana},
  {Valenziano}, {Valiviita}, {Van Tent}, {Vielva}, {Villa}, {Wade}, {Wandelt},
  {Wehus}, {White}, {White}, {Wilkinson}, {Yvon}, {Zacchei}, \&
  {Zonca}}]{2016A&A...594A..13P}
{Planck Collaboration}, {Ade}, P. A.~R., {Aghanim}, N., {et~al.} 2016, \aap,
  594, A13, \dodoi{10.1051/0004-6361/201525830}

\bibitem[{{Portegies Zwart} \& {McMillan}(2000)}]{2000ApJ...528L..17P}
{Portegies Zwart}, S.~F., \& {McMillan}, S. L.~W. 2000, \apj, 528, L17,
  \dodoi{10.1086/312422}

\bibitem[{{Portegies Zwart} {et~al.}(2010){Portegies Zwart}, {McMillan}, \&
  {Gieles}}]{2010ARA&A..48..431P}
{Portegies Zwart}, S.~F., {McMillan}, S. L.~W., \& {Gieles}, M. 2010, Annual
  Review of Astronomy and Astrophysics, 48, 431,
  \dodoi{10.1146/annurev-astro-081309-130834}

\bibitem[{{Portegies Zwart} \& {Yungelson}(1998)}]{1998A&A...332..173P}
{Portegies Zwart}, S.~F., \& {Yungelson}, L.~R. 1998, \aap, 332, 173.
\newblock \doarXiv{astro-ph/9710347}

\bibitem[{{Postnov} \& {Kuranov}(2019)}]{2019MNRAS.483.3288P}
{Postnov}, K.~A., \& {Kuranov}, A.~G. 2019, \mnras, 483, 3288,
  \dodoi{10.1093/mnras/sty3313}

\bibitem[{{Postnov} \& {Yungelson}(2014)}]{2014LRR....17....3P}
{Postnov}, K.~A., \& {Yungelson}, L.~R. 2014, Living Reviews in Relativity, 17,
  3, \dodoi{10.12942/lrr-2014-3}

\bibitem[{{Qin} {et~al.}(2018){Qin}, {Fragos}, {Meynet}, {Andrews},
  {S{\o}rensen}, \& {Song}}]{2018A&A...616A..28Q}
{Qin}, Y., {Fragos}, T., {Meynet}, G., {et~al.} 2018, \aap, 616, A28,
  \dodoi{10.1051/0004-6361/201832839}

\bibitem[{{Qin} {et~al.}(2019){Qin}, {Marchant}, {Fragos}, {Meynet}, \&
  {Kalogera}}]{2019ApJ...870L..18Q}
{Qin}, Y., {Marchant}, P., {Fragos}, T., {Meynet}, G., \& {Kalogera}, V. 2019,
  \apj, 870, L18, \dodoi{10.3847/2041-8213/aaf97b}

\bibitem[{{Quinlan} \& {Shapiro}(1987)}]{1987ApJ...321..199Q}
{Quinlan}, G.~D., \& {Shapiro}, S.~L. 1987, \apj, 321, 199,
  \dodoi{10.1086/165624}

\bibitem[{{Racine}(2008)}]{2008PhRvD..78d4021R}
{Racine}, {\'E}. 2008, \prd, 78, 044021, \dodoi{10.1103/PhysRevD.78.044021}

\bibitem[{{Rakavy} \& {Shaviv}(1967)}]{1967ApJ...148..803R}
{Rakavy}, G., \& {Shaviv}, G. 1967, \apj, 148, 803, \dodoi{10.1086/149204}

\bibitem[{{Rodriguez} {et~al.}(2018{\natexlab{a}}){Rodriguez}, {Amaro-Seoane},
  {Chatterjee}, {Kremer}, {Rasio}, {Samsing}, {Ye}, \&
  {Zevin}}]{2018PhRvD..98l3005R}
{Rodriguez}, C.~L., {Amaro-Seoane}, P., {Chatterjee}, S., {et~al.}
  2018{\natexlab{a}}, \prd, 98, 123005, \dodoi{10.1103/PhysRevD.98.123005}

\bibitem[{{Rodriguez} {et~al.}(2018{\natexlab{b}}){Rodriguez}, {Amaro-Seoane},
  {Chatterjee}, \& {Rasio}}]{2018PhRvL.120o1101R}
{Rodriguez}, C.~L., {Amaro-Seoane}, P., {Chatterjee}, S., \& {Rasio}, F.~A.
  2018{\natexlab{b}}, \prl, 120, 151101, \dodoi{10.1103/PhysRevLett.120.151101}

\bibitem[{{Rodriguez} \& {Antonini}(2018)}]{2018ApJ...863....7R}
{Rodriguez}, C.~L., \& {Antonini}, F. 2018, \apj, 863, 7,
  \dodoi{10.3847/1538-4357/aacea4}

\bibitem[{{Rodriguez} {et~al.}(2016{\natexlab{a}}){Rodriguez}, {Chatterjee}, \&
  {Rasio}}]{2016PhRvD..93h4029R}
{Rodriguez}, C.~L., {Chatterjee}, S., \& {Rasio}, F.~A. 2016{\natexlab{a}},
  \prd, 93, 084029, \dodoi{10.1103/PhysRevD.93.084029}

\bibitem[{{Rodriguez} {et~al.}(2016{\natexlab{b}}){Rodriguez}, {Haster},
  {Chatterjee}, {Kalogera}, \& {Rasio}}]{2016ApJ...824L...8R}
{Rodriguez}, C.~L., {Haster}, C.-J., {Chatterjee}, S., {Kalogera}, V., \&
  {Rasio}, F.~A. 2016{\natexlab{b}}, \apj, 824, L8,
  \dodoi{10.3847/2041-8205/824/1/L8}

\bibitem[{{Rodriguez} \& {Loeb}(2018)}]{2018ApJ...866L...5R}
{Rodriguez}, C.~L., \& {Loeb}, A. 2018, \apj, 866, L5,
  \dodoi{10.3847/2041-8213/aae377}

\bibitem[{{Rodriguez} {et~al.}(2015){Rodriguez}, {Morscher}, {Pattabiraman},
  {Chatterjee}, {Haster}, \& {Rasio}}]{2015PhRvL.115e1101R}
{Rodriguez}, C.~L., {Morscher}, M., {Pattabiraman}, B., {et~al.} 2015, \prl,
  115, 051101, \dodoi{10.1103/PhysRevLett.115.051101}

\bibitem[{{Rodriguez} {et~al.}(2016{\natexlab{c}}){Rodriguez}, {Zevin},
  {Pankow}, {Kalogera}, \& {Rasio}}]{2016ApJ...832L...2R}
{Rodriguez}, C.~L., {Zevin}, M., {Pankow}, C., {Kalogera}, V., \& {Rasio},
  F.~A. 2016{\natexlab{c}}, \apj, 832, L2, \dodoi{10.3847/2041-8205/832/1/L2}

\bibitem[{{Roulet} \& {Zaldarriaga}(2019)}]{2019MNRAS.tmp..231R}
{Roulet}, J., \& {Zaldarriaga}, M. 2019, \mnras, 231,
  \dodoi{10.1093/mnras/stz226}

\bibitem[{{Sadowski} {et~al.}(2008){Sadowski}, {Belczynski}, {Bulik},
  {Ivanova}, {Rasio}, \& {O'Shaughnessy}}]{2008ApJ...676.1162S}
{Sadowski}, A., {Belczynski}, K., {Bulik}, T., {et~al.} 2008, \apj, 676, 1162,
  \dodoi{10.1086/528932}

\bibitem[{{Samsing} {et~al.}(2014){Samsing}, {MacLeod}, \&
  {Ramirez-Ruiz}}]{2014ApJ...784...71S}
{Samsing}, J., {MacLeod}, M., \& {Ramirez-Ruiz}, E. 2014, \apj, 784, 71,
  \dodoi{10.1088/0004-637X/784/1/71}

\bibitem[{{Sasaki} {et~al.}(2016){Sasaki}, {Suyama}, {Tanaka}, \&
  {Yokoyama}}]{2016PhRvL.117f1101S}
{Sasaki}, M., {Suyama}, T., {Tanaka}, T., \& {Yokoyama}, S. 2016, \prl, 117,
  061101, \dodoi{10.1103/PhysRevLett.117.061101}

\bibitem[{{Sathyaprakash} {et~al.}(2012){Sathyaprakash}, {Abernathy},
  {Acernese}, {Ajith}, {Allen}, {Amaro-Seoane}, {Andersson}, {Aoudia}, {Arun},
  {Astone}, {Krishnan}, {Barack}, {Barone}, {Barr}, {Barsuglia}, {Bassan},
  {Bassiri}, {Beker}, {Beveridge}, {Bizouard}, {Bond}, {Bose}, {Bosi},
  {Braccini}, {Bradaschia}, {Britzger}, {Brueckner}, {Bulik}, {Bulten},
  {Burmeister}, {Calloni}, {Campsie}, {Carbone}, {Cella}, {Chalkley},
  {Chassande-Mottin}, {Chelkowski}, {Chincarini}, {Di Cintio}, {Clark},
  {Coccia}, {Colacino}, {Colas}, {Colla}, {Corsi}, {Cumming}, {Cunningham},
  {Cuoco}, {Danilishin}, {Danzmann}, {Daw}, {De Salvo}, {Del Pozzo}, {Dent},
  {De Rosa}, {Di Fiore}, {Emilio}, {Di Virgilio}, {Dietz}, {Doets}, {Dueck},
  {Edwards}, {Fafone}, {Fairhurst}, {Falferi}, {Favata}, {Ferrari}, {Ferrini},
  {Fidecaro}, {Flaminio}, {Franc}, {Frasconi}, {Freise}, {Friedrich}, {Fulda},
  {Gair}, {Galimberti}, {Gemme}, {Genin}, {Gennai}, {Giazotto}, {Glampedakis},
  {Gossan}, {Gouaty}, {Graef}, {Graham}, {Granata}, {Grote}, {Guidi}, {Hallam},
  {Hammond}, {Hannam}, {Harms}, {Haughian}, {Hawke}, {Heinert}, {Hendry},
  {Heng}, {Hennes}, {Hild}, {Hough}, {Huet}, {Husa}, {Huttner}, {Iyer},
  {Jones}, {Jones}, {Kamaretsos}, {Kant Mishra}, {Kawazoe}, {Khalili}, {Kley},
  {Kokeyama}, {Kokkotas}, {Kroker}, {Kumar}, {Kuroda}, {Lagrange}, {Lastzka},
  {Li}, {Lorenzini}, {Losurdo}, {L{\"u}ck}, {Majorana}, {Malvezzi}, {Mandel},
  {Mandic}, {Marka}, {Marin}, {Marion}, {Marque}, {Martin}, {McLeod},
  {Mckechan}, {Mehmet}, {Michel}, {Minenkov}, {Morgado}, {Morgia}, {Mosca},
  {Moscatelli}, {Mours}, {M{\"u}ller-Ebhardt}, {Murray}, {Naticchioni},
  {Nawrodt}, {Nelson}, {O' Shaughnessy}, {Ott}, {Palomba}, {Paoli}, {Parguez},
  {Pasqualetti}, {Passaquieti}, {Passuello}, {Perciballi}, {Piergiovanni},
  {Pinard}, {Pitkin}, {Plastino}, {Plissi}, {Poggiani}, {Popolizio}, {Porter},
  {Prato}, {Prodi}, {Punturo}, {Puppo}, {Rabeling}, {Racz}, {Rapagnani}, {Re},
  {Read}, {Regimbau}, {Rehbein}, {Reid}, {Ricci}, {Richard}, {Robinson},
  {Rocchi}, {Romano}, {Rowan}, {R{\"u}diger}, {Samblowski}, {Santamar{\'\i}a},
  {Sassolas}, {Schilling}, {Schmidt}, {Schnabel}, {Schutz}, {Schwarz}, {Scott},
  {Seidel}, {Sintes}, {Somiya}, {Sopuerta}, {Sorazu}, {Speirits}, {Storchi},
  {Strain}, {Strigin}, {Sutton}, {Tarabrin}, {Taylor}, {Th{\"u}rin},
  {Tokmakov}, {Tonelli}, {Tournefier}, {Vaccarone}, {Vahlbruch}, {van den
  Brand}, {Van Den Broeck}, {van der Putten}, {van Veggel}, {Vecchio},
  {Veitch}, {Vetrano}, {Vicere}, {Vyatchanin}, {We{\ss}els}, {Willke},
  {Winkler}, {Woan}, {Woodcraft}, \& {Yamamoto}}]{2012CQGra..29l4013S}
{Sathyaprakash}, B., {Abernathy}, M., {Acernese}, F., {et~al.} 2012, Classical
  and Quantum Gravity, 29, 124013, \dodoi{10.1088/0264-9381/29/12/124013}

\bibitem[{{Savchenko} {et~al.}(2017){Savchenko}, {Ferrigno}, {Kuulkers},
  {Bazzano}, {Bozzo}, {Brandt}, {Chenevez}, {Courvoisier}, {Diehl}, {Domingo},
  {Hanlon}, {Jourdain}, {von Kienlin}, {Laurent}, {Lebrun}, {Lutovinov},
  {Martin- Carrillo}, {Mereghetti}, {Natalucci}, {Rodi}, {Roques}, {Sunyaev},
  \& {Ubertini}}]{2017ApJ...848L..15S}
{Savchenko}, V., {Ferrigno}, C., {Kuulkers}, E., {et~al.} 2017, \apj, 848, L15,
  \dodoi{10.3847/2041-8213/aa8f94}

\bibitem[{{Scheepmaker} {et~al.}(2007){Scheepmaker}, {Haas}, {Gieles},
  {Bastian}, {Larsen}, \& {Lamers}}]{2007A&A...469..925S}
{Scheepmaker}, R.~A., {Haas}, M.~R., {Gieles}, M., {et~al.} 2007, \aap, 469,
  925, \dodoi{10.1051/0004-6361:20077511}

\bibitem[{{Schmidt} {et~al.}(2015){Schmidt}, {Ohme}, \&
  {Hannam}}]{2015PhRvD..91b4043S}
{Schmidt}, P., {Ohme}, F., \& {Hannam}, M. 2015, \prd, 91, 024043,
  \dodoi{10.1103/PhysRevD.91.024043}

\bibitem[{{Shu} \& {Lubow}(1981)}]{1981ARA&A..19..277S}
{Shu}, F.~H., \& {Lubow}, S.~H. 1981, Annual Review of Astronomy and
  Astrophysics, 19, 277, \dodoi{10.1146/annurev.aa.19.090181.001425}

\bibitem[{{Sigurdsson} \& {Hernquist}(1993)}]{1993Natur.364..423S}
{Sigurdsson}, S., \& {Hernquist}, L. 1993, \nat, 364, 423,
  \dodoi{10.1038/364423a0}

\bibitem[{{Skilling}(2004)}]{2004AIPC..735..395S}
{Skilling}, J. 2004, in American Institute of Physics Conference Series, ed.
  R.~{Fischer}, R.~{Preuss}, \& U.~V. {Toussaint}, Vol. 735, 395--405.
\newblock \url{https://doi.org/10.1063/1.1835238}

\bibitem[{{Soares-Santos} {et~al.}(2017){Soares-Santos}, {Holz}, {Annis},
  {Chornock}, {Herner}, {Berger}, {Brout}, {Chen}, {Kessler}, {Sako}, {Allam},
  {Tucker}, {Butler}, {Palmese}, {Doctor}, {Diehl}, {Frieman}, {Yanny}, {Lin},
  {Scolnic}, {Cowperthwaite}, {Neilsen}, {Marriner}, {Kuropatkin}, {Hartley},
  {Paz-Chinch{\'o}n}, {Alexander}, {Balbinot}, {Blanchard}, {Brown}, {Carlin},
  {Conselice}, {Cook}, {Drlica-Wagner}, {Drout}, {Durret}, {Eftekhari}, {Farr},
  {Finley}, {Foley}, {Fong}, {Fryer}, {Garc{\'\i}a-Bellido}, {Gill}, {Gruendl},
  {Hanna}, {Kasen}, {Li}, {Lopes}, {Louren{\c{c}}o}, {Margutti}, {Marshall},
  {Matheson}, {Medina}, {Metzger}, {Mu{\\textasciitilde n}oz}, {Muir},
  {Nicholl}, {Quataert}, {Rest}, {Sauseda}, {Schlegel}, {Secco}, {Sobreira},
  {Stebbins}, {Villar}, {Vivas}, {Walker}, {Wester}, {Williams}, {Zenteno},
  {Zhang}, {Abbott}, {Abdalla}, {Banerji}, {Bechtol}, {Benoit-L{\'e}vy},
  {Bertin}, {Brooks}, {Buckley-Geer}, {Burke}, {Carnero Rosell}, {Carrasco
  Kind}, {Carretero}, {Castander}, {Crocce}, {Cunha}, {D'Andrea}, {da Costa},
  {Davis}, {Desai}, {Dietrich}, {Doel}, {Eifler}, {Fernandez}, {Flaugher},
  {Fosalba}, {Gaztanaga}, {Gerdes}, {Giannantonio}, {Goldstein}, {Gruen},
  {Gschwend}, {Gutierrez}, {Honscheid}, {Jain}, {James}, {Jeltema}, {Johnson},
  {Johnson}, {Kent}, {Krause}, {Kron}, {Kuehn}, {Kuhlmann}, {Lahav}, {Lima},
  {Maia}, {March}, {McMahon}, {Menanteau}, {Miquel}, {Mohr}, {Nichol}, {Nord},
  {Ogando}, {Petravick}, {Plazas}, {Romer}, {Roodman}, {Rykoff}, {Sanchez},
  {Scarpine}, {Schubnell}, {Sevilla-Noarbe}, {Smith}, {Smith}, {Suchyta},
  {Swanson}, {Tarle}, {Thomas}, {Thomas}, {Troxel}, {Vikram}, {Wechsler},
  {Weller}, {Dark Energy Survey}, \& {Dark Energy Camera GW-EM
  Collaboration}}]{2017ApJ...848L..16S}
{Soares-Santos}, M., {Holz}, D.~E., {Annis}, J., {et~al.} 2017, \apj, 848, L16,
  \dodoi{10.3847/2041-8213/aa9059}

\bibitem[{{Spera} \& {Mapelli}(2017)}]{2017MNRAS.470.4739S}
{Spera}, M., \& {Mapelli}, M. 2017, \mnras, 470, 4739,
  \dodoi{10.1093/mnras/stx1576}

\bibitem[{{Spera} {et~al.}(2015){Spera}, {Mapelli}, \&
  {Bressan}}]{2015MNRAS.451.4086S}
{Spera}, M., {Mapelli}, M., \& {Bressan}, A. 2015, \mnras, 451, 4086,
  \dodoi{10.1093/mnras/stv1161}

\bibitem[{{Spruit}(2002)}]{2002A&A...381..923S}
{Spruit}, H.~C. 2002, \aap, 381, 923, \dodoi{10.1051/0004-6361:20011465}

\bibitem[{{Stevenson} {et~al.}(2017{\natexlab{a}}){Stevenson}, {Berry}, \&
  {Mandel}}]{2017MNRAS.471.2801S}
{Stevenson}, S., {Berry}, C. P.~L., \& {Mandel}, I. 2017{\natexlab{a}}, \mnras,
  471, 2801, \dodoi{10.1093/mnras/stx1764}

\bibitem[{{Stevenson} {et~al.}(2015){Stevenson}, {Ohme}, \&
  {Fairhurst}}]{2015ApJ...810...58S}
{Stevenson}, S., {Ohme}, F., \& {Fairhurst}, S. 2015, \apj, 810, 58,
  \dodoi{10.1088/0004-637X/810/1/58}

\bibitem[{{Stevenson} {et~al.}(2017{\natexlab{b}}){Stevenson},
  {Vigna-G{\'o}mez}, {Mandel}, {Barrett}, {Neijssel}, {Perkins}, \& {de
  Mink}}]{2017NatCo...814906S}
{Stevenson}, S., {Vigna-G{\'o}mez}, A., {Mandel}, I., {et~al.}
  2017{\natexlab{b}}, Nature Communications, 8, 14906,
  \dodoi{10.1038/ncomms14906}

\bibitem[{{Stone} {et~al.}(2017){Stone}, {Metzger}, \&
  {Haiman}}]{2017MNRAS.464..946S}
{Stone}, N.~C., {Metzger}, B.~D., \& {Haiman}, Z. 2017, \mnras, 464, 946,
  \dodoi{10.1093/mnras/stw2260}

\bibitem[{{Sukhbold} {et~al.}(2016){Sukhbold}, {Ertl}, {Woosley}, {Brown}, \&
  {Janka}}]{2016ApJ...821...38S}
{Sukhbold}, T., {Ertl}, T., {Woosley}, S.~E., {Brown}, J.~M., \& {Janka}, H.~T.
  2016, \apj, 821, 38, \dodoi{10.3847/0004-637X/821/1/38}

\bibitem[{{Talbot} \& {Thrane}(2017)}]{2017PhRvD..96b3012T}
{Talbot}, C., \& {Thrane}, E. 2017, \prd, 96, 023012,
  \dodoi{10.1103/PhysRevD.96.023012}

\bibitem[{{Talbot} \& {Thrane}(2018)}]{2018ApJ...856..173T}
---. 2018, \apj, 856, 173, \dodoi{10.3847/1538-4357/aab34c}

\bibitem[{{Talon} \& {Charbonnel}(2005)}]{2005A&A...440..981T}
{Talon}, S., \& {Charbonnel}, C. 2005, \aap, 440, 981,
  \dodoi{10.1051/0004-6361:20053020}

\bibitem[{{Talon} \& {Charbonnel}(2008)}]{2008A&A...482..597T}
---. 2008, \aap, 482, 597, \dodoi{10.1051/0004-6361:20078620}

\bibitem[{{Taracchini} {et~al.}(2014){Taracchini}, {Buonanno}, {Pan},
  {Hinderer}, {Boyle}, {Hemberger}, {Kidder}, {Lovelace}, {Mrou{\'e}},
  {Pfeiffer}, {Scheel}, {Szil{\'a}gyi}, {Taylor}, \&
  {Zenginoglu}}]{2014PhRvD..89f1502T}
{Taracchini}, A., {Buonanno}, A., {Pan}, Y., {et~al.} 2014, \prd, 89, 061502,
  \dodoi{10.1103/PhysRevD.89.061502}

\bibitem[{{Tauris} {et~al.}(2017){Tauris}, {Kramer}, {Freire}, {Wex}, {Janka},
  {Langer}, {Podsiadlowski}, {Bozzo}, {Chaty}, {Kruckow}, {van den Heuvel},
  {Antoniadis}, {Breton}, \& {Champion}}]{2017ApJ...846..170T}
{Tauris}, T.~M., {Kramer}, M., {Freire}, P. C.~C., {et~al.} 2017, \apj, 846,
  170, \dodoi{10.3847/1538-4357/aa7e89}

\bibitem[{{Thorne}(1974)}]{1974ApJ...191..507T}
{Thorne}, K.~S. 1974, \apj, 191, 507, \dodoi{10.1086/152991}

\bibitem[{{Thorne}(1983)}]{1983grr..proc....1T}
{Thorne}, K.~S. 1983, in Gravitational Radiation, 1--57

\bibitem[{{Tiwari}(2018)}]{2018CQGra..35n5009T}
{Tiwari}, V. 2018, Classical and Quantum Gravity, 35, 145009,
  \dodoi{10.1088/1361-6382/aac89d}

\bibitem[{{Tiwari} {et~al.}(2018){Tiwari}, {Fairhurst}, \&
  {Hannam}}]{2018ApJ...868..140T}
{Tiwari}, V., {Fairhurst}, S., \& {Hannam}, M. 2018, \apj, 868, 140,
  \dodoi{10.3847/1538-4357/aae8df}

\bibitem[{{Troja} {et~al.}(2017){Troja}, {Piro}, {van Eerten}, {Wollaeger},
  {Im}, {Fox}, {Butler}, {Cenko}, {Sakamoto}, {Fryer}, {Ricci}, {Lien}, {Ryan},
  {Korobkin}, {Lee}, {Burgess}, {Lee}, {Watson}, {Choi}, {Covino}, {D'Avanzo},
  {Fontes}, {Gonz{\'a}lez}, {Khandrika}, {Kim}, {Kim}, {Lee}, {Lee}, {Kutyrev},
  {Lim}, {S{\'a}nchez-Ram{\'\i}rez}, {Veilleux}, {Wieringa}, \&
  {Yoon}}]{2017Natur.551...71T}
{Troja}, E., {Piro}, L., {van Eerten}, H., {et~al.} 2017, \nat, 551, 71,
  \dodoi{10.1038/nature24290}

\bibitem[{{Tutukov} \& {Yungelson}(1993)}]{1993MNRAS.260..675T}
{Tutukov}, A.~V., \& {Yungelson}, L.~R. 1993, \mnras, 260, 675,
  \dodoi{10.1093/mnras/260.3.675}

\bibitem[{{Ugliano} {et~al.}(2012){Ugliano}, {Janka}, {Marek}, \&
  {Arcones}}]{2012ApJ...757...69U}
{Ugliano}, M., {Janka}, H.-T., {Marek}, A., \& {Arcones}, A. 2012, \apj, 757,
  69, \dodoi{10.1088/0004-637X/757/1/69}

\bibitem[{{Valsecchi} {et~al.}(2010){Valsecchi}, {Glebbeek}, {Farr}, {Fragos},
  {Willems}, {Orosz}, {Liu}, \& {Kalogera}}]{2010Natur.468...77V}
{Valsecchi}, F., {Glebbeek}, E., {Farr}, W.~M., {et~al.} 2010, \nat, 468, 77,
  \dodoi{10.1038/nature09463}

\bibitem[{{Van Den Broeck}(2014)}]{2014JPhCS.484a2008V}
{Van Den Broeck}, C. 2014, in Journal of Physics Conference Series, Vol. 484,
  Journal of Physics Conference Series, 012008.
\newblock \url{https://doi.org/10.1088/1742-6596/484/1/012008}

\bibitem[{{Vanbeveren}(2009)}]{2009NewAR..53...27V}
{Vanbeveren}, D. 2009, New Astronomy Reviews, 53, 27,
  \dodoi{10.1016/j.newar.2009.03.001}

\bibitem[{Veitch {et~al.}(2017)Veitch, {Del Pozzo}, Cody, \&
  Pitkin}]{john_veitch_2017_835874}
Veitch, J., {Del Pozzo}, W., Cody, \& Pitkin, M. 2017, johnveitch/cpnest: Minor
  optimisation, \dodoi{10.5281/zenodo.835874}.
\newblock \url{https://doi.org/10.5281/zenodo.835874}

\bibitem[{{Veitch} {et~al.}(2015){Veitch}, {Raymond}, {Farr}, {Farr}, {Graff},
  {Vitale}, {Aylott}, {Blackburn}, {Christensen}, {Coughlin}, {Del Pozzo},
  {Feroz}, {Gair}, {Haster}, {Kalogera}, {Littenberg}, {Mandel},
  {O'Shaughnessy}, {Pitkin}, {Rodriguez}, {R{\"o}ver}, {Sidery}, {Smith}, {Van
  Der Sluys}, {Vecchio}, {Vousden}, \& {Wade}}]{2015PhRvD..91d2003V}
{Veitch}, J., {Raymond}, V., {Farr}, B., {et~al.} 2015, \prd, 91, 042003,
  \dodoi{10.1103/PhysRevD.91.042003}

\bibitem[{{Vink} \& {de Koter}(2005)}]{2005A&A...442..587V}
{Vink}, J.~S., \& {de Koter}, A. 2005, \aap, 442, 587,
  \dodoi{10.1051/0004-6361:20052862}

\bibitem[{{Vink} {et~al.}(2001){Vink}, {de Koter}, \&
  {Lamers}}]{2001A&A...369..574V}
{Vink}, J.~S., {de Koter}, A., \& {Lamers}, H. J. G. L.~M. 2001, \aap, 369,
  574, \dodoi{10.1051/0004-6361:20010127}

\bibitem[{{Vitale} \& {Farr}(2018)}]{2018arXiv180800901V}
{Vitale}, S., \& {Farr}, W.~M. 2018, arXiv e-prints, arXiv:1808.00901.
\newblock \doarXiv{1808.00901}

\bibitem[{{Vitale} {et~al.}(2017){Vitale}, {Lynch}, {Sturani}, \&
  {Graff}}]{2017CQGra..34cLT01V}
{Vitale}, S., {Lynch}, R., {Sturani}, R., \& {Graff}, P. 2017, Classical and
  Quantum Gravity, 34, 03LT01, \dodoi{10.1088/1361-6382/aa552e}

\bibitem[{{Voss} \& {Tauris}(2003)}]{2003MNRAS.342.1169V}
{Voss}, R., \& {Tauris}, T.~M. 2003, \mnras, 342, 1169,
  \dodoi{10.1046/j.1365-8711.2003.06616.x}

\bibitem[{{Wong} {et~al.}(2012){Wong}, {Valsecchi}, {Fragos}, \&
  {Kalogera}}]{2012ApJ...747..111W}
{Wong}, T.-W., {Valsecchi}, F., {Fragos}, T., \& {Kalogera}, V. 2012, \apj,
  747, 111, \dodoi{10.1088/0004-637X/747/2/111}

\bibitem[{{Woosley}(2017)}]{2017ApJ...836..244W}
{Woosley}, S.~E. 2017, \apj, 836, 244, \dodoi{10.3847/1538-4357/836/2/244}

\bibitem[{{Woosley} {et~al.}(2007){Woosley}, {Blinnikov}, \&
  {Heger}}]{2007Natur.450..390W}
{Woosley}, S.~E., {Blinnikov}, S., \& {Heger}, A. 2007, \nat, 450, 390,
  \dodoi{10.1038/nature06333}

\bibitem[{{Wysocki} {et~al.}(2018){Wysocki}, {Lange}, \&
  {O'Shaughnessy}}]{2018arXiv180506442W}
{Wysocki}, D., {Lange}, J., \& {O'Shaughnessy}, R. 2018, arXiv e-prints,
  arXiv:1805.06442.
\newblock \doarXiv{1805.06442}

\bibitem[{Wysocki \& O'Shaughnessy(2018)}]{WysockiTechDoc}
Wysocki, D., \& O'Shaughnessy, R. 2018, Rescaling VT factors to match tabulated
  VT averages, Tech. Rep. {LIGO}-T1800427, LIGO Scientific Collaboration and
  Virgo Collaboration

\bibitem[{{Zevin} {et~al.}(2017){Zevin}, {Pankow}, {Rodriguez}, {Sampson},
  {Chase}, {Kalogera}, \& {Rasio}}]{2017ApJ...846...82Z}
{Zevin}, M., {Pankow}, C., {Rodriguez}, C.~L., {et~al.} 2017, \apj, 846, 82,
  \dodoi{10.3847/1538-4357/aa8408}

\bibitem[{{Ziosi} {et~al.}(2014){Ziosi}, {Mapelli}, {Branchesi}, \&
  {Tormen}}]{2014MNRAS.441.3703Z}
{Ziosi}, B.~M., {Mapelli}, M., {Branchesi}, M., \& {Tormen}, G. 2014, \mnras,
  441, 3703, \dodoi{10.1093/mnras/stu824}

\end{thebibliography}

\appendix

\section{Systematics} \label{sec:systematics}
\changes{In this section, we discuss the systematic uncertainties that affect our analysis, and show that they are subdominant to statistical uncertainties. We focus on two major sources of systematic uncertainty. The first of these is introduced by the waveform models that are used to extract the parameters of individual events, and the second is in the estimation of the detection efficiency.} 

\subsection{Waveform systematics}

\begin{figure}[b!]
  \centering
  \includegraphics[width=\columnwidth]{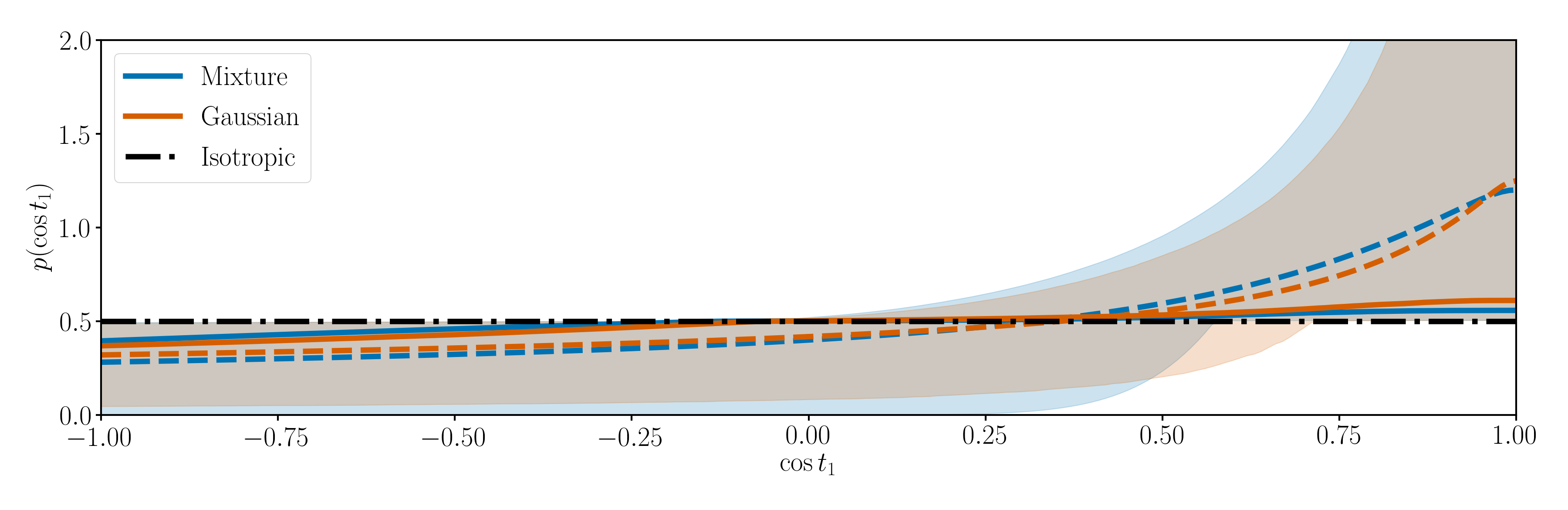}
  \caption{
Inferred distribution of cosine spin tilt for the more massive black hole for two choices of prior (see Section~\ref{sec:spin_models}) with the {\tt SEOBNRv3} waveform model, with the same definitions as Figure \ref{fig:parametric_spin_tilts}.}
\label{fig:parametric_spin_tilts_seob}
\end{figure}

In~\cite{o2catalog}, two waveform families are used to extract the parameters of individual events: {\tt SEOBNRv3}~\citep{2014PhRvD..89h4006P,2017PhRvD..95b4010B} and {\tt IMRPhenomPv2}~\citep{2014PhRvL.113o1101H,2016PhRvD..93d4007K,2016PhRvD..93d4006H}. While both families capture a wide variety of physical effects, including simple precession and other spin effects, they do not match each other exactly over the whole of the parameter space. Differences between the waveforms can therefore lead to slight biases in the inference of individual events' parameters, and thereby impact the inferred population distributions.To directly assess the impact of these
uncertainties on our results, we have repeated our calculations using parameter estimates based on {\tt SEOBNRv3} (the results in the main text all use {\tt IMRPhenomPv2}). \changes{See Table \ref{tab:systematics_vt_wf} --- } we find that the two waveform models produce at most modestly different inferences about key parameters.
For example, the standard Model B mass and spin distribution analysis with {\tt SEOBNRv3} leads us to infer that the 90\% upper bound of  $a_1$ is
$\APRIMARYimrpNoSpinVTQuadBSEOBboundninety$  
and credible intervals on $\mmax$ and ${\cal R}$ are
$\MMAXimrpNoSpinVTQuadBSEOBbound M_\odot$ 
and $\RATEimrpNoSpinVTQuadBSEOBboundU$,
which is consistent with the {\tt IMRPhenomPv2} Model B estimates of
$\APRIMARYimrpNoSpinVTQuadBboundninety$,
$\MMAXimrpNoSpinVTQuadBbound M_\odot$, 
and $\RATEimrpNoSpinVTQuadBboundU$ presented in the main text.
Similarly, in the redshift evolution analysis, we infer that the redshift evolution parameter $\lambda = \lambdaintervalSEOB$ under the {\tt SEOBNRv3} waveform compared to $\lambda = \lambdaintervalIMRP$ under the {\tt IMRPhenomPv2} waveform.
\changes{In summary, for the mass and rate part of the distributions, we find there is no significant change whatsoever. The most significant change is to the parameter controlling the primary tilt angle. The {\tt SEOBNRv3} waveform predicts that this parameter is smaller by 50\%, along with a smaller reduction in the secondary tilt angle parameter. Compare Figure \ref{fig:parametric_spin_tilts_seob}, produced with {\tt SEOBNRv3} derived event posteriors, with Figure \ref{fig:parametric_spin_tilts}. {\tt SEOBNRv3} produces a distribution of tilts which is closer to isotropic than its {\tt IMRPhenomPv2} counterpart. However, the models are compatible to within their uncertainties over the distribution $p(\cos t_1)$.}

\subsection{Selection Effects and Sensitive Volume}

In this subsection we detail the various assumptions and possible systematics that enter into our calculation of the detection efficiency.
The detectability of a BBH merger in GWs depends on the distance and orientation of the binary along with its intrinsic parameters, especially its component masses.
In order to model the underlying population and determine the BBH  merger rate, we must properly model the mass, redshift and spin-dependent selection effects,
and incorporate them into our population analysis according to Equation~\ref{eq:hyperparameter-likelihood}.
One way to infer the sensitivity of the detector network to a given population of BBH mergers is by carrying out large-scale simulations in which synthetic GW waveforms are injected into the detector 
data and subsequently searched for. 
The parameters of the injected waveforms can be drawn directly from the fixed population of interest, or alternatively, the injections can be placed to more broadly cover parameter space and reweighed to match the properties of the population \citep{2018CQGra..35n5009T}.
Such injection campaigns were carried out in~\cite{o2catalog} to measure the total sensitive spacetime volume $\VT$ and the corresponding merger rate for two fixed-parameter populations (power-law and flat-in-log).
However, it is computationally expensive to carry out an injection campaign that sufficiently covers the multi-dimensional population hyper-parameter space considered in this work. 
For this reason, for the parametric population studies in this work, we employ a semi-analytic method to estimate the fraction of found detections as a function of masses, spins and redshift (or equivalently, distance).

Our estimates of the network sensitivity are based on the semi-analytic method that was used to infer the BBH mass distribution from the first four GW detections~\citep{2016PhRvX...6d1015A,2017PhRvL.118v1101A}.
This method assumes that a BBH system is detectable if and only if it produces an SNR $\rho \geq \rho_{\rm th}$ in a single detector, where the threshold SNR, $\rho_{\rm th}$, is typically chosen to be 8.
Given a BBH system with known component masses, spins, and cosmological redshift, and a detector with stationary Gaussian noise characterized by a given power spectral density (PSD), one can calculate the optimal SNR, $\rho_{\rm opt}$, of the signal emitted by the BBH merger.
The optimal SNR corresponds to the SNR of the signal produced by a face-on, directly overhead BBH merger with the same masses, spins and redshift.
Given $\rho_{\rm opt}$, the distribution of single-detector SNRs $\rho$ -- corresponding to sources with random orientations with respect to the detector --  can be calculated using the analytic distribution of angular factors $\Theta \equiv \rho / \rho_{\rm opt}$~\citep{1993PhRvD..47.2198F}.
Under these assumptions, the probability of detecting a system of given masses, spins and redshift, $P_\mathrm{det}(m_1, m_2, \mathbf{\chi}_1, \mathbf{\chi}_2, z)$, is given by the probability that $\rho \geq \rho_{\rm th}$, or equivalently, that a randomly drawn $\Theta \geq \rho_{\rm th} / \rho_{\rm opt}(m_1, m_2, \mathbf{\chi}_1, \mathbf{\chi}_2, z)$.
$P_\mathrm{det}$ referred to in this section is equivalent to the $f(z \mid \xi)$ that appears in Equation~\ref{eqn:sens_vol} of Section~\ref{sec:notation}.

The semi-analytic calculation relies on two main simplifying assumptions: the detection threshold $\rho_\mathrm{th}$, and the choice of PSD for characterizing the detector noise.
When fitting the mass distribution to the first four BBH events in~\cite{2017PhRvL.118v1101A}, we assumed that the PSD in each LIGO interferometer could be approximated by the Early High Sensitivity curve in~\cite{2018LRR....21....3A} during O1 and the first few months of O2, and we fixed $\rho_\mathrm{th} = 8$.
We refer to the sensitivity estimate under these assumptions as the raw semi-analytic calculation.
In reality, the detector PSD fluctuates throughout the observing period. 
Additionally, the fixed detection threshold on SNR does not directly account for the empirical distributions of astrophysical and noise triggers,
and does not have a direct correspondence with the detection statistic used by the GW searches to rank significance of triggers~\citep{2017ApJ...849..118N, 2017PhRvD..95d2001M, o2catalog}.
Consequently, the sensitive spacetime volume of a population estimated using an SNR threshold may differ from the one obtained using injections, which threshold on the pipeline-dependent detection statistic.

We therefore pursue two modifications to the raw semi-analytic calculation in order to reduce the bias in our sensitivity estimates and the resulting population estimates. 
\changes{We emphasize that these modifications do not noticeably affect the inferred \emph{shape} of the population, e.g. the mass power-law slope, but do lead to different rate estimates, reflecting a systematic uncertainty in the inferred merger rate and its evolution with redshift that, given the small number of events and uncertainty in the phenomenological population models, remains subdominant to the statistical uncertainty. This is explicitly shown in the remainder of this section.}

In the first modification, which we employ throughout the mass distribution analysis (Section~\ref{sec:mass_distr}), we calibrate the raw semi-analytic method to the injection campaign in~\cite{o2catalog}.
The calibration takes the form of mass-dependent calibration factors, calculated by \changes{least squares regression as
described below; see ~\cite{WysockiTechDoc} for relevant data products}.
\changes{Specifically, we use injections to evaluate $\E{VT}_i \equiv \int d\xi p(\xi|\theta_i) VT_{true}(\xi)$ for a set of reference
hyperparameters $\theta_i$ (here, mass distribution models with different exponents $\alpha$ and maximum masses $\mmax$), where $\xi$ denotes all
binary parameters.}
To calculate $\E{VT}_i$ from injections into the PyCBC detection pipeline, we consider injections to be ``detected" if they have a detection statistic $\varrho \geq 8$, where $\varrho$ is the statistic used in the PyCBC analysis of O2 data~\citep{2017ApJ...849..118N,o2catalog}.
This is comparable to the detection statistic $\varrho = 8.7$ of the lowest-significance GW event included in our analysis, GW170729.
Note that, as discussed in Section~\ref{sec:rates}, because we adopt a fixed detection threshold, our analysis differs from the rate analysis in~\cite{o2catalog}, which does not fix a detection threshold, instead assigning to each trigger a probability of astrophysical origin~\citep{2015PhRvD..91b3005F}.
\changes{Once we have computed $\E{VT}_i$, we correct the raw semi-analytic model $VT_\mathrm{raw}$ described above by a factor $f(\xi)$ which is a
low-order polynomial in $\xi$: $f(\xi) = \sum_\alpha \lambda_\alpha F_\alpha(\xi)$, with $F_\alpha$ the relevant basis
polynomials.  We  minimize the mean-square difference between $\E{VT}$ as computed by injections and $\int d\xi f(\xi)
\VT_{raw}(\xi) p(\xi|\theta)$.    If  $H$ is the precomputed matrix of weight ``moments''
$
H_{k,\alpha} = \int d\xi p(\xi|\theta_k) VT(\xi) F_{\alpha}(\xi)
$
then the coefficients of this least-squares expression can be computed analytically as
$
\lambda = (H^T\gamma H)^{-1}H^T \gamma \E{VT}
$
where $\gamma$ is a diagonal inverse covariance matrix characterizing the Monte Carlo integration errors of each
individual $\E{VT}_i$.
}
This procedure yields the mass-calibrated sensitive volume $\VT_\mathrm{cal}$.

\begin{figure}
\includegraphics[width=\columnwidth]{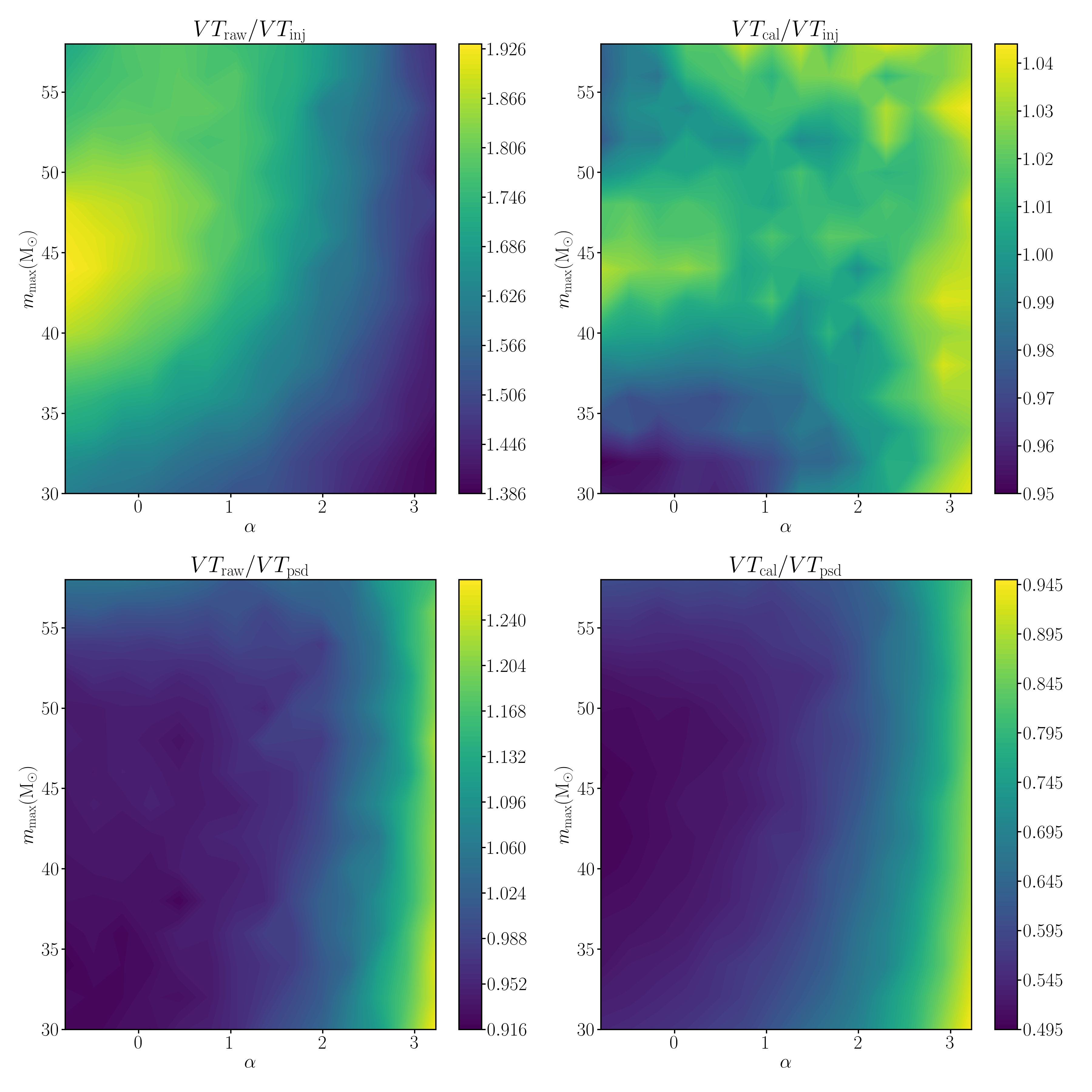}
\caption{Ratio between the raw semi-analytic computation of $\VT$ to the $\VT$ computed by injections into the PyCBC search pipeline (\changes{top} left panel), and the same ratio for the mass-calibrated $\VT$ (\changes{top} right panel),
for different mass distributions described by the two-parameter Model A. 
\changes{The bottom panel shows the same ratios, but this time comparing the $\VT$ derived with the time-varying PSDs applied to the semi-analytic calculation $VT_\mathrm{psd}$ to the raw and the calibrated $\VT$.}
The $\VT$ for the injections is calculated for a threshold of $\varrho = 8.0$, where $\varrho$ is the signal-noise model statistic used in the PyCBC analysis of O2 data.
This threshold roughly matches the detection statistic of the lowest significance detection, GW170729, which has $\varrho = 8.7$ in PyCBC.
We use the mass-calibrated $VT_\mathrm{cal}$ for the parametric mass- and spin-distribution analyses in Section~\ref{sec:mass_distr} and~\ref{sec:spin_distr} in order to better match the injection results, while the redshift evolution analysis uses $VT_\mathrm{psd}$ in order to carefully track the sensitivity at high redshift.
\changes{The discrepancy between the methods may be due to the limited number of high-redshift injections.}
However, the difference between all three methods is relatively constant as a function of the mass population, particularly where posterior support for the mass distribution hyper-parameters is high, indicating that systematic uncertainties in the $\VT$ estimation do not have a large impact on our results.}
\label{fig:VTcompare}
\end{figure}

\changes{The top panel of} Figure~\ref{fig:VTcompare} shows the comparison between the raw semi-analytic $\VT$, the calibrated $\VT$, and the injection $\VT$ across the two-dimensional hyperparameter space of Model A for the mass distribution.
We have repeated our mass distribution analysis with different choices of the $\VT$ calibration, and found that the effect on the shape of the mass distribution and the overall merger rate $\mathcal{R}$
are much smaller than the differences between Models A, B and C and the statistical errors associated with a small sample of 10 events.

\changes{As shown in Figure~\ref{fig:VTcompare}, the main effect of this calibration is to decrease $\VT$ by a factor of $\sim1.6$. Over the relevant part of parameter-space (i.e. the regions of the $\alpha$--$\mmax$ plane that have likelihood support), this factor remains fairly constant, implying that the inferred shape of the mass distribution is not affected by applying the $\VT$ calibration, although the overall rate is increased by about a factor of $\sim 1.6$ compared to the raw semi-analytic calculation. We have verified this explicitly by repeating the analysis with and without calibrated $\VT$.} 

\begin{figure}
\includegraphics[width=\columnwidth]{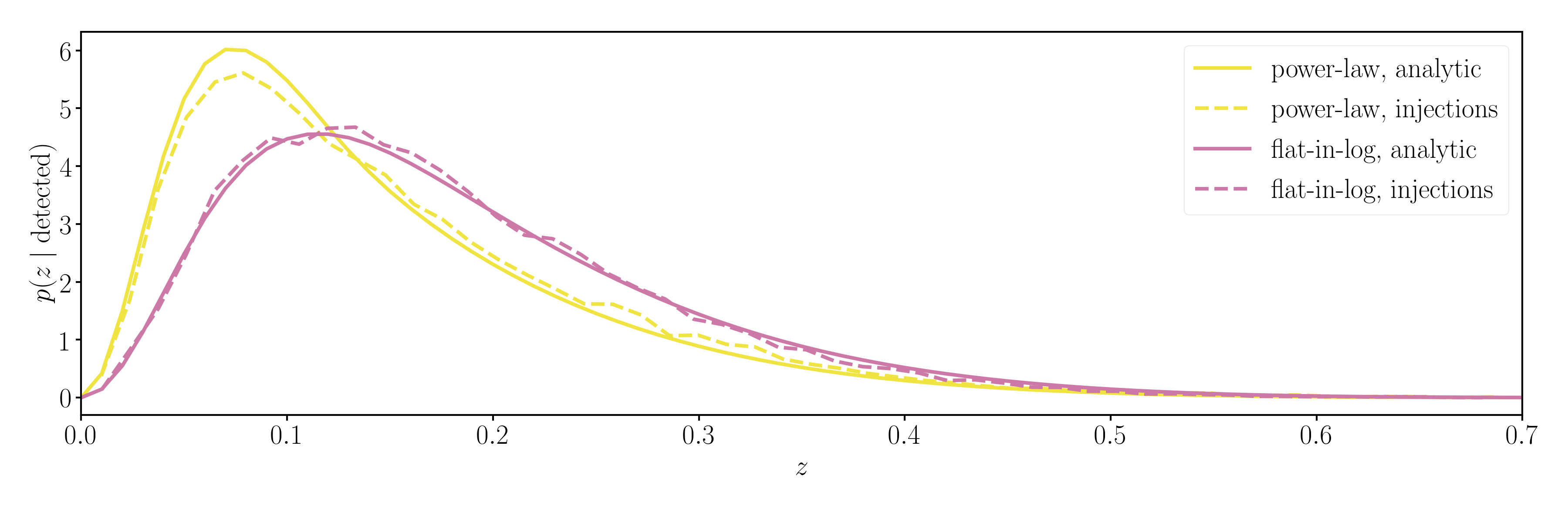}
\caption{Redshift distribution of injections recovered with a false alarm rate (FAR) less than 0.1 yr$^{-1}$ by the search pipeline GstLAL for the two fixed-parameter injection sets, power-law (red) and flat-in-log (green) compared to the expectation from the semi-analytic calculation used for the redshift evolution analysis, as described in the text. 
The underlying redshift distribution of the injected populations are assumed to follow a uniform in comoving volume and source-frame time distribution.
The FAR threshold of 0.1 / year nearly matches the FAR of the lowest-significance GW event, GW170729, with a FAR of 0.18 / year in the GstLAL pipeline.
The semi-analytic calculation closely predicts the redshift distribution of the found injections.}
\label{fig:pzdet}
\end{figure}

For the redshift evolution analysis (Section~\ref{sec:rates}), it is not sufficient to calibrate the mass-dependence of the detection probability; we must verify that the semi-analytic calculation reproduces the proper redshift-dependence. Therefore,
we pursue an alternative modification to the raw semi-analytic calculation. In this modification, we replace the single PSD of the raw semi-analytic calculation with a different PSD calculated for the Livingston detector for each five-day chunk of observing time in O1 and O2.
We find that this assumption correctly reproduces the redshift-dependent sensitivity empirically determined by the injection campaigns into the GstLAL pipeline for two fixed mass distributions (see Figure~\ref{fig:pzdet}), whereas adopting different assumptions, such as using the PSDs calculated for the Hanford detector instead of the Livingston detector, or changing the single-detector SNR threshold away from 8, yields curves in Figure~\ref{fig:pzdet} that deviate noticeably from the distribution of recovered injections. 
\changes{This modification to the sensitivity calculation is necessary in the redshift analysis because the detection probability can fluctuate significantly at high redshifts $z > 0.5$, where there is a very small probability of detection but considerable physical volume. Due to computational cost, the number of detections available at high redshift is insufficient to directly calibrate the redshift-dependent detection probability to injections as we did in the mass distribution section.}

\changes{We find that between the two methods we use to estimate detection efficiency, the effect on the inferred mass distribution is negligible. However, the second time-varying approach employed in the redshift analysis underpredicts the overall merger rate by $\sim 70 \%$ compared to the first calibrated approach (see the bottom right panel of Figure~\ref{fig:VTcompare}). 
This reflects a systematic uncertainty in the high-redshift detection efficiency and the implied merger rate. When additional detections lead to improved statistical constraints on the merger rate across redshift, it will become increasingly necessary to place a very large number of injections at high redshift and closely spaced in time in order to accurately estimate the high-redshift sensitivity.}

\begin{table}[htbp]
\centering
\begin{tabular}{|c|ccc|} \hline
Parameter/Model & Reference & spin $\VT$ & using SEOBNR \\ \hline
\multicolumn{4}{|c|}{Mass} \\ \hline
$\alpha$ (Model A) & \ALPHAMimrpNoSpinVTQuadApm  & \ALPHAMimrpSpinVTQuadApm  & \ALPHAMimrpNoSpinVTQuadASEOBpm  \\
$\alpha$ (Model B) & \ALPHAMimrpNoSpinVTQuadBpm  & \ALPHAMimrpSpinVTQuadBpm  & \ALPHAMimrpNoSpinVTQuadBSEOBpm  \\
$\alpha$ (Model C) & \ALPHAMimrpNoSpinVTQuadCpm  & - & \ALPHAMseobNoSpinVTQuadCpm  \\
$\beta_q$ (Model B) & \BETAQimrpNoSpinVTQuadBpm  & \BETAQimrpSpinVTQuadBpm  & \BETAQimrpNoSpinVTQuadBSEOBpm  \\
$\beta_q$ (Model C) & \BETAQimrpNoSpinVTQuadCpm  & - & \BETAQseobNoSpinVTQuadCpm  \\
$m_{\text{max}}$ (Model A) & \MMAXimrpNoSpinVTQuadApm  & \MMAXimrpSpinVTQuadApm  & \MMAXimrpNoSpinVTQuadASEOBpm  \\
$m_{\text{max}}$ (Model B) & \MMAXimrpNoSpinVTQuadBpm  & \MMAXimrpSpinVTQuadBpm  & \MMAXimrpNoSpinVTQuadBSEOBpm  \\
$m_{\text{max}}$ (Model C) & \MMAXimrpNoSpinVTQuadCpm  & - & \MMAXseobNoSpinVTQuadCpm  \\
$m_{\text{min}}$ (Model B) & \MMINimrpNoSpinVTQuadBpm  & \MMINimrpSpinVTQuadBpm  & \MMINimrpNoSpinVTQuadBSEOBpm  \\
$m_{\text{min}}$ (Model C) & \MMINimrpNoSpinVTQuadCpm  & - & \MMINseobNoSpinVTQuadCpm  \\
$\lambda_m$ (Model C) & \LAMBDAMimrpNoSpinVTQuadCpm  & - & \LAMBDAMseobNoSpinVTQuadCpm  \\
$\mu_m$ (Model C) & \MUMimrpNoSpinVTQuadCpm  & - & \MUMseobNoSpinVTQuadCpm  \\
$\sigma_m$ (Model C) & \SIGMAMimrpNoSpinVTQuadCpm  & - & \SIGMAMseobNoSpinVTQuadCpm  \\
\hline


\multicolumn{4}{|c|}{Rate} \\ \hline
$R$ (Model A) & \RATEimrpNoSpinVTQuadApm  & \RATEimrpSpinVTQuadApm  & \RATEimrpNoSpinVTQuadASEOBpm  \\
$R$ (Model B) & \RATEimrpNoSpinVTQuadBpm  & \RATEimrpSpinVTQuadBpm  & \RATEimrpNoSpinVTQuadBSEOBpm  \\
$R$ (Model C) & \RATEimrpNoSpinVTQuadCpm  & - & \RATEseobNoSpinVTQuadCpm  \\
\hline

\multicolumn{4}{|c|}{Spin} \\ \hline
$\cos t_1$ (Model A) & \COSTILTPRIMARYimrpNoSpinVTQuadApm  & \COSTILTPRIMARYimrpSpinVTQuadApm  & \COSTILTPRIMARYimrpNoSpinVTQuadASEOBpm  \\
$\cos t_1$ (Model B) & \COSTILTPRIMARYimrpNoSpinVTQuadBpm  & \COSTILTPRIMARYimrpSpinVTQuadBpm  & \COSTILTPRIMARYimrpNoSpinVTQuadBSEOBpm  \\
$\cos t_1$ (Model C) & \COSTILTPRIMARYimrpNoSpinVTQuadCpm  & - & \COSTILTPRIMARYseobNoSpinVTQuadCpm  \\
$\cos t_2$ (Model A) & \COSTILTSECONDARYimrpNoSpinVTQuadApm  & \COSTILTSECONDARYimrpSpinVTQuadApm  & \COSTILTSECONDARYimrpNoSpinVTQuadASEOBpm  \\
$\cos t_2$ (Model B) & \COSTILTSECONDARYimrpNoSpinVTQuadBpm  & \COSTILTSECONDARYimrpSpinVTQuadBpm  & \COSTILTSECONDARYimrpNoSpinVTQuadBSEOBpm  \\
$\cos t_2$ (Model C) & \COSTILTSECONDARYimrpNoSpinVTQuadCpm  & - & \COSTILTSECONDARYseobNoSpinVTQuadCpm  \\
\hline

\end{tabular}
\caption{\changes{Summary of intervals for each of the parameters considered in the models of Sections \ref{sec:mass_distr}, \ref{sec:rates} \ref{sec:spin_distr}. The reference uses the posteriors derived from the {\tt IMRPhenomPv2} waveform model and without spin effects included in $\VT$. The second column allows for spin effects in $\VT$ estimation. Finally, the third column shows the population model parameters inferred when the {\tt SEOBNRv3} waveform model is used to derive the event posteriors. Spin enabled $\VT$ is only available for Models A and B, but we expect that Model C would exhibit similar trends. Broadly, the mass and rate parameters are nearly the same and well within their respective uncertainties with and without spin effects in $\VT$, as well as considering the {\tt SEOBNRv3} waveform model. The most notable difference comes from the parameterized spin distribution. The differences are primarily related to the spin tilt distribution, and, for clarity, we suppress the spin magnitude distribution and mixture parameters which are nearly identical.}}
\label{tab:systematics_vt_wf}
\end{table}

\changes{Another difference between the mass distribution analysis presented in Section~\ref{sec:mass_distr} and the redshift evolution analysis of Section~\ref{sec:rates} is in the treatment of BBH spins. Section~\ref{sec:mass_distr} marginalizes over the spin distribution and includes first-order spin effects in the calculation of $\VT$ while the redshift analysis of Section~\ref{sec:rates} does not.}
\changes{From Table \ref{tab:systematics_vt_wf},} we find that including first-order spin effects in the calculation of $P_\mathrm{det}$ and the corresponding sensitive spacetime volume $\VT$ results in mostly indistinguishable population estimates compared to neglecting spin entirely. 
\changes{Similarly, fixing the spin distribution does not appreciably affect the inferred mass distribution.
Therefore, for simplicity, we neglect the effect of spin distribution in the redshift evolution analysis.} 
\changes{Meanwhile, the effects of spin on the sensitive volume $\VT$ do have a moderate influence on inferences about the spin tilt angles, presented in~\ref{sec:spin_param_infer}. When considering the effects of spin with $\VT$, there is about a 10\% shift in the median spin tilt angle parameters inferred, but this is well within the much wider credible interval.}
Therefore, such effects does not change our overall \changes{astrophysical} conclusions, and their influence on the results shown is comparable to what would result from different priors on the population parameters (for example, choosing a different prior range of $\sigma_i$ as compared to~\ref{tab:spin_priors}).

We also note that all our calculations of the detection efficiency are based on the {\tt IMRPhenomPv2} waveform. Differences between the phasing, and more importantly, the amplitude of the waveform can lead to different SNRs and detection statistics for the same sets of physical parameters. 
To bound the significance of this effect, we carry out the injection-based $\VT$ estimation for both the {\tt IMRPhenomPv2} and {\tt SEOBNRv2} waveforms and find that for populations described by the two-parameter mass Model A, the waveforms produce $\VT$ estimates consistent to 10\% across the relevant region of hyperparameter space with high posterior probability.
Therefore, compared to the statistical uncertainties, the choice in waveform does not contribute a significant systematic uncertainty for the $\VT$ estimation. 

Finally, an additional systematic uncertainty we have neglected in the $\VT$ and parametric rates calculations is the calibration uncertainty.
While the event posterior samples have incorporated a marginalization over uncertainties on
the calibration~\cite{SplineCalMarg-T1400682} for both strain amplitude and phase, the $\VT$ estimation here does not.
The amplitude calibration uncertainty results in an 18\% volume uncertainty~\citep{o2catalog}, which is currently below the level of statistical uncertainty in our population-averaged merger rate estimate.

\section{Alternative Spin Models} \label{sec:alt_spins}

\changes{We perform here a number of complementary analyses to reinforce the robustness of the results in Section~\ref{sec:spin_distr}, and gauge the effect of fixed parameter choices on spin inferences. Instead of a parameterized model such as those used in Section \ref{sec:spin_distr}, we focus on a few discrete choices of model parameters to reinforce the conclusions in that Section. These choices provide a complementary view to the results presented earlier and also display our current ability (or inability) to measure features in differing parts of the mass and spin parameter space.}

\subsection{Model Selection}\label{ap:sub:model-selection}
We choose a set of specific realizations of the general model described in Section \ref{ssec:model-features}, building on~\cite{2017Natur.548..426F,2018ApJ...868..140T}.
\changes{Four discrete spin magnitude models are considered, the first three being special cases of Equation~\ref{eq:spin_mag}:}

\begin{itemize}
	\item Low (L): $p(a) = 2 (1 - a)$, i.e., $\alpha_a = 1$, $\beta_a = 2$.
	\item Flat (F): $p(a) = 1$, i.e., $\alpha_a = 1$, $\beta_a = 1$.
	\item High (H): $p(a) = 2 a$, i.e., $\alpha_a = 2$, $\beta_a = 1$.
	\item Very low (V): $p(a) \propto e^{-(a/0.2)}$
\end{itemize}

Such magnitude distributions are chosen as simple representations of low, moderate and highly spinning individual black holes. 
The very low (V) population is added to capture the features of an even lower spinning population --- this is motivated by the features at low spin of the parametric distribution \changes{displayed} in Figure~\ref{fig:parametric_spin}.

For spin orientations we consider three fixed models representing extreme cases of Equation~\ref{eq:spin-alignment}:
\begin{itemize}
    \item Isotropic (I): $p(\cos t_i) = 1 / 2$; $-1 < \cos t_i < 1$, i.e., $\zeta = 0$.
    \item Aligned (A): $p(\cos t_i) = \delta(\cos t_i - 1)$, i.e., $\zeta = 1$, $\sigma_i = 0$.
    \item Restricted (R): $p(\cos t_i) = 1$; $0 < \cos t_i < 1$, this is the same as I, except the spins are restricted to point above the orbital plane.   
\end{itemize}

\changes{The isotropic distribution is motivated by dynamical or similarly disordered assembly scenarios, while the aligned one better capture a population of isolated binaries, under the simplifying assumption that the stars remain perfectly aligned throughout their evolution. 
In order to assess any preferences in the data for binaries with $\chi_{\rm eff}>0$, we introduce the restricted model: it resembles the isotropic distribution, but limits tilt angles to be positive.}
While we have mathematically defined the R model by assuming tilted spins, the same $\chi_{\rm eff}$ distribution can be generated with nonprecessing spins.

\begin{figure}[htbp]
\centering
\includegraphics[width=\columnwidth]{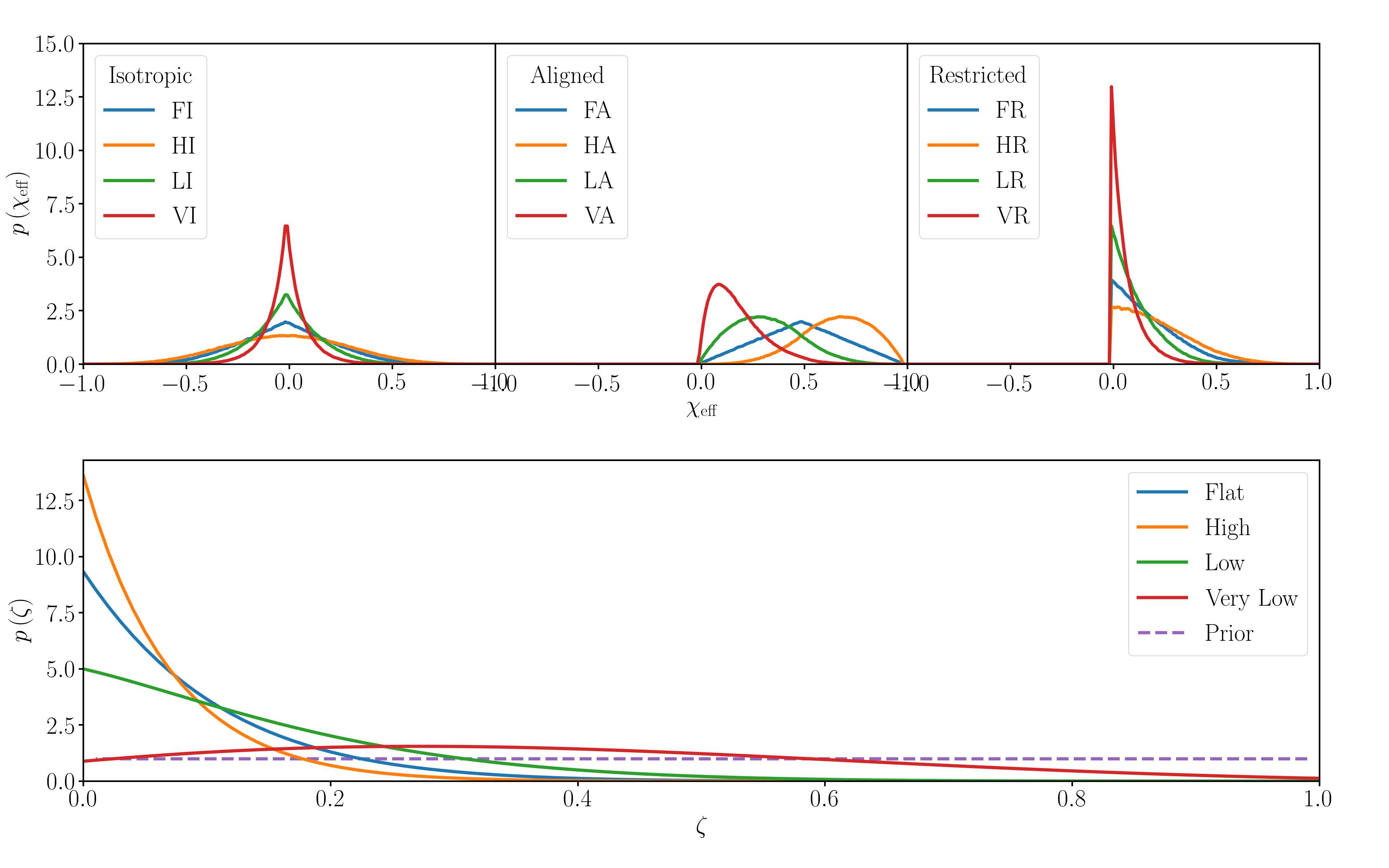}
\caption{
  Upper row: $p(\chi_{\textrm{eff}})$ under various model
  assumptions. Labels in each subpanel legend correspond to the tilt
  and magnitude models defined in
  \ref{ap:sub:model-selection}. Isotropic models (left) provide
  support for both negative and positive
  $\chi_{\textrm{eff}}$. Aligned models (center) assume perfect
  alignment for each of the four magnitudes distributions. Restricted
  models (right) have the same shape as the Isotropic ones, with
  support over $\chi_{\textrm{eff}}>0$ only. However they can be
  generated with nonprecessing spins.
  Bottom row: Posterior on the mixture fraction $\zeta$ between
  isotropic and aligned distributions.  $\zeta=0$ corresponds to a
  completely isotropic distribution.
}
\label{fig:chieff_toymodels}
\label{fig:mix_frac}
\end{figure}

\changes{Here we perform our inference entirely through $\chieff$, whose 12 different distributions are illustrated in Figure \ref{fig:chieff_toymodels}. Since we do not have conclusive results on $\beta_q$ from Figure \ref{fig:dndm:hyperparameters_ab}, we cannot make a single simplifying assumption on the mass model, which the $\chieff$ distribution depends on. We therefore consider three limiting cases: two of these fix the mass ratio to fiducial values, $q=1$ and $q=0.5$. The third corresponds to a fixed parameter model with $\alpha=1, \mmin=5, \mmax=50$.}
Figure \ref{fig:chieff_toymodels} illustrates the $\chieff$ distributions implied by each of these scenarios.

\begin{table}[htbp]

\begin{center}
\begin{tabular}{|l|c|c|c|c|}
\hline
$q=1$	& Very low & Low & Flat	& High \\ \hline
Isotropic & 1.29 & 0.0 & -1.04 & -2.25 \\ \hline
Restricted	& 3.5 & 3.22 & 1.06 & -0.2 \\ \hline
Aligned & 1.39 & -4.57 & -13.62 & -33.13 \\ \hline
\hline
$q=0.5$& Very low & Low & Flat	& High \\ \hline
Isotropic & 1.32 & 0.0 & -1.12 & -2.6 \\ \hline
Restricted	& 3.55 & 3.23 & 1.0 & -0.58 \\ \hline
Aligned & 1.52 & -4.15 & -12.86 & -31.6 \\ \hline
\hline
fixed param. & Very low & Low &  Flat	& High    \\ \hline
Isotropic & 0.64 & 0.0 & -2.0 & -3.85 \\ \hline
Restricted	& 1.83 & 0.8 & -2.3 & -5.0 \\ \hline
Aligned & -2.69 & -11.98 & -21.98 & -44.6 \\ \hline
\end{tabular}
\end{center}
\caption{\label{tab:odds_ratio} \textbf{Natural log Bayes factors for various spin distributions}.
The orientation models are described in Section~\ref{sec:notation}.
We find modest evidence for small spins. When spins are small, we cannot make strong statements about the distribution of spin orientations.
}
\end{table}

\changes{Following \cite{2017Natur.548..426F,2018ApJ...868..140T}, we calculate the evidence and compute the Bayes factors for each of the zero dimensional spin models. 
Results are provided in Table \ref{tab:odds_ratio}, with the low and isotropic distribution (LI) as the reference.}

Because of degeneracies in the GW waveform between mass ratio and $\chieff$, the choice of mass distribution impacts inferences about spins.
This effect explains the significant difference in Bayes factors for the third row in the table.
We find again our result moderately favors small black hole spins.
The restricted models with $\chi_{\rm eff}$ strictly positive consistently produce the highest Bayes factors. 
For the small-spin magnitude models we cannot make strong statements about the distribution of spin orientations.
Models containing highly spinning components are significantly disfavored, with high or flat aligned spins particularly selected against (e.g., FA and HA are disfavored with Bayes factors ranging in $\left[10^{-11},10^{-6}\right]$ and $\left[10^{-21},10^{-13}\right]$, respectively).
As a bracket for our uncertainty on the mass and mass ratio distribution, we evaluated the Bayes factors for the fixed parameter model $\alpha=2.3$, $\mmin=5$, $\mmax=50$. They differ from the third mass model in Table \ref{tab:odds_ratio} by a factor comparable to unity.

\subsection{Spin Mixture Models}

The models considered for model selection in Table~\ref{tab:odds_ratio} all assume a fixed set of spin magnitudes and tilts.
There is no reason to believe, however, that the Universe produces from only one of these distributions.
A natural extension is to allow for a mixing fraction describing the relative abundances of perfectly-aligned and isotropically distributed black holes spins.

We assume that the aligned and isotropic components follow the same spin magnitude distribution.
It is possible that black holes with a different distribution of spin orientations would have a different distribution of spin magnitudes, but given our weaker constraints on spin magnitudes, we focus on spin tilts sharing the same magnitude distribution.

We compute the posterior on the fraction of aligned binaries $\zeta$ in the population as per Equation \ref{eq:spin-alignment} in the limit ($\sigma_i \rightarrow 0$).
The models here are subsets of the Mixture distribution, with a purely isotropic being $\zeta=0$, and completely aligned being $\zeta=1$.
The prior on the mixing fraction is flat.

All of the models which contain a completely aligned component favor isotropy over alignment.
This ability to distinguish a mixing fraction diminishes with smaller spin magnitudes.
This is because such spin magnitudes yield populations which are not distinguishable to within measurement uncertainty of \chieff{}.
We do not include the most-favored restricted (R) configuration, but expect that the results would be similar.
Coupled with the model selection results in the previous section, this implies that the mixing fraction is not well determined when fixed to the models (low and very low) which are favored by the data (see Figure~\ref{fig:mix_frac}).
As stated above, in this case our ability to measure the mixing fraction is negligible.

\subsection{Three-bin Analysis of \chieff} \label{sec:three_bin}

\begin{figure}[htbp]
\includegraphics[width=\textwidth]{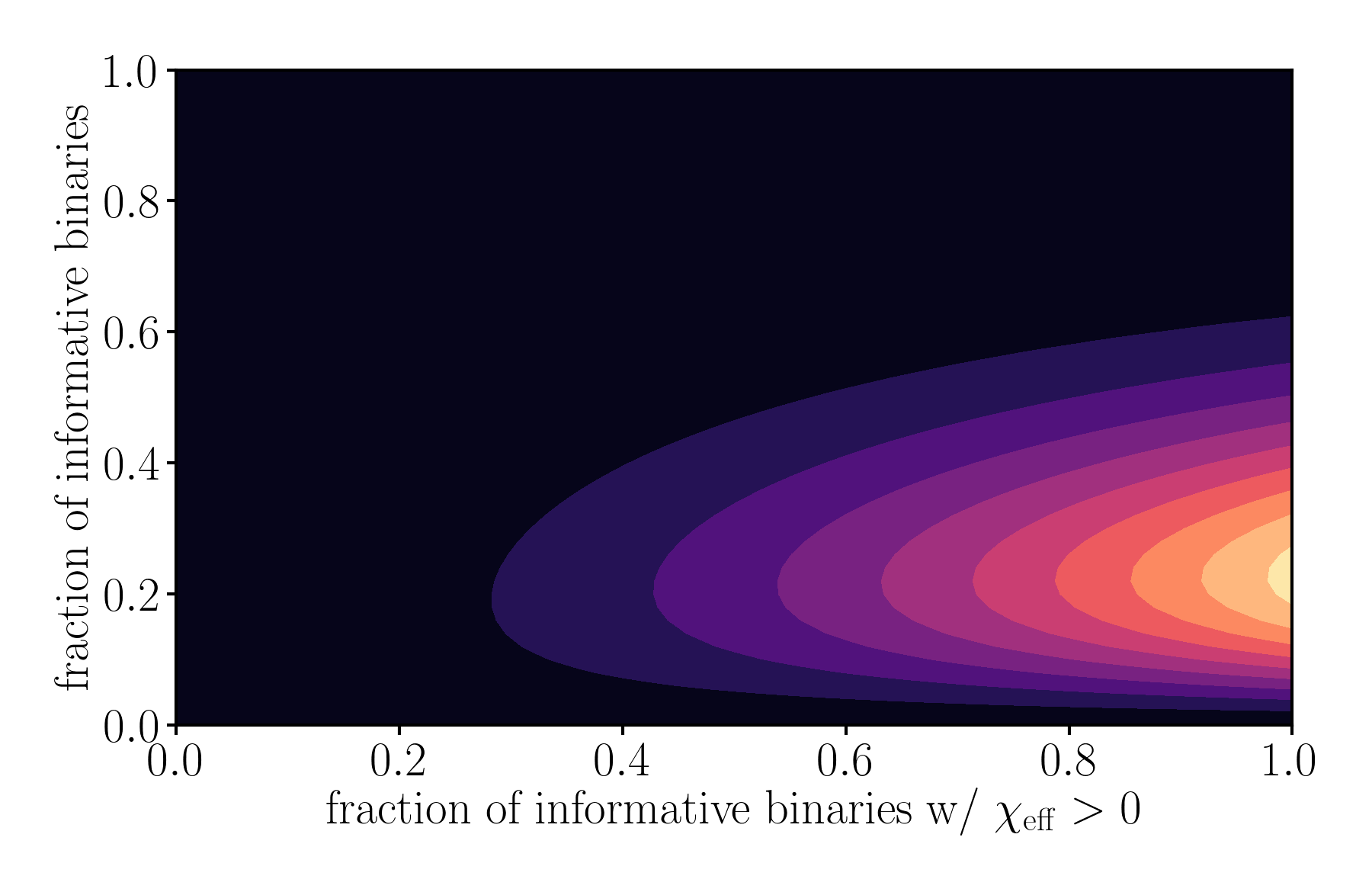}
\caption{Posterior distribution for the fraction of informative binaries (i.e., $|\chieff|>0.05$), and the fraction of those informative binaries with positive $\chieff$ (i.e., $\chieff > 0.05$).}
\label{fig:chieff_bins}
\end{figure}

We illustrate here how \chieff{} measurements can provide insights into discerning spin orientation distributions. Following~\cite{2018ApJ...854L...9F}, we split the range of $\chieff$ into three bins. One encompasses the fraction of uninformative binaries with \chieff{} consistent with zero ($|\chieff| \leq 0.05$); the vertical axis of Figure~\ref{fig:chieff_bins} shows the fraction of binaries lying outside of this bin. The other two capture significantly positive ($\chieff > 0.05$), and significantly negative ($\chieff < -0.05$) binaries. The width $0.05$ is chosen to be of the order of the uncertainty in a typical event posterior.

The aligned spin scenario is preferred in the posterior support on the right half of Figure~\ref{fig:chieff_bins}: the small fraction of binaries which are informative tend to possess \chieff{} greater than zero. Conversely, if the spins are isotropic, there would be no preference for positive or negative $\chieff$, and the posterior in Figure~\ref{fig:chieff_bins} would peak towards the middle. 
However, of the ten observed binaries, eight are consistent with zero \chieff{} and only two are informative, thus demonstrating our ability to distinguish between the two scenarios is weak.

\section{Importance Resampling The Single-Event Likelihood} \label{sec:reweighting}

Our hierarchical population analysis uses the individual-event likelihood for
each event $n = 1, \ldots, N$, $\mathcal{L}\left( d_n \mid \xi, z \right)$ (see Section
\ref{sec:notation}, Eq. \eqref{eq:hyperparameter-likelihood}).  Individual-event
analyses report posterior samples drawn a density that is proportional to this
likelihood times a prior \citep{2015PhRvD..91d2003V,o2catalog}.  The prior
density used is uniform in \emph{detector frame} masses and proportional to the
square of the luminosity distance \citep{2015PhRvD..91d2003V}; in terms of the
\emph{source frame} masses and redshift, the prior is
\begin{equation}
  p\left( \msource{1}, \msource{2}, z \right) \propto d_L^2(z) \frac{\partial \mdet{1}}{\partial \msource{1}} \frac{\partial \mdet{2}}{\partial \msource{2}} \frac{\partial d_L}{\partial z} = d_L^2(z) \left( 1 + z \right)^2 \frac{\partial d_L}{\partial z}.
\end{equation}
The derivative of the luminosity distance in a spatially flat universe
\citep{1999astro.ph..5116H} is
\begin{equation}
  \frac{\partial d_L}{\partial z} = \frac{d_L}{1+z} + \left( 1 + z \right) \frac{d_H}{E(z)},
\end{equation}
where $d_H = c / H_0$ is the Hubble distance and
\begin{equation}
  E(z) \equiv \sqrt{\Omega_M \left( 1 + z \right)^2 + \Omega_\Lambda}
\end{equation}
in a $\Lambda$CDM universe.

Given a set of posterior samples as described above, we can transform them to
samples from the likelihood over source frame masses and redshift by importance
resampling with weights that are the inverse prior
\begin{equation}
    w\left( \msource{1}, \msource{2}, z \right) = \frac{1}{p\left( \msource{1}, \msource{2}, z \right)}.
\end{equation}
The integral in Eq.\ \eqref{eq:hyperparameter-likelihood} may then be approximated as
\begin{equation}
  \int \dd \xi \, \dd z \, \mathcal{L}\left( d_n \mid \xi, z \right) \frac{\dd N}{\dd \xi \dd z} \propto \sum_{i = 1}^{N_\mathrm{samp}} w\left( \msource{1,i}, \msource{2,i}, z_i \right) \frac{\dd N}{\dd \xi_i \dd z_i},
\end{equation}
where the sum is taken over all posterior samples, and we assume that the
population distribution is expressed in terms of source-frame masses (i.e.\ $\xi
= \left( \msource{1}, \msource{2}, \ldots \right)$).  Alternately, we can
construct a random resampling of the set of existing posterior samples, with
sample $i$ appearing in the resampling with probability proportional to $w\left(
\msource{1,i}, \msource{2,i}, z_i \right)$; the integral is then proportional to
the average value of $\dd N/\dd \xi \dd z$ over the resampled set.  The
(unknown) constant of proportionality is related to the Bayesian evidence for
event $n$; as long as a consistent method (weighted sum or resampling) is used
to compare different population models, this constant is irrelevant to computing
Bayes factors between models.

\end{document}